\pdfoutput=1

\documentclass[11pt,a4paper]{article}
\usepackage[T1]{fontenc}
\usepackage[utf8]{inputenc}
\usepackage{lmodern,amsmath,amssymb,amsthm,mathtools}
\usepackage[margin=1.08in]{geometry}
\usepackage{booktabs,array,enumitem,microtype,xcolor,needspace}
\usepackage{tikz}
\usepackage[hypcap=false]{caption}
\usepackage[colorlinks=true,linkcolor=blue!45!black,citecolor=blue!45!black,urlcolor=blue!45!black]{hyperref}
\setlist{nosep}
\newtheorem{theorem}{Theorem}[section]
\newtheorem{lemma}[theorem]{Lemma}
\newtheorem{proposition}[theorem]{Proposition}
\newtheorem{corollary}[theorem]{Corollary}
\newtheorem*{theoremA}{Theorem A (spatial perfection)}
\newtheorem*{theoremB}{Theorem B (perfect play by von Neumann type)}
\newtheorem*{theoremD}{Theorem D (the entanglement spectrum)}
\newtheorem*{theoremC}{Theorem C (zero-error communication)}
\newtheorem*{backgroundlemma}{Background lemma}
\theoremstyle{definition}
\newtheorem{definition}[theorem]{Definition}
\theoremstyle{remark}
\newtheorem*{remark}{Remark}
\DeclareMathOperator{\Tr}{Tr}
\DeclareMathOperator{\Cat}{Cat}
\DeclareMathOperator{\ran}{ran}
\DeclareMathOperator{\supp}{supp}
\DeclareMathOperator{\spec}{spec}
\DeclareMathOperator{\Ad}{Ad}
\DeclareMathOperator{\Rea}{Re}
\newcommand{\HH}{\mathcal H}
\newcommand{\KK}{\mathcal K}
\newcommand{\WW}{\mathcal W}

\newcommand{\perf}{\mathrm{PERF}}
\newcommand{\qs}{\mathrm{qs}}
\newcommand{\qc}{\mathrm{qc}}
\newcommand{\finS}{\mathrm{finS}}
\newcommand{\sfhost}{\mathrm{sf}}
\newcommand{\finhost}{\mathrm{fin}}
\newcommand{\co}{\mathrm{co}}
\newcommand{\ZE}{\mathrm{ZE}}
\newcommand{\II}{\mathrm{II}}
\newcommand{\III}{\mathrm{III}}
\newcommand{\vN}{\bar\otimes}
\newcommand{\Mat}{\mathrm{Mat}}
\newcommand{\Rdec}{\mathcal R_{\rm dec}}
\newcommand{\Adec}{\mathcal A_{\rm dec}}
\newcommand{\Gclk}{G^{\rm clk}}
\newcommand{\Gsel}{G^{\rm sel}}
\newcommand{\Gsh}{G_{\rm sh}}
\newcommand{\Glin}{G_{\rm lin}}
\newcommand{\Gthree}{G_{\III}}
\newcommand{\Gsep}{G_{\rm sep}}
\newcommand{\hcat}{h_{\mathrm{Cat}}}
\newcommand{\kcat}{k_{\mathrm{Cat}}}
\newcommand{\Ccat}{C_{\mathrm{Cat}}}
\newcommand{\Lq}{\mathbf q}
\newcommand{\LE}{\mathbf E}
\newcommand{\LI}{\mathbf I}
\newcommand{\LJ}{\mathbf J}
\newcommand{\LU}{\mathbf U}
\newcommand{\LC}{\mathbf C}
\newcommand{\Fin}{\mathrm{Fin}}
\newcommand{\btrans}{\mathsf U}
\newcommand{\Csp}{C_{\rm sp}}
\newcommand{\Zcomb}{Z_{\rm comb}}
\newcommand{\rhoreg}{\rho_{\rm reg}}
\newcommand{\rhocomb}{\rho_{\rm comb}}
\newcommand{\tailbd}{t^{\rm bd}}
\newcommand{\Pact}{E^{\rm act}}
\newcommand{\codearchive}{\url{https://github.com/Apsiape/perfect-play-arithmetical-hierarchy}}
\newcommand{\precursordisclosure}{This paper consolidates and supersedes our preprints \cite{SpatialPrecursor,TypePrecursor,CommunicationPrecursor}. Theorem D and the completion of the semifinite-but-not-spatial difference classification are new here. The commuting $\Pi^0_1$ classification is known \cite{Slofstra,MNY}; Theorem B identifies its language with that of every fixed type-III subtype.}
\hypersetup{pdftitle={A trace costs one quantifier: the exact complexity of perfect play in nonlocal games},pdfauthor={Seth Douglas and Nidhal Mghirbi}}
\title{A trace costs one quantifier:\\ the exact complexity of perfect play in nonlocal games}
\author{Seth Douglas \and Nidhal Mghirbi}
\date{November 2026}
\begin{document}
\maketitle
\begin{abstract}
Given a finite nonlocal game, can the players win with certainty? We prove that the existence of a perfect normal spatial strategy is $\Sigma^0_2$-complete, even when restricted to games of value one, so it cannot be decided even with an oracle for the halting problem. This is strictly harder than perfect play in the finite-dimensional and commuting-operator models, which are $\Sigma^0_1$-complete and $\Pi^0_1$-complete. Within the commuting model, the existence of a perfect normal strategy with a finite or semifinite von Neumann host is $\Sigma^0_2$-complete, whereas for every fixed type $\mathrm{III}_\lambda$ subtype it has the same $\Pi^0_1$-complete language as unrestricted commuting play. A trace thus costs one existential number quantifier: the hardness constructions force every perfect strategy to choose a finite clock period or a finite tree depth. Spatial $\Sigma^0_2$-completeness also holds when finite entanglement entropy is required, and both forms are formally verified in Lean. For every computable $0<\alpha<1$, perfect spatial play with $\mathrm{Tr}\,h^\alpha<\infty$ for the reduced density $h$ is $\Sigma^0_2$-complete as well; these conditions form a strict ladder between finite rank and finite entropy. We also prove $\Sigma^0_2$-completeness for one-shot zero-error communication with normal entanglement, and construct a finite channel with zero-error codes of a given size each of which needs infinite entanglement entropy.
\end{abstract}
\begin{center}\small Opus 5.5, GPT-6.1 Sol and GPT-6 assisted with this work; the authors take responsibility for every claim (see ``How this work was done''). Programs and finite checks: \codearchive. Archive: \url{https://doi.org/10.5281/zenodo.23962889}.\end{center}
\section{Introduction}\label{sec:intro}
In a nonlocal game a referee sends each of two separated players a question from a finite list. Each player answers from a finite list, and a fixed rule decides whether they win. Shared entanglement can raise the winning probability. When can the players win with certainty?

For a fixed game this is a yes-or-no question. It has a precise answer only once one says what a quantum strategy is, and only once one separates approaching certainty from attaining it: the supremum of the winning probability can be one while no strategy reaches it. The standard models are finite-dimensional tensor products, normal states on infinite-dimensional tensor products (the spatial model), and commuting operator algebras on one Hilbert space. Inside the commuting model one can further ask which von Neumann algebra hosts the players. This paper finds, for each of these models, the exact level of the arithmetical hierarchy at which perfect play is decided.

The answers are not monotone in the model. As sets of correlations the spatial model sits between the finite-dimensional and the commuting one, yet perfect spatial play is strictly harder to decide than both: it is $\Sigma^0_2$-complete, while the other two are $\Sigma^0_1$- and $\Pi^0_1$-complete. A $\Sigma^0_2$ question asks whether some integer passes every one of infinitely many finite tests; no algorithm decides such questions in general, even with an oracle for the halting problem. Completeness means that perfect spatial play has this form and that every question of this form reduces to it. Our constructions make the integer concrete: a perfect strategy must select a finite clock period or a finite tree depth, and then pass every test attached to that choice. Spatial and semifinite perfection combine an unbounded finite choice with an infinite verification.

\subsection{Why a trace costs a quantifier}\label{sec:picture}
Commuting perfection is $\Pi^0_1$ because the commuting state space is compact. Perfection then equals commuting value one, which a convergent hierarchy of relaxations refutes in finitely many steps whenever it fails. Spatial play has no such compactness: Schmidt mass can escape to infinitely many small coefficients. Our upper bound keeps one Schmidt coefficient from disappearing. Once that coefficient is bounded below and scaled to one, the amplitude operator lies in a norm-compact family, and the losing probabilities become homogeneous zero equations that survive limits. So perfect spatial play has the form
\[
\exists\kappa\ \forall d\ \ F(G,\kappa,d),
\]
with $F$ decidable: for each integer $\kappa$, the existence of a perfect strategy whose largest Schmidt probability is at least $1/\kappa$ is a $\Pi^0_1$ property (Theorem~\ref{thm:strata}).

The same integer appears for every semifinite host. A trace gives a finite corner whose overlap with the state can be bounded below; in type I a rank-one corner recovers the largest Schmidt amplitude (Section~\ref{sec:cornerupper}). A type-III factor has no finite corner, and its perfect-play language collapses to the commuting one. In this sense a trace costs exactly one existential quantifier. Finite dimension is the degenerate case: there the whole space is a finite corner, and once the dimension is fixed, perfection is a question of real algebra that Tarski's method decides \cite{BPR}. Only the search over the dimension remains, which is why that row is $\Sigma^0_1$. An operational split between semifinite and type III algebras was known for embezzlement of entanglement \cite{VLSWW}; here the split appears as a quantifier.

The hardness proofs show that the quantifier is needed. In each case the existential witness is a finite object that every perfect strategy must choose. In the spatial model it is the order of a clock phase. The Weyl unitaries of the clock commute with the trace-class density, so they act on its eigenspaces, which are finite-dimensional, and there a determinant forces the commutator phase to be a root of unity. In a diffuse semifinite algebra no such determinant exists, but a trace forces a critical binary tree to end at a finite depth. A finite game then checks an infinite sequence of decidable conditions about the chosen object. No computable function of the game bounds a sufficient clock order (Section~\ref{sec:reduction}) or a sufficient Schmidt stratum (Theorem~\ref{thm:strata}).

\subsection{Results}\label{sec:results}
Table~\ref{tab:landscape} collects the answers, and four theorems fill it.

\begin{table}[ht]
\centering\small
\begin{tabular}{@{}p{.58\linewidth}lp{.17\linewidth}@{}}
\toprule
Decision problem & Complete class & Source\\
\midrule
Perfect finite-dimensional strategy & $\Sigma^0_1$ & \cite{MIPRE}\\
Perfect commuting strategy & $\Pi^0_1$ & \cite{Slofstra,MNY}\\
Finite-dimensional value equals one & $\Pi^0_2$ & \cite{MNY20,MNY}\\
\addlinespace
Perfect spatial strategy; also with finite entropy & $\Sigma^0_2$ & Theorem~A\\
Perfect strategy with finite $\alpha$-moment, fixed computable $0<\alpha<1$ & $\Sigma^0_2$ & Theorem~D\\
Perfect play in type $\II_1$, $\II_\infty$ or any semifinite algebra & $\Sigma^0_2$ & Theorem~B\\
Perfect play in a type $\III_\lambda$ factor, any fixed $\lambda\in[0,1]$ & $\Pi^0_1$ & Theorem~B\\
Commuting perfect play but no semifinite perfect play & $\Pi^0_2$ & Theorem~\ref{thm:Brefine}\\
Semifinite perfect play without spatial (or finite-entropy) perfection & $D_2(\Sigma^0_2)$ & Theorem~\ref{thm:Brefine}\\
\addlinespace
One-shot zero-error code, normal entanglement; also finite entropy & $\Sigma^0_2$ & Theorem~C\\
One-shot zero-error code, commuting operators & $\Pi^0_1$ & Theorem~\ref{thm:qc}\\
\bottomrule
\end{tabular}
\caption{Finite rational Boolean games and channels; computable many-one reductions. A value is a supremum. A perfect strategy attains one. The first three rows are known; the others are proved here.}\label{tab:landscape}
\end{table}

The first theorem completes the standard row of models.

\Needspace{5\baselineskip}
\begin{theoremA}\label{thm:A}
Deciding whether a finite rational $0/1$ game has a perfect normal spatial strategy is $\Sigma^0_2$-complete. The same holds when a perfect strategy with finite marginal entropy is required. Hardness holds under the promise $\omega_q(G)=1$, for a family in which no instance has a perfect finite-dimensional strategy.
\end{theoremA}

The second sorts the commuting model by the type of the players' algebra. A host is an ambient von Neumann algebra containing Alice's measurements, with Bob's measurements in its commutant. Both the host and its normal state may be chosen; the state need not be tracial, and the measurements need not generate the host. Write $\perf(X)$ for the set of games with a perfect strategy whose host is of class $X$; finite-host perfection uses a $\sigma$-finite finite von Neumann algebra, equivalently one admitting a faithful normal tracial state (Section~\ref{sec:prelim}).

\Needspace{5\baselineskip}
\begin{theoremB}\label{thm:B}
Perfect normal play with a finite von Neumann host, a type $\II_1$ host, or a semifinite host is $\Sigma^0_2$-complete, and so are the type $\II_\infty$ and type $\II$ existence languages. For every fixed $\lambda\in[0,1]$, the type $\III_\lambda$ language equals commuting perfection and is $\Pi^0_1$-complete.
\end{theoremB}

Theorem~\ref{thm:Brefine} completes the type classification: commuting perfection without semifinite perfection is $\Pi^0_2$-complete, the semifinite-but-not-spatial differences are $D_2(\Sigma^0_2)$-complete, and every nonempty difference among six models is complete for its class.

The third shows that the same level governs a communication task. A sender uses a finite classical channel once and must convey one of $m$ messages with zero error, helped by shared entanglement. Write $\ZE_t(N,m)$ for the existence of such a code for the channel $N$ in model $t$ (Section~\ref{sec:normal}).

\Needspace{5\baselineskip}
\begin{theoremC}\label{thm:C}
For finite rational classical channels, $\ZE_{\qs}$ and $\ZE_{\finS}$ are $\Sigma^0_2$-complete. Moreover, there are a finite rational classical channel $N$ and an integer $m$ for which $\ZE_{\qs}(N,m)$ holds, and every normal spatial $m$-message zero-error resource has infinite marginal entropy on both sides.
\end{theoremC}

The fourth measures how fast the entanglement spectrum of a perfect strategy must decay. For a perfect spatial strategy with reduced density $h$, finite rank, finite $\alpha$-moments $\Tr h^\alpha<\infty$ and finite entropy are successively weaker resource conditions.

\Needspace{5\baselineskip}
\begin{theoremD}\label{thm:D}
For every fixed computable $0<\alpha<1$, perfect play with $\Tr h^\alpha<\infty$, denoted $\mathcal P_\alpha$, is $\Sigma^0_2$-complete. With $\mathcal P_0$ denoting finite rank, $\mathcal P_1$ finite entropy and $\mathcal P_{\rm normal}$ normal perfection, they form the strict ladder $\mathcal P_0\subsetneq\mathcal P_a\subsetneq\mathcal P_b\subsetneq\mathcal P_1\subsetneq\mathcal P_{\rm normal}$ for $0<a<b<1$, and all moment orders greater than one collapse to normal perfection.
\end{theoremD}

Theorem~\ref{thm:momentrefine} adds a single hardness family for all rungs, the completeness of the differences between rungs, and an exact description of the possible critical indices.

\paragraph{A running example.}
Consider the rooted tree of finite words $w$ over four letters; $|w|$ is the word length, and appending a letter selects one of four branches. Give the basis vector $|w\rangle$ density weight $(3/4)16^{-|w|}$, so that each child has one sixteenth of its parent's weight. The weights sum to one and define a normal state of finite entropy. Its half-moment, the sum of the square roots of all weights, diverges: the four child square roots sum to the parent's square root at every level. A \emph{bridge} is a finite game gadget that doubles a density weight between two selected subspaces; four bridges give the factor $16$. A fifth changes the ratio to $1/32$ and makes the half-moment converge, because $4/\sqrt{32}=1/\sqrt2$. Figure~\ref{fig:running} shows the tree and its bridges. This tree is the root of the spatial hardness game (Section~\ref{sec:tree}), it separates the moment rungs (Section~\ref{sec:renyi}), and it sits exactly at the zero-error threshold (Section~\ref{sec:communication}).

\begin{figure}[tb]
\centering
\colorlet{rxaccent}{orange!80!black}%
\begin{tikzpicture}[x=1mm,y=1mm,font=\footnotesize,
  rxbar/.style={fill=black!10,draw=black!80,line width=0.25pt},
  rxhi/.style={fill=rxaccent!45,draw=rxaccent,line width=0.6pt},
  rxedge/.style={draw=black!55,line width=0.25pt},
  rxgrid/.style={draw=black!15,line width=0.3pt},
  rxarrow/.style={-stealth,line width=0.55pt,shorten <=0.9mm,shorten >=0.9mm}]
%
\node[anchor=base west] at (3,7) {(a) one step down the tree};
\node[rotate=90,font=\scriptsize] at (48.6,-15.25) {weight, relative to $\xi$};
\foreach \g in {0,...,4}{
  \draw[rxgrid] ({12+7*\g},-0.6) -- ({12+7*\g},-29.9);
  \node[anchor=north] at ({12+7*\g},-30.6) {$e_{\g}$};
}
\foreach \rxj/\lab in {0/1,1/\frac12,2/\frac14,3/\frac18,4/\frac1{16}}{
  \draw[rxgrid] (10.4,{-2.25-6.5*\rxj}) -- (41.6,{-2.25-6.5*\rxj});
  \node[anchor=west,font=\scriptsize] at (42.6,{-2.25-6.5*\rxj}) {$\lab$};
}
\foreach \g in {1,...,4}{
  \draw[rxarrow] ({12+7*(\g-1)},{-2.25-6.5*(5-\g)}) -- ({12+7*\g},{-2.25-6.5*(4-\g)});
  \node[inner sep=0pt] at ({12+7*\g-3.5+1.75},{-2.25-6.5*(4-\g)-3.25-1.9}) {$\btrans_{\g}$};
}
\draw[rxarrow] (12,-2.25) -- (40,-2.25);
\node[anchor=south] at (26,-2.0) {$a_\alpha$};
\draw[rxarrow,draw=rxaccent,line width=0.9pt] (12,-2.25) -- (12,-28.25);
\node[anchor=east,text=rxaccent] at (11.2,-15.25) {$A_\alpha$};
\foreach \g in {1,...,3}{ \fill ({12+7*\g},{-2.25-6.5*(4-\g)}) circle (0.5); }
\fill (40,-2.25) circle (0.55);
\filldraw[fill=rxaccent,draw=rxaccent] (12,-2.25) circle (0.65);
\filldraw[fill=rxaccent,draw=rxaccent] (12,-28.25) circle (0.65);
\node[anchor=east] at (11.1,-2.25) {$\xi$};
\node[anchor=south] at (40,-1.3) {$a_\alpha\xi$};
\node[anchor=east] at (11.1,-28.25) {$A_\alpha\xi$};
%
\node[anchor=base west] at (53,7) {(b) four bridges};
\node[anchor=base west,font=\scriptsize] at (53,2.2) {weight $\tfrac34\,16^{-m}$ per word};
\foreach \m in {1,2,3}{
  \pgfmathtruncatemacro{\rxN}{4^\m-1}
  \pgfmathsetmacro{\rxsc}{43/4^\m}
  \pgfmathsetmacro{\rxsp}{43/4^(\m-1)}
  \foreach \rxk in {0,...,\rxN}{
    \pgfmathtruncatemacro{\rxp}{floor(\rxk/4)}
    \draw[rxedge] ({53+(\rxp+0.5)*\rxsp},{-8*(\m-1)-4.5}) -- ({53+(\rxk+0.5)*\rxsc},{-8*\m});
  }
}
\foreach \m in {0,...,3}{
  \draw[rxbar] (53,{-8*\m}) rectangle (96,{-8*\m-4.5});
}
\foreach \m in {1,2,3}{
  \pgfmathtruncatemacro{\rxN}{4^\m-1}
  \pgfmathsetmacro{\rxsc}{43/4^\m}
  \foreach \rxk in {1,...,\rxN}{
    \draw[draw=black!80,line width=0.25pt] ({53+\rxk*\rxsc},{-8*\m}) -- ({53+\rxk*\rxsc},{-8*\m-4.5});
  }
}
\draw[rxhi] (53,0) rectangle (96,-4.5);
\draw[rxhi] (63.75,-8) rectangle (74.5,-12.5);
\node[font=\scriptsize] at (74.5,-2.25) {$\emptyset$};
\foreach \rxk/\w in {0/00,1/01,2/10,3/11}{
  \node[font=\scriptsize] at ({53+(\rxk+0.5)*10.75},-10.25) {$\w$};
}
\node at (74.5,-31.2) {$\vdots$};
\node[anchor=base] at (74.5,-37.5) {$\Tr\sqrt{h_{F,1/16}}=\infty$};
%
\node[anchor=base,font=\scriptsize] at (100.5,2.2) {$m$};
\foreach \m in {0,...,3}{
  \node[font=\scriptsize] at (100.5,{-8*\m-2.25}) {$\m$};
}
%
\pgfmathsetmacro{\rxW}{43*sqrt(7/6)}
\node[anchor=base west] at (105,7) {(c) five bridges};
\node[anchor=base west,font=\scriptsize] at (105,2.2) {weight $\tfrac78\,32^{-m}$ per word};
\foreach \m in {1,2,3}{
  \pgfmathtruncatemacro{\rxN}{4^\m-1}
  \pgfmathsetmacro{\rxsc}{\rxW/4^\m}
  \pgfmathsetmacro{\rxsp}{\rxW/4^(\m-1)}
  \foreach \rxk in {0,...,\rxN}{
    \pgfmathtruncatemacro{\rxp}{floor(\rxk/4)}
    \draw[rxedge] ({105+(\rxp+0.5)*\rxsp},{-8*(\m-1)-4.5}) -- ({105+(\rxk+0.5)*\rxsc},{-8*\m});
  }
}
\foreach \m in {0,...,3}{
  \pgfmathtruncatemacro{\rxN}{4^\m-1}
  \pgfmathsetmacro{\rxsc}{\rxW/4^\m}
  \pgfmathsetmacro{\rxbw}{\rxW/(sqrt(32)^\m)}
  \foreach \rxk in {0,...,\rxN}{
    \draw[rxbar] ({105+(\rxk+0.5)*\rxsc-0.5*\rxbw},{-8*\m}) rectangle ({105+(\rxk+0.5)*\rxsc+0.5*\rxbw},{-8*\m-4.5});
  }
}
\pgfmathsetmacro{\rxbwone}{\rxW/sqrt(32)}
\draw[rxhi] (105,0) rectangle ({105+\rxW},-4.5);
\draw[rxhi] ({105+1.5*\rxW/4-0.5*\rxbwone},-8) rectangle ({105+1.5*\rxW/4+0.5*\rxbwone},-12.5);
\node[font=\scriptsize] at ({105+0.5*\rxW},-2.25) {$\emptyset$};
\foreach \rxk/\w in {0/00,1/01,2/10,3/11}{
  \node[font=\scriptsize] at ({105+(\rxk+0.5)*\rxW/4},-10.25) {$\w$};
}
\node at ({105+0.5*\rxW},-31.2) {$\vdots$};
\node[anchor=base] at ({105+0.5*\rxW},-37.5) {$\Tr\sqrt{h_{F,1/32}}<\infty$};
\end{tikzpicture}%
\input{figures/fig-running-example-caption}
\end{figure}

Theorem~C comes from a correspondence between games and channels due to Man\v{c}inska \cite{Mancinska}, which we extend from finite dimension to trace-class operators. A perfect strategy gives a code exactly when its Schmidt amplitudes are summable. The compiled games of Theorem~A sit exactly at this threshold: with four bridges, the amplitude sum diverges for every perfect strategy, and five bridges restore summability. A Catalan version of the construction gives a channel whose game has finite-entropy perfect strategies, while every code of the required size has infinite entropy.

\paragraph{Reading guide.} For the trace mechanism, read Sections~\ref{sec:prelim} and~\ref{sec:upper} (in particular Section~\ref{sec:cornerupper}), then Sections~\ref{sec:selector}, \ref{sec:typeIII} and~\ref{sec:sep}, treating the compiler as a black box with the input and output contract stated at the start of Section~\ref{sec:compiler}. For the spatial classification, read Section~\ref{sec:upper}, that contract, and Sections~\ref{sec:tree}--\ref{sec:reduction}; the implementation in the rest of Section~\ref{sec:compiler} can be deferred, and Section~\ref{sec:reduction} shows how the clock period becomes the existential witness. For communication, read Section~\ref{sec:prelim} and Sections~\ref{sec:normal}--\ref{sec:catalan}.

\subsection{Further results}
The body also proves refinements of Theorems~A--D. The largest-Schmidt strata are $\Pi^0_1$, and no total $0'$-computable function bounds a sufficient stratum (Theorem~\ref{thm:strata} and Section~\ref{sec:entropybudgets}). Under a supplied entropy budget, full correlation tables are compact; the entropy busy beaver has Turing degree $0''$ and eventually dominates every total $0'$-computable function (Section~\ref{sec:entropybudgets}). Exact attainment of a rational Bell bound, of an unspecified supremum, or of a specified value is $\Sigma^0_2$-complete (Theorem~\ref{thm:bell}). The $I_{3322}$ inequality is an explicit instance. P\'al and V\'ertesi computed its supremum numerically, with finite-dimensional strategies converging to it, and conjectured that no finite dimension attains it \cite{PalVertesi}, and Krohn-Grimberghe has claimed a computer-assisted exact identification of the value \cite{KrohnGrimberghe}; the supremum is not attained in finite dimension \cite{Pauwels,Coladangelo3322,Douglas3322} but is attained by an infinite-dimensional spatial strategy \cite{Coladangelo3322,Douglas3322}. The type-I and type-$\II_1$ languages are incomparable, and both are proper subsets of the type-$\II_\infty$ language (Section~\ref{sec:sep}). Theorem~\ref{thm:differences} classifies every difference ``perfect in $X$ but not in $Y$'' among six models. Finally, one-shot zero-error feasibility with finite-dimensional entanglement is $\Sigma^0_1$-complete, a corollary of \cite{Mancinska} and \cite{MIPRE} (Section~\ref{sec:fdrow}).

\subsection{Prior work and attribution}
The finite-dimensional row of Table~\ref{tab:landscape} uses perfect completeness in Theorem~12.6 of Ji, Natarajan, Vidick, Wright and Yuen \cite{MIPRE}, not just a gap statement about a supremum; upper membership follows by a dimension search and real quantifier elimination. Slofstra's embedding theorem \cite{Slofstra} and the account of Mousavi, Nezhadi and Yuen \cite[Theorem~6.4]{MNY} give the commuting row. The value-one row was first proved by Mousavi, Nezhadi and Yuen \cite{MNY20}; see also \cite[Theorem~6.9]{MNY}. The commuting upper bound uses the convergent moment hierarchy of Navascu\'es, Pironio and Ac\'in \cite{NPA}. The operator-solution framework for binary constraint system games originates with Cleve and Mittal \cite{CM}; Cleve, Liu and Slofstra gave the commuting-operator extension \cite{CLS}. Slofstra also established nonclosure of finite-dimensional quantum correlations \cite{SlofstraNonclosed}. Lin's $\mathsf{MIP}^{co}=\mathsf{coRE}$ theorem concerns gap approximation in the commuting model \cite{Lin}; it does not classify spatial attainment.

Infinite-dimensional spatial correlations were constructed by Coladangelo and Stark \cite{CS}. Kim, Paulsen and Schafhauser prove that perfect spatial play collapses to perfect finite-dimensional play for synchronous games \cite[Corollary 3.11]{KPS}, and Lupini and coauthors prove the corresponding equivalence for imitation games \cite[Corollary~5.4]{Imitation}. Our games keep nonsynchronous bridge questions, so no YES instance of the spatial reduction is synchronous or an imitation game.

Man\v{c}inska's flagged magic-square construction supplies the bridge geometry \cite{Mancinska}; the magic square itself is due to Mermin and Peres \cite{Mermin,Peres}. The normal bridge, shared-label refinement, graph products and Boolean rational-row tests were developed in our entropy-game paper \cite[Lemmas~4.1 and 4.2, Theorem~B, Eq.~(32)]{Entropy}; its Catalan game \cite[Theorem~C]{Entropy} is used here as a separator, and a four-bridge version of its Catalan assembly gives the infinite-entropy channel. Section~\ref{sec:compiler} restates and proves the versions needed here and extends them to $n$ bridges, arbitrary finite lists of endpoint labels and normal commuting hosts. The standard-form reduction, the commuting bridge and expected-inclusion transfer come from our earlier paper \cite[Lemmas~1.6, 2.2 and 3.1]{ReadType}; the type rows of Table~\ref{tab:landscape} answer the question left open in \cite[Theorem~7.6]{ReadType}. The separation $\perf(\II_1)\not\subseteq\perf(\mathrm I)$ uses the tracial mechanism of \cite[Proposition~7.4]{ReadType} and Slofstra's nonclosure construction; the other direction uses a new one-bridge separator. Our earlier preprints \cite{Gamma} and \cite{Hyperfinite} gave single games with, respectively, finite-energy spatial perfection that no finite-dimensional strategy attains, and a perfect type $\III_{1/2}$ strategy with no injective host.

Group embeddings originate with Higman \cite{Higman}; the effective form follows Chiodo \cite[Theorem 2.2]{Chiodo}, with Craig's conversion \cite{Craig} and the HNN normal form \cite{HNN}. Subtype transfer uses Connes--Takesaki flow-of-weights theory \cite{ConnesTakesaki}, Takesaki's expectation theorem \cite{Takesaki}, Connes's invariants \cite{Connes}, and the argument inside Haagerup--Musat \cite[Lemma 3.8]{HM}. The nearest operational work on host type is the type classification of embezzling states by van Luijk, Stottmeister, Werner and Wilming \cite{VLSWW}, the commuting self-test for exact embezzlement of Paddock and Schmidt \cite{PS}, and Harris's Cuntz-algebra self-testing and factor classification \cite{HarrisEmbezzlement}. These concern embezzlement rather than the arithmetical classification of host-existence languages.

For communication, Cubitt, Leung, Matthews and Winter established the operator characterization of entanglement-assisted zero-error codes and the pseudo-telepathy construction \cite{CLMW}. Man\v{c}inska proved finite-dimensional clique saturation in Lemma~5, the converse and square-root state map in Theorem~6, and the entanglement-assisted independence-number characterization in Corollary~7 of \cite{Mancinska}; her zero-error results assume a finite-dimensional shared state. Her Corollary~8 gives the fifty-input flagged-magic-square channel with ten entangled messages and at most nine messages from finite-dimensional maximally entangled resources. Man\v{c}inska, Roberson and Varvitsiotis studied perfect strategies and independent-set game reductions \cite{MRV}; Man\v{c}inska and Roberson developed quantum graph homomorphisms \cite{MR}; Harris proved universality results for synchronous graph homomorphism games \cite{HarrisGraph}. Bhattacharyya, Mehta and Zhao proved undecidability for a restricted entanglement-assisted one-shot parameter of classical-to-quantum channels \cite{BMZ}; their parameter uses maximal entanglement and projective measurements \cite[Definition~2.3]{BMZ}. The finite-dimensional row here removes these restrictions for classical channels and a one-shot threshold; it does not resolve the asymptotic capacity question. For fixed-scenario correlation membership, studied by Fu, Miller and Slofstra \cite{FMS}, see Section~\ref{sec:discussion}.

\precursordisclosure

\subsection{Organization}
Section~\ref{sec:prelim} fixes the models, languages and notation. Section~\ref{sec:upper} proves the $\Sigma^0_2$ upper bounds for spatial, finite-entropy, finite and semifinite perfection; Section~\ref{sec:momentupper} treats moment rungs and Section~\ref{sec:codeupper} reduces codes to games. Section~\ref{sec:compiler} builds the finite compiler shared by all hardness proofs. Sections~\ref{sec:tree}--\ref{sec:reduction} prove the spatial hardness of Theorem~A, and Section~\ref{sec:bell} the Bell variants. Sections~\ref{sec:renyi}--\ref{sec:programmable} prove Theorem~D. Sections~\ref{sec:selector}--\ref{sec:typeIII}, together with Section~\ref{sec:iionly}, prove Theorem~B, and Section~\ref{sec:sep} compares the type languages. Sections~\ref{sec:normal}--\ref{sec:catalan} prove Theorem~C and the other communication rows. Section~\ref{sec:discussion} discusses open problems.

{\small\tableofcontents}
\clearpage
\section{Preliminaries}\label{sec:prelim}
\subsection{Games, strategies and models}
A finite rational Boolean game $G$ is a finite list of questions and answers, a rational distribution $\pi(x,y)$, and a predicate $V(a,b\mid x,y)\in\{0,1\}$. Perfection means zero probability for every losing event of positive question weight. A spatial strategy consists of positive operator-valued measurements (POVMs) $E_{a|x}$ and $F_{b|y}$ and a normal state, represented by a positive trace-class density $\rho$ of trace one, on $\HH_A\otimes\HH_B$. Its winning probability is
\[
 \sum_{xyab}\pi(x,y)V(a,b\mid x,y)\Tr\rho(E_{a|x}\otimes F_{b|y}).
\]
We write $C_q,C_{qs},C_{qc}$ for finite-dimensional, normal spatial, and commuting correlations. For a strategy $\mathcal S$, write $\omega(G,\mathcal S)$ for the displayed winning probability and $\operatorname{loss}_G(\mathcal S)=1-\omega(G,\mathcal S)$. The value $\omega_q$ is the supremum over $C_q$. Finite Schmidt truncations show that $C_q$ and $C_{qs}$ have the same closure and the same supremum for every Bell functional.

The set $\perf_q$ consists of games with a perfect finite-dimensional strategy. The set $\perf_{\qs}$ consists of games with a perfect normal spatial strategy. Its stratum $\perf_{\qs}(\kappa)$ requires a perfect pure strategy whose largest Schmidt probability is at least $1/\kappa$. The set $\perf_{\finS}$ requires a perfect strategy with finite marginal entropy. The one-marginal existence convention gives the same game language as requiring both reduced entropies finite, by purification onto the other party and equality of the pure state's marginal spectra; the conventions can differ on a fixed mixed strategy. Commuting perfection, $\perf_\co$, uses two commuting families of POVMs on one Hilbert space and a state.

\paragraph{Hosts.}
For a class $X$ of von Neumann algebras, $\perf(X)$ denotes existence of a perfect strategy normal on an ambient von Neumann algebra of class $X$; Bob acts in its commutant. The ambient algebra need not be generated by the measurements. The state may be chosen along with the host. Finite-host perfection permits non-tracial normal states; this is consistent with $\omega_{\rm tr}(G_{\rm FMS})<1$ in \cite[Theorem~3]{Mancinska}. These are existence languages: they do not test a prescribed algebra, state or full correlation table. Write $\perf_{\finhost}$ for hosting in a $\sigma$-finite finite von Neumann algebra, equivalently one admitting a faithful normal tracial state, and $\perf_{\sfhost}$ for hosting in a semifinite von Neumann algebra. Factor hosting gives the same existence languages as hosting in algebras purely of the indicated type.

We use separable hosts when decomposing. For commuting input this follows from the joint GNS representation of the finitely generated measurement $C^*$-algebra. For finite or semifinite input the finite-corner overlap construction of Section~\ref{sec:cornerupper} gives a separable host in the same class: generate the two unital $C^*$-algebras by the finitely many effects and the finite band projection, and take the closure of their orbit of the trace vector. It is separable; Lemma~\ref{lem:overlap} preserves a nonzero perfect component and proves its represented host finite or semifinite. One may then standardize Bob and decompose over the center. Every losing probability is an integral of nonnegative probabilities, so the finitely many zero integrals vanish on a common conull set of factor fibres. A finite or semifinite fibre remains in that class; the amplifications in \eqref{eq:languageequal} give the required pure type-II factor. For type I use its direct-integral decomposition into spatial factor strategies; in particular the type-I language $\perf(\mathrm I)$ is the spatial language $\perf_\qs$. The type-III constructions of Section~\ref{sec:typeIII} start with an unrestricted commuting factor fibre.

When several languages are compared at once we abbreviate
\[
 \Lq=\perf_q,\quad \LE=\perf_{\finS},\quad \LI=\perf(\mathrm I),
\]
\[
 \LJ=\perf(\II_1),\quad \LU=\perf_{\sfhost},\quad \LC=\perf_{\co}.
\]
Here $\Lq$ requires attainment in finite dimension; it is not the value-one language. The inclusions are $\Lq\subseteq\LE\subseteq\LI\subseteq\LU\subseteq\LC$ and $\Lq\subseteq\LJ\subseteq\LU$; every displayed inclusion is strict (Section~\ref{sec:sep}). We keep $\perf$ strictly for attainment. Nezhadi's 2025 thesis \cite{Nezhadi} uses ``perfectly winnable'' for the value-one $\Pi^0_2$ problem of \cite{MNY}, so its terminology has a different scope.

\subsection{Arithmetical classes}
The set $\Sigma^0_2$ consists of predicates $\exists j\,\forall t\,\Rdec(e,j,t)$ with decidable $\Rdec$. Completeness means that a problem is itself of this form and every set of this form reduces computably to it. The classes $\Sigma^0_1$ and $\Pi^0_1$ use a single existential or universal number quantifier, respectively, and $\Pi^0_2$ consists of complements of $\Sigma^0_2$ sets. We also use the class of differences
\[
 D_2(\Sigma^0_2)=\{A\setminus B:A,B\in\Sigma^0_2\}.
\]
It is a class of differences and should not be confused with the intersection of the pointclasses $\Sigma^0_2$ and $\Pi^0_2$. The canonical complete language is $\{(e,f):e\in\mathrm{FIN},\ f\notin\mathrm{FIN}\}$, where $\mathrm{FIN}$ is the $\Sigma^0_2$-complete set of indices of finite c.e.\ sets. Indeed reduce the two $\Sigma^0_2$ predicates separately to FIN and pair the resulting indices.

All encodings and reductions in this paper are effective finite encodings and computable many-one reductions. Completeness is over all finite scenarios; the numbers of questions and answers are not fixed in advance.

\subsection{Conventions}
Entropy uses logarithms to base two: $S(h)=-\Tr(h\log_2h)$ and $H(p)=-\sum_i p_i\log_2p_i$, with $0\log_2 0=0$. A projection-valued measurement (PVM) is a POVM consisting of orthogonal projections. Write $|\Phi_d\rangle=d^{-1/2}\sum_{j=0}^{d-1}|j,j\rangle$ for a finite maximally entangled vector. Vectorization of an operator $T$ in fixed orthonormal bases is
\[
 |T\rangle\!\rangle=\sum_{i,j}T_{ij}|i\rangle\otimes|j\rangle,
 \qquad \|\,|T\rangle\!\rangle\|^2=\Tr T^*T,
\]
when this sum is square summable. A purification is a unit vector with the given state as its partial trace over an added environment. Normal means represented by a positive trace-class density of trace one; on a spatial tensor product this includes pure vectors and countable mixtures. The centralizer of a density consists of the bounded operators commuting with it; for a faithful normal state $\varphi$ on a von Neumann algebra $M$ it is $M_\varphi=\{x\in M:\varphi(xy)=\varphi(yx)\ \forall y\in M\}$. For a game written in a code or compiler, YES and NO instances are those with and without the specified perfect strategy or code. $\Mat_r$ denotes the scalar $r$-by-$r$ matrix algebra.

\subsection{Supports and synchronization}\label{sec:support}
The compiler of Section~\ref{sec:compiler} uses the following elementary facts, in a spatial and in a standard-form version.

\paragraph{Pure components.}
If a mixed normal strategy is perfect, write $\rho=\sum_j\lambda_j|\psi_j\rangle\langle\psi_j|$. Each losing event is positive. Its expectation vanishes in every component with $\lambda_j>0$, so every such component is perfect with the same measurements. Trace-class marginals have separable support. Thus all arguments may be made on separable state supports.

\paragraph{Matching and support.}
For effects on opposite sides, perfect binary matching gives, on a purification $\Omega$,
\[
 0=\langle E+F-2EF\rangle
 =\langle(E-F)^2+(E-E^2)+(F-F^2)\rangle.
\]
Each positive term annihilates $\Omega$. If $T$ is the coefficient operator of the pure vector on the chosen Schmidt cut, matching gives $ET=T F^{\mathsf T}$; dense coefficient range first makes the local state support reducing. On that support the effects are projections and commute with the marginal density. A positive operator $B$ with $\Tr(hB)=0$ for faithful $h$ is zero, since $B^{1/2}h^{1/2}=0$ and $h^{1/2}$ has dense range.

\paragraph{Sharp compressions and joint measurements.}
If $PEP$ is a projection for an effect $E$ and support projection $P$, positivity shows that $E$ acts as identity on its range and zero on its kernel inside $P\HH$. Hence $E$ reduces the support. A POVM with sharp coarse marginals is their joint spectral measurement: each fine effect is supported in the corresponding marginal ranges, and summation gives the commuting products. These facts apply after compression before projectivity is used.

\paragraph{Synchronization and forbidden events.}
We call a projection synchronized when its binary question is perfectly matched to the corresponding opposite question. If synchronized projections satisfy $P_A^{\rm sync}\Omega=P_B^{\rm sync}\Omega$ and a forbidden event gives $Q_AP_B^{\rm sync}\Omega=0$, dense Schmidt range gives the local identity $Q_AP_A^{\rm sync}=0$. The arguments work on either party and on purifications of mixed states.

\paragraph{Standard form.}
A standard-form strategy for a faithful normal state $\varphi$ on $M$ uses its natural-cone vector $\xi_\varphi$ in $L^2(M)$, Alice effects in $M$, and Bob effects in $M'=\mathcal JM\mathcal J$, where $\mathcal J$ is the modular conjugation and $\sigma_t^\varphi$ the modular group. Nonfaithful states are interpreted on their support corners. A general normal-marginal commuting strategy has this form without changing probabilities: after joint GNS, compress Bob to $\overline{M\psi}$ using the isometry $x\xi_\varphi\mapsto x\psi$. The compression lies in the standard commutant and is still a POVM. Semifiniteness passes to corners. This is \cite[Lemma 1.6]{ReadType}; the argument includes mixed states.

Perfectly matched binary effects are sharp centralizer projections with mirror $\mathcal JE\mathcal J$. Indeed different matched classes satisfy $E_cF_d\xi_\varphi=0$. Summing gives $E_c\xi_\varphi=F_c\xi_\varphi$ and $E_cE_d\xi_\varphi=0$ for $c\ne d$. Separation gives orthogonality, and completeness gives idempotence. The self-adjoint mirror gives $\varphi(E_cx)=\varphi(xE_c)$. A fine positive effect below sharp marginals is their joint projection. These facts replace any finite-dimensional rigidity assumption. On faithful support a synchronized projection is therefore a centralizer projection with its standard commutant mirror.

\subsection{Notation}
Table~\ref{tab:notation} collects the notation used across sections. Distinct fonts separate question names (sans serif), languages (bold), automaton symbols (typewriter) and operators. Integer indices such as $i,j,k,\ell,m,n,r,t$ are local to the statement or proof in which they appear, except that $n$ counts bridges in the compiler and in Sections~\ref{sec:communication}--\ref{sec:catalan}. Each root presentation (the Fock clock of Sections~\ref{sec:tree}--\ref{sec:reduction}, the selector of Section~\ref{sec:selector}, the shift separator of Section~\ref{sec:sep} and the Catalan presentation of Section~\ref{sec:catalan}) has its own labels, introduced where it is defined.

\newpage
\begin{center}\footnotesize
\captionof{table}{Notation.}\label{tab:notation}
\begin{tabular}{@{}>{\raggedright\arraybackslash}p{.33\linewidth}>{\raggedright\arraybackslash}p{.60\linewidth}@{}}
\toprule Symbol & Meaning\\\midrule
\multicolumn{2}{@{}l}{\emph{Games, models and languages}}\\
$G;\ \pi,V$ & finite rational Boolean game; question distribution and predicate\\
$\mathcal S,\omega(G,\mathcal S),\omega_q$ & strategy, its winning probability, finite-dimensional value\\
$\perf_q,\perf_\qs$; $\perf_\finS,\perf_\co$ & finite-dimensional, spatial; finite-entropy, commuting perfection\\
$\perf(X),\perf_\finhost,\perf_\sfhost$ & hosting in class $X$, in a finite, in a semifinite von Neumann algebra\\
$\Lq,\LE,\LI,\LJ,\LU,\LC$ & abbreviations for six existence languages\\
$\kappa;\ \perf_\qs(\kappa),\perf_\finS(m)$ & Schmidt stratum integer; stratum and integer entropy-budget stratum\\
$F_\qs,F_\sfhost,F_{\rm Bell}$ & decidable stage predicates of the upper bounds\\
$\Rdec,\Adec$ & decidable predicate and its $\Sigma^0_2$ set\\
$\Gclk_e,G^{(n)}_e,\Gsel_e$ & Weyl-clock reduction (four bridges), its $n$-bridge version, selector reduction\\
$\Gsh,\Glin,G^{(2)}_{\Cat},\Gthree$ & fixed separating games of Section~\ref{sec:sep}\\
\multicolumn{2}{@{}l}{\emph{States and hosts}}\\
$D_{\rm Sch},D$ & largest-amplitude-normalized and $\Tr D^2=1$ Schmidt operators\\
$M,\varphi,\sigma_t^\varphi;\ \mathcal J,\xi_\varphi$ & host, faithful state, modular group; standard conjugation and state vector\\
$\tau,h$ & trace and density\\
$R_{\rm hyp},R_\infty,R_\lambda$ & injective $\II_1$, $\III_1$, $\III_\lambda$ factors\\
\multicolumn{2}{@{}l}{\emph{The compiler}}\\
$\mathsf R_i,\mathsf C_j,\mathsf T_x,\mathsf B_y;\ f,\perp$ & flagged magic-square questions; flag and inactive answers\\
$c_x,r_y;\ P_{\rm mag},m_{ab}$ & selected column and row triples; magic support and its column-zero projections\\
$\mathfrak x,\mathfrak z;\ \btrans$ & Pauli axis reflections; calibrated flag-to-magic transport\\
$P^{\rm lab}$ & synchronized label projection in a refinement\\
$F_i,F_{ij},G_{ij},P_\triangle$ & qutrit frame, pair support, graph and triangle projections\\
$P^{\rm par},J_{ab},s_{\rm par}$ & bound parameter, its joint partition and reflection label\\
$\Delta,L,V_j;\ k_{\rm row}$ & rational row, denominator, signed reflections; scalar qubits\\
$\Pi^\pm_j,\Pi^{\rm row}_t,\Pi^{\rm ref}_d,\mathcal I_t$ & scalar parity events, row partition, reference projection, column isometries\\
$d_{\max};\ \varphi_{\rm old},h_{\rm old}$ & row-ancilla dimension; state and marginal before the row ancilla\\
$A_\alpha,P_\circ;\ \nu$ & creation contractions and root projection; modular ratio $2^{-n}$\\
$e_0,e_g,m_{g,ab},m_g$ & source and selected corners, magic corners and supports of bridge $g$\\
$\Pact_g$ & activity $e_{g-1}+m_g$ of bridge $g$\\
$\btrans_g,v;\ \mathfrak c_n,\mu_n$ & bridges and their chain; scaffold cover constant and weights\\
$\Lambda,u_{r,\ell},V_{g,\ell},L_r$ & transported encodings, corner copies, block unitaries, decoded copies\\
$a_\alpha;\ L_{\rm end},R_{\rm end}$ & lifted centralizer contractions; endpoint coefficients\\
\bottomrule\end{tabular}\end{center}
\newpage
\begin{center}\footnotesize
\captionof*{table}{Table~\ref{tab:notation}: Notation (continued).}
\begin{tabular}{@{}>{\raggedright\arraybackslash}p{.33\linewidth}>{\raggedright\arraybackslash}p{.60\linewidth}@{}}
\toprule Symbol & Meaning\\\midrule
\multicolumn{2}{@{}l}{\emph{The Fock clock}}\\
$\WW,A_w,\KK,h_{F,\nu},h_F,\rho_{\rm spec}$ & four-letter words, word isometries, root multiplicity, tree densities, spectator\\
$U_x,U_y;P_x,P_y$ & cyclic coordinate shifts; zero-column and zero-row projections\\
$\Gamma_{\rm CA},P_c;\ \mathtt s,\mathtt b,\mathtt f,\mathtt z,\mathtt H$ & automaton alphabet and its projections; start, blank, front, mark, halt\\
$J_{\rm W},K_{\rm W},Z_{\rm clk},\zeta,d_{\rm clk}$ & Weyl pair, central clock phase, its value and order\\
$W,C_{\rm ctr},P_{\rm ret},T_{\rm inv}$ & coordinate clock, counter, return pulse and bounded reader inverse\\
$h^{[m]},P_m,t_m$ & truncated density, top-depth projection and its mass\\
\multicolumn{2}{@{}l}{\emph{The quarter-plane register and comb (Section~\ref{sec:programmable})}}\\
$S,T;\ \Pi_x,\Pi_t;\ C_x,C_t$ & shifts, boundary projections and dyadic counters\\
$Z,J,K,W;\ E_{\rm ret},R_{\rm ret}$ & clock labels, return projection and bounded reader\\
$d_{\rm clk};\ \rho_{\rm spec},\rhoreg$ & clock order; spectator and geometric register densities\\
$V_{\rm sp},B_0,B_1;\ Q,P,E_{\rm guard}$ & spine and branch maps; spine, root and guard projections\\
$\Csp,L;\ E_{\Csp},E_L,F_{\Csp},F_L$ & comb counters, equality projections and their readers\\
$k_m,\rhocomb,\Zcomb$ & bush depths, comb density and its normalizer\\
$\mathcal P_\beta,\alpha_c(h),c(G);\ d_{\rm cut}$ & moment language, witness and game critical indices; cutoff dimension\\
\multicolumn{2}{@{}l}{\emph{The nested assembly (Appendix~\ref{nested:sec:nested})}}\\
$X_i;\ a_i,a_i'$ & root modular contractions; inner and transported lifts\\
$f,e,\Pact_{\rm in},E;\ f',e'$ & inner corners and activity, outer corner; transported corners\\
$u,\btrans;\ \btrans_0,\btrans_1;\ t,\ell$ & inner/outer transports, outer restrictions, equal-density map and quadratic coefficient\\
$U_k,V_k,V^{\rm blk}_k$ & inner encodings, outer copies and whole-activity block unitaries\\
$R_{\rm tar};\ K_L,K_M,K_R$ & target centralizer expression; arbitrary polynomial coefficients\\
$\epsilon_{xy},\varsigma;\ J_{\rm sgn},Z^{\rm gr}_{ij}$ & FMS signs, scalar axes; sign grading and graph reflection\\
\multicolumn{2}{@{}l}{\emph{Communication}}\\
$N,m;\ \ZE_t(N,m)$ & finite rational channel and message count; one-shot feasibility\\
$\Gamma(G);\ \mathcal Q_A,\mathcal Q_B;\ k$ & tagged confusability graph; question sets; number of tagged questions\\
$R_{i,x},\rho,\rho_{\rm code};\ \mathcal C_c$ & assemblage and code marginal; a clique\\
$\hat\mu_n;\ c^{\rm amp}_n$ & normalized square roots of $\mu_n$; amplitude constant\\
$\hcat,\kcat,\Ccat$ & critical Catalan density, its normalized square, Catalan height counter\\
\bottomrule\end{tabular}\end{center}

\section{Upper bounds: one existential quantifier}\label{sec:upper}
Each tracial language in this paper is a union over one integer of $\Pi^0_1$ sets. For spatial play the integer bounds the inverse of the largest Schmidt probability; for semifinite play it bounds the inverse overlap of the state with a finite tracial corner. We first treat the spatial case and prove the following theorem.

\Needspace{5\baselineskip}
\begin{theorem}[Schmidt strata]\label{thm:strata}
Uniformly for integers $\kappa\geq1$, the existence of a perfect pure spatial strategy with largest Schmidt probability at least $1/\kappa$ is a $\Pi^0_1$ property. Their union is $\perf_{\qs}\in\Sigma^0_2$. No total computable function $\kappa(G)$ bounds a sufficient stratum for every spatially perfect game. If $\omega_q(G)=1$ and no perfect spatial strategy exists, the largest Schmidt probability tends to zero along every sequence of pure spatial strategies whose winning probabilities tend to one.
\end{theorem}

\subsection{Normalize one Schmidt coefficient}
For a pure strategy write $\psi=\sum_i s_i|i,i\rangle$ with $s_1\geq s_2\geq\cdots$ and $\sum_i s_i^2=1$. Set $d_i=s_i/s_1$ and $D_{\rm Sch}=\operatorname{diag}(d_i)$. For some integer $\kappa$,
\begin{equation}\label{eq:DK}
 d_1=1,\qquad d_i\downarrow0,\qquad \Tr D_{\rm Sch}^2\leq \kappa.
\end{equation}
For positive effects $E,F$, vectorization gives
\[
 p(E,F)=\frac{\|\sqrt E\,D_{\rm Sch}\sqrt{F^{\mathsf T}}\|_{\rm HS}^2}{\Tr D_{\rm Sch}^2},
 \qquad p(E,F)=0\ \Longleftrightarrow\ ED_{\rm Sch}F^{\mathsf T}=0.
\]
To prove the reverse implication, $ED_{\rm Sch}F^{\mathsf T}=0$ sends $D_{\rm Sch}\ran F^{\mathsf T}$ into $\ker E$. The closures of the ranges of $F^{\mathsf T}$ and $\sqrt{F^{\mathsf T}}$ agree, and $\ker E=\ker\sqrt E$. Continuity gives the square-root sandwich identity. The forward implication follows by multiplying the zero square-root sandwich by the square roots.

Let $P_{d_{\rm cut}}$ be the first ${d_{\rm cut}}$ coordinate projection. Monotonicity in \eqref{eq:DK} gives
\begin{equation}\label{eq:tail}
 \|D_{\rm Sch}-P_{d_{\rm cut}}D_{\rm Sch}P_{d_{\rm cut}}\|=d_{{d_{\rm cut}}+1}\leq\sqrt{\kappa/({d_{\rm cut}}+1)}.
\end{equation}
Consequently the family in \eqref{eq:DK} is compact in operator norm: take coordinatewise limits and use this common tail estimate; Fatou preserves the squared-sum bound. Finite families of POVMs on $\ell^2$ are compact in the weak operator topology. Positivity and the finite sum-to-identity relations survive these limits.

\begin{lemma}[Compact sandwiches]\label{lem:sandwich}
If $D_{{\rm Sch},j}\to D_{\rm Sch}$ in norm with compact $D_{\rm Sch}$, and uniformly bounded $E_j,F_j$ converge weakly to $E,F$, then $E_jD_{{\rm Sch},j}F_j$ converges weakly to $ED_{\rm Sch}F$.
\end{lemma}
\begin{proof}
For fixed $v$, compactness makes $D_{\rm Sch}F_jv\to D_{\rm Sch}Fv$ in norm. Pair this with the weakly convergent vectors $E_j^*u$. The replacement of $D_{\rm Sch}$ by $D_{{\rm Sch},j}$ has norm error tending to zero.
\end{proof}

\subsection{An effective finite predicate}
Let $\mathcal L_G$ be the losing tuples with $\pi(x,y)>0$. Define $F_\qs(G,\kappa,{d_{\rm cut}})$ by the existence of ${d_{\rm cut}}$-dimensional POVMs and real numbers $1=d_1\geq\cdots\geq d_{d_{\rm cut}}\geq0$, $\sum_i d_i^2\leq \kappa$, such that
\begin{equation}\label{eq:stage}
 |(E_{a|x}D_{{\rm Sch},{d_{\rm cut}}}F_{b|y}^{\mathsf T})_{ij}|^2\leq \kappa/({d_{\rm cut}}+1)
 \quad((x,y,a,b)\in\mathcal L_G;\ i,j\leq {d_{\rm cut}}).
\end{equation}
This is semialgebraic over the rationals. Splitting complex entries into real and imaginary parts and using principal minors for positivity makes real quantifier elimination a decision procedure \cite{BPR}. Decidability is more than the upper bound needs. All variables lie in $[-1,1]$, where each constraint is Lipschitz in the maximum norm, so $F_\qs$ fails exactly when, at some grid resolution, every grid point violates some constraint by more than its Lipschitz constant times the spacing. Thus $F_\qs$ is $\Pi^0_1$ by compactness alone; Appendix~\ref{app:leanA} checks this route.

A perfect strategy with $s_1^2\geq1/\kappa$ supplies all these witnesses. Compress its POVMs. Since $P_{d_{\rm cut}}$ commutes with $D_{\rm Sch}$ and $ED_{\rm Sch}F^{\mathsf T}=0$, the finite sandwich is $-P_{d_{\rm cut}}E(D_{\rm Sch}-D_{{\rm Sch},{d_{\rm cut}}})F^{\mathsf T}P_{d_{\rm cut}}$, with norm bounded by \eqref{eq:tail}.

Conversely choose a witness for every ${d_{\rm cut}}$. Extend $D_{{\rm Sch},{d_{\rm cut}}}$ by zero and each POVM by a fixed deterministic POVM on the complement. Pass to a subsequence with $D_{{\rm Sch},{d_{\rm cut}}}\to D_{\rm Sch}$ in norm and all effects weakly convergent. For every fixed matrix entry, \eqref{eq:stage} and Lemma~\ref{lem:sandwich} give $ED_{\rm Sch}F^{\mathsf T}=0$. We retain
\[
 1\leq\Tr D_{\rm Sch}^2\leq \kappa,\qquad d_1=1.
\]
The normalized vectorization of $D_{\rm Sch}$ is a perfect normal strategy with largest Schmidt probability $1/\Tr D_{\rm Sch}^2\geq1/\kappa$. Therefore
\begin{equation}\label{eq:hierarchy}
 G\in\perf_{\qs}(\kappa)\ \Longleftrightarrow\ \forall {d_{\rm cut}}\ F_\qs(G,\kappa,{d_{\rm cut}}),
 \qquad \perf_{\qs}=\bigcup_{\kappa\geq1}\perf_{\qs}(\kappa).
\end{equation}
This proves the upper membership in Theorems~A and \ref{thm:strata}.

\subsection{No effective choice of stratum; vanishing largest weight}
If a total computable $\kappa(G)$ were sufficient for every YES game, \eqref{eq:hierarchy} would put spatial perfection in $\Pi^0_1$. Perfect completeness of \cite[Theorem 12.6]{MIPRE} already makes it $\Sigma^0_1$-hard. NO instances of that map have spatial value at most $1/2$, because the finite-dimensional and spatial suprema agree. Thus the same map reduces the halting set to spatial perfection. The halting set cannot be $\Pi^0_1$, so such a function does not exist.

Now suppose pure spatial strategies approach perfect play while $p_{\max}\geq1/\kappa$ on a subsequence. Normalize their largest Schmidt amplitudes. For a losing event,
\[
 \|ED_{\rm Sch}F^{\mathsf T}\|_{\rm HS}
 \leq\|\sqrt E D_{\rm Sch}\sqrt{F^{\mathsf T}}\|_{\rm HS}
 =\sqrt{\Tr D_{\rm Sch}^2\,p(E,F)}.
\]
Positive question weights force every such probability to zero. The same compact limit gives a perfect spatial strategy. Contraposition proves the final assertion of Theorem~\ref{thm:strata}, including sequences of finite-dimensional pure strategies. Equivalently their Schmidt min-entropy tends to infinity.

\subsection{Entropy budgets}\label{sec:finS}\label{sec:entropybudgets}
This subsection treats entropy budgets and their busy-beaver profiles; readers heading for the trace mechanism may skip to Section~\ref{sec:cornerupper}.

Fix a finite Bell scenario and let $K_H$ be its correlations realized by
normal spatial strategies with $S(\rho_A)\le H$, for $0\le H<\infty$.
Purify onto Bob, the unconstrained party: Alice's marginal and every
correlation are preserved. Thus pure strategies suffice, with the same
budget, for entire correlation tables as well as winning probabilities.

\begin{proposition}[Entropy-budget compactness]\label{prop:entropycompact}
The set $K_H$ is compact. It is an effectively closed ($\Pi^0_1$) subset of $[0,1]^t$ uniformly in the scenario
and $H$: for rational budgets this is an ordinary effective presentation;
for real budgets it is uniform relative to a name of $H$.
\end{proposition}
\begin{proof}
For sorted Schmidt probabilities $p_i$, $p_i\le1/i$, so
\begin{equation}\label{eq:entropytail}
 \delta_n=\sum_{i>n}p_i\le H/\log_2(n+1).
\end{equation}
This gives tightness in $\ell^1$, and finite partial entropy sums give
lower semicontinuity. Hence the sorted distributions of entropy at most
$H$ are compact in $\ell^1$. Their Schmidt vectors are compact in $\ell^2$,
since $\|\sqrt p-\sqrt{p'}\|_2^2\le\|p-p'\|_1$.
Compress the POVMs to the Schmidt supports and absorb the basis changes
into them, obtaining common copies of $\ell^2$ (if the Schmidt rank is finite, extend each POVM by a fixed deterministic answer on the complement in $\ell^2$). Finite POVM tuples are
weak-operator sequentially compact there. Tensor products of convergent
effects converge weakly, first on simple tensors and then by boundedness
on all vectors. The norm-convergent Schmidt vectors therefore give
convergence of every correlation entry. This is compactness of
\emph{correlations}; the original states on fixed local spaces need not
form a compact set. This argument is the bounded-entropy compactness
of \cite[Section~9.1, ``Divergence at bounded entropy'']{Entropy}.

Here are explicit outer approximations. Fix an integer $\bar H\ge1$ and work
with $0\le H\le\bar H$. Choose positive rational
$\eta_{\bar H,n}>\sqrt{\bar H/\log_2(n+1)}$ tending to zero. Compute rational
polynomials $h_n$ with
\[
 \sup_{p\in\mathsf S_n}|h_n(p)-S(p)|\le1/n,
 \qquad \mathsf S_n=\{p_i\ge0:\textstyle\sum_i p_i=1\}.
\]
For example, Bernstein approximation to $-t\log_2t$ on $[0,1]$, with
rationally rounded coefficients and error at most $n^{-2}$ per summand,
gives these polynomials. Let $A^{\bar H}_{H,n}$ consist of tables $c\in[0,1]^t$
($t$ is the number of entries) for which there are sorted $p\in\mathsf S_n$,
nonnegative $s_i$ with $s_i^2=p_i$, and $n$-dimensional POVMs (extended by a fixed deterministic answer on the complement in $\ell^2$) satisfying
\begin{equation}\label{eq:entropyouter}
 h_n(p)\le H+1/n,\qquad
 |c_{ab|xy}-\langle\psi_s,E_{a|x}\otimes F_{b|y}\psi_s\rangle|
 \le\eta_{\bar H,n},\quad \psi_s=\sum_i s_i|i,i\rangle.
\end{equation}
Split complex entries into real and imaginary parts, impose positivity
by principal minors, and impose the POVM sums exactly. These are compact
semialgebraic outer sets, uniformly decidable by real quantifier
elimination \cite[Chapter~14]{BPR} for rational input data. The joint sets in $(H,c)$
are semialgebraic over the rationals as well.

The normalized first-$n$ truncation majorizes $p$, so its entropy does not
increase. Its pure-state trace distance from the original state is
$\sqrt{\delta_n}$; compressing the POVMs thus changes each entry by at most
this amount. Consequently $K_H\subseteq A^{\bar H}_{H,n}$ for every $n$.
Conversely witnesses in \eqref{eq:entropyouter} have entropy at most
$H+2/n$. Along any sequence $n\to\infty$, the preceding tail and POVM
argument gives a subsequential limit with entropy at most $H$, and the
entrywise slack vanishes. Hence
\[
 K_H=\bigcap_{n\ge1}A^{\bar H}_{H,n}.
\]
Compact semialgebraic sets with rational data are effectively closed;
the displayed intersection proves the uniform $\Pi^0_1$ presentation.
It also gives a joint effectively closed set on $[0,\bar H]\times[0,1]^t$,
which explains the assertion for real $H$.
\end{proof}

\begin{corollary}[Exact tables and the vanishing-loss entropy limit]\label{cor:entropybudget}
Membership of a prescribed rational correlation table in the class
realizable with some finite marginal entropy is in $\Sigma^0_2$,
uniformly over finite scenarios. The same holds for real-algebraic tables.
For a finite rational game $G$, set
\[
 E_G(\varepsilon)=\inf\{S(\rho_A):\omega(G,\mathcal S)\ge1-\varepsilon\},
 \qquad H_{\rm ex}(G)=\inf\{S(\rho_A):\omega(G,\mathcal S)=1\},
\]
where the infima range over normal spatial strategies and
$\inf\varnothing=\infty$. In particular $E_G(\varepsilon)=\infty$ when
$\varepsilon<1-\omega_q(G)$. Then
\[
 \lim_{\varepsilon\downarrow0}E_G(\varepsilon)=H_{\rm ex}(G),\qquad
 G\in\perf_{\finS}\ \Longleftrightarrow\
 \sup_{\varepsilon>0}E_G(\varepsilon)<\infty.
\]
If $H_{\rm ex}(G)$ is finite, a perfect strategy attains it.
\end{corollary}
\begin{proof}
Exact membership in $K_m$ is uniformly $\Pi^0_1$ for integer $m\ge0$;
take the union over $m$. The perfect-value constraint is closed too,
so the games $\perf_{\finS}(m)$ with perfect play of entropy at most $m$
form a uniform $\Pi^0_1$ stratum, and $\perf_{\finS}\in\Sigma^0_2$.
Purification preserves the supplied marginal
entropy, and a pure state's two marginal entropies agree.

The increasing limit $L=\lim_{\varepsilon\downarrow0}E_G(\varepsilon)$
is at most $H_{\rm ex}(G)$. If $L<\infty$, choose strategies of loss at most
$1/j$ and entropy at most $L+1/j$. Proposition~\ref{prop:entropycompact},
with its lower-semicontinuity argument, gives a perfect limit of entropy
at most $L$. Thus $H_{\rm ex}(G)\le L$, and equality follows, also in the
extended reals. A sequence of perfect strategies whose entropies decrease
to a finite $H_{\rm ex}(G)$ gives attainment by the same argument.
\end{proof}

\begin{lemma}[Entropy of clock witnesses]\label{lem:clockentropy}
Suppose $e\in\Adec$, and let $j_*$ be the least witness to $\forall t\,\Rdec(e,j,t)$; if $e\notin\Adec$, Theorem~\ref{thm:reduction} gives no perfect strategy.
Every perfect normal spatial strategy for $\Gclk_e$ has full
Alice-marginal entropy at least $\log_2 p_{j_*+1}$.
\end{lemma}
\begin{proof}
This lemma and the next corollary use the hardness construction of
Sections~\ref{sec:compiler}--\ref{sec:reduction}, which does not use them.
Purify onto Bob as in (N1) of Section~\ref{sec:compiler}; this preserves
Alice's full marginal $h$ and all correlations. Work on its support.
Lemma~\ref{lem:four} extracts a faithful normalized source density
$h_0=121e_0he_0$. By \eqref{eq:fockdensity},
$h_0=h_F\otimes\rho_{\rm spec}$. Every positive eigenspace of
$\rho_{\rm spec}$ splits into $Z_{\rm clk}$ phase sectors. The soundness
proof of Theorem~\ref{thm:reduction} applies to each nonzero sector:
its word descendants persist at every depth, so an invalid order or a
false check would contradict $P_{\mathtt H}=0$. Thus each phase order
is $p_{j+1}$ for a valid witness $j\ge j_*$. Equation~\eqref{eq:det}
makes the dimension of that sector divisible by its order. Consequently
every positive eigenvalue of $\rho_{\rm spec}$ has multiplicity at least
$p_{j_*+1}$, and hence is at most $1/p_{j_*+1}$. Therefore
$S(\rho_{\rm spec})\ge\log_2p_{j_*+1}$.

To pass to the full marginal, use the entire synchronized scaffold
cover, rather than an entropy inequality for a compression. Set
$v_0=e_0$ and $v_g={\btrans}_g\cdots {\btrans}_1$ for $1\le g\le4$.
Let $t_{g,ab}=\mathfrak x_g^a\mathfrak z_g^b e_g$ be the synchronized
magic matrix units, and put $T_{g,ab}=t_{g,ab}v_g$. These partial
isometries all have initial projection $e_0$ and final projections
$m_{g,ab}$, respectively. Together with $e_0$, their final projections
are orthogonal and sum to the identity on the full support. Moreover
$ht_{g,ab}=t_{g,ab}h$ and $hv_g=2^gv_gh$, so
\[
 T_{g,ab}^*hT_{g,ab}=2^g e_0he_0.
\]
Write $T_0=e_0$. Since each corner commutes with $h$ and the corners
are orthogonal, $T_k^*hT_l=0$ for distinct members $k\ne l$ of the
seventeen-element list $(T_0;T_{g,ab})$. Hence
$U(\delta_k\otimes\xi)=T_k\xi$ is a unitary from
$\mathbb C^{17}\otimes e_0P_A\HH_A$ onto $P_A\HH_A$, and by \eqref{eq:121}
\[
 h\simeq\operatorname{diag}(\mu_4)\otimes h_0,
 \qquad \mu_4=121^{-1}(1;2^{\oplus4};4^{\oplus4};8^{\oplus4};16^{\oplus4}).
\]
Here $h_0$ is the density of the extracted realization on the whole source
corner $e_0P_A\HH_A$; by \eqref{eq:fockdecomp} any further tensor factor
present there lies in $\KK$ and is counted in $\rho_{\rm spec}$. Nonnegative spectral sums now give,
also when an entropy is infinite,
\[
 S(h)=S(\operatorname{diag}(\mu_4))+S(h_F)+S(\rho_{\rm spec})
 \ge\log_2p_{j_*+1}.
\]
\end{proof}

\begin{corollary}[The entropy busy beaver]\label{cor:entropybeaver}
Fix a binary encoding of finite rational games, with decidable validity,
and write $|G|$ for its bit length. Put
\[
 \begin{gathered}
 m_*(G)=\min\{m\in\mathbb N:G\in\perf_{\finS}(m)\},\\
 B(n)=\max\bigl(\{m_*(G):|G|\le n,\ m_*(G)<\infty\}\cup\{0\}\bigr),
 \end{gathered}
\]
where $\mathbb N$ includes zero and an empty minimum is infinity.
For finite $m_*(G)$, attainment gives $m_*(G)=\lceil H_{\rm ex}(G)\rceil$.
There is no total $0'$-computable upper bound on $B$, and
\[
 B\equiv_T0'',\qquad B\oplus0'\equiv_T0''.
\]
Moreover $B$ eventually strictly dominates every total $0'$-computable function,
and hence exceeds each such function infinitely often. This is the profile of Rad\'o's busy beaver function \cite{Rado} relativized to $0'$, the ``super Busy Beaver'' $\mathrm{BB}_1$ of \cite[Section~3]{AaronsonBB}, with Beeping Busy Beavers giving an earlier natural example of this growth \cite[Section~5.10]{AaronsonBB}; the content of the corollary is that a resource of perfect play realizes it.
\end{corollary}
\begin{proof}
The fixed-budget test is uniformly $\Pi^0_1$, hence decidable in $0'$;
at budget zero it reduces to the decidable product-state test.
If a total $0'$-computable $b(n)$ bounded $B(n)$, testing the single budget
$b(|G|)$ would decide $\perf_{\finS}$ in $0'$, contradicting its
$\Sigma^0_2$-completeness in Theorem~A.
Using $0''$, decide which of the finitely many games of size at most $n$
lie in $\perf_{\finS}$, then search their budgets using $0'$ to compute
$B(n)$. Conversely $B\oplus0'$ decides $\perf_{\finS}$ by testing budget
$B(|G|)$, so computes $0''$. Finally, an eventual upper bound could be
patched at finitely many inputs to a total $0'$-computable upper bound,
which has just been excluded.

For the reverse reduction $0'\le_T B$, use the clock games for the
decidable predicate ``the $e$th machine on input $e$ halts within $j$
steps.'' If it halts at time $T_e$, the valid witnesses are exactly $j\ge T_e$, and Lemma~\ref{lem:clockentropy} gives
\[
 T_e<p_{T_e+1}\le2^{m_*(\Gclk_e)}\le2^{B(|\Gclk_e|)}.
\]
Thus simulating $2^{B(|\Gclk_e|)}$ steps decides the halting set from $B$: a computation that has not halted by then never halts.
Combining this with $B\oplus0'\equiv_T0''$ proves $B\equiv_T0''$.

For eventual domination, fix a computable array $g(a,n,s)$ obtained by
finite-stage simulations of oracle machines with the standard finite
approximations to $0'$. Every total $0'$-computable function occurs as
$n\mapsto\lim_s g(a,n,s)$ for some $a$. Compile the decidable predicates
\[
 \Rdec(a,n,\langle s_0,v\rangle,t)
 \ \Longleftrightarrow\
 \forall s\in[s_0,t]\ g(a,n,s)=v
\]
into clock games $G_{a,n}$, with hardwired input $e=\langle a,n\rangle$, using a computable pairing with
$\langle s_0,v\rangle\ge v$. Pad by harmless always-accepted questions
so that $|G_{a,n}|\ge n+1$. A perfect strategy for a padded game restricts
to one for the unpadded game with the same state, and conversely extends
by arbitrary answers; padding therefore changes neither $h$ nor the
bound of Lemma~\ref{lem:clockentropy}. Define the computable, nondecreasing,
unbounded size envelope
\[
 \ell(n)=\max_{a,k\le n}|G_{a,k}|.
\]
For a given total $0'$-computable $F:\mathbb N\to\mathbb N$, the limit
lemma \cite{Shoenfield} supplies an index $a_0$ with
\[
 \lim_s g(a_0,n,s)=f(n),\qquad
 f(n)=2^{1+\max_{m\le \ell(n+1)}F(m)}.
\]
The envelope was defined using all indices before selecting $a_0$,
so there is no circular size definition. Each $G_{a_0,n}$ is a YES
instance, and every witness index is at least $f(n)$. For $n\ge a_0$
and $\ell(n)\le m<\ell(n+1)$, Lemma~\ref{lem:clockentropy} therefore gives
\[
 B(m)\ge m_*(G_{a_0,n})\ge\log_2p_{f(n)+1}
 >\log_2f(n)>F(m).
\]
The intervals cover every sufficiently large integer since $\ell$ is
nondecreasing and unbounded. This proves eventual strict domination.
\end{proof}

\begin{remark}[Finite strata and jump bounds]
Let $n\ge2$ and suppose a $\Sigma^0_n$-complete attainment language has
$e\in L\iff\exists m\,R(e,m)$ with $R$ uniformly $\Pi^0_{n-1}$.
No total $0^{(n-2)}$-computable function bounds a sufficient stratum index
on all YES inputs; in fact no total $0^{(n-1)}$-computable function does.
Indeed the latter oracle decides each stratum test, so a bounded finite
search would put $L$ in $\Delta^0_n$, contradicting completeness.
The entropy busy beaver is the case $n=2$. The Schmidt resource has the full profile as well.
\end{remark}

\begin{corollary}[The Schmidt busy beaver]\label{cor:schmidtbeaver}
For the same encoding, put
\[
 \begin{gathered}
 \kappa_{\min}(G)=\min\{\kappa\ge1:G\in\perf_{\qs}(\kappa)\},\\
 K(n)=\max\bigl(\{\kappa_{\min}(G):|G|\le n,\ G\in\perf_{\qs}\}\cup\{0\}\bigr).
 \end{gathered}
\]
An empty minimum is infinity. For each YES clock game,
\[
 \kappa_{\min}(\Gclk_e)\ge\frac{121}{12}p_{j_*+1}.
\]
Moreover $K\equiv_T0''$, and $K$ eventually strictly dominates every
total $0'$-computable function.
\end{corollary}
\begin{proof}
For every perfect pure strategy, the full-marginal equivalence in
Lemma~\ref{lem:clockentropy} and its spectral-multiplicity argument give
\[
 \|h\|\le\|\operatorname{diag}(\mu_4)\|\,\|h_F\|\,
 \|\rho_{\rm spec}\|\le\frac{16}{121}\frac34\frac1{p_{j_*+1}}
 =\frac{12}{121p_{j_*+1}}.
\]
The largest Schmidt probability is $\|h\|$. A strategy in stratum
$\kappa$ therefore requires $\kappa\ge(121/12)p_{j_*+1}$, proving the bound.
Every spatially perfect game lies in some finite stratum by
Theorem~\ref{thm:strata}. Using $0''$, decide which of the finitely many
valid games of length at most $n$ are spatially perfect, then use $0'$
to test their uniform $\Pi^0_1$ strata and search for each least one.
This computes $K$. Conversely $K\oplus0'$ decides spatial perfection
by testing the single stratum $K(|G|)$ (or stratum one if that value is
zero), and hence computes $0''$ by Theorem~A.

To remove the extra $0'$ oracle, use the halting clock games from the
proof of Corollary~\ref{cor:entropybeaver}. If machine $e$ halts at time
$T_e$, its least valid witness is $T_e$, so
\[
 T_e<p_{T_e+1}\le\frac{12}{121}\kappa_{\min}(\Gclk_e)
 \le\frac{12}{121}K(|\Gclk_e|)\le K(|\Gclk_e|).
\]
Simulating $K(|\Gclk_e|)$ steps therefore decides whether it halts:
if it has not halted by then, it never halts. Thus $0'\le_T K$, and
the preceding reductions give $K\equiv_T0''$.

Finally use the same computable array, padded games and nondecreasing
unbounded envelope $\ell$ as in Corollary~\ref{cor:entropybeaver}; padding
preserves the state and hence the Schmidt probabilities. Given total
$0'$-computable $F$, the limit lemma supplies $a_0$ for
$f(n)=2^{1+\max_{r\le \ell(n+1)}F(r)}$. Every witness for $G_{a_0,n}$
is at least $f(n)$. For $n\ge a_0$ and $\ell(n)\le m<\ell(n+1)$,
\[
 K(m)\ge\kappa_{\min}(G_{a_0,n})
 \ge\frac{121}{12}p_{f(n)+1}>f(n)>F(m).
\]
The intervals cover all sufficiently large integers, proving eventual
strict domination.
\end{proof}

\begin{remark}[Least entropy and the least witness]
For a YES clock game, let $d_{\max}(e)$ be the computable maximum row
ancilla dimension in its compiler. Then
\[
 H_{\rm ex}(\Gclk_e)=\log_2p_{j_*+1}+H(\mu_4)+S(h_F)+\delta_e,
 \qquad 0\le\delta_e\le\log_2(3d_{\max}(e)).
\]
Indeed the full-marginal identity and multiplicity bound in
Lemma~\ref{lem:clockentropy} give the lower bound
$S(h)\ge H(\mu_4)+S(h_F)+\log_2p_{j_*+1}$ for every perfect
normal strategy, including mixed strategies by purification onto Bob.
For the upper bound choose the least witness in the explicit pure
strategy of Theorem~\ref{thm:reduction}: its marginal is the tensor
product of $\operatorname{diag}(\mu_4)$, $h_F$,
$I_{p_{j_*+1}}/p_{j_*+1}$, $I_3/3$ and
$I_{d_{\max}(e)}/d_{\max}(e)$. Entropy additivity gives exactly the
same lower-bound expression plus $\log_2(3d_{\max}(e))$.
Taking the infimum proves the display. In particular the least entropy
determines $\log_2p_{j_*+1}$ up to a computable additive error.
\end{remark}

\subsection{A finite tracial corner}\label{sec:cornerupper}
The spatial upper bound fixes a nonzero Schmidt coefficient. For a general semifinite host, the corresponding parameter is a nonzero overlap with a finite tracial corner. The existential integer bounds this overlap, and the quantifier pattern is the same.

\begin{lemma}[Finite-corner overlap]\label{lem:overlap}
Let $\mathcal A,\mathcal B$ be commuting unital $C^*$-algebras containing a game's effects. Suppose unit vectors $\psi,\eta$ and $e=e^2=e^*\in\mathcal A$ satisfy: $\psi$ is perfect, $e\eta=\eta$, $\langle\eta,\psi\rangle\ne0$, and the vector state of $\eta$ is tracial on $e\mathcal A e$. Then the game has a perfect normal semifinite strategy. If $e=1$, it has a perfect normal strategy in a finite von Neumann algebra.
\end{lemma}
\begin{proof}
Let $\mathcal H_\eta=\overline{\operatorname{span}\mathcal A\mathcal B\eta}$. It reduces both algebras because they are self-adjoint and commute. Put $M=(\mathcal A|_{\mathcal H_\eta})''$ and $p=e|_{\mathcal H_\eta}$. The vector functional of $\eta$ is normal on $M$ and tracial on $pMp$: Kaplansky density approximates each corner element by a bounded net from $p\mathcal A p$, and separate ultraweak continuity extends $\tau_\eta(xy)=\tau_\eta(yx)$ one variable at a time. Moreover
\begin{equation}\label{eq:cornercyclic}
 p\mathcal H_\eta=\overline{\operatorname{span}\mathcal B(pMp)\eta}.
\end{equation}
Indeed $pab\eta=bpap\eta$, and bounded strong approximation supplies the reverse closure. If $x\geq0$ in $pMp$ has $\tau_\eta(x)=0$, then, for $a\in pMp$,
\[
 \|x^{1/2}a\eta\|^2=\tau_\eta(a^*xa)
 =\tau_\eta(x^{1/2}aa^*x^{1/2})\leq\|a\|^2\tau_\eta(x)=0.
\]
Since $x$ commutes with $\mathcal B$, \eqref{eq:cornercyclic} gives $x=0$. Thus $pMp$ has a faithful normal finite trace. If a central projection $z$ of $M$ annihilates $p$, it annihilates $\eta$, and hence $\mathcal A\mathcal B\eta$ and $\mathcal H_\eta$; so $z=0$. Thus $p$ has full central support. A type-III summand has no nonzero finite projection, so its intersection with $p$ is zero and fullness excludes that summand. Consequently $M$ is semifinite.

The orthogonal projection onto $\mathcal H_\eta$ commutes with both parties. $\zeta=P_{\mathcal H_\eta}\psi$ is nonzero by its overlap with $\eta$. Each losing positive product annihilates $\psi$ and hence $\zeta$. Normalize $\zeta$. Its state is normal on $M$ and Bob's compressed effects commute with $M$. If $e=1$, the faithful finite trace is on $M$.
\end{proof}

Conversely standardize a normal semifinite strategy and write $\psi=h^{1/2}\in L^2(M,\tau)$, with $\tau(h)=1$. Some band $e=1_{[a,b]}(h)$, $0<a<b<\infty$, is nonzero. It has $0<\tau(e)\leq1/a$. For $\eta=e/\sqrt{\tau(e)}$,
\begin{equation}\label{eq:overlap}
 \langle\eta,\psi\rangle
 =\frac{\tau(eh^{1/2}e)}{\sqrt{\tau(e)}}
 \geq\sqrt a\sqrt{\tau(e)}>0.
\end{equation}
Its corner state is tracial. In a finite von Neumann algebra take a faithful normalized trace and $e=1$; then $\tau(h^{1/2})>0$. Adjoin $e$ to Alice's algebra if necessary.

\begin{theorem}\label{thm:upper}
$\perf_{\finhost}$ and $\perf_{\sfhost}$ belong to $\Sigma^0_2$.
\end{theorem}
\begin{proof}
For $k\geq1$ seek moments $m_{ij}(w)=\langle\xi_i,w\xi_j\rangle$, where $\xi_0=\psi,\xi_1=\eta$. Letters are the self-adjoint effects and Alice's extra projection $e$. Impose positivity for all polynomial vector lists, localizing positivity for each effect and its complement, POVM normalization, Alice--Bob commutation \emph{including $e$}, $e^2=e$, $e\eta=\eta$, unit norms, zero losing expectations on $\psi$, and
\[
 \Rea\langle\eta,\psi\rangle\geq1/k,\qquad
 \langle\eta,eue\,eve\,\eta\rangle
 =\langle\eta,eve\,eue\,\eta\rangle
\]
for all Alice words $u,v$, including $e$. A vector equality means vanishing squared norm. These form a recursive list of closed rational polynomial constraints. Impose $|m_{ij}(w)|\leq1$ directly. All genuine contraction moments obey this bound, and the ambient product of closed disks is compact. Localizing positivity gives bounded GNS effects $0\leq E\leq1$. One can also use the certificate $E-E^2=(I-E)E(I-E)+E(I-E)E$.

At stage $m$ take the first $m$ constraints and their coordinates. Feasibility $F_{\sfhost}(G,k,m)$ is a non-strict rational semialgebraic question, decidable by real quantifier elimination \cite{BPR}. Its solution set is a closed cylinder in the compact product of all moment disks. The cylinders are nested because stages retain previous constraints. If every stage is feasible, their intersection is nonempty: otherwise their open complements have a finite subcover, already making one stage empty. A point of the intersection gives a positive matrix-valued functional. Quotient polynomial pairs by its null space and complete. Left multiplication yields the bounded GNS effects and the two vectors. Relations hold on the polynomial dense space and extend by boundedness. Corner traciality holds on the linear span of $eue$ and then in norm on $e\mathcal A e$. Apply Lemma~\ref{lem:overlap}. Conversely \eqref{eq:overlap} supplies some $k$ for every semifinite strategy. Thus
\begin{equation}\label{eq:sigma2sf}
 G\in\perf_{\sfhost}\iff\exists k\,\forall m\,F_{\sfhost}(G,k,m).
\end{equation}
Set $e=1$ for finite von Neumann algebras.
\end{proof}

Tensoring a finite factor host with $R_{\rm hyp}$ gives a $\II_1$ host. Tensoring a semifinite factor host with $R_{\rm hyp}\vN B(\ell^2)$ gives a $\II_\infty$ host. Product-state expectations, or standard-form tensoring, preserve the probabilities. Factor decomposition as in Section~\ref{sec:prelim} therefore gives
\begin{equation}\label{eq:languageequal}
 \perf_{\finhost}=\perf(\II_1),\qquad
 \perf_{\sfhost}=\perf(\II)=\perf(\II_\infty).
\end{equation}
These are equalities of existence languages.

\begin{remark}[One parameter for both corners]
In a type I factor $B(\HH)$ with its trace, the state vector is $\psi=h^{1/2}$. If $e$ is a rank-one subprojection of the spectral projection of $h$ for its largest eigenvalue, then $\langle e,\psi\rangle=\Tr(eh^{1/2})=s_1$ is the largest Schmidt amplitude. This rank-one choice recovers the largest Schmidt amplitude from the corner overlap. General finite-corner overlaps need not equal it, and the parameters $\kappa$ and $k$ need not agree. Both upper bounds put one existential quantifier on that amount and one universal quantifier on finite stages. Without a trace there is no such corner. Commuting perfection is in $\Pi^0_1$ by the convergent hierarchy of Navascu\'es, Pironio and Ac\'in \cite{NPA}: a NO instance eventually has an infeasible perfect-value relaxation, and feasibility at every order gives a commuting strategy by compactness. Section~\ref{sec:typeIII} shows that every type-III subtype has exactly this commuting language.
\end{remark}

\section{A finite compiler for normal strategies}\label{sec:compiler}
\subsection*{From a presentation to a game}
\addcontentsline{toc}{subsection}{\protect\numberline{}From a presentation to a game}
The spatial, semifinite and communication hardness proofs compile operator relations into one finite game. This section builds that compiler once, for normal spatial and for normal commuting strategies; the finite predicates are the same in both models. The proofs follow in Sections~\ref{sec:bridge}--\ref{sec:approxcompiler}.

\emph{Input.} An admissible presentation (Section~\ref{sec:admissible}) is a finite rational list: four contraction symbols $A_\alpha$, a root projection $P_\circ$, finitely many typed centralizer labels and finitely many rational $*$-polynomial equations. Its intended solutions are normal realizations: bounded operators on a Hilbert space $\HH_0$ with a faithful trace-class density $h_0$, such that the labels commute with $h_0$, the equations hold, and $h_0A_\alpha=A_\alpha h_0/16$.

\emph{Output.} A finite game with a rational question distribution and a $0/1$ predicate. Its perfect normal spatial strategies and the normal realizations of the presentation correspond in both directions (Lemma~\ref{lem:four}).

\emph{Construction.}
\begin{enumerate}
\item \emph{Labels become unitaries.} A projection $P$ is encoded by $2P-I$, a positive contraction $T_+$ by a unitary $u_+$ with $T_+=(2I+u_++u_+^*)/4$, and a general contraction by an average of two unitaries.
\item \emph{A scaffold carries the factor $16$.} A synchronized event commutes with the marginal density, so it cannot change a density eigenvalue. Conditional flagged magic-square bridges, whose sign questions are not synchronized, supply that change: seventeen synchronized corners and four such bridges certify partial isometries ${\btrans}_g$ with $h{\btrans}_g=2{\btrans}_gh$ (Lemma~\ref{lem:bridge}). Their product $v$ maps the source corner $e_0$ onto $e_4$ with $hv=16vh$, and the cover fixes the source mass $1/121$.
\item \emph{Labels travel through the bridges.} Corner copies, block unitaries and shared-question refinements (Lemma~\ref{lem:label}) force $L_4v=vL_0$ for every transported label $L$.
\item \emph{The modular operators disappear.} Each creation symbol is written $A_\alpha=v^*a_\alpha$ with a centralizer contraction $a_\alpha$ from $e_0$ to $e_4$. Every application equation becomes an equation among centralizer labels in which $v$ does not occur, so every word passed to the graph compiler lies in the density centralizer.
\item \emph{Words become projections.} A qutrit frame, graph projections and triangle projections certify actual products of unitary labels. A rational equation among words becomes a rational linear equation among synchronized projections (Section~\ref{sec:projection}).
\item \emph{Linear equations become Boolean tests.} Each such equation is enforced by finitely many orthogonality tests against a certified scalar matrix algebra (Lemma~\ref{lem:rows}).
\item \emph{Assembly.} All questions are kept, the question sets are padded, and question pairs are sampled uniformly; unspecified pairs are accepted.
\end{enumerate}

\emph{Soundness.} In a perfect normal strategy every losing event of positive weight has probability zero. This makes each synchronized event a projection on the state support that commutes with the marginal density, and it turns each imposed equation into an operator identity on that support. The bridges then give the realization on the source corner, with density $h_0=121e_0he_0$.

\emph{Completeness.} A realization is copied into the scaffold density \eqref{eq:scaffold}. Contractions are written as averages of unitaries inside the finite-dimensional positive eigenspaces of that density. A qutrit frame and a finite maximally entangled row ancilla are appended, and Bob uses transposes in an eigenbasis of the density. Every test then holds exactly. Only finite-dimensional factors are added, so finite entropy is preserved.

\emph{Size and approximation.} Section~\ref{sec:approxcompiler} bounds the loss when the application equations hold only approximately. Section~\ref{sec:reduction} bounds the size of the reduction game and counts the questions of one full instance. Appendix~\ref{sec:repro} describes a small compiled instance with four bridges, one modular isometry and no transported labels.

\subsection*{Normal states, supports and exact cancellation}
\addcontentsline{toc}{subsection}{\protect\numberline{}Normal states, supports and exact cancellation}
We first restate the bridge, label and rational-row tools from \cite[Lemmas~4.1 and 4.2, Theorem~B, Eq.~(32)]{Entropy}, with their proofs, and extend them from spatial strategies to normal commuting strategies, where separation in standard form replaces Schmidt-range cancellation. These arguments concern arbitrary normal supports and finite word lists. We then give the $n$-bridge assembly.

The proofs use only the following facts, which hold on arbitrary, possibly nonseparable, Hilbert spaces. When a marginal density has infinite rank, its inverse on its range is unbounded. Cancellations therefore go through the dense range of the marginal: by (N3), an operator that kills the state vector vanishes on the whole support.

\paragraph{(N1) Normal states and purifications.}
A normal state on $\HH_A\otimes\HH_B$ is a positive trace-class operator $\rho$ of trace one. By the spectral theorem for compact operators, $\rho=\sum_j\lambda_j|\psi_j\rangle\langle\psi_j|$ with countably many $\lambda_j>0$. Let $J$ be the countable index set, put $\HH_{\rm env}=\ell^2(J)$, and put $\Omega=\sum_j\lambda_j^{1/2}\psi_j\otimes|j\rangle$. Then
\[
 \Tr\rho(E\otimes F)=\langle\Omega,(E\otimes F\otimes I_{\rm env})\Omega\rangle
\]
for all bounded $E,F$, and no measurement acts on $\HH_{\rm env}$. The marginals $h_A=\Tr_B\rho$ and $h_B=\Tr_A\rho$ are positive trace-class operators of trace one. Write $P_A,P_B$ for their support projections, the projections onto the closures of their ranges; these closures are separable. The restriction of $h_A$ to $P_A\HH_A$ is injective with dense range. Its positive eigenspaces are finite-dimensional and span a dense subspace of $P_A\HH_A$. The same holds for $h_B$.

\paragraph{(N2) Coefficient operators.}
Put $\HH_{B,{\rm env}}=\HH_B\otimes\HH_{\rm env}$. Fix orthonormal bases $(b_j)$ of $\HH_B$ and $(e_\ell)$ of
$\HH_{\rm env}$, and use their tensor-product basis
$(f_{j\ell})=(b_j\otimes e_\ell)$ of $\HH_{B,{\rm env}}$, and let $Y^{\mathsf T}=CY^*C$ be the transpose of $Y\in B(\HH_{B,{\rm env}})$, where $C$ is the complex conjugation fixing every $f_{j\ell}$. Bob's transpose $F^{\mathsf T}$ is defined in the basis $(b_j)$. Transposition preserves norms, adjoints, positivity and projections, reverses products, and satisfies $(F\otimes I_{\rm env})^{\mathsf T}=F^{\mathsf T}\otimes I_{\rm env}$. The map $T\mapsto\sum_{j,\ell}Tf_{j\ell}\otimes f_{j\ell}$ is a unitary from the Hilbert--Schmidt operators $\HH_{B,{\rm env}}\to\HH_A$ onto $\HH_A\otimes\HH_{B,{\rm env}}$. If it sends $T_\Omega$ to $\Omega$, it sends $XT_\Omega Y^{\mathsf T}$ to $(X\otimes Y)\Omega$, and $T_\Omega T_\Omega^*=h_A$. We call $T_\Omega$ the coefficient operator. Identities such as $ET=TF^{\mathsf T}$ below are coefficient forms of vector identities, and we state both forms where it helps. For a mixed state this is the cut $A\mid B\,{\rm env}$, and every Bob operator enters as $F\otimes I_{\rm env}$.

\paragraph{(N3) Exact cancellation.}
For $X\in B(\HH_A)$,
\[
 \|(X\otimes I)\Omega\|^2=\Tr(h_AX^*X)=\|Xh_A^{1/2}\|_{\rm HS}^2 .
\]
Hence $(X\otimes I)\Omega=0$, equivalently $XT_\Omega=0$, holds exactly when $XP_A=0$. Indeed the range of $h_A^{1/2}$ is dense in $P_A\HH_A$, and a bounded operator that vanishes on a dense subspace vanishes on its closure. Likewise $(I\otimes F\otimes I_{\rm env})\Omega=0$, equivalently $T_\Omega(F\otimes I_{\rm env})^{\mathsf T}=0$, holds exactly when $FP_B=0$, because its squared norm is $\Tr(h_BF^*F)$. These are the dense-range cancellations used throughout. The same statements hold for any vector in place of $\Omega$, with the supports of its own marginals.

\paragraph{(N4) Mirrors.}
Call $X\in B(\HH_A)$ and $Y\in B(\HH_B)$ \emph{mirrors} if $(X\otimes I)\Omega=(I\otimes Y\otimes I_{\rm env})\Omega$, that is, $XT_\Omega=T_\Omega(Y\otimes I_{\rm env})^{\mathsf T}$.
\begin{enumerate}[label=(\alph*)]
\item If $X_k$ and $Y_k$ are mirrors for $k=1,\ldots,n$, then $X_1\cdots X_n$ and $Y_n\cdots Y_1$ are mirrors. Operators on different tensor factors commute, so $(X_1X_2\otimes I)\Omega=(X_1\otimes Y_2\otimes I)\Omega=(I\otimes Y_2Y_1\otimes I)\Omega$; induct. Linear combinations of mirror pairs are mirror pairs.
\item If $X$ and $X'$ are both mirrors of $Y$, then $(X-X')P_A=0$ by (N3). If $Y$ and $Y'$ are both mirrors of $X$, then $(Y-Y')P_B=0$.
\item Suppose $X,Y$ are mirrors and $X^*,Y^*$ are mirrors. This holds for self-adjoint mirrors, and for unitary mirrors because applying $X^*\otimes Y^*$ to their mirror identity gives $(I\otimes Y^*\otimes I)\Omega=(X^*\otimes I)\Omega$. Then $Xh_A=h_AX$ and $Yh_B=h_BY$. For $Z\in B(\HH_A)$,
\[
 \Tr(h_AZX)=\langle(I\otimes Y^*\otimes I)\Omega,(Z\otimes I)\Omega\rangle
 =\langle(X^*\otimes I)\Omega,(Z\otimes I)\Omega\rangle=\Tr(h_AXZ).
\]
Cyclicity of the trace, valid because $h_A$ is trace class and $X,Z$ are bounded, gives $\Tr((Xh_A-h_AX)Z)=0$. Taking $Z$ of rank one shows that the trace-class operator $Xh_A-h_AX$ is zero. Bob's statement is symmetric.
\item An operator commuting with $h_A$ commutes with $P_A$ and with every spectral projection of $h_A$. In particular it maps each finite-dimensional positive eigenspace of $h_A$ into itself.
\end{enumerate}

\paragraph{(N5) Perfect matching.}
Let $E$ be an Alice effect and $F$ a Bob effect whose two disagreement events $E\otimes(I-F)$ and $(I-E)\otimes F$ have probability zero. By the matching identity, the sum of these two probabilities is the expectation of
\[
 (E\otimes I-I\otimes F)^2+(E-E^2)\otimes I+I\otimes(F-F^2),
\]
a sum of three positive operators. Each expectation vanishes, so
\[
 (E\otimes I)\Omega=(I\otimes F\otimes I)\Omega,\qquad
 ((E-E^2)^{1/2}\otimes I)\Omega=0,\qquad
 (I\otimes(F-F^2)^{1/2}\otimes I)\Omega=0 .
\]
Thus $E$ and $F$ are mirrors. Next, $E$ reduces $P_A$. Since $((I-P_A)\otimes I)\Omega=0$,
\[
 ((I-P_A)E\otimes I)\Omega=(I\otimes F\otimes I)((I-P_A)\otimes I)\Omega=0,
\]
so $(I-P_A)EP_A=0$ by (N3), and self-adjointness gives $EP_A=P_AEP_A=P_AE$. Then (N3) gives $(E-E^2)P_A=0$, so $EP_A$ is a projection, and $E$ commutes with $h_A$ by (N4c). The same holds for $F$ on Bob's side. We call $EP_A$ a \emph{synchronized projection} and mean this support-restricted operator in all local identities. Reduction of the support is proved before projectivity is used.

\paragraph{(N6) Compressions and joint measurements.}
\begin{enumerate}[label=(\alph*)]
\item If $0\leq Y\leq Z$ with $Z$ a projection, then $Y=ZYZ$. Indeed $0\leq(I-Z)Y(I-Z)\leq(I-Z)Z(I-Z)=0$, so $Y^{1/2}(I-Z)=0$.
\item \emph{Projective compression.} If $P$ is a projection, $E$ an effect and $PEP$ a projection, then $E$ reduces $P\HH$. For $u\in\ran(PEP)$, $\langle u,(I-E)u\rangle=0$ and positivity give $Eu=u$. For $u\in P\HH$ with $PEPu=0$, $\langle u,Eu\rangle=0$ gives $Eu=0$. Thus $E$ maps $P\HH$ into itself, and self-adjointness gives reduction.
\item \emph{Only compressions matter.} Since $(P_A\otimes I)\Omega=\Omega=(I\otimes P_B\otimes I)\Omega$, every probability equals $\langle\Omega,(P_AEP_A\otimes P_BFP_B\otimes I)\Omega\rangle$. Identities may therefore be proved first for compressed effects. Changing effects outside the supports, while keeping them POVMs, changes no probability.
\item \emph{Sharp marginals.} Let $(E_c)$ be a finite POVM indexed by tuples $c=(c_1,\ldots,c_m)$ whose compressions to a reducing subspace $P\HH$ have projections as coarse marginals $Q^k_s=\sum_{c:\,c_k=s}E_c$. By (a), each compressed $E_c$ satisfies $E_c=Q^k_{c_k}E_cQ^k_{c_k}$, hence $E_cQ^k_s=\delta_{s,c_k}E_c$. Summing, $Q^1_{s_1}Q^2_{s_2}=\sum_{c_1=s_1,\,c_2=s_2}E_c$, which is self-adjoint, so the marginals commute. Iterating, $Q^1_{s_1}\cdots Q^m_{s_m}$ is the compressed effect of the outcome $(s_1,\ldots,s_m)$, or zero if there is no such outcome. Each compressed fine effect is thus a product of commuting projections, hence a projection, and (b) makes the original fine effect reduce $P\HH$.
\end{enumerate}

\paragraph{(N7) Forbidden events.}
Let $Q$ be an Alice effect and $R$ a Bob effect whose compression $R_B=P_BRP_B$ is a synchronized projection with Alice mirror $R_A$. If the event $Q\otimes R$ has probability zero, then $QR_AP_A=0$. If $Q$ also reduces $P_A$, then $R_AQP_A=0$ as well. Indeed $0=\langle\Omega,(Q\otimes R_B\otimes I)\Omega\rangle=\|(Q^{1/2}\otimes R_B\otimes I)\Omega\|^2$, and $(Q^{1/2}\otimes R_B\otimes I)\Omega=(Q^{1/2}R_A\otimes I)\Omega$, so (N3) gives $Q^{1/2}R_AP_A=0$. The second claim is the adjoint of the first, because $P_A$ commutes with $Q$ and $R_A$.

\paragraph{(N8) Canonical purifications for completeness.}
Let $h$ be a faithful positive trace-class operator of trace one on $\HH$. By (N1), $\HH$ is separable and has an orthonormal basis $(\varphi_i)$ with $h\varphi_i=\lambda_i\varphi_i$ and $\lambda_i>0$. Put $\Omega_h=\sum_i\lambda_i^{1/2}\varphi_i\otimes\varphi_i$, and let $X^{\mathsf T}$ denote the transpose in $(\varphi_i)$. Both marginals of $\Omega_h$ equal $h$. If $X$ commutes with $h$, it commutes with $h^{1/2}$, and the rule of (N2) gives
\[
 (I\otimes X^{\mathsf T})\Omega_h=(X\otimes I)\Omega_h,
\]
since both vectors correspond to $h^{1/2}X=Xh^{1/2}$. The map $X\mapsto X^{\mathsf T}$ is linear, reverses products, and preserves adjoints, positivity and projections; and $X^{\mathsf T}$ commutes with $h^{\mathsf T}=h$. Hence a centralizer projection $X$ for Alice and $X^{\mathsf T}$ for Bob are perfectly matched: their disagreement probability is $\langle\Omega_h,((X(I-X)+(I-X)X)\otimes I)\Omega_h\rangle=0$. For centralizer effects $Q$ of Alice and $R^{\mathsf T}$ of Bob, the joint probability is $\Tr(hQR)$, so an event with $QR=0$ has probability zero.

\subsection{The normal flagged magic-square bridge}\label{sec:bridge}
The magic-square relations originate with Mermin and Peres \cite{Mermin,Peres}. We use Man\v{c}inska's flagged magic-square (FMS) geometry \cite{Mancinska}. We spell out the predicate and prove its normal extension because the published finite-dimensional statement does not suffice here. All FMS question and answer indices start at zero; frame indices below start at one.

\subsubsection{Questions, calibration and exact bindings}
Alice has $\mathsf R_0,\mathsf R_1,\mathsf R_2,\mathsf T_0,\mathsf T_1$ and Bob has $\mathsf C_0,\mathsf C_1,\mathsf C_2,\mathsf B_0,\mathsf B_1$. A row question has a flag answer $f$ or a triple of product $+1$. A column question has $f$ or a triple of product $+1,+1,-1$ according to its column. The magic square is
\begin{equation}\label{eq:square}
\begin{pmatrix}XI&IX&XX\\ IZ&ZI&ZZ\\ XZ&ZX&YY\end{pmatrix}.
\end{equation}
Here $X,Y,Z$ denote the Pauli matrices, $I$ the $2\times2$ identity, and adjacent letters their tensor product. The displayed matrices describe a canonical realization; the predicate only checks the corresponding signs. Select column-zero outcomes $c_0=(+++)$, $c_1=(+--)$ and row-zero outcomes $r_0=(+++)$, $r_1=(+--)$. Question $\mathsf T_x$ has answers $+,-$ and the three column-zero triples other than $c_x$; $\mathsf B_y$ analogously retains all row-zero triples other than $r_y$.

\begin{figure}[ht]
\centering
\setlength{\unitlength}{1mm}
\begin{picture}(114,35)
\put(0,15){\framebox(22,14){flag $f$}}
\put(23,22){\vector(1,0){20}}
\put(24,25){${\btrans}$}
\put(44,0){\framebox(68,34){}}
\put(47,27){column-zero magic directions}
\put(47,18){$m_{00}=|c_0\rangle\langle c_0|\quad m_{01}=|c_1\rangle\langle c_1|$}
\put(47,10){$m_{10},\ m_{11}:\ \text{other two directions}$}
\put(47,3){$r_y=(c_0+(-1)^yc_1)/\sqrt2$}
\end{picture}
\caption{Selected column and row directions. The flag transports to $c_0$;
$m_{ab}$ denote projections.}\label{fig:fms}
\end{figure}
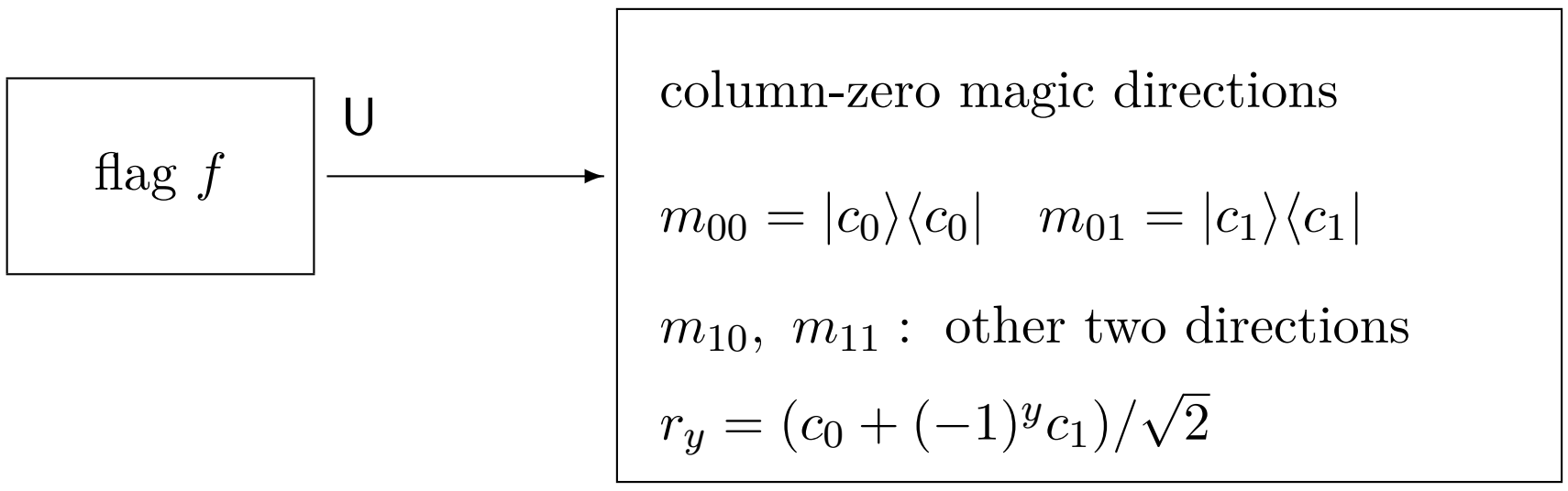

Table~\ref{tab:fms} and Figure~\ref{fig:fms} summarize the rules. On $\mathsf R_i,\mathsf C_j$, reject flag/nonflag mismatch and otherwise check the intersecting cell; two flags are accepted. On $\mathsf T_x,\mathsf C_0$, a sign is allowed only against $f$ or $c_x$; a retained triple must equal the opposite triple. On $\mathsf R_0,\mathsf B_y$ use the analogous rule with $r_y$. On $\mathsf T_x,\mathsf B_y$, reject two signs $a,b\in\{+1,-1\}$ unless $ab=(-1)^{xy}$; all pairs with a nonsign answer are accepted. Other question pairs are accepted. For a conditional instance add an inactive answer $\perp$, accept two inactive answers, and reject exactly one inactive answer. Bind each noninactive event to its synchronized active projection and all row/column flags to the synchronized flag.

Let $P_{\rm mag}$ be the nonflag magic support. Its four column-zero projections $m_{00},m_{01},m_{10},m_{11}$ correspond respectively to $(+++),(+--),(-+-),(--+)$. On Alice, $m_{00}$ is read from the retained $(+++)$ answer of $\mathsf T_1$ and the other three from $\mathsf T_0$. On Bob all four come from $\mathsf C_0$. Each of these chosen occurrences is matched to its opposite synchronized binary question. The positive nonflag $ZI$ event is read from Alice's $\mathsf R_1$ component 1 and Bob's $\mathsf C_1$ component 1. The positive $IX$ event comes from Alice's $\mathsf R_0$ component 1 and Bob's $\mathsf C_1$ component 0. Bind both positive nonflag events to synchronized binary questions, denoted $E_{ZI,+}$ and $E_{IX,+}$. Put
\[
 \mathfrak x=I-2P_{\rm mag}+2E_{ZI,+},\quad \mathfrak z=I-2P_{\rm mag}+2E_{IX,+},\quad
 t_{ab}=\mathfrak x^a\mathfrak z^bm_{00}.
\]
These are reflections extended by identity outside $P_{\rm mag}$, and $t_{ab}^*t_{ab}=m_{00}$, $t_{ab}t_{ab}^*=m_{ab}$.

\begin{table}[ht]
\centering\small
\begin{tabular}{@{}p{.21\linewidth}p{.26\linewidth}p{.44\linewidth}@{}}
\toprule Questions & Answers & Tested rule\\\midrule
$\mathsf R_i,\mathsf C_j$ & flag or parity triple & flags match; nonflags agree at $(i,j)$\\
$\mathsf T_x,\mathsf C_0$ & sign or retained triple & signs pair only with $f,c_x$; retained triples match\\
$\mathsf R_0,\mathsf B_y$ & triple or sign & symmetric calibration at $r_y$\\
$\mathsf T_x,\mathsf B_y$ & signs or triples & two signs satisfy $ab=(-1)^{xy}$\\
Conditional copies & also inactive $\perp$ & activity matches; flags and events bind to named projections\\
\bottomrule
\end{tabular}
\caption{Flagged magic-square questions, answers and rules. Unspecified pairs are accepted.}\label{tab:fms}
\end{table}

\Needspace{5\baselineskip}
\subsubsection{Normal bridge theorem}
\begin{lemma}\label{lem:bridge}
A perfect normal conditional FMS strategy has on its active supports the decomposition
\[
 \HH_{\rm active}=\HH_f\oplus(\mathbb C^4\otimes\mathcal H_{\rm mag}),
 \qquad h=h_f\oplus(I_4\otimes h_{\rm mag}).
\]
The calibrated transports ${\btrans}_A,{\btrans}_B$ from the flag to the selected $c_0$ corner satisfy ${\btrans}^*{\btrans}=f$, ${\btrans}{\btrans}^*=m_{00}$ and $h{\btrans}=2{\btrans}h$. On a purification their matched components satisfy
\begin{equation}\label{eq:bridgevector}
 \Omega_{c_0c_0}=\sqrt2({\btrans}_A\otimes {\btrans}_B\otimes I_{\rm env})\Omega_{ff}.
\end{equation}
The flag answer and its projection are both denoted $f$. The other three magic directions obey the transported versions of this equation.
\end{lemma}
The proof first extracts the magic matrix algebra from the synchronized cells. Calibration then identifies the two binary sign supports. The four sign equations force an off-diagonal unitary between the flag and a magic corner, and commutation with the calibrated marginal gives the density ratio two.

\begin{proof}
Fix a purification $\Omega$ as in (N1). The predicate makes the following pairs of events perfectly matched in the sense of (N5): an inactive answer on one side and the inactive answer on the other; each noninactive event and the opposite synchronized activity projection; each row or column flag and the opposite synchronized flag; and, for every cell $(i,j)$ and sign $s$, Alice's event ``$\mathsf R_i$ is not a flag and its entry $j$ equals $s$'' and Bob's event ``$\mathsf C_j$ is not a flag and its entry $i$ equals $s$''. The last pair is matched because the rule on $\mathsf R_i,\mathsf C_j$ rejects a flag against a nonflag and checks the intersecting cell otherwise. By (N5), each coarse effect in this list reduces its party's support, its compression is a projection commuting with that party's marginal density, and matched compressions are mirrors. Let $a_A$ and $f_A$ be Alice's activity and flag projections, put $N_A=a_A-f_A$, and let $Q^s_{ij}$ be the compression of the cell event above, so that $Q^+_{ij}+Q^-_{ij}=N_A$. The cell observables $\mathsf M^A_{ij}=Q^+_{ij}-Q^-_{ij}$ are self-adjoint unitaries on $N_AP_A\HH_A$. Each is a mirror of Bob's operator $\mathsf M^B_{ij}$, built in the same way from $\mathsf C_j$.

For a nonflag answer $t=(t_0,t_1,t_2)$ of $\mathsf R_i$, the compressed effect lies below $Q^{t_j}_{ij}$ for each $j$, since the other nonflag effects of $\mathsf R_i$ with entry $j$ equal to $t_j$ are positive. By (N6d) it equals $Q^{t_0}_{i0}Q^{t_1}_{i1}Q^{t_2}_{i2}$, the three cell observables of row $i$ commute, and the original fine effect reduces $P_A$. Summing over the triples of product $+1$ gives $\mathsf M^A_{i0}\mathsf M^A_{i1}\mathsf M^A_{i2}=N_A$ on $P_A\HH_A$. The same argument for Bob's columns gives commuting $\mathsf M^B_{0j},\mathsf M^B_{1j},\mathsf M^B_{2j}$ with product $\epsilon_jN_B$, where $\epsilon_0=\epsilon_1=1$ and $\epsilon_2=-1$. By (N4a), $\mathsf M^A_{0j}\mathsf M^A_{1j}\mathsf M^A_{2j}$ is a mirror of $\mathsf M^B_{2j}\mathsf M^B_{1j}\mathsf M^B_{0j}=\epsilon_jN_B$, which is a mirror of $\epsilon_jN_A$. Then (N4b) gives $\mathsf M^A_{0j}\mathsf M^A_{1j}\mathsf M^A_{2j}=\epsilon_jN_A$ on $P_A\HH_A$. In the same way $\mathsf M^A_{ij}\mathsf M^A_{i'j}$ and $\mathsf M^A_{i'j}\mathsf M^A_{ij}$ are mirrors of the equal operators $\mathsf M^B_{i'j}\mathsf M^B_{ij}$ and $\mathsf M^B_{ij}\mathsf M^B_{i'j}$, so they agree on $P_A\HH_A$. The nine cell observables therefore satisfy every row and column relation of \eqref{eq:square} as operator identities on $N_AP_A\HH_A$. The symmetric argument gives the same for Bob on $N_BP_B\HH_B$.

Next comes the Pauli-pair calculation. Write $a=\mathsf M^A_{00},b=\mathsf M^A_{01},c=\mathsf M^A_{10},d=\mathsf M^A_{11}$ for the four top-left cell observables of the square. Context relations give
\[
 [a,b]=[c,d]=[a,c]=[b,d]=0,\qquad
 (ac)(bd)=-(ab)(cd).
\]
Cancel the outside $a,d$ to obtain $cb=-bc$. Commutation of $ac$ and $bd$ then gives $da=-ad$. Thus $(a,d)$ and $(b,c)$ are anticommuting pairs, and each commutes with the other pair. Put $\mathcal N_A=N_AP_A\HH_A$, and let $\mathcal N_\pm$ be the $\pm1$ eigenspaces of $a$, so that $\mathcal N_A=\mathcal N_+\oplus\mathcal N_-$. Since $da=-ad$, $d$ maps $\mathcal N_\pm$ into $\mathcal N_\mp$, and $d^2=I$ makes $d|_{\mathcal N_+}$ unitary onto $\mathcal N_-$. This identifies $\mathcal N_A$ with $\mathbb C^2\otimes\mathcal N_+$, with $a=Z\otimes I$ and $d=X\otimes I$. The operators $b,c$ commute with $a$ and $d$, hence with the algebra $\Mat_2\otimes I$ they generate, so $b=I\otimes b'$ and $c=I\otimes c'$ for anticommuting self-adjoint unitaries $b',c'$. Repeating the argument for $c',b'$ gives $\mathcal N_+=\mathbb C^2\otimes\mathcal H_{\rm mag}$. The space $\mathcal H_{\rm mag}$ may be infinite-dimensional; it is separable because it lies in $P_A\HH_A$.

For later use we record the identification through the units $t_{ab}$. The projection $m_{00}$ is the joint $(+1,+1)$ eigenprojection of the column-zero cells $a=\mathsf M_{00}$ and $c=\mathsf M_{10}$, and $m_{ab}$ is their joint $((-1)^a,(-1)^b)$ eigenprojection; this is the stated correspondence with $(+++),(+--),(-+-),(--+)$. On $\mathcal N_A$ the reflection $\mathfrak x$ equals $d=\mathsf M_{11}$, which anticommutes with $a$ and commutes with $c$, and $\mathfrak z$ equals $b=\mathsf M_{01}$, which commutes with $a$ and anticommutes with $c$. Hence $t_{ab}=\mathfrak x^a\mathfrak z^bm_{00}$ maps $m_{00}\mathcal N_A$ isometrically onto $m_{ab}\mathcal N_A$, and
\[
 U_A:\mathbb C^4\otimes\mathcal H^A_{\rm mag}\to\mathcal N_A,\qquad
 U_A(u_{ab}\otimes\xi)=t_{ab}\xi,\qquad \mathcal H^A_{\rm mag}=m_{00}\mathcal N_A,
\]
is unitary. With $u_{ab}=|a\rangle\otimes|b\rangle$, these coordinates give $m_{ab}=|u_{ab}\rangle\langle u_{ab}|\otimes I$, $a=Z\otimes I$, $c=I\otimes Z$, $d=X\otimes I$ and $b=I\otimes X$. The operators $\mathfrak x,\mathfrak z,m_{00}$ are built from synchronized projections, so they commute with $h_A$ by (N5), and so does each $t_{ab}$. Therefore $U_A^*h_AU_A=I_4\otimes h^A_{\rm mag}$, where $h^A_{\rm mag}$ is the restriction of $h_A$ to $\mathcal H^A_{\rm mag}$. Together with $[f_A,h_A]=0$ this is the decomposition in the lemma, with $\mathcal H_f=f_AP_A\HH_A$. Both blocks are faithful, because $h_A$ is faithful on $P_A\HH_A$ and they are its restrictions to reducing subspaces. The same construction on Bob's side gives $U_B$ and $h^B_{\rm mag}$.

Now the coefficient blocks. Flags and activity are matched, so $(f_A\otimes I)\Omega=(I\otimes f_B\otimes I)\Omega$ and $(N_A\otimes I)\Omega=(I\otimes N_B\otimes I)\Omega$. Put $\Omega_{ff}=(f_A\otimes f_B\otimes I)\Omega$ and $\Omega_{NN}=(N_A\otimes N_B\otimes I)\Omega$. The mixed components vanish; for example $(f_A\otimes N_B\otimes I)\Omega=(f_AN_A\otimes I)\Omega=0$. Each of $\mathfrak x,\mathfrak z,m_{00}$ is a linear combination of matched events, hence a mirror of its Bob counterpart by (N4a); for $\mathfrak x$ the events are the activity, the flag and the positive $ZI$ event, which the rule on $\mathsf R_1,\mathsf C_1$ matches. The reflections are self-adjoint, so (N4a) makes $t^A_{ab}$ a mirror of $(t^B_{ab})^*$, and the matrix unit $t^A_{ab}(t^A_{a'b'})^*$ a mirror of $t^B_{a'b'}(t^B_{ab})^*$. In the coordinates $U_A\otimes U_B$ write $\Omega_{NN}=\sum_{k,l}u_k\otimes u_l\otimes\omega_{kl}$, where $k,l$ range over the four labels $ab$ and $\omega_{kl}\in\mathcal H^A_{\rm mag}\otimes\mathcal H^B_{\rm mag}\otimes\HH_{\rm env}$. The mirror identity for the units $|u_k\rangle\langle u_l|$ and $|u_l\rangle\langle u_k|$ reads
\[
 \sum_{l'}u_k\otimes u_{l'}\otimes\omega_{ll'}=\sum_{k'}u_{k'}\otimes u_l\otimes\omega_{k'k},
\]
so $\omega_{ll'}=0$ for $l'\ne l$ and $\omega_{ll}=\omega_{kk}$. Hence
\[
 \Omega_{NN}=\sum_ku_k\otimes u_k\otimes\omega_{\rm mag}
\]
for a single vector $\omega_{\rm mag}$, whose two marginals are $h^A_{\rm mag}$ and $h^B_{\rm mag}$. In coefficient form, the coefficient operator of $\Omega_{NN}$ is $I_4\otimes T_{\rm mag}$, where $T_{\rm mag}$ is that of $\omega_{\rm mag}$, and $T_f$ is that of $\Omega_{ff}$. Here $T_{\rm mag}:\mathcal H^B_{\rm mag}\otimes\HH_{\rm env}
\to\mathcal H^A_{\rm mag}$ and
$T_f:\mathcal H^B_f\otimes\HH_{\rm env}\to\mathcal H^A_f$
are the coefficient maps in the fixed product coordinates.
They satisfy $T_{\rm mag}T_{\rm mag}^*=h^A_{\rm mag}$ and $T_fT_f^*=h^A_f$, and have dense range. All cancellations below use (N3) on the corresponding original local support.

Now the sign questions. For a column-zero triple $c$, let $\Pi_c$ be the joint spectral projection of Alice's three cell reflections in column zero, and $\Pi_r$ the joint projection of Bob's row-zero cells; put $\epsilon_{xy}=(-1)^{xy}$. The rule on $\mathsf T_x,\mathsf C_0$ matches Alice's coarse sign event of $\mathsf T_x$ with Bob's event $\{f,c_x\}$ of $\mathsf C_0$, and each retained triple of $\mathsf T_x$ with the same triple of $\mathsf C_0$. Bob's $\mathsf C_0$ projection for a triple is the joint projection of his column-zero cells, and by (N4a) its mirror is Alice's joint projection $\Pi_c$ of her cells $\mathsf M^A_{00},\mathsf M^A_{10},\mathsf M^A_{20}$. Thus (N4b) gives $S^A_x=f_A+\Pi_{c_x}$ on $P_A\HH_A$, and the retained effects of $\mathsf T_x$ are the other column-zero projections. Symmetrically, the rule on $\mathsf R_0,\mathsf B_y$ gives $S^B_y=f_B+\Pi_{r_y}$, where $\Pi_{r_y}$ is Bob's joint projection of his row-zero cells. In the coordinates above, $c_0=u_{00}$ and $c_1=u_{01}$, while $r_y=(u_{00}+(-1)^yu_{01})/\sqrt2$, because row zero consists of $a=Z\otimes I$, $b=I\otimes X$ and $ab$. Hence $\langle c_x,r_y\rangle=\epsilon_{xy}/\sqrt2$, and
\[
 \Omega_{xy}:=(S^A_x\otimes S^B_y\otimes I)\Omega
 =\Omega_{ff}+\frac{\epsilon_{xy}}{\sqrt2}\,c_x\otimes r_y\otimes\omega_{\rm mag}.
\]
The maps $V^A_x(\xi\oplus\eta)=\xi+c_x\otimes\eta$ and $V^B_y(\xi'\oplus\eta')=\xi'+r_y\otimes\eta'$ identify $\mathcal H_f\oplus\mathcal H^A_{\rm mag}$ and $\mathcal H^B_f\oplus\mathcal H^B_{\rm mag}$ with the two sign supports. Under them $\Omega_{xy}$ becomes $\Omega_{ff}+\epsilon_{xy}\omega_{\rm mag}/\sqrt2$, with the two terms in the flag-flag and the magic-magic blocks; its coefficient operator is
\[
 W_{xy}=\operatorname{diag}(T_f,\epsilon_{xy}T_{\rm mag}/\sqrt2).
\]
Its Alice marginal is $h_f\oplus h^A_{\rm mag}/2$ and its Bob marginal is $h^B_f\oplus h^B_{\rm mag}/2$; both are faithful.

Let $E'_a=P_AE_aP_A$ and $F'_b=P_BF_bP_B$ be the compressed sign effects of $\mathsf T_x$ and $\mathsf B_y$. Then $0\leq E'_a\leq S^A_x$ and $0\leq F'_b\leq S^B_y$, so $E'_a=E'_aS^A_x$ and $F'_b=F'_bS^B_y$ by (N6a). If $ab\ne\epsilon_{xy}$, the pair is rejected, so
\[
 0=\langle\Omega,(E'_a\otimes F'_b\otimes I)\Omega\rangle=\|(E'^{1/2}_a\otimes F'^{1/2}_b\otimes I)\Omega\|^2,
\]
whence $(E'_a\otimes F'_b\otimes I)\Omega=0$; in coefficient form, $E_a^{1/2}W_{xy}(F_b^T)^{1/2}=0$ and $E_aW_{xy}F_b^T=0$. Writing $\epsilon=\epsilon_{xy}$ and summing the complete binary effects,
\[
 (E'_a\otimes I)\Omega_{xy}=(E'_a\otimes S^B_y\otimes I)\Omega=(E'_a\otimes F'_{\epsilon a}\otimes I)\Omega=(S^A_x\otimes F'_{\epsilon a}\otimes I)\Omega=(I\otimes F'_{\epsilon a}\otimes I)\Omega_{xy}.
\]
Next, $(E'_aE'_{-a}\otimes I)\Omega_{xy}=(E'_a\otimes F'_{-\epsilon a}\otimes I)\Omega_{xy}=(E'_a\otimes F'_{-\epsilon a}\otimes I)\Omega=0$. Since $E'_{-a}=S^A_x-E'_a$, this says $((E'_a-E'^2_a)\otimes I)\Omega_{xy}=0$. The Alice marginal of $\Omega_{xy}$ is faithful on $S^A_xP_A\HH_A$, so (N3), applied to $\Omega_{xy}$, gives $E'^2_a=E'_a$. This is where dense range enters. Bob's case is symmetric. By (N6b) the original sign effects reduce the supports. The differences $\mathcal X_x=E'_+-E'_-$ and $\mathcal Y_y=F'_+-F'_-$ are self-adjoint unitaries on $S^A_xP_A\HH_A$ and $S^B_yP_B\HH_B$, and
\[
 (\mathcal X_x\otimes I)\Omega_{xy}=\epsilon_{xy}(I\otimes\mathcal Y_y\otimes I)\Omega_{xy}.
\]
In coefficient form this reads
\begin{equation}\label{eq:sign}
 \mathcal X_xW_{xy}=\epsilon_{xy}W_{xy}\mathcal Y_y^T.
\end{equation}

Set $T_{\rm cal}=\operatorname{diag}(T_f,T_{\rm mag}/\sqrt2)$ and ${J_{\rm sgn}}=\operatorname{diag}(I,-I)$, so that $W_{xy}=T_{\rm cal}$ for $(x,y)\ne(1,1)$ and $W_{11}={J_{\rm sgn}}T_{\rm cal}$. Put $\widehat{\mathcal X}_x=V^{A*}_x\mathcal X_xV^A_x$ and $\widehat{\mathcal Y}_y=V^{B*}_y\mathcal Y_yV^B_y$, and let $\Omega_{\rm cal}=\Omega_{ff}+\omega_{\rm mag}/\sqrt2$, whose coefficient operator is $T_{\rm cal}$. For $(x,y)\ne(1,1)$ the vector $\Omega_{xy}$ becomes $\Omega_{\rm cal}$, and for $(1,1)$ it becomes $({J_{\rm sgn}}\otimes I)\Omega_{\rm cal}=(I\otimes{J_{\rm sgn}}\otimes I)\Omega_{\rm cal}$. The cases $(0,0)$ and $(1,0)$ give $((\widehat{\mathcal X}_0-\widehat{\mathcal X}_1)\otimes I)\Omega_{\rm cal}=0$, and the cases $(0,0)$ and $(0,1)$ give $(I\otimes(\widehat{\mathcal Y}_0-\widehat{\mathcal Y}_1)\otimes I)\Omega_{\rm cal}=0$. Both marginals of $\Omega_{\rm cal}$, namely $h_f\oplus h^A_{\rm mag}/2$ and $h^B_f\oplus h^B_{\rm mag}/2$, are faithful, so (N3) gives $\widehat{\mathcal X}_0=\widehat{\mathcal X}_1=\mathcal X$ and $\widehat{\mathcal Y}_0=\widehat{\mathcal Y}_1=\mathcal Y$. These are applications of (N3) to the two original local marginals. The case $(1,1)$ then gives
\[
 (\mathcal X{J_{\rm sgn}}\otimes I)\Omega_{\rm cal}=-(I\otimes\mathcal Y\otimes I)({J_{\rm sgn}}\otimes I)\Omega_{\rm cal}
 =-({J_{\rm sgn}}\mathcal X\otimes I)\Omega_{\rm cal},
\]
so $\mathcal X{J_{\rm sgn}}+{J_{\rm sgn}}\mathcal X=0$ by (N3). A self-adjoint unitary anticommuting with ${J_{\rm sgn}}$ has zero diagonal blocks, so
\[
 \mathcal X=\begin{pmatrix}0&\beta^*\\\beta&0\end{pmatrix},
\]
and $\mathcal X^2=I$ gives $\beta^*\beta=I_{\mathcal H_f}$ and $\beta\beta^*=I_{\mathcal H_{\rm mag}}$: the block $\beta:\HH_f\to\mathcal H_{\rm mag}$ is unitary. The operators $\mathcal X,\mathcal Y$ are self-adjoint mirrors for $\Omega_{\rm cal}$, so (N4c) gives $[\mathcal X,h_f\oplus h^A_{\rm mag}/2]=0$. In coefficient form: from $\mathcal XT_{\rm cal}=T_{\rm cal}\mathcal Y^T$, the operator $\mathcal XT_{\rm cal}T_{\rm cal}^*=T_{\rm cal}\mathcal Y^TT_{\rm cal}^*$ is self-adjoint, and $T_{\rm cal}T_{\rm cal}^*=h_f\oplus h^A_{\rm mag}/2$ with $h_f=T_fT_f^*$ and $h^A_{\rm mag}=T_{\rm mag}T_{\rm mag}^*$. Its off-diagonal block is $\beta h_f=\tfrac12h^A_{\rm mag}\beta$, so $h^A_{\rm mag}\beta=2\beta h_f$: the corner transport has the required ratio. On Bob's side the same argument gives a unitary $\beta_B$ with $h^B_{\rm mag}\beta_B=2\beta_Bh^B_f$.

The actual Alice extraction is ${\btrans}_A=\Pi_{c_0}\mathcal X_0f$. Bob's $r_0$ projection is the retained $(+++)$ answer of $\mathsf B_1$, equivalently the sign support of $\mathsf B_0$ minus its flag. His extraction to the same $c_0$ direction is
\[
 {\btrans}_B=\sqrt2\Pi_{c_0}\Pi_{r_0}\mathcal Y_0f.
\]
The coefficient $\sqrt2$ enters only the extraction; the predicate stays rational.

In the coordinates $V^A_0$, ${\btrans}_A$ is the map $\xi\mapsto c_0\otimes\beta\xi$ from $\mathcal H_f$ into $m_{00}P_A\HH_A$. Thus ${\btrans}_A^*{\btrans}_A=f_A$ and ${\btrans}_A{\btrans}_A^*=m_{00}$ on the support, and $h_A{\btrans}_A=2{\btrans}_Ah_A$, because $h_A$ acts as $h_f$ on the flag and as $h^A_{\rm mag}$ on $c_0\otimes\mathcal H^A_{\rm mag}$. On Bob's side $\mathcal Y_0f_B$ is $\xi'\mapsto r_0\otimes\beta_B\xi'$, and $\sqrt2\Pi_{c_0}\Pi_{r_0}$ restricts to the isometry $r_0\otimes\eta\mapsto c_0\otimes\eta$ from $\Pi_{r_0}P_B\HH_B$ onto $m_{00}P_B\HH_B$, since $\langle c_0,r_0\rangle=2^{-1/2}$. So ${\btrans}_B=c_0\otimes\beta_B$ has the same three properties. The components in $\mathcal H^A_{\rm mag}\otimes\mathcal H^B_f\otimes\HH_{\rm env}$ of the two sides of $(\mathcal X\otimes I)\Omega_{\rm cal}=(I\otimes\mathcal Y\otimes I)\Omega_{\rm cal}$ give $(\beta\otimes I)\Omega_{ff}=(I\otimes\beta_B^*\otimes I)\omega_{\rm mag}/\sqrt2$, that is, $\omega_{\rm mag}=\sqrt2(\beta\otimes\beta_B\otimes I)\Omega_{ff}$. Since $\Omega_{c_0c_0}=c_0\otimes c_0\otimes\omega_{\rm mag}$, this is \eqref{eq:bridgevector}. The formula for $\Omega_{NN}$ also gives $\Omega_{m_{ab}m_{ab}}=(t^A_{ab}\otimes t^B_{ab}\otimes I)\Omega_{c_0c_0}$, and the components for two different magic directions vanish. These are the transported versions for the other three directions.

Mixed states are covered by the same calculation. It used the purification $\Omega$ on the cut $A\mid B\,{\rm env}$, with every Bob operator of the form $F\otimes I_{\rm env}$. The synchronized opposite effects reduce the Schmidt support, and the sign effects do so once their compressed sharpness is proved, so every compressed product above is the restriction of an original Bob-only product tensored with $I_{\rm env}$. Bob's identities come from the cut $B\mid A\,{\rm env}$, the Bob form of (N3). The supports $P_A,P_B$ are those of $\Tr_B\rho$ and $\Tr_A\rho$, so every identity holds on the original marginal supports, and \eqref{eq:bridgevector} holds on the original purification with ${\btrans}_A\otimes {\btrans}_B\otimes I_{\rm env}$. The partial trace over $\HH_{\rm env}$ of the rank-one operators of its two sides gives $(m_{00}\otimes m_{00})\rho(m_{00}\otimes m_{00})=2({\btrans}_A\otimes {\btrans}_B)(f\otimes f)\rho(f\otimes f)({\btrans}_A\otimes {\btrans}_B)^*$.
\end{proof}

\Needspace{5\baselineskip}
\subsubsection{The commuting form}\label{sec:bridgecomm}
On a normal commuting host the same predicate calibrates a modular eigenoperator. We recall the arbitrary-normal bridge argument of \cite[Lemma 2.2]{ReadType}. Matching makes the cells sharp centralizer effects. Write the upper-left four cell reflections as $a,b,c,d$. The row/column relations give $[a,b]=[a,c]=[b,d]=[c,d]=0$, $(ac)(bd)=-(ab)(cd)$, and $[ac,bd]=0$. Canceling the outside $a,d$ gives $cb=-bc$; substitution into the last commutator gives $ad=-da$. The two commuting Pauli pairs therefore generate centralizer $\Mat_4$. In $\mathbb C^2\otimes\mathbb C^2$ the calibrated unit vectors are $\chi_x=|+\rangle\otimes|x\rangle$ and $\rho_y^{\rm dir}=|+\rangle\otimes|(-1)^y\rangle$, where $|+\rangle=(|0\rangle+|1\rangle)/\sqrt2$ and $|(-1)^y\rangle=(|0\rangle+(-1)^y|1\rangle)/\sqrt2$. Choose a minimal projection $m_{\rm min}$ of the magic matrix algebra $\Mat_4$ and centralizer partial isometries $u^A_x,u^B_y$ with initial projection $m_{\rm min}$ and ranges the projections onto $\chi_x,\rho_y^{\rm dir}$ (tensored with the multiplicity identity). Choose their phases so that their overlap is $(u^A_x)^*u^B_y=(-1)^{xy}m_{\rm min}/\sqrt2$. If $\Sigma_x^A,\Sigma_y^B$ are the original sign differences, their pull-backs on the standard corner $(f+m_{\rm min})\mathcal J(f+m_{\rm min})\mathcal J L^2(M)$ are
\[
 \mathcal X_x=(f+(u^A_x)^*)\Sigma_x^A(f+u^A_x),\qquad
 \mathcal Y_y=\mathcal J(f+(u^B_y)^*)\mathcal J\Sigma_y^B \mathcal J(f+u^B_y)\mathcal J.
\]
They are commuting self-adjoint contractions. Their joint-support vector is
\[
 \theta_{xy}=(f+(-1)^{xy}m_{\rm min}/\sqrt2)\xi_\varphi .
\]
Forbidden sign events give $\mathcal X_x\theta_{xy}=(-1)^{xy}\mathcal Y_y\theta_{xy}$ and $\mathcal X_x\mathcal Y_y\theta_{xy}=(-1)^{xy}\theta_{xy}$. The multiplier $f\pm m_{\rm min}/\sqrt2$ is bounded invertible in the corner centralizer, so $\theta_{xy}$ separates both corner algebras. Commuting the opposite contraction through the first identity gives $\mathcal X_x^2\theta_{xy}=\theta_{xy}$ and $\mathcal Y_y^2\theta_{xy}=\theta_{xy}$; separation makes them reflections. Put $c_{\rm cal}=f+m_{\rm min}/\sqrt2$. The three positive cells give $\mathcal X c_{\rm cal}\xi_\varphi=\mathcal Y c_{\rm cal}\xi_\varphi$ and identify $\mathcal X_0=\mathcal X_1=\mathcal X$ and $\mathcal Y_0=\mathcal Y_1=\mathcal Y$ by separation on $c_{\rm cal}\xi_\varphi$. The fourth vector is ${J_{\rm sgn}} c_{\rm cal}\xi_\varphi$, ${J_{\rm sgn}}=f-m_{\rm min}$. Its equation is $\mathcal X{J_{\rm sgn}} c_{\rm cal}\xi_\varphi=-\mathcal Y{J_{\rm sgn}} c_{\rm cal}\xi_\varphi=-{J_{\rm sgn}}\mathcal Y c_{\rm cal}\xi_\varphi=-{J_{\rm sgn}}\mathcal X c_{\rm cal}\xi_\varphi$, since opposite algebras commute. Separation gives $\mathcal X{J_{\rm sgn}}=-{J_{\rm sgn}}\mathcal X$. Thus $\mathcal X={\btrans}+{\btrans}^*$, ${\btrans}=m_{\rm min}\mathcal Xf$, ${\btrans}^*{\btrans}=f$, ${\btrans}{\btrans}^*=m_{\rm min}$. The mirror equation means that $\mathcal X$ centralizes $\omega(\cdot)=\varphi(d_{\rm cal}\,\cdot)$, $d_{\rm cal}=f+m_{\rm min}/2$. The bounded centralizer perturbation formula of Pedersen--Takesaki \cite[Lemma 4.5]{PT} $\sigma_t^\omega(x)=d_{\rm cal}^{it}\sigma_t^\varphi(x)d_{\rm cal}^{-it}$ gives on the $m_{\rm min}$--$f$ block
\begin{equation}\label{eq:bridgecomm}
 \sigma_t^\varphi({\btrans})=2^{it}{\btrans},\qquad \varphi(m_{\rm min})=2\varphi(f).
\end{equation}
The last equality is KMS. Multiplication by the magic matrix units gives all four transports. If one sector vanished, the off-diagonal reflection would kill the other; hence on a nonzero active corner both sectors are nonzero.

\subsubsection{A shared question fixes transported labels}
\begin{lemma}[Label refinement]\label{lem:label}
Retain the original FMS questions. For a synchronized projection $P^{\rm lab}$, add Alice's $\mathsf T_0[P^{\rm lab}]$ with answers $(a,z)$. Ignore $z$ against the unchanged $\mathsf B_0,\mathsf B_1,\mathsf C_0$ and use the original predicates; match $z$ to Bob's binary $P^{\rm lab}$. Add the symmetric $\mathsf B_0[P^{\rm lab}]$ against unchanged $\mathsf T_0,\mathsf T_1,\mathsf R_0$. Also retain activity matching. In every perfect normal strategy the coarse signs equal the original signs and commute with $P^{\rm lab}$ on their state supports. Conversely commuting labels give these refinements. Distinct labels require no mutual commutation.
\end{lemma}
\begin{proof}
Let $q_a=\sum_zq_{a,z}$ be the compressed coarse effects of $\mathsf T_0[P^{\rm lab}]$. Against $\mathsf C_0$ the original rule is applied to $a$, so the matching argument in the proof of Lemma~\ref{lem:bridge} gives, by (N5) and (N4b), $q_t=\Pi_t$ for each retained triple $t$ and $q_++q_-=R_A=f_A+\Pi_{c_0}$. Against $\mathsf B_0,\mathsf B_1$ the forbidden sign pairs are the original ones. The calculation in that proof which gave $(E'_a\otimes I)\Omega_{0y}=(I\otimes F'_a\otimes I)\Omega_{0y}$ and $E'^2_a=E'_a$ used only these pairs and the projection $R_A$, so it applies verbatim to $q_\pm$. With $\mathcal X'=q_+-q_-$ it gives
\[
 (\mathcal X'\otimes I)\Omega_{00}=(I\otimes\mathcal Y_0\otimes I)\Omega_{00}=(\mathcal X_0\otimes I)\Omega_{00},
\]
the vector form of $\mathcal X'W_{00}=W_{00}\mathcal Y_0^T$. The Alice marginal of $\Omega_{00}$ is faithful on $R_AP_A\HH_A$, so (N3) gives $\mathcal X'=\mathcal X_0$ and $q_\pm=(R_A\pm\mathcal X_0)/2$.

The $z$-marginal $\sum_aQ_{a,1}$ is perfectly matched with Bob's binary question for $P^{\rm lab}$. By (N5) its compression is a projection and a mirror of Bob's synchronized projection, whose Alice mirror is $P^{\rm lab}_A$; (N4b) gives equality. For every $a$, positivity gives $0\leq q_{a,1}\leq P^{\rm lab}_A$ and $0\leq q_{a,0}\leq P_A-P^{\rm lab}_A$. By (N6a), $q_{a,1}=P^{\rm lab}_Aq_{a,1}P^{\rm lab}_A$ and $q_{a,0}=(P_A-P^{\rm lab}_A)q_{a,0}(P_A-P^{\rm lab}_A)$, so $P^{\rm lab}_Aq_a=q_{a,1}=q_aP^{\rm lab}_A$. All of these are identities on $P_A\HH_A$. For the converse, if the original effects $E_a$ commute with $P^{\rm lab}$, then $E_aP^{\rm lab}$ and $E_a(I-P^{\rm lab})$ are positive. Summing over $z$ gives $E_a$, and summing the $z=1$ effects over $a$ gives $P^{\rm lab}$. Every probability involving only $a$ is an original probability, and the $z$-test is the original perfect matching of $P^{\rm lab}$. By (N6c) one may first replace $P^{\rm lab}$ by its support-restricted projection. No two different labels are measured jointly, so no commutation between them enters either direction.
\end{proof}

If a label is block diagonal with flag block $P^{\rm lab}_{f}$ and identical magic blocks $P^{\rm lab}_{m}$, the commutator says $\beta P^{\rm lab}_{f}=P^{\rm lab}_{m}\beta$. In particular, for a graph projection of a block unitary it certifies transport of that same unitary by the calibrated ${\btrans}$. It does not supply a fresh bridge with an independent gauge.
Indeed, with respect to $R_A=f\oplus\Pi_{c_0}$ the commutator $[\mathcal X_0,P^{\rm lab}]$ has the $(c_0,f)$ block $\beta P^{\rm lab}_f-P^{\rm lab}_m\beta$.

\begin{proof}[Proof in the commuting model]
Calibration fixes the retained effects and the sign support. Pull the new coarse sign difference $\mathcal X'$ to the same corner. Against the unchanged opposite sign it satisfies $\mathcal X'\theta_{0y}=\mathcal Y_y\theta_{0y}=\mathcal X_0\theta_{0y}$. The separating vector of Section~\ref{sec:bridgecomm} gives $\mathcal X'=\mathcal X_0$. Thus its coarse effects $E_a$ are the old projections. For the fine effects $Q_{a,z}$, label matching gives $\sum_aQ_{a,1}=P^{\rm lab}$ and $\sum_aQ_{a,0}=I-P^{\rm lab}$ on faithful support. Positivity gives $0\leq Q_{a,1}\leq E_a,P^{\rm lab}$ and $0\leq Q_{a,0}\leq E_a,I-P^{\rm lab}$. Hence $E_aP^{\rm lab}=Q_{a,1}=P^{\rm lab}E_a$, using completeness of each marginal. This proves commutation, and the symmetric argument handles Bob. Conversely use the joint products $E_aP^{\rm lab},E_a(I-P^{\rm lab})$. Each label has its own refined question with the same unchanged opposite signs, so there is no cross-label commutation constraint.
\end{proof}

\subsection{Rational projection equations that certify products}\label{sec:projection}
All projections in this section are synchronized. We give the finite rules used in the compiler. The projection-partition viewpoint is standard \cite{Fritz}; the formulas here provide the needed actual multiplication identities. See also the graph-projection encoding in \cite[Section 4]{CZ}.

\subsubsection{Graphs and triangles}
Let $F_i,F_j$ be orthogonal, put $F_{ij}=F_i+F_j$, and introduce projections $G,H,G^\perp,H^\perp$ with
\begin{equation}\label{eq:G1}
 G+G^\perp=F_{ij},\quad H+H^\perp=F_{ij},\quad 5H+F_{ij}=3F_i+4G.
\end{equation}
On $F_{ij}$, put ${Z_{ij}^{\rm gr}}=F_i-F_j$, $\Xi_{ij}=2G-F_{ij}$. These are reflections, and $2H-F_{ij}=(3{Z_{ij}^{\rm gr}}+4\Xi_{ij})/5$. Squaring forces ${Z_{ij}^{\rm gr}}\Xi_{ij}+\Xi_{ij}{Z_{ij}^{\rm gr}}=0$. Thus
\[
 \Xi_{ij}=\begin{pmatrix}0&u_{ij}\\u_{ij}^*&0\end{pmatrix},\qquad
 u_{ij}^*u_{ij}=F_j,\quad u_{ij}u_{ij}^*=F_i.
\]
The equations thus certify the graph blocks. Conversely any such unitary corner map gives the four projections in \eqref{eq:G1}.

For a partition $F_1+F_2+F_3=I$ impose the graph tests on all three pairs and add a projection $P_\triangle$ with
\begin{equation}\label{eq:G2}
 3P_\triangle=2(G_{12}+G_{23}+G_{13})-I.
\end{equation}
Its block matrix has diagonal $I/3$ and off-diagonal $u_{ij}/3$. The $(1,3)$ block of $P_\triangle^2=P_\triangle$ gives $2u_{13}+u_{12}u_{23}=3u_{13}$, hence $u_{13}=u_{12}u_{23}$. Conversely this product identity makes the entire matrix a projection.

Start with a reference triangle for identity. Its scalar matrix algebra is called the \emph{frame}. It gives scalar matrix units and $\HH=\mathbb C^3\otimes\mathcal H_{\rm base}$. For every unitary label $\Theta$, use graphs in all three positions. Triangles $(\Theta,1,\Theta)$ and $(1,\Theta,\Theta)$ identify the placements. A triangle $(\Theta,\Theta_2,\Theta_3)$ then certifies $\Theta_3=\Theta\Theta_2$; a triangle with product identity certifies inverse labels. Build each word from its proper prefix and final token. Finally a real rational equation $\sum_wq_w\Theta_w=0$ is precisely
\begin{equation}\label{eq:G3}
 2\sum_wq_wG_{12}(\Theta_w)-\Bigl(\sum_wq_w\Bigr)(F_1+F_2)=0.
\end{equation}
Its two off-diagonal blocks are the equation and its adjoint. All products therefore refer to actual common operators. Synchronization makes the extracted unitaries commute with the base marginal density.

\subsubsection{Binding prescribed parameters}
A prescribed synchronized projection $P^{\rm par}$ is linked to the frame by projections $P^{\rm par}F_i,NF_i$ with
\[
 P^{\rm par}F_i+NF_i=F_i,\quad \sum_iP^{\rm par}F_i=P^{\rm par},\quad\sum_iNF_i=I-P^{\rm par}.
\]
These symbols are names for projections before the equations identify their products; the names $NF_i$ denote complements of $P^{\rm par}F_i$ inside $F_i$. For every pair $ij$, introduce $J_{11},J_{10},J_{01},J_{00}$ and impose
\begin{align*}
 J_{11}+J_{10}&=P^{\rm par}F_i+P^{\rm par}F_j,&
 J_{00}+J_{01}&=NF_i+NF_j,\\
 J_{11}+J_{01}&=G_{ij}(1),&
 G_{ij}(s_{\rm par})&=J_{11}+J_{00}.
\end{align*}
Projection sums give the indicated joint partitions. The symbol $s_{\rm par}$ is a unitary parameter label. These equations force $P^{\rm par}$ to commute with the frame and identify the extracted $s_{\rm par}$ with $2e-I$, where $P^{\rm par}=I_3\otimes e$. In particular each parameter is the specified FMS event itself. The four-bridge assembly uses twenty-five such parameters.

These rules are a deterministic finite compiler for the finite word list. Cancellation of adjacent inverse tokens is performed before taking proper prefixes. All coefficients stay rational. Conversely a centralizer unitary assignment with the prescribed parameter projections gives every graph, complement, $H$, triangle and joint projection simultaneously by these formulas.

\paragraph{Scope in infinite dimensions.}
All projections in this section are support-restricted synchronized projections on Alice's support $P_A\HH_A$, as in (N5), and every displayed equation holds there as an operator identity by Lemma~\ref{lem:rows}. The derivations are block-matrix calculations for orthogonal decompositions of $P_A\HH_A$. Projections whose sum is a projection are mutually orthogonal, and a self-adjoint unitary anticommuting with $\operatorname{diag}(I,-I)$ has the off-diagonal form used for $\Xi_{ij}$. The frame identification is the unitary
\[
 \mathbb C^3\otimes\mathcal H_{\rm base}\to P_A\HH_A,\qquad e_i\otimes\xi\mapsto u_{i1}\xi,\qquad \mathcal H_{\rm base}=F_1P_A\HH_A,
\]
where $u_{11}=F_1$, $u_{21}=u_{12}^*$, $u_{31}=u_{13}^*$, and $u_{ij}=F_i(2G_{ij}(1)-F_i-F_j)F_j$. The frame projections and the maps $u_{ij}$ commute with $h_A$ by (N5), so $h_A=I_3\otimes h_{\rm base}$ in these coordinates. An extracted label $\Theta$ is defined by $F_1(2G_{12}(\Theta)-F_1-F_2)F_2=e_{12}\otimes\Theta$, and it commutes with $h_{\rm base}$.

The binding equations are solved in the same way. Write $\pi^{\rm par}_i=P^{\rm par}F_i$ and $\eta^{\rm par}_i=NF_i$. These are projections below $F_i$ with $\pi^{\rm par}_i+\eta^{\rm par}_i=F_i$, and $\sum_i\pi^{\rm par}_i=P^{\rm par}$, so $P^{\rm par}F_i=\pi^{\rm par}_i=F_iP^{\rm par}$ and $P^{\rm par}$ commutes with every $F_i$. For a pair $ij$ the four projections $J_{ab}$ sum to $F_{ij}$ and are therefore mutually orthogonal. Both $\pi^{\rm par}=\pi^{\rm par}_i+\pi^{\rm par}_j$ and $G_{ij}(1)$ are sums of them, so they commute and $\pi^{\rm par} G_{ij}(1)=J_{11}$. In frame coordinates write $\pi^{\rm par}_i=e_{ii}\otimes p_i$. Commutation of $\pi^{\rm par}$ with $\Xi_{ij}(1)=(e_{ij}+e_{ji})\otimes I$ gives $p_i=p_j$; over all pairs, $P^{\rm par}=I_3\otimes e$. Finally $J_{00}=(F_{ij}-\pi^{\rm par})(F_{ij}-G_{ij}(1))$, so
\[
 G_{ij}(s_{\rm par})=\pi^{\rm par} G_{ij}(1)+(F_{ij}-\pi^{\rm par})(F_{ij}-G_{ij}(1))=\tfrac12\bigl(F_{ij}+(e_{ij}+e_{ji})\otimes(2e-I)\bigr),
\]
which identifies $s_{\rm par}$ with $2e-I$.

\subsection{Rational rows as Boolean winning conditions}\label{sec:rows}
We implement each rational row by a subspace orthogonality test. This is the general rational-row theorem of \cite[Theorem~B]{Entropy}, restated here to specify the extension and its quantitative loss.

\Needspace{5\baselineskip}
\begin{lemma}[Boolean rational rows]\label{lem:rows}
Let a finite strategy specification name finitely many synchronized binary projections and finitely many rational linear identities among them. There is a finite extension by Boolean tests with the following properties.

\emph{Soundness:} in every perfect normal spatial strategy for the extended tests, the identities hold as operator identities for the named support-restricted projections on each local state support.

\emph{Completeness:} any normal spatial strategy with these synchronized projections, satisfying the identities on its supports, extends to satisfy all the added tests perfectly after adjoining a finite maximally entangled state. The old measurements act trivially on the new factor. The rest of the old strategy, including nonsynchronous questions, is retained; if its old tests were perfect, the whole extension is perfect.
\end{lemma}
\begin{remark}
The same proof, paragraph ``The commuting form of rational rows'', gives the commuting form used in Section~\ref{sec:chain}.
\end{remark}

\subsubsection{A certified scalar matrix algebra}
Choose $d_{\max}=2^{2r}$ large enough for every row construction below. Use $r$ synchronized magic squares, binding each context's cell marginals to synchronized binary cells. Add four-answer joint questions for all pairs of axes from distinct squares, joint questions between every named old projection and every scalar axis, and global $Z$ and global $X$ measurements with all bit outcomes, binding each bit marginal to its axis. These steps concern only synchronized projections, never an FMS sign.

The Pauli-pair calculation in the proof of Lemma~\ref{lem:bridge}, without flags, gives two anticommuting pairs of axes per square. A four-answer joint question for axes $\varsigma,\varsigma'$ of distinct squares has its two binary marginals bound to synchronized cells, so the axes commute on $P_A\HH_A$ by (N6d). Iterating the eigenspace argument of that proof over the $2r$ mutually commuting anticommuting pairs identifies $P_A\HH_A$ with $\mathbb C^{d_{\max}}\otimes\mathcal H'$, the axes generating $\Mat_{d_{\max}}\otimes I$; the space $\mathcal H'$ may be infinite-dimensional. The axes commute with $h_A$ by (N5), so $h_A$ lies in $(\Mat_{d_{\max}}\otimes I)'=I\otimes B(\mathcal H')$: the marginal is $I_{d_{\max}}/d_{\max}\otimes h$ for a density $h$ on $\mathcal H'$. A joint question between a named projection $E_i$ and an axis makes them commute by (N6d). The axes generate $\Mat_{d_{\max}}\otimes I$ as an algebra, so $E_i$ lies in its commutant, and $E_i=I_{d_{\max}}\otimes e_i$ for a projection $e_i$ on $\mathcal H'$. The global $Z$ and $X$ measurements have sharp bit marginals, so by (N6d) they are the joint spectral measures of the commuting $Z$ axes and of the commuting $X$ axes.

For a row dimension $d=2^{k_{\rm row}}\leq d_{\max}$ use the first ${k_{\rm row}}$ qubits; the rest are spectators. For each nonzero $j\in\mathbb F_2^{k_{\rm row}}$, choose a basis $\ell_1,\ldots,\ell_{{k_{\rm row}}-1}$ of $j^\perp$. Add a joint context for $X^j,Z^{\ell_1},\ldots,Z^{\ell_{{k_{\rm row}}-1}}$, binding its marginals to the corresponding global parity events. These commute because $\ell\cdot j=0$. The zero-$Z$-syndrome outcomes with $X^j$ sign $\pm$ are
\[
 \Pi_j^\pm=\tfrac12(|0\rangle\pm|j\rangle)(\langle0|\pm\langle j|)
                       \otimes I_{\rm spectator}.
\]
Write $|+_d\rangle=d^{-1/2}\sum_{j=0}^{d-1}|j\rangle$. The reference $\Pi^{\rm ref}_d=|+_d\rangle\langle+_d|\otimes I_{\rm spectator}$ is the event that the first ${k_{\rm row}}$ global $X$ bits are positive. It includes all spectator outcomes.

\subsubsection{Signed reflections and column graphs}
For one row $\Delta=\alpha_0I+\sum_i\alpha_ie_i$, let $L$ clear all coefficient denominators and set
\[
 s_i=2e_i-I,\quad c_i=L\alpha_i,\quad
 c_0=2L\alpha_0+\sum_iL\alpha_i.
\]
Then $2L\Delta=c_0I+\sum_ic_is_i$. Expand this into signed reflections with integer repetitions, add a canceling $I,-I$ pair with $V_0=I$, and pad by further such pairs to a power of four $d$. The list has even length because its coefficient sum is even. Thus
\begin{equation}\label{eq:slots}
 \sum_{j=0}^{d-1}V_j=2L\Delta,\qquad V_0=I,\qquad
 V_j\in\{\pm I,\ \pm(2e_i-I)\}.
\end{equation}
For each $j>0$ use a six-answer joint question of $e_i$ and the three-way scalar measurement $\Pi_j^+,\Pi_j^-,I-\Pi_j^+-\Pi_j^-$. For $V_j=\pm I$, bind its binary bit to a synchronized deterministic anchor: its $1$-projection is $I$ and its $0$-projection is zero. Matching both marginals identifies all joint effects. Appropriate coarse events are the two-coordinate graph projections
\[
 G_j^\pm=\frac12\begin{pmatrix}I&\pm V_j\\\pm V_j&I\end{pmatrix}
\]
on coordinates $0,j$, zero elsewhere. A negative reflection sign exchanges the two coarse events.

Add a synchronized $d$-answer question with projections $\Pi^{\rm row}_t$, indexed by $t\in\mathbb F_2^{k_{\rm row}}$. Put $\chi_t(j)=(-1)^{t\cdot j}$. Forbid $\Pi^{\rm row}_t$ together with $G_j^-$ when $\chi_t(j)=1$, or $G_j^+$ when $\chi_t(j)=-1$. Define
\[
 \mathcal I_t\xi=d^{-1/2}\bigl(\chi_t(j)V_j\xi\bigr)_{j=0}^{d-1},
 \qquad \mathcal P_t=\mathcal I_t\mathcal I_t^*.
\]
Unitarity and character orthogonality give $\mathcal I_t^*\mathcal I_u=\delta_{tu}I$ and $\sum_t\mathcal P_t=I$. Neither equation requires different $V_j$ to commute. Finally forbid $\Pi^{\rm row}_0$ jointly with the opposite $\Pi^{\rm ref}_d$.

For soundness, the forbidden pairs are handled by (N7). Take $Q$ to be the coarse event $G^\mp_j$ of the party that asks it and $R$ the opposite synchronized $\Pi^{\rm row}_t$. Both reduce the support, so $G^\mp_j\Pi^{\rm row}_t=\Pi^{\rm row}_tG^\mp_j=0$ on that party's support. The events $G^\pm_j$ and $\Pi^{\rm row}_t$ are products of mirrored projections, so (N4a) and (N3) transfer this identity to the other party. In particular it holds on $P_A\HH_A$. Write a vector of $P_A\HH_A=\mathbb C^d\otimes\mathbb C^{d_{\max}/d}\otimes\mathcal H'$ as $(v_j)_{j\in\mathbb F_2^{k_{\rm row}}}$, and regard each $V_j$ as a support-restricted reflection, a self-adjoint unitary on $\mathbb C^{d_{\max}/d}\otimes\mathcal H'$. If $\chi_t(j)=1$, the equation $G^-_jv=0$ says $v_j=V_jv_0$. If $\chi_t(j)=-1$, the equation $G^+_jv=0$ says $v_j=-V_jv_0$. Thus $\ran\Pi^{\rm row}_t\subseteq\ran\mathcal I_t$, that is, $\Pi^{\rm row}_t\leq\mathcal P_t$. The two families are complete, so $\sum_t(\mathcal P_t-\Pi^{\rm row}_t)=0$ with positive terms, and $\Pi^{\rm row}_t=\mathcal P_t$. The forbidden pair $(\Pi^{\rm row}_0,\Pi^{\rm ref}_d)$ gives $\Pi^{\rm ref}_d\mathcal P_0=0$ by (N7), and $\Pi^{\rm ref}_d\mathcal I_0=\Pi^{\rm ref}_d\mathcal P_0\mathcal I_0=0$. Finally $(\langle+_d|\otimes I)\mathcal I_0=d^{-1}\sum_jV_j=(2L/d)\Delta$, so
\[
 \|\Pi^{\rm ref}_d\mathcal I_0\xi\|^2=\frac{4L^2}{d^2}\|\Delta\xi\|^2
\]
for every $\xi$ in the support. Hence $\Pi^{\rm ref}_d\mathcal I_0=0$ holds exactly when $\Delta$ vanishes on the support. This proves soundness on arbitrary normal supports.

For completeness, let the old strategy have purification $\Omega$ and supports $P_A,P_B$. By (N6c) we may replace every named synchronized effect by its support-restricted projection, completed by a fixed outcome off the support; no probability changes. Adjoin $\Phi_{d_{\max}}$, so that the new purification is $\Omega'=\Omega\otimes\Phi_{d_{\max}}$ and the new state is again normal. Alice's new effects are the displayed operators: scalar Paulis, joint products $E_i\varsigma$ and $E_i\Pi_j^\pm$ of commuting projections, and the column projections $\mathcal P_t=\mathcal I_t\mathcal I_t^*$, all defined on the support and completed by a fixed outcome off it. For an Alice operator $\sum_kz_k\otimes p_k(e)$, with scalar factors $z_k$ and products $p_k(e)$ of named projections, Bob uses $\sum_kz_k^{\mathsf T}\otimes p^{\rm rev}_k(e^B)$, where $p^{\rm rev}_k(e^B)$ is the reversed product of his matched projections; equivalently, Bob uses transposes in canonical Schmidt coordinates. The order matters when the $e_i$ do not commute. By (N4a) and $(z\otimes I)\Phi_{d_{\max}}=(I\otimes z^{\mathsf T})\Phi_{d_{\max}}$, it is a mirror of the Alice operator. This assignment commutes with adjoints, so the counterpart $Y$ of an Alice projection $P$ is self-adjoint. Moreover $(I\otimes(Y^2-Y)\otimes I)\Omega'=((P^2-P)\otimes I)\Omega'=0$, so $Y$ is a projection on Bob's support by (N3), and the counterparts of the column projections sum to the identity. Every matching test therefore has disagreement probability zero. Every forbidden pair has probability zero because the corresponding Alice product vanishes, and the row-zero pair does so because $\Delta=0$ on the support. The old measurements act as $E\otimes I_{d_{\max}}$, so all old tests, including nonsynchronous ones, keep their probabilities. Since the argument runs on a purification, it covers mixed original states.

Write $h_{\rm old}$ for the trace-one marginal before the row ancilla is appended (including the qutrit frame when present). For the honest column construction, without assuming $\Delta=0$, the sole row-zero event has probability
\begin{equation}\label{eq:rowloss}
 p(\Pi^{\rm row}_0,\Pi^{\rm ref}_d)=\frac{4L^2}{d^3}\Tr(h_{\rm old}\Delta^2).
\end{equation}
All column-graph and matching tests remain exact even when the row is nonzero. This formula is \cite[Eq.~(32)]{Entropy} in the general row setting. To prove it, note that Bob's reference effect is the transpose of $\Pi^{\rm ref}_d$ on the scalar factor and is a mirror of $\Pi^{\rm ref}_d$. The probability is therefore $\|(\mathcal P_0\otimes (\Pi^{\rm ref}_d)^{\mathsf T})\Omega'\|^2=\Tr(h'\Pi^{\rm ref}_d\mathcal P_0\Pi^{\rm ref}_d)$, where $h'=(I_d/d)\otimes(d/d_{\max})I_{\rm spectator}\otimes h_{\rm old}$. Since $(\langle+_d|\otimes I)\mathcal I_0=(2L/d)\Delta$, we have $\Pi^{\rm ref}_d\mathcal P_0\Pi^{\rm ref}_d=|+_d\rangle\langle+_d|\otimes I_{\rm spectator}\otimes(4L^2/d^2)\Delta^2$. The trace against $h'$ is \eqref{eq:rowloss}: the factor $1/d$ is $\langle+_d|I_d/d|+_d\rangle$, and the spectator trace is one. The calculation uses only that the $V_j$ are self-adjoint unitaries, so it holds for every row, zero or not.

\paragraph{The commuting form of rational rows.}
On a normal commuting host the same finite extension is used. The Pauli calculation of Section~\ref{sec:bridgecomm} identifies the tensor matrix algebra, and old projections are $I\otimes e_i$ with centralizer scalar matrix units. A forbidden pair with mirror $\mathcal JG\mathcal J$ gives $\Pi^{\rm row}_tG\xi_\varphi=0$, since $\mathcal JG\mathcal J\xi_\varphi=G\xi_\varphi$; separation gives $\Pi^{\rm row}_tG=0$. The remaining soundness argument is unchanged. Conversely amplify by $\Mat_{d_{\max}}$ with normalized trace. Use the actual scalar Paulis, joint products and projections $\mathcal P_t$. All are centralizer projections. Bob uses $\mathcal J_{\rm new}X\mathcal J_{\rm new}$ for each projection $X$. For a centralizer word $xy$, $\mathcal J(xy)\mathcal J\xi_\varphi=(xy)^*\xi_\varphi$; adjunction supplies the reversed order. Thus Bob's products are the actual transferred products, even when the old projections do not commute. Every structural forbidden event is zero, and the final event is zero precisely for a zero row. Old measurements act trivially on the new matrix factor, preserving all old probabilities. Finite matrix amplification preserves all indicated factor types. This proves Lemma~\ref{lem:rows}.

\subsection{Admissible presentations}\label{sec:admissible}
\begin{definition}[Admissible presentation; normal realization]\label{def:admissible}
Fix an integer $n\geq3$ and put $\nu=2^{-n}$. The four-isometry normal model requires $4\nu<1$; the generic chained-bridge statement of Lemma~\ref{lem:chain} separately permits $n\geq1$. An admissible presentation is a finite rational list with four contraction symbols $A_\alpha$ ($\alpha\in\{0,1\}^2$), a centralizer projection label $P_{\circ}$ for the root, and finitely many centralizer labels, each typed as a unitary, projection, positive contraction ($0\leq T\leq I$), or general contraction ($\|T\|\leq1$). Its application equations (the identities enforced by the final row tests) are rational $*$-polynomials in the centralizer labels, the creation and range equations
\[
 A_\alpha^*A_\beta=\delta_{\alpha\beta}I,\qquad
 \sum_\alpha A_\alpha A_\alpha^*=I-P_{\circ},
\]
and endpoint equations $L_{\rm end}A_\alpha=A_\beta R_{\rm end}$ or $L_{\rm end}A_\alpha=0$. Finite sums $\sum_\iota L_\iota A_{\alpha_\iota}L'_\iota=0$ are allowed. Here $L_{\rm end},R_{\rm end},L_\iota,L'_\iota$ are centralizer polynomial expressions. Apart from the displayed creation and range equations, no monomial contains more than one creation symbol. Every label used on a left endpoint is included in the transported unitary list $\Lambda$ defined in the proof of Lemma~\ref{lem:four}, and the root projection is transported even when it occurs in no endpoint equation.

A normal realization consists of a nonzero Hilbert space $\HH_0$, a faithful positive trace-class operator $h_0$ with $\Tr h_0=1$, and bounded operators of the declared types on $\HH_0$. Every centralizer label, including $P_{\circ}$, commutes with $h_0$. All application equations hold as operator identities on $\HH_0$, and
\begin{equation}\label{eq:covariance}
 h_0A_\alpha=\nu A_\alpha h_0.
\end{equation}
\end{definition}
Thus all identities are imposed on the state support, where the density is faithful. A positive contraction need not be bounded below. Finite entropy of a realization means $S(h_0)<\infty$. The ratio is imposed by the assembly, rather than by a real coefficient in the Boolean game. The spatial hardness proof of Sections~\ref{sec:tree}--\ref{sec:bell} uses $n=4$, so $\nu=1/16$; Section~\ref{sec:communication} uses $n=5$.

\subsection{The \texorpdfstring{$n$}{n}-bridge compiler}\label{sec:fourbridge}
Each FMS bridge forces a density ratio of two between consecutive selected corners (Figure~\ref{fig:chain}). A chain of $n$ bridges therefore gives the ratio $2^n$ from source to far corner, and transporting a source-to-far creation operator back to the source gives $\nu=2^{-n}$. Set
\begin{equation}\label{eq:weights}
 \mathfrak c_n=1+4\sum_{g=1}^n2^g=2^{n+3}-7,\qquad
 \mu_n=\mathfrak c_n^{-1}(1;2^{\oplus4};4^{\oplus4};\ldots;(2^n)^{\oplus4}).
\end{equation}
The notation $t^{\oplus4}$ denotes four entries equal to $t$.

\begin{lemma}[$n$-bridge compiler]\label{lem:four}
Every admissible presentation effectively yields a finite rational Boolean game $G^{(n)}$. Every perfect normal spatial strategy yields a normal realization in the sense of Definition~\ref{def:admissible}, on the nonzero source corner with faithful trace-one density $h_0=\mathfrak c_ne_0he_0$, where $h$ is the marginal density. Every such normal realization extends to a perfect normal spatial strategy after adjoining the scaffold and finite ancillary factors (scaffold, qutrit and row factors). Finite marginal entropy is preserved by the completeness construction.\end{lemma}
The assembly parameters are
\begin{center}
\begin{tabular}{@{}ll@{}}\toprule
quantity&value\\\midrule
orthogonal scaffold corners&$4n+1$\\
frame-bound parameters&$6n+1$\\
cover constant&$\mathfrak c_n=2^{n+3}-7$\\
chain covariance&$hv=2^n vh$\\
normalized source density&$\mathfrak c_n e_0he_0$\\
far-corner weight relative to source&$2^n$\\\bottomrule
\end{tabular}
\end{center}
For $n=4$ there are seventeen corners and twenty-five parameters, $\mathfrak c_4=121$, $hv=16vh$ and $h_0=121e_0he_0$.

\begin{proof}
We give the assembly and both directions.

\paragraph{Corners and parameters.}
Use $4n+1$ orthogonal synchronized corners
\[
 e_0;\quad m_{g,00},m_{g,01},m_{g,10},m_{g,11}\quad(1\leq g\leq n).
\]
Put $e_g=m_{g,00}$ and $m_g=\sum_{ab}m_{g,ab}$, and impose the cover $e_0+\sum_gm_g=I$. Conditional bridge $g$ has flag $e_{g-1}$, magic support $m_g$ and activity $\Pact_g=e_{g-1}+m_g$. Use all event bindings described above in both orientations, including the flag/four-magic/inactive partition. Bind to the qutrit frame the $4n+1$ corners and the $2n$ positive $ZI,IX$ events. These are the $6n+1$ frame-bound parameters. Introduce $2n$ axis labels $\mathfrak x_g,\mathfrak z_g$ with the explicit equations
\[
 \mathfrak x_g=I-2m_g+2E_{g,ZI,+},\qquad
 \mathfrak z_g=I-2m_g+2E_{g,IX,+},
\]
where $E_{g,ZI,+},E_{g,IX,+}$ are the positive nonflag events of bridge $g$, bound to synchronized binary questions as above.

Let a perfect normal spatial strategy be given, with purification and supports as in (N1), and write $h=h_A$. By (N5) every corner, frame and graph question yields a synchronized projection on $P_A\HH_A$ commuting with $h$, and by Section~\ref{sec:projection} and Lemma~\ref{lem:rows} every imposed equation, including the cover, holds on $P_A\HH_A$. Projections with sum $I$ are mutually orthogonal, so the $4n+1$ corners partition $P_A\HH_A$. Bridge $g$ is a perfect conditional FMS strategy with activity $e_{g-1}+m_g$, flag $e_{g-1}$ and magic support $m_g$, whose magic directions are the bound corners $m_{g,ab}$. Lemma~\ref{lem:bridge} gives calibrated partial isometries ${\btrans}_g$ with ${\btrans}_g^*{\btrans}_g=e_{g-1}$, ${\btrans}_g{\btrans}_g^*=e_g$ and $h{\btrans}_g=2{\btrans}_gh$ on the support; the other three magic corners are identified with $e_g$ by synchronized Pauli matrix units and have the same density block. Moving $h$ through the $n$ factors gives
\begin{equation}\label{eq:fourv}
 v={\btrans}_n\cdots {\btrans}_1,\quad v^*v=e_0,\quad vv^*=e_n,\quad hv=2^nvh.
\end{equation}
For the masses, $h$ is trace class and ${\btrans}_g$ is bounded, so cyclicity gives
\[
 \Tr(he_g)=\Tr(h{\btrans}_g{\btrans}_g^*)=\Tr({\btrans}_g^*h{\btrans}_g)=2\Tr({\btrans}_g^*{\btrans}_gh)=2\Tr(he_{g-1}).
\]
Each $t_{ab}$ of bridge $g$ commutes with $h$, so $\Tr(hm_{g,ab})=\Tr(he_g)$: the four magic masses are twice the preceding flag mass. Summing over the cover, with $\Tr h=1$, gives
\begin{equation}\label{eq:121}
 \Tr(he_0)=1/\mathfrak c_n;\qquad 121=\mathfrak c_4=1+4(2+4+8+16).
\end{equation}
In particular the source corner is nonzero.

\begin{figure}[ht]
\centering\small
\[
 e_0\xrightarrow{{\btrans}_1}\boxed{m_1\supset e_1}
 \xrightarrow{{\btrans}_2}\boxed{m_2\supset e_2}
 \xrightarrow{{\btrans}_3}\boxed{m_3\supset e_3}
 \xrightarrow{{\btrans}_4}\boxed{m_4\supset e_4}
\]
\begin{tabular}{@{}c|ccccc@{}}
corner & $e_0$&$e_1$&$e_2$&$e_3$&$e_4$\\
$121\Tr(he_g)$&1&2&4&8&16\\
$121\Tr(hm_g)$&--&8&16&32&64
\end{tabular}
\caption{Four calibrated bridges ($n=4$). Each $m_g$ has four equal-mass magic corners; its selected $e_g$ is the next flag. For $n$ bridges the selected masses are $2^g/\mathfrak c_n$, and returning along the chain gives the creation ratio $\nu=2^{-n}$.}\label{fig:chain}
\end{figure}

\paragraph{Every transported label.}
Encode a unitary directly, a projection $P$ by $2P-I$, a positive contraction $T_+$ by a unitary $u_+$ with $T_+=(2I+u_++u_+^*)/4$, and a general contraction by an average of two unitaries. Let $\Lambda$ be the resulting finite unitary label list that must be transported; it contains the encoding of $P_\circ$ and all unitary encodings required at left endpoints. For every $\ell\in\Lambda$, $r=0,\ldots,n$, introduce a unitary $u_{r,\ell}$ and impose
\begin{equation}\label{eq:supportcopies}
 (u_{r,\ell}-I)(I-e_r)=0.
\end{equation}
The unitary reduces $e_r$ and is identity on its complement. For a typed label $L$, its corner copy $L_r$ is explicitly decoded as
\[
 \begin{array}{c|c}
 \text{type of }L&L_r\\\hline
 \text{unitary}&e_ru_{r,L}\\
 \text{projection}&e_r(I+u_{r,L})/2,\quad u_{r,L}^2=I\\
 \text{positive contraction}&e_r(2I+u_{r,L}+u_{r,L}^*)/4\\
 \text{general contraction}&e_r(u_{r,L,+}+u_{r,L,-})/2
 \end{array}
\]
Thus $(P_{\circ})_n$ denotes the decoded root projection at $e_n$. The two encodings of a general contraction are transported separately when that label occurs at a left endpoint.

For each $g=1,\ldots,n$ and every $\ell\in\Lambda$ introduce a unitary $V_{g,\ell}$ with
\begin{equation}\label{eq:blockcopies}
 V_{g,\ell}=I-\Pact_g+e_{g-1}u_{g-1,\ell}
 +\sum_{a,b}\mathfrak x_g^a\mathfrak z_g^b e_gu_{g,\ell}\mathfrak z_g^{-b}\mathfrak x_g^{-a}.
\end{equation}
For this bridge, apply Lemma~\ref{lem:label} to the six projections
\[
 F_1,F_2,F_3,G_{12}(1),G_{13}(1),G_{23}(1)
\]
and the $|\Lambda|$ graph projections $G_{12}(V_{g,\ell})$. Thus there are $6+|\Lambda|$ refinements of each retained sign question per bridge, on each party. Each label has its own duplicated question; the unchanged opposite questions are always retained. There are $(n+1)|\Lambda|$ support equations and $n|\Lambda|$ block equations.

We show that the commutator of the original sign with $G_{12}(V_{g,\ell})$, in its $(c_0,f)$ block, is precisely
\begin{equation}\label{eq:transportcopies}
 {\btrans}_gu_{g-1,\ell}=u_{g,\ell}{\btrans}_g,
\end{equation}
with the same actual ${\btrans}_g$ for every label; distinct transported labels need not commute with each other. Equation~\eqref{eq:supportcopies} says $u_{r,\ell}(I-e_r)=I-e_r$. Multiplying by $u_{r,\ell}^*$ gives $u_{r,\ell}^*(I-e_r)=I-e_r$, and the adjoint of this is $(I-e_r)u_{r,\ell}=I-e_r$. Hence $u_{r,\ell}$ commutes with $e_r$ and is the identity on $(I-e_r)P_A\HH_A$. On the active space of bridge $g$, the $(a,b)$ term of \eqref{eq:blockcopies} is $t_{ab}(e_gu_{g,\ell}e_g)t_{ab}^*$, because $\mathfrak x_g,\mathfrak z_g$ are reflections and $\mathfrak x_g^a\mathfrak z_g^be_g=t_{ab}$. So $V_{g,\ell}$ is block diagonal for $e_{g-1}\oplus\bigoplus_{ab}m_{g,ab}$, with the block $u_{g-1,\ell}$ on the flag and transported copies of $u_{g,\ell}$ on the magic corners; in particular it reduces $e_{g-1}$ and $e_g$. The six frame refinements make the sign $\mathcal X_0$ of bridge $g$ commute with $F_1,F_2,F_3$ and $G_{ij}(1)$. These generate $\Mat_3\otimes I$ in the frame coordinates of Section~\ref{sec:projection}, whose commutant is $I_3\otimes B(\mathcal H_{\rm base})$, and the corners are frame-bound; hence ${\btrans}_g=I_3\otimes\widetilde {\btrans}_g$. The refinement by $G_{12}(V_{g,\ell})=\frac12(F_1+F_2+e_{12}\otimes V_{g,\ell}+e_{21}\otimes V_{g,\ell}^*)$ makes $\mathcal X_0$ commute with it, and the $(1,2)$ frame block gives $[\widetilde{\mathcal X}_0,V_{g,\ell}]=0$ on the base. With respect to $e_{g-1}\oplus e_g$,
\[
 \widetilde{\mathcal X}_0=\begin{pmatrix}0&\beta^*\\\beta&0\end{pmatrix},\qquad
 V_{g,\ell}=\begin{pmatrix}u_{g-1,\ell}&0\\0&u_{g,\ell}\end{pmatrix},
\]
and the lower-left block of the commutator is $\beta u_{g-1,\ell}-u_{g,\ell}\beta$. This is \eqref{eq:transportcopies}. Induction over $g$ gives $u_{n,\ell}v=vu_{0,\ell}$. Multiplying on the left by $u_{n,\ell}^*$ and on the right by $u_{0,\ell}^*$ gives $u_{n,\ell}^*v=vu_{0,\ell}^*$. The decoded copies in the table are polynomials in $e_r,u_{r,L},u_{r,L}^*$ with the same coefficients for every $r$, and $e_nv=v=ve_0$. Hence $L_nv=vL_0$ and $L_n^*v=vL_0^*$ for every decoded transported label, where $L_r$ is the decoded copy at $e_r$ with corner identity $e_r$, and the same holds for sums and products of decoded labels.

\paragraph{Extraction and exact translation.}
Introduce four centralizer contractions $a_\alpha=(a_{\alpha,+}+a_{\alpha,-})/2$ and impose $a_\alpha=e_na_\alpha e_0$. On the root corner extract $A_\alpha=v^*a_\alpha$. Supply the following polynomials to the graph compiler:
\begin{center}\small
\begin{tabular}{@{}ll@{}}
\toprule
Root identity & Centralizer identity\\
\midrule
$A_\alpha^*A_\beta=\delta_{\alpha\beta}I$ & $a_\alpha^*a_\beta=\delta_{\alpha\beta}e_0$\\
$\sum_\alpha A_\alpha A_\alpha^*=I-P_{\circ}$ & $\sum_\alpha a_\alpha a_\alpha^*=e_n-(P_{\circ})_n$\\
$L_{\rm end}A_\alpha=A_\beta R_{\rm end}$ & $(L_{\rm end})_na_\alpha=a_\beta (R_{\rm end})_0$\\
$L_{\rm end}A_\alpha=0$ & $(L_{\rm end})_na_\alpha=0$\\
$\sum_\iota L_\iota A_{\alpha_\iota}L'_\iota=0$ & $\sum_\iota (L_\iota)_na_{\alpha_\iota}(L'_\iota)_0=0$\\
Root-only polynomial & Same polynomial in source copies\\
\bottomrule
\end{tabular}
\end{center}
Expand the finite words, bind their parameters, apply graphs and triangles, and then Lemma~\ref{lem:rows}. Retain all FMS and refinement questions. Pad the two question sets if necessary and sample uniformly from all pairs, accepting every unspecified pair. The result has rational question distribution and a Boolean predicate. The support arguments prove soundness on a nonzero corner, using Alice alone; mirror equations are unnecessary.

Put $\HH_0=e_0P_A\HH_A$ and let $h_0$ be the restriction of $\mathfrak c_ne_0h_Ae_0$ to $\HH_0$. It is positive and trace class, it is faithful because $h_A$ is faithful on $P_A\HH_A$ and commutes with $e_0$, and it has trace one by \eqref{eq:121}; in particular $\HH_0\ne0$. Every source copy $L_0$ maps $\HH_0$ into itself and commutes with $h_0$, because $u_{0,L}$ commutes with $e_0$ and with $h_A$. The declared types hold: $e_0u_{0,L}$ is unitary on $\HH_0$; $e_0(I+u_{0,L})/2$ is a projection when $u_{0,L}^2=I$; $e_0(2I+u+u^*)/4$ is a positive contraction; and an average of two unitaries is a contraction. The contraction $a_\alpha=e_na_\alpha e_0$ commutes with $h_A$, so $A_\alpha=v^*a_\alpha$ maps $\HH_0$ into itself, and
\[
 h_0A_\alpha=\mathfrak c_n\,e_0h_Av^*a_\alpha=2^{-n}\mathfrak c_n\,v^*h_Aa_\alpha=2^{-n}\mathfrak c_n\,v^*a_\alpha e_0h_Ae_0=2^{-n}A_\alpha h_0,
\]
using $e_0v^*=v^*$, the adjoint $h_Av^*=2^{-n}v^*h_A$ of \eqref{eq:fourv}, $a_\alpha=a_\alpha e_0$ and $[e_0,h_A]=0$; this is $h_0A_\alpha=\nu A_\alpha h_0$. The translations are exact. For the creation equations, $A_\alpha^*A_\beta=a_\alpha^*vv^*a_\beta=a_\alpha^*e_na_\beta=a_\alpha^*a_\beta$. For the range sum, $\sum_\alpha A_\alpha A_\alpha^*=v^*(\sum_\alpha a_\alpha a_\alpha^*)v$. If $\sum_\alpha a_\alpha a_\alpha^*=e_n-(P_\circ)_n$, this equals $e_0-v^*(P_\circ)_nv=e_0-(P_\circ)_0$; conversely, conjugation by $v$ recovers the centralizer form, because $vv^*=e_n$ and $a_\alpha=e_na_\alpha$. For an endpoint equation, $v^*L_n=L_0v^*$ gives $L_0A_\alpha-A_\beta R_0=v^*(L_na_\alpha-a_\beta R_0)$. The bracket has range in $e_nP_A\HH_A$, where $v^*$ is isometric, so either side vanishes exactly when the other does. Sums $\sum_\iota L_\iota A_{\alpha_\iota}L'_\iota$ translate in the same way, with $(L_\iota)_n$ on the left. Root-only polynomials involve only source copies. Every application equation therefore holds as an operator identity on $\HH_0$.

\paragraph{One simultaneous realization.}
Let $(\HH_0,h_0)$ be a normal realization, and put $\HH_{\rm sc}=\mathbb C^{4n+1}\otimes\HH_0$ with the density
\begin{equation}\label{eq:scaffold}
 h_{\rm sc}=\operatorname{diag}(\mu_n)\otimes h_0
 =\frac1{\mathfrak c_n}\operatorname{diag}
 (h_0;\,2h_0^{\oplus4};\,4h_0^{\oplus4};\,\ldots;\,2^nh_0^{\oplus4}).
\end{equation}
It is faithful and trace class, of trace one by the definition of $\mathfrak c_n$. Use identity corner identifications in the bridges and $a_\alpha=|e_n\rangle\langle e_0|\otimes A_\alpha$; its commutation with $h_{\rm sc}$ is exactly the modular relation. The positive eigenspaces of $h_{\rm sc}$ are spanned by corner vectors tensored with positive eigenvectors of $h_0$, so they are finite-dimensional, and a bounded operator commuting with $h_{\rm sc}$ maps each of them into itself by (N4d). On such a block $\mathcal E$ the contraction $T=a_\alpha|_{\mathcal E}$ has a polar decomposition $T=V|T|$. Because $\dim\ker T=\dim\ker T^*$ in finite dimension, $V$ extends to a unitary $V_{\rm pol}$ with $T=V_{\rm pol}|T|$, and $|T|\pm i(I-|T|^2)^{1/2}$ are commuting unitaries, so
\[
 T=(U_++U_-)/2,\qquad
 U_\pm=V_{\rm pol}(|T|\pm i\sqrt{I-|T|^2}).
\]
The direct sums of the resulting unitaries over all blocks are unitaries commuting with $h_{\rm sc}$; they give centralizer encodings for all four $a_\alpha$ at once. Finite-dimensional blocks are needed here: a nonunitary isometry, such as a root isometry $A_\alpha$, is an extreme point of the unit ball and is therefore not an average of two unitaries. That is why $A_\alpha$ is never encoded directly and the contraction completion takes place on the scaffold. Encode each source contraction inside the source density blocks, choose $2P-I$ for projections, and use
\[
 u_+=2T_+-I+2i\sqrt{T_+(I-T_+)}
\]
for a positive contraction $T_+$. These encodings are functions of operators commuting with $h_0$, so they commute with $h_0$. Copy these same choices through the corner identifications and extend by identity outside the corners: the copies $u_{r,\ell}=|e_r\rangle\langle e_r|\otimes U_\ell+(I-|e_r\rangle\langle e_r|)\otimes I$ satisfy \eqref{eq:supportcopies}, the block unitaries \eqref{eq:blockcopies} act as $U_\ell$ on every corner of bridge $g$ and as $I$ elsewhere, and ${\btrans}_g=|e_g\rangle\langle e_{g-1}|\otimes I$ gives \eqref{eq:transportcopies}. Thus the block equations hold simultaneously. The translated equations hold because, for example, $L_na_\alpha-a_\beta R_0=|e_n\rangle\langle e_0|\otimes(LA_\alpha-A_\beta R)$. Append the scalar qutrit frame, use the graph formulas, and take actual joint products for all refinements; their block unitaries commute with the coarse FMS effects, and separate labels need not commute. Finally append the finite maximally entangled row factor of dimension $d_{\max}$. The scaffold and ancillary dimensions are finite, so they add only finite entropy.

All tests hold with one state and one assignment; the strategy is explicit. Take $\HH_A=\HH_B=\mathbb C^3\otimes\HH_{\rm sc}\otimes\mathbb C^{d_{\max}}$, the density $h=(I_3/3)\otimes h_{\rm sc}\otimes(I_{d_{\max}}/d_{\max})$ and the pure state $\Omega_h$ of (N8), formed in a product eigenbasis. Every synchronized question of Alice measures a projection commuting with $h$: corner, frame, graph, triangle and binding projections built from the centralizer unitaries by the formulas of Section~\ref{sec:projection}; the FMS cell, activity and corner projections; and the row projections of Lemma~\ref{lem:rows}. Bob measures their transposes. By (N8) every matching test passes. Every forbidden pair between centralizer events has probability $\Tr(hQR)=0$, because the defining identities give $QR=0$. The FMS sign questions are not in the centralizer. For bridge $g$, Alice's sign observable for $\mathsf T_x$ is the reflection $|f\rangle\langle c_x|+|c_x\rangle\langle f|$ on the span of the flag corner $e_{g-1}$ and the corner $c_x$, tensored with the identity of $\HH_0$. Bob's sign observable for $\mathsf B_y$ is $|f\rangle\langle r_y|+|r_y\rangle\langle f|$, with $r_y=(c_0+(-1)^yc_1)/\sqrt2$. Their retained answers are the remaining column-zero and row-zero projections. If $\mu$ is the mass of $e_{g-1}$, each magic corner has mass $2\mu$, and $\langle c_x,r_y\rangle=\epsilon_{xy}/\sqrt2$. So, up to the frame and row factors, the compressed state is $\Omega_{xy}=\mu^{1/2}(|f,f\rangle+\epsilon_{xy}|c_x,r_y\rangle)\otimes\Omega_{h_0}$. Substitution gives $(\mathcal X_x\otimes I)\Omega_{xy}=\epsilon_{xy}(I\otimes\mathcal Y_y)\Omega_{xy}$, so every forbidden sign pair has probability zero; the calibrated rules against $\mathsf C_0$ and $\mathsf R_0$ hold by construction. The FMS effects are corner operators tensored with the identity of $\HH_0$. They commute with every block unitary $V_{g,\ell}$ and with the frame, on Alice's side and, after transposition in the product eigenbasis, on Bob's. The refinements $E_aP^{\rm lab}$ of Lemma~\ref{lem:label} are therefore POVMs. The marginal entropy is $\log_23+S(h_{\rm sc})+\log_2d_{\max}$, where $S(h_{\rm sc})=H(\mu_n)+S(h_0)$.\qedhere

\end{proof}

\subsection{Chained bridges on normal commuting hosts}\label{sec:chain}
The same assembly works on arbitrary normal commuting supports. Here the edges may be any finite family and the range equation may carry a transported defect.

\Needspace{5\baselineskip}
\begin{lemma}[Chained bridges]\label{lem:chain}
For any fixed $n\geq1$, the conditional FMS predicates, individual label refinements and rational rows above compile the following finite root specifications on arbitrary normal commuting supports. The centralizer labels have prescribed types (projection, positive contraction, contraction or unitary). Each edge $A_\alpha$ has modular ratio $2^{-n}$, with isometry and range equations and endpoint relations $L_{\rm end}A_\alpha=A_\beta R_{\rm end}$ or $L_{\rm end}A_\alpha=0$; $L_{\rm end},R_{\rm end}$ are real rational polynomials in the typed centralizer labels (for example $R_{\rm end}=C^2/2$). Other relations are root centralizer polynomials with real rational coefficients. Every perfect strategy extracts that root specification on a nonzero corner. Every faithful normal root realization extends simultaneously to a perfect strategy after finite amplification, preserving its factor type and, for a spatial finite-entropy model, finiteness of entropy. Only labels occurring at a left endpoint (including a range-equation label) need transport; transporting further labels is permitted and changes neither direction.
\end{lemma}
\begin{proof}
Use the synchronized partition
\[
 e_0+\sum_{g=1}^n m_g=I,\qquad
 m_g=\sum_{\epsilon,\delta\in\{0,1\}}m_{g,\epsilon\delta},\qquad e_g=m_{g,00}.
\]
For bridge $g$ use a conditional FMS copy with flag $e_{g-1}$, magic support $m_g$, and activity $\Pact_g=e_{g-1}+m_g$. The inactive effect is $I-\Pact_g$. Match activity on every conditional question and bind the flag and four chosen magic events to their named binary projections, in both party orientations. The flag/four-magic/inactive column partition then has precisely these six marginals. Different copies have disjoint question names; no agreement between their nonsynchronous signs is imposed.

The commuting bridge of Section~\ref{sec:bridgecomm} applies in each $\Pact_gM\Pact_g$. Indeed $\Pact_g$ is centralizer, so its normalized restriction has the restricted modular group. It gives ${\btrans}_g^*{\btrans}_g=e_{g-1}$, ${\btrans}_g{\btrans}_g^*=e_g$ and $\sigma_t^\varphi({\btrans}_g)=2^{it}{\btrans}_g$. These equations extend by zero off $\Pact_g$ and imply
\[
 v={\btrans}_n\cdots {\btrans}_1,\quad
 v^*v=e_0,\quad
 vv^*=e_n,\quad
 \sigma_t^\varphi(v)=2^{nit}v.
\]
This is an extracted operator, never a centralizer word passed to the graph compiler. KMS and the magic matrix units give
\[
 \varphi(m_{g,\epsilon\delta})=2^g\varphi(e_0),\qquad
 \varphi(e_0)=\mathfrak c_n^{-1}.
\]
If an active corner vanishes its flag and all four magic corners vanish together; otherwise the local bridge makes all five nonzero. The recurrence therefore holds also before nonvanishing is known, and the cover proves that $e_0\ne0$.

Bind every corner and every positive Pauli parameter to the same qutrit frame. Let $\Lambda$ be the unitary encodings of the labels to be transported, including every label used on a left endpoint: use a unitary directly, a reflection for a projection, the displayed functional calculus for a positive contraction, and two unitaries for a general contraction. Introduce the axis labels $\mathfrak x_g,\mathfrak z_g$ of each bridge as in Lemma~\ref{lem:four}. For each $\ell\in\Lambda$ and $r=0,\ldots,n$, impose $(u_{r,\ell}-I)(I-e_r)=0$, and introduce the unitary label $V_{g,\ell}$ of \eqref{eq:blockcopies}. The orthogonal block supports make this a unitary. At bridge $g$ apply Lemma~\ref{lem:label} separately to the six frame projections and to each $G_{12}(V_{g,\ell})$. There is one refinement of each retained sign question per label per bridge, on each party, always against the unchanged opposite questions. There is no joint refinement of distinct labels. The frame refinements give ${\btrans}_g=I_3\otimes\widetilde {\btrans}_g$. In the selected magic--flag block the graph commutator is exactly
\[
 {\btrans}_gu_{g-1,\ell}=u_{g,\ell}{\btrans}_g.
\]
Induction gives $(L_{\rm end})_nv=v(L_{\rm end})_0$ for every decoded transported label and its adjoints and words. Labels used only at the source need one source encoding and no refinement.

For each edge introduce a centralizer contraction $a_\alpha=e_na_\alpha e_0$ and extract $A_\alpha=v^*a_\alpha$. Compile the centralizer equations
\[
 a_\alpha^*a_\beta=\delta_{\alpha\beta}e_0,\qquad
 \sum_\alpha a_\alpha a_\alpha^*=e_n-(P_\circ)_n,
 \qquad (L_{\rm end})_na_\alpha=a_\beta (R_{\rm end})_0
\]
for the corresponding root equations; the zero endpoint uses zero on the right. The equations with a specified initial defect use that defect in the first right side. More generally a range equation $\sum_\alpha A_\alpha A_\alpha^*=R_{\rm ran}$, with $R_{\rm ran}$ a transported centralizer polynomial, is compiled as $\sum_\alpha a_\alpha a_\alpha^*=(R_{\rm ran})_n$. For the Catalan input of \cite[Section 6]{Entropy} this is $R_{\rm ran}=I-BB^*-P$; its centralizer contraction $B$ and projection $P$ are already in the transported list. Multiplying by $v$ or its adjoint proves both directions, since its initial and final projections are $e_0,e_n$. Every root-only polynomial stays at $e_0$. In particular the edge has modular ratio $2^{-n}$ and all labels refer to the same extracted root operators. Graphs, triangles and Lemma~\ref{lem:rows} enforce these finitely many centralizer polynomials without changing the conditional tests. This proves commuting soundness.

For completeness take a faithful normal root $(M_0,\varphi_0)$ and on $\Mat_{1+4n}\vN M_0$ put
\[
 \varphi_{\rm sc}=\operatorname{diag}(\mu_n)\otimes\varphi_0
 =\mathfrak c_n^{-1}\operatorname{diag}
 (1;\,2^{\oplus4};\,4^{\oplus4};\,\ldots;\,(2^n)^{\oplus4})
 \otimes\varphi_0.
\]
Choose the actual matrix corner maps for ${\btrans}_g$, the ideal real FMS effects on $\Pact_g$, and $I-\Pact_g$ for every inactive answer. Take $a_\alpha=|e_n\rangle\langle e_0|\otimes A_\alpha$. Its modular ratio is $2^n2^{-n}=1$, so it lies in the finite centralizer. Its polar partial isometry therefore has a unitary completion, giving its two-unitary encoding. Choose the root centralizer encodings once and copy them identically to the required corners. All $V_{g,\ell}$ then commute with the FMS effects at their own bridge, even though adjacent activities overlap at $e_g$. Use their actual joint products for the separate refinements. Append the common scalar frame and row factors and use standard commutant mirrors as in Lemma~\ref{lem:rows}. All old conditional probabilities and every added row are simultaneously correct. Only finite matrix factors have been appended. Their state entropy is finite; in the spatial case it adds to the root entropy. This proves completeness and the stated preservation properties.
\end{proof}

\subsection{A dimension-independent completeness estimate}\label{sec:approxcompiler}
The exact construction gives a useful quantitative statement for the four-bridge compiler ($n=4$). Keep the label types, centralizer conditions, contractivity and modular covariance exact, but allow residuals $r_j$ in the application equations of Definition~\ref{def:admissible}. Write $\|T\|_{2,h_0}^2=\Tr(h_0T^*T)$. The constructed strategy satisfies
\begin{equation}\label{eq:generalestimate}
 1-\omega(G,\mathcal S)\leq C_{\rm comp}\sum_j\|r_j\|_{2,h_0}^2,
\end{equation}
where $C_{\rm comp}$ depends only on the finite presentation and the compiler.

Here is the proof, including the density factors. The same scaffold, exact unitary encodings and actual products make all structural tests perfect. Only the final application row-zero tests may lose. Let $p_j$ be the lifted polynomial residual. For a source or root-polynomial row it is the source copy of $r_j$; for an endpoint row it is $vr_j$; for a range row it is $vr_jv^*$. Consequently
\[
 \|p_j\|_{2,h_{\rm sc}}^2=\frac{\beta_j}{121}\|r_j\|_{2,h_0}^2,
 \qquad \beta_j=\begin{cases}16,&\text{range row at }e_4,\\1,&\text{otherwise}.\end{cases}
\]
For the endpoint case use $v^*v=e_0$. For the range case use $v^*h_{\rm sc}v=16h_0/121$.

The qutrit graph row $\Delta_j$ has off-diagonal blocks $p_j,p_j^*$. Since $p_j$ commutes with $h_{\rm sc}$,
\[
 \Tr(h_{\rm old}\Delta_j^2)
 =\frac13\Tr[h_{\rm sc}(p_j^*p_j+p_jp_j^*)]
 =\frac{2\beta_j}{363}\|r_j\|_{2,h_0}^2.
\]
Apply \eqref{eq:rowloss}, with $L_j$ and $d_j$ the denominator and slot dimension of row $j$. An orientation of the row-zero test is a choice of which party receives the row question, and each orientation has probability \eqref{eq:rowloss}. If the forbidden pair is imposed in both ordered question pairs, as in the formal compiler of Appendix~\ref{app:leanA}, both contributions count. For uniform $Q$-by-$Q$ sampling one may therefore take
\begin{equation}\label{eq:generalconstant}
 C_{\rm comp}=\max_j\frac{16L_j^2\beta_j}{363d_j^3Q^2}.
\end{equation}
With one oriented row-zero test per row, the convention of \eqref{eq:valueconstant}, the coefficient $16$ may be replaced by $8$. Summing event probabilities gives an upper bound even if rejecting rules overlap. The dimension, clock order, density spectrum and spectator dimensions do not enter this constant. The estimate applies to the constructed strategies, whose structural tests are exact.

\section{The rooted tree and computation squares}\label{sec:tree}
Sections~\ref{sec:tree}--\ref{sec:reduction} prove the hardness half of Theorem~A. The game is the four-bridge compilation of one admissible presentation, the Fock clock. Its root carries a four-ary tree, each depth of the tree is a finite square, and a deterministic computation runs on the squares. A Weyl clock supplies the existential witness in Section~\ref{sec:clock}.

\subsection{Normality removes every rootless sector}
Require four isometries and a root projection with
\begin{equation}\label{eq:fockrelations}
 A_{ab}^*A_{cd}=\delta_{ac}\delta_{bd}I,\qquad
 \sum_{ab}A_{ab}A_{ab}^*=I-P_{\circ},\qquad h_0A_{ab}=A_{ab}h_0/16.
\end{equation}
Here $h_0$ is a faithful density on the extracted root-corner Hilbert space. For $w=w_1\cdots w_m$ set $A_w=A_{w_1}\cdots A_{w_m}$ (the empty product is $I$) and put $Q_m=\sum_{|w|=m}A_wA_w^*$. Orthogonality and trace cyclicity give $\Tr(h_0Q_m)=4^{-m}$. The projections $Q_m$ decrease. Their strong limit has zero $h_0$-weight, hence is zero by faithfulness. Also $Q_m-Q_{m+1}=\sum_{|w|=m}A_wP_{\circ}A_w^*$. Therefore
\begin{equation}\label{eq:fockdecomp}
 \HH=\bigoplus_w A_w\KK\simeq\ell^2(\WW)\otimes\KK,
 \qquad \KK=P_{\circ}\HH\ne0.
\end{equation}
Here $\WW$ denotes finite words in the four-letter alphabet $\{0,1\}^2$. The modular equation makes $h_0$ commute with every $A_wP_{\circ}A_w^*$. Since $\Tr(h_0P_{\circ})=3/4$, transporting its root block gives
\begin{equation}\label{eq:fockdensity}
 h_0=h_F\otimes\rho_{\rm spec},\qquad
 h_F=\frac34\sum_w16^{-|w|}|w\rangle\langle w|,
 \qquad \Tr\rho_{\rm spec}=1.
\end{equation}
In particular
\begin{equation}\label{eq:fockentropy}
 S(h_F)=\log_2(4/3)+4/3.
\end{equation}
Indeed the depth-$m$ mass is $(3/4)4^{-m}$ and its mean depth is $1/3$; the log probability at each word is $\log_2(4/3)+4m$. This proves the formula and excludes a rootless normal summand. Section~\ref{sec:generaltree} repeats this calculation at a general ratio; there $h_F=h_{F,1/16}$.

\subsection{Each depth is a finite square}
For $w=(a_1,b_1)\cdots(a_m,b_m)$ define
\[
 x(w)=\sum_{j=1}^m a_j2^{j-1},\qquad
 y(w)=\sum_{j=1}^m b_j2^{j-1}.
\]
The first letter is the least significant bit. Prefixing $A_{ab}$ maps $(x,y)$ to $(2x+a,2y+b)$. Depth $m$ is a $2^m\times2^m$ square. Require centralizer unitaries $U_x,U_y$ and projections $P_x,P_y$ satisfying, for $a,b\in\{0,1\}$,
\begin{align}
 U_xA_{0b}&=A_{1b},&U_xA_{1b}&=A_{0b}U_x,&U_xP_{\circ}&=P_{\circ},\label{eq:U_x}\\
 U_yA_{a0}&=A_{a1},&U_yA_{a1}&=A_{a0}U_y,&U_yP_{\circ}&=P_{\circ},\label{eq:U_y}\\
 P_xA_{0b}&=A_{0b}P_x,&P_xA_{1b}&=0,&P_xP_{\circ}&=P_{\circ},\label{eq:P_x}\\
 P_yA_{a0}&=A_{a0}P_y,&P_yA_{a1}&=0,&P_yP_{\circ}&=P_{\circ}.\label{eq:P_y}
\end{align}
Induction on word length identifies $U_x$ and $U_y$ as cyclic increments of $x$ and $y$ respectively, acting identically on $\KK$. The same induction makes $P_x,P_y$ the indicators of $x=0,y=0$. In particular $U_x,U_y$ commute, and
\[
 P_{x,\rm last}=U_x^*P_xU_x,\qquad P_{y,\rm last}=U_y^*P_yU_y
\]
mark the final column and row. These identities hold on the full root multiplicity space.

\subsection{A finite deterministic tableau}\label{sec:tableau}
Choose a deterministic radius-one cellular automaton on the half-line. Its finite alphabet $\Gamma_{\rm CA}$ includes start $\mathtt s$, blank $\mathtt b$ and halt $\mathtt H$, and its update is $\Phi(\ell,c,r,\varepsilon)$ with left boundary symbol $\#$ and external pulse $\varepsilon\in\{0,1\}$. The initial row is $\mathtt s,\mathtt b,\mathtt b,\ldots$. We use these conventions:
\begin{itemize}
\item $\Phi(\mathtt b,\mathtt b,\mathtt b,\varepsilon)=\mathtt b$ for both pulse values.
\item The initial start state ignores the pulse at time zero. Later waiting-front states respond to positive-time pulses.
\item Cells passed by the waiting front keep their unary marks. Frozen cells and all states of the later input-checking computation ignore subsequent pulses.
\item A transition using the pulse at time $y$ produces row $y+1$. Thus the pulse at time ${d_{\rm clk}}$ records the front at coordinate ${d_{\rm clk}}$ in that next row.
\end{itemize}
Here is an explicit front layer, with alphabet $\{\mathtt s,\mathtt b,\mathtt f,\mathtt z,\mathtt H\}$; $\mathtt f$ is the front state. The right neighbour is unused in these six rules:
\[
\begin{array}{c|l}
\text{condition}&\text{next centre}\\\hline
c=\mathtt s&\mathtt b\\
c=\mathtt f&\mathtt z\text{ if }\varepsilon=1,\ \mathtt b\text{ if }\varepsilon=0\\
c=\mathtt b,\ \ell=\mathtt s&\mathtt f\\
c=\mathtt b,\ \ell=\mathtt f&\mathtt f\text{ if }\varepsilon=0,\ \mathtt b\text{ if }\varepsilon=1\\
c=\mathtt z&\mathtt z\\
c=\mathtt H&\mathtt H
\end{array}
\]
All remaining blank centres stay blank. At time $y\geq1$, until a return, the unique front is at $x=y$. The first positive pulse at ${d_{\rm clk}}$ writes $\mathtt z$ at $({d_{\rm clk}},{d_{\rm clk}}+1)$ and suppresses the front's successor. The frozen mark's distance from the left boundary is the unary input ${d_{\rm clk}}$; blank cells between the boundary and the mark represent its unary positions. Pulse-dependence is confined to neighbourhoods containing the front. In particular a blank right of $\mathtt f$ may read the pulse.

Add a persistent unary bit $u\in\{0,1\}$, initially zero, with $u'=1$ when the old front centre is $\mathtt s$ or $\mathtt f$, and $u'=u$ otherwise. Thus passed cells retain their unary marks on a separate track, even though their front-layer component is blank. Take the product alphabet $\{\mathtt s,\mathtt b,\mathtt f,\mathtt z,\mathtt H\}\times\{0,1\}\times\mathcal T_{\rm TM}$, with a finite TM alphabet $\mathcal T_{\rm TM}=\mathcal A_{\rm tape}\times(Q_{\rm TM}\cup\{\mathrm{no\ head}\})$ for the fixed one-tape half-line program in Section~\ref{sec:reduction}. All TM cells start blank without a head. The transition $\mathtt f\to \mathtt z$ initializes its head in the start state at that cell. Thereafter a centre carrying $(a,p)$ writes the symbol prescribed by $\delta(p,a,u)$ and retains a head only for a stationary move. A head moving right (left) is received by the right (left) neighbour, whose old tape symbol is retained. A centre receiving no head retains its symbol without a head. Invalid multiple incoming heads may be sent to a fixed error state; they never occur from this initialization. The left boundary $\#$ is read as a tape end marker and no transition moves left of it. The frozen front track is read-only input for this TM. A halting head writes a designated halt symbol $\mathtt H$ in the combined alphabet and keeps it. These prescriptions give a finite radius-one lookup table effectively from the finite TM transition table; all subsequent TM transitions ignore the pulse. Denote this finite product alphabet, with the halt convention, by $\Gamma_{\rm CA}$. In the combined alphabet, $\mathtt s$ and $\mathtt b$ abbreviate the start and fully blank product states. Collapse any product state with a halting head to the single absorbing symbol $\mathtt H$.

Blank quiescence holds before a front or head arrives. The $\mathtt s$ rule ignores the time-zero pulse and seeds $\mathtt f$ at time one via its right neighbour. The mark is permanent and later pulses cannot restart a head. Finally, all disturbances start at the origin, or at the marked front within its radius-one cone, so no nonblank cell can occur at $x>y$. These checks give all four conventions and the blank right ghost used below.

Let $(P_c)_{c\in\Gamma_{\rm CA}}$ be a centralizer PVM and $P_{\rm ret}$ a centralizer projection, with $P_{{\rm ret},0}=I-P_{\rm ret}$, $P_{{\rm ret},1}=P_{\rm ret}$. Require
\begin{equation}\label{eq:initial}
 P_{\mathtt s}P_xP_y=P_xP_y,\qquad P_{\mathtt b}(I-P_x)P_y=(I-P_x)P_y,\qquad P_{\mathtt H}=0.
\end{equation}
Define centralizer polynomials
\begin{align*}
 L_\#&=P_x,& L_c&=U_xP_cU_x^*(I-P_x),\\
 N_r&=U_x^*P_rU_x(I-P_{x,\rm last})+\mathbf1_{\{r=\mathtt b\}}P_{x,\rm last}.
\end{align*}
For every $a\in\Gamma_{\rm CA}$ impose
\begin{equation}\label{eq:update}
 \left[U_y^*P_aU_y-\sum_{\Phi(\ell,c,r,\varepsilon)=a}
 L_\ell P_cN_rP_{{\rm ret},\varepsilon}\right](I-P_{y,\rm last})=0.
\end{equation}
The factors have the displayed order. They need not commute globally. The factor $I-P_{y,\rm last}$ removes a spurious transition from the final row to row zero. The ghost to the right of the last column is blank.

\begin{lemma}[Definite computation]\label{lem:tableau}
Suppose a root vector $\xi$ belongs to a sector where the clock pulse on every coordinate has a specified scalar value. Then every $P_c$ acts on $A_w\xi$ as the indicator of the genuine classical computation at $(x(w),y(w))$.
\end{lemma}
\begin{proof}
Work at a fixed depth $m$ and induct on time, for every spatial coordinate at once. Equation~\eqref{eq:initial} gives row zero. On an induction vector every factor $L_\ell,P_c,N_r,P_{{\rm ret},\varepsilon}$ acts by its classical zero or one. Their ordered product therefore selects exactly one update. Equation~\eqref{eq:update} gives $U_y^*P_aU_y$ with the correct scalar on that vector. Multiplying by $U_y$ gives the next-row assertion.

The half-line and the square agree throughout this induction. At time $y$ no disturbance has travelled beyond coordinate $y$. In particular the ghost cell $2^m$ is blank for all times $y\leq2^m-1$, by radius one and quiescence under either pulse. The left ghost is always $\#$. This proves the genuine-prefix assertion. A halting event at finite coordinates and time occurs in every sufficiently large square and contradicts $P_{\mathtt H}=0$.
\end{proof}

The tableaux at distinct depths need not intertwine with the creation operators. Determinism forces each of them to the same finite prefix; soundness needs only one sufficiently large square. Every depth has positive normal weight by \eqref{eq:fockdensity}.

\subsection{A worked front and mask example}\label{sec:toy}
Table~\ref{tab:toy} shows the depth-$3$ square for a clock of order three; the verification software regenerates it (Appendix~\ref{sec:repro}). Its $\mathtt d$ denotes a passed cell and $\mathtt m$ a frozen mark. These persistent annotations differ from the minimal front layer above, whose passed cells are blank and whose mark is $\mathtt z$; both put the mark at $(3,4)$. The start state $\mathtt s$ ignores the pulse at time zero. The pulse at $y=3$ produces the mark in row 4; the later pulse at $y=6$ has no effect.
\begin{table}[ht]
\centering\small
\begin{tabular}{@{}cc|*{8}{>{\ttfamily}c}|c@{}}
\toprule $y$&pulse&0&1&2&3&4&5&6&7&update mask\\\midrule
0&1&s&b&b&b&b&b&b&b&1\\
1&0&d&f&b&b&b&b&b&b&1\\
2&0&d&d&f&b&b&b&b&b&1\\
3&1&d&d&d&f&b&b&b&b&1\\
4&0&d&d&d&m&b&b&b&b&1\\
5&0&d&d&d&m&b&b&b&b&1\\
6&1&d&d&d&m&b&b&b&b&1\\
7&0&d&d&d&m&b&b&b&b&0\\
\bottomrule
\end{tabular}
\caption{Depth $m=3$, period $q=3$: the decoder's $8\times8$ toy tableau.
$P_x$ marks column 0, $P_{x,\rm last}$ column 7, $P_y$ row 0 and
$P_{y,\rm last}$ row 7. The last-column neighbour is replaced by blank;
the last-row mask suppresses the toroidal $7\to0$ update.}\label{tab:toy}
\end{table}

\section{A finite-order clock and its period reader}\label{sec:clock}
\subsection{A determinant supplies a finite witness}
Add centralizer unitaries $J_{\rm W},K_{\rm W},Z_{\rm clk}$ with
\begin{equation}\label{eq:weyl}
\begin{gathered}
 J_{\rm W}K_{\rm W}=Z_{\rm clk}K_{\rm W}J_{\rm W},\qquad
 [Z_{\rm clk},J_{\rm W}]=[Z_{\rm clk},K_{\rm W}]=0,\\
 [J_{\rm W},A_{ab}]=[K_{\rm W},A_{ab}]=[Z_{\rm clk},A_{ab}]=0.
\end{gathered}
\end{equation}
A unitary commuting with an isometry also commutes with its adjoint: its inverse commutes with the isometry, and taking adjoints gives the assertion. Thus these labels reduce $\KK=P_{\circ}\HH$ and commute with $\rho_{\rm spec}$. Their action on every word sector is their root action.

Every positive eigenspace of $\rho_{\rm spec}$ is finite-dimensional. It reduces $J_{\rm W},K_{\rm W},Z_{\rm clk}$ and decomposes into $Z_{\rm clk}$ eigenspaces. On a nonzero such eigenspace of dimension $d$, with phase $\zeta$, the Weyl relation gives
\begin{equation}\label{eq:det}
 \det(J_{\rm W}K_{\rm W})=\zeta^d\det(K_{\rm W}J_{\rm W}),\qquad \zeta^d=1.
\end{equation}
Since the positive eigenspaces exhaust $\KK$, $Z_{\rm clk}$ has pure point spectrum there and every phase is a root of unity. At least one sector is nonzero. Conversely a ${d_{\rm clk}}$-dimensional Weyl pair realizes every primitive ${d_{\rm clk}}$th root, with density $I_{d_{\rm clk}}/{d_{\rm clk}}$.

\subsection{The clock and a counter without a positive lower bound}
Take a centralizer unitary $W$ and a positive centralizer contraction $C_{\rm ctr}$ with
\begin{align}
 WP_{\circ}&=P_{\circ},&WA_{ab}&=A_{ab}Z_{\rm clk}^bW^2,\label{eq:W}\\
 C_{\rm ctr}P_{\circ}&=P_{\circ},&C_{\rm ctr}A_{ab}&=2^{-b}A_{ab}C_{\rm ctr}^2.\label{eq:C}
\end{align}
Also require $Z_{\rm clk}$ to commute with $U_x,U_y,P_x,P_y,W,C_{\rm ctr},P_{\rm ret},T_{\rm inv}$ and all $P_c$. Induction on word length, using the coordinate recurrence, gives
\begin{equation}\label{eq:coordinateclock}
 W|_{(x,y)}=Z_{\rm clk}^y,\qquad C_{\rm ctr}|_{(x,y)}=2^{-y}I.
\end{equation}
In particular $C_{\rm ctr}$ is injective even though its spectrum accumulates at zero.

For a centralizer projection $P_{\rm ret}$ and contraction $T_{\rm inv}$, impose
\begin{equation}\label{eq:reader}
 (W-I)P_{\rm ret}=0,\qquad T_{\rm inv}(W-I)=\tfrac12 C_{\rm ctr}(I-P_{\rm ret}).
\end{equation}
The first equation puts $\ran P_{\rm ret}$ in $\ker(W-I)$. For $\xi$ in that kernel the second gives $C_{\rm ctr}(I-P_{\rm ret})\xi=0$. Injectivity of $C_{\rm ctr}$ gives $P_{\rm ret}\xi=\xi$. Thus
\begin{equation}\label{eq:return}
 P_{\rm ret}=\operatorname{proj}\ker(W-I).
\end{equation}
On a root phase sector of order ${d_{\rm clk}}$, the pulse is exactly $\mathbf1_{\{{d_{\rm clk}}\mid y\}}$ at every spatial coordinate. It cannot vanish spuriously or postpone the first return.

\subsection{A bounded inverse on every prime clock}
\begin{lemma}[A good primitive phase]\label{lem:goodphase}
For every prime ${d_{\rm clk}}$ there is a primitive ${d_{\rm clk}}$th root $\zeta$ such that
\[
 |\zeta^t-1|\geq2^{-t-1}\quad(t\geq1,\ {d_{\rm clk}}\nmid t).
\]
\end{lemma}
\begin{proof}
For $1\leq r<{d_{\rm clk}}$,
\[
 |e^{2\pi ir/{d_{\rm clk}}}-1|\geq4\min(r,{d_{\rm clk}}-r)/{d_{\rm clk}}.
\]
Fix $1\leq t<{d_{\rm clk}}$. Multiplication by $t$ permutes nonzero residues. A phase $e^{2\pi ik/{d_{\rm clk}}}$ violating the required inequality must satisfy $\min(tk\bmod {d_{\rm clk}},{d_{\rm clk}}-(tk\bmod {d_{\rm clk}}))<{d_{\rm clk}}2^{-t-3}$. There are at most ${d_{\rm clk}}2^{-t-2}$ such residues. Summing over $t$ excludes fewer than ${d_{\rm clk}}/4$ choices, while ${d_{\rm clk}}/4<{d_{\rm clk}}-1$. One of the ${d_{\rm clk}}-1$ nontrivial phases works. For $t\geq {d_{\rm clk}}$ reduce to a nonzero residue $r<{d_{\rm clk}}$; its lower bound $2^{-r-1}$ is at least $2^{-t-1}$.
\end{proof}

For this phase set
\begin{equation}\label{eq:D}
 T_{\rm inv}|_{(x,y)}=
 \begin{cases}
 2^{-y-1}(\zeta^y-1)^{-1}I,&{d_{\rm clk}}\nmid y,\\
 0,&{d_{\rm clk}}\mid y.
 \end{cases}
\end{equation}
Lemma~\ref{lem:goodphase} gives $\|T_{\rm inv}\|\leq1$. Equations~\eqref{eq:coordinateclock} and \eqref{eq:return} then verify \eqref{eq:reader}. The phase belongs to the strategy, so the game coefficients remain rational.

\section{Spatial hardness: the reduction and value one}\label{sec:reduction}
Fix any decidable predicate $\Rdec(e,j,t)$ and the set $\Adec=\{e:\exists j\,\forall t\ \Rdec(e,j,t)\}$. We effectively choose the cellular automaton in Section~\ref{sec:tree} to run the following program, with $e$ hardwired.

\paragraph{Read a positive-time return.}
At time $y$ a waiting front is at position $y$. Ignore the time-zero pulse. At the first positive-time pulse, freeze the front and retain its coordinate as unary input ${d_{\rm clk}}$. With the transition convention above the mark is written into row ${d_{\rm clk}}+1$ at position ${d_{\rm clk}}$. Test ${d_{\rm clk}}$. If ${d_{\rm clk}}=1$ or ${d_{\rm clk}}$ is composite, halt. Otherwise compute the unique $j\geq0$ with ${d_{\rm clk}}=p_{j+1}$, the $(j+1)$st prime. Then evaluate $\Rdec(e,j,0),\Rdec(e,j,1),\ldots$ and halt at the first false answer. All later computation states ignore the external pulse.

For a $\Sigma^0_2$-complete choice of $\Adec$, there is no computable bound on a sufficient clock order as a function of the input game: such a bound would leave only finitely many possible $j$ in this reduction, making $\exists j\,\forall t\ \Rdec(e,j,t)$ a $\Pi^0_1$ predicate. Indeed one tests, for every finite time cutoff, whether some one of those finitely many candidates survives all checks so far; if none survives forever, all have failed by a common finite cutoff. This contradicts $\Sigma^0_2$-hardness.

This finite program has an effective finite radius-one simulation. The initial-front rule, frozen unary track and ordinary tape/head tracks give the stated boundary conventions. We do not require a bound on its running time. Each completed false check causes the halt symbol after finitely many updates.

\paragraph{The clock presentation and its labels.}
Include the Fock, coordinate, clock, reader and tableau equations above, all typing relations, and the stated $Z_{\rm clk}$ commutators. They form an admissible finite presentation with $n=4$. In the four-bridge compiler of Lemma~\ref{lem:four} the transported typed list is exactly
\begin{equation}\label{eq:labelset}
 P_{\circ},\ U_x,\ U_y,\ P_x,\ P_y,\ W,\ C_{\rm ctr},\ Z_{\rm clk},\ J_{\rm W},\ K_{\rm W}.
\end{equation}
Table~\ref{tab:labels} records the label types. Here $P_{\circ},P_x,P_y$ are projections, $C_{\rm ctr}$ is positive, and the other six are unitary. Thus one unitary encoding per typed label suffices, so $|\Lambda|=10$. There are fifty support equations, forty block equations and sixteen refinements per bridge per party. Source-only labels $P_{\rm ret},T_{\rm inv},P_c$ do not occur at left endpoints of an $A_\alpha$ relation and need no transport. For example the transported polynomials include
\[
 W_4a_{ab}=a_{ab}Z_{{\rm clk},0}^bW_0^2,\qquad
 C_{{\rm ctr},4}a_{ab}=2^{-b}a_{ab}C_{{\rm ctr},0}^2,\qquad
 J_{{\rm W},4}a_{ab}=a_{ab}J_{{\rm W},0}.
\]
The source-only return equations and tableau equations are centralizer polynomials. Lemma~\ref{lem:four} therefore computes a finite rational Boolean game $\Gclk_e$.

\begin{table}[ht]
\centering\small
\begin{tabular}{@{}llll@{}}
\toprule Label & Type & Transported? & Use\\\midrule
$P_{\circ}$ & projection & yes & Fock root\\
$U_x,U_y$ & unitary & yes & coordinate increments\\
$P_x,P_y$ & projection & yes & boundaries\\
$W$ & unitary & yes & clock\\
$C_{\rm ctr}$ & positive contraction & yes & counter\\
$Z_{\rm clk},J_{\rm W},K_{\rm W}$ & unitary & yes & Weyl clock\\
$P_{\rm ret}$ & projection & no & pulse\\
$T_{\rm inv}$ & general contraction & no & reader inverse\\
$P_c$ & projection & no & tableau\\
$a_\alpha$ & general contraction & scaffold only & lifted creation\\
\bottomrule
\end{tabular}
\caption{Typed application labels of the clock presentation. Transported labels are copied through all bridges for use at a left endpoint.}\label{tab:labels}
\end{table}

\begin{figure}[ht]
\centering\small
\begin{tabular}{@{}c@{}}
Presentation $\xrightarrow{\text{Lemma~\ref{lem:four}}}$ finite game $\Gclk_e$\\[4pt]
Perfect normal strategy $\longrightarrow$ normal realization\\[2pt]
$\longrightarrow$ Fock tree $\longrightarrow$ finite clock sector\\[2pt]
$\longrightarrow$ deterministic tableau $\longrightarrow\forall t\ \Rdec(e,j,t)$\\[4pt]
Witness $j$ $\longrightarrow$ Fock/Weyl model\\[2pt]
$\longrightarrow$ scaffold, frame, row ancilla $\longrightarrow$ perfect strategy
\end{tabular}
\caption{Dependencies of the spatial hardness proof. Compactness and Bell attainment have separate upper-bound arguments.}\label{fig:dependencies}
\end{figure}

\begin{theorem}\label{thm:reduction}
The map $e\mapsto \Gclk_e$ satisfies
\[
 \Gclk_e\in\perf_{\qs}\ \Longleftrightarrow\
 \Gclk_e\in\perf_{\finS}\ \Longleftrightarrow\
 \exists j\,\forall t\ \Rdec(e,j,t).
\]
No $\Gclk_e$ has a perfect finite-dimensional strategy.
\end{theorem}
\begin{proof}
A perfect normal strategy extracts a nonzero faithful root realization. The determinant argument supplies a nonzero root phase sector of finite order ${d_{\rm clk}}$. By \eqref{eq:weyl} its word descendants persist at all depths. Equations~\eqref{eq:W}--\eqref{eq:return} force the exact return sequence, with its first positive return at ${d_{\rm clk}}$. Lemma~\ref{lem:tableau} needs only that the initial root vector lie in this $Z_{\rm clk}$-eigenspace: the clock pulse is then scalar on all its word descendants. The imposed $[Z_{\rm clk},P_c]=0$ relations are redundant for this induction. An invalid ${d_{\rm clk}}$ halts. For valid ${d_{\rm clk}}=p_{j+1}$ any false $\Rdec(e,j,t)$ also halts. A sufficiently large square would contain that halt, contrary to $P_{\mathtt H}=0$. Hence the extracted $j$ witnesses membership in $\Adec$.

Conversely take a witness $j$ and set ${d_{\rm clk}}=p_{j+1}$. Use \eqref{eq:fockdensity} with $\rho_{\rm spec}=I_{d_{\rm clk}}/{d_{\rm clk}}$, a good primitive phase and its Weyl pair, the canonical coordinate permutations, \eqref{eq:coordinateclock}, \eqref{eq:return}, \eqref{eq:D}, and the classical nonhalting tableau. Every required operator is bounded and commutes with the density when so specified. This is a simultaneous realization. On $\mathbb C^{d_{\rm clk}}$ one may take $J_{\rm W}|i\rangle=\zeta^i|i\rangle$, $K_{\rm W}|i\rangle=|i+1\bmod {d_{\rm clk}}\rangle$, and $Z_{\rm clk}=\zeta I$. Its root entropy and largest Schmidt probability are
\begin{equation}\label{eq:witnessentropy}
 S(h)=\log_2{d_{\rm clk}}+\log_2(4/3)+4/3,\qquad p_{\max}=3/(4{d_{\rm clk}}).
\end{equation}
The compiler adds finite factors, and the resulting perfect strategy can be written explicitly. Let $\mathcal H_F=\ell^2(\WW)$ and let $\mathcal H_{\rm sc}=\mathbb C^{17}$ index the scaffold corners. Take
\[
 \mathcal H_A=\mathcal H_B=
 \mathcal H_{\rm sc}\otimes\mathcal H_F\otimes\mathbb C^{d_{\rm clk}}
 \otimes\mathbb C^3\otimes\mathbb C^{d_{\max}}.
\]
With $\mu_4=(1,2,2,2,2,4,4,4,4,8,8,8,8,16,16,16,16)/121$, use the pure vector
\[
 |\psi\rangle=|\sqrt{\mu_4}\rangle\!\rangle\otimes
 |\sqrt{h_F}\rangle\!\rangle\otimes|\Phi_{d_{\rm clk}}\rangle
 \otimes|\Phi_3\rangle\otimes|\Phi_{d_{\max}}\rangle,
 \quad |\sqrt{\mu_4}\rangle\!\rangle=\sum_i\sqrt{\mu_{4,i}}|i,i\rangle.
\]
The Fock factor is $\sum_w\sqrt{(3/4)16^{-|w|}}|w,w\rangle$. Alice uses the scaffold FMS effects, actual graph projections, refinements and column partitions of Lemma~\ref{lem:four}; Bob uses the prescribed opposite FMS effects and the transposed transferred centralizer products. The row factor is the finite maximally entangled ancilla of Lemma~\ref{lem:rows}. These simultaneous effects were checked in the compiler proof. The marginal entropy is $H(\mu_4)+S(h_F)+\log_2{d_{\rm clk}}+\log_2 3+\log_2d_{\max}<\infty$.

Finally finite-dimensional perfect play would extract four isometries with mutually orthogonal ranges on one nonzero finite-dimensional root space. Their range dimensions would sum to four times the dimension of that space, which is impossible. The argument applies to every input, whether or not it has a witness.
\end{proof}

Together with Section~\ref{sec:upper}, this proves both completeness statements in Theorem~A. The value-one promise is established next. For the promise version, the reduction maps every input into that promise; the unrestricted $\Sigma^0_2$ membership predicate restricts to it, giving promise-problem completeness. The construction can be made syntactically nonsynchronous and non-imitation on every input by adding a fresh binary question to each side, accepting every pair involving it. Either answer then has the same nonempty allowed-answer support against every opposite question, which violates the imitation separation condition. This addition does not change perfection or the value-one property.

\paragraph{Size of the finite game.}
Here is a coarse order-of-magnitude bound, independent of the witness ${d_{\rm clk}}$. Let $|\Gamma_{\rm CA}|$ be the alphabet size, let $N^{\rm proj}_e$ count the named graph projections, rows and word prefixes after endpoint translation, and let $b_e$ bound the bit length of their rational coefficients. The lookup table has at most $2(|\Gamma_{\rm CA}|+1)|\Gamma_{\rm CA}|^2$ entries, so unshared explicit polynomials give $N^{\rm proj}_e=\operatorname{poly}(|\Gamma_{\rm CA}|)$ for this fixed list of operator relations. Denominator clearing and signed repetitions give row dimension $d_{\max}\leq2^{O(N^{\rm proj}_e(b_e+\log(N^{\rm proj}_e+1)))}$. The scalar contexts, column questions and marginal bindings are polynomial in $N^{\rm proj}_e,d_{\max}$, so the padded question count $Q_e$ and maximum answer count obey the same coarse exponential bound. An explicit predicate table is polynomial in those two counts and has the same exponential order. Thus the effective compilation depends on the hardwired program, never on its chosen prime or running time. For scale, the small compiled example of Appendix~\ref{sec:repro}, with one modular isometry and no transported labels, already has $748$ rows and $27\,991$ questions per party. For the full ten-label instance with $e=0$ we counted the questions without writing out the game. Expanding the tableau polynomials literally gives $6\,191\,638\,087$ questions per party. Sharing bounded-circuit gates among the eight update rows gives $8\,516\,580$; this encoding adds source-only gates and keeps the soundness and completeness translation. Both counts use an explicit eight-symbol scan automaton for $e=0$, with alphabet $\mathtt s,\mathtt b,\mathtt f,\mathtt d,\mathtt{z0},\mathtt z,\mathtt h,\mathtt H$, which halts unless the clock order is two; it replaces the generic product alphabet of Section~\ref{sec:tableau}.

\subsection{Every constructed game has value one}\label{sec:value}
The finite-dimensional approximations do not need a witness for $e$. Truncate the Fock tree to depths $0,\ldots,m$ and choose a prime ${d_{\rm clk}}>2^m$ with a good phase (Lemma~\ref{lem:goodphase}). Use the normalized density
\[
 h^{[m]}=\frac{\sum_{|w|\leq m}16^{-|w|}|w\rangle\langle w|}
 {\sum_{i=0}^m4^{-i}}\otimes\frac{I_{d_{\rm clk}}}{{d_{\rm clk}}}.
\]
Every represented time is less than ${d_{\rm clk}}$. The decoder therefore remains in the waiting stage in every square and never halts. Let $P_m$ denote the top-depth projection. Its mass is
\begin{equation}\label{eq:topmass}
 t_m=\Tr(h^{[m]}P_m)=\frac{3}{4^{m+1}-1}.
\end{equation}

Every endpoint identity $L_{\rm end}A_\alpha=A_\beta R_{\rm end}$ remains exact. Below top depth it is the canonical identity; on top depth both sides vanish because the endpoint expression $R_{\rm end}$ preserves depth. The modular equations, the range-sum identity and all off-diagonal $A_\alpha^*A_\beta=0$ also remain exact. The only failed original relations are the four diagonal isometry equations:
\begin{equation}\label{eq:fourdefects}
 A_\alpha^*A_\alpha-I=-P_m\quad(\alpha=00,01,10,11).
\end{equation}
All graph and triangle projections are defined from the actual products. Every structural graph, synchrony, binding, FMS, refinement and column test stays exact.

Here is the full row accounting. For each $\alpha$ let $\ell_\alpha$ be the denominator-clearing integer and $d_\alpha$ the column dimension of the row obtained by translating $a_\alpha^*a_\alpha-e_0=0$ into graph projections. They depend only on the finite presentation, so they are independent of $m$ and ${d_{\rm clk}}$. In the scalar qutrit frame its defect $\Delta_\alpha$ has off-diagonal blocks $-(P_m)_0$ and adjoint and zero diagonal blocks. The frame density is $I_3/3$ and the source scaffold mass is $1/121$, so
\begin{equation}\label{eq:defectweight}
 \Tr(h_{\rm old}\Delta_\alpha^2)=\frac23\frac1{121}t_m.
\end{equation}
The same calculation applies separately to each of the four rows. Their row-zero events have distinct column questions and occur in no other failed rule. If the game has $Q_e$ equally padded questions per party, \eqref{eq:rowloss} gives
\begin{align}
 1-\omega(\Gclk_e,\mathcal S_m)
 &\leq C_e\frac{3}{4^{m+1}-1},\label{eq:valueloss}\\
 C_e&=\frac{1}{Q_e^2}\frac{2}{363}
 \sum_{\alpha\in\{00,01,10,11\}}\frac{4\ell_\alpha^2}{d_\alpha^3}.\label{eq:valueconstant}
\end{align}
If the four row constructions have common $\ell,d$, this is $C_e=32\ell^2/(363Q_e^2d^3)$. The constant counts one oriented row-zero test per row; testing both orientations would double it.

Since the density factor $I_{d_{\rm clk}}/{d_{\rm clk}}$ has trace one, no factor of ${d_{\rm clk}}$ enters the loss. Thus $\omega_{\mathrm{q}}(\Gclk_e)=1$ for every $e$, proving the promise in Theorem~A.

\section{Exact Bell attainment}\label{sec:bell}
The same compactness decides when a rational Bell functional attains a value exactly. A Bell functional generalizes a game: its coefficients may be any rationals, and attainment of its supremum replaces perfection.

\Needspace{5\baselineskip}
\begin{theorem}[Bell attainment]\label{thm:bell}
For rational Bell functionals, each of the following is $\Sigma^0_2$-complete in the normal spatial model: attainment of a supplied rational global upper bound; attainment of the supremum, with no bound supplied; and attainment of a supplied rational scalar value. In the first problem, validity of the supplied bound can either be promised or required as part of the decision problem.
\end{theorem}

The Bell functional $I_{3322}$ is an explicit instance of the second problem in which finite dimension does not suffice. P\'al and V\'ertesi computed its supremum numerically, with finite-dimensional strategies converging to it, and conjectured that no finite dimension attains it \cite{PalVertesi}, and Krohn-Grimberghe has claimed a computer-assisted exact identification of the value \cite{KrohnGrimberghe}. The supremum is attained by no finite-dimensional strategy \cite{Pauwels,Coladangelo3322,Douglas3322}, and it is attained by an infinite-dimensional spatial strategy \cite{Coladangelo3322,Douglas3322}.

Let $L(p)=\sum_{xyab}c_{xyab}p(a,b\mid x,y)$ have rational coefficients and let $C_{\rm abs}=\sum|c_{xyab}|$. Its value on any strategy lies in $[-C_{\rm abs},C_{\rm abs}]$. Finite Schmidt truncation shows that its supremum over spatial strategies equals its supremum over finite-dimensional strategies.

\subsection{A specified global bound}
Assume first that $\beta\in\mathbb Q$ is a global upper bound. For fixed measurements the Bell operator
\[
 B(E,F)=\sum c_{xyab}E_{a|x}\otimes F_{b|y}
\]
satisfies $B\leq\beta I$. A normal state attains $\beta$ only if every positive-weight pure component does so. For a pure vector this is equivalent to $B\psi=\beta\psi$, because $\beta I-B$ is positive. Compress to its Schmidt supports and normalize the largest amplitude as before. Vectorization yields the homogeneous equation
\begin{equation}\label{eq:bellequation}
 \sum c_{xyab}E_{a|x}D_{\rm Sch}F_{b|y}^{\mathsf T}=\beta D_{\rm Sch}.
\end{equation}
At stage ${d_{\rm cut}}$ use the variables of \eqref{eq:stage} and require every squared entry of the residual in \eqref{eq:bellequation} to be at most $C_{\rm abs}^2\kappa/({d_{\rm cut}}+1)$. Compression of a true witness satisfies this bound: only the compact tails of the sandwiched terms contribute, with coefficient sum $C_{\rm abs}$. Conversely the compact-sandwich lemma gives an exact nonzero limit solution. Its vectorization is an attaining strategy. Thus attainment is $\exists \kappa\,\forall {d_{\rm cut}}$ of a decidable predicate.

If validity is not promised, impose also $\omega_q(L)\leq\beta$. This is $\Pi^0_1$: for each finite dimension, the universal assertion that all states and POVMs give value at most $\beta$ is decidable by real quantifier elimination. Finite-dimensional density states suffice for this validity test. The conjunction remains $\Sigma^0_2$.

\subsection{No bound supplied}
The quantifier pattern is explicitly $\exists \kappa\geq1\,\forall {d_{\rm cut}}\geq1\ F_{\rm Bell}(L,\kappa,{d_{\rm cut}})$, with $F_{\rm Bell}$ decidable as follows. At stage ${d_{\rm cut}}$ introduce also a real variable $\beta\in[-C_{\rm abs},C_{\rm abs}]$, use the same residual bound, and require $\beta\geq\omega_{d_{\rm cut}}(L)$, where $\omega_{d_{\rm cut}}$ is the maximum in local dimension ${d_{\rm cut}}$. This requirement is a universal semialgebraic formula in $\beta$, so no numerical value of $\omega_{d_{\rm cut}}(L)$ is needed. Full real quantifier elimination decides the stage formula with its existential and universal real variables.

An attaining strategy supplies all stages. Conversely pass to a subsequence with $\beta_{d_{\rm cut}}\to\beta_*$, $D_{{\rm Sch},{d_{\rm cut}}}\to D_{\rm Sch}$ in norm and effects weakly convergent. For every fixed dimension $d$, all sufficiently late stages have $\beta_{d_{\rm cut}}\geq\omega_d(L)$. Hence $\beta_*\geq\sup_d\omega_d(L)$. The limit equation \eqref{eq:bellequation} gives a unit vector attaining $\beta_*$. Its expectation cannot exceed that supremum, so equality holds. This proves $\Sigma^0_2$ membership for attainment of the unspecified supremum.

\subsection{A specified scalar value}
Let $t\in\mathbb Q$ be supplied without a bound promise. A value strictly between the infimum and supremum has finite-dimensional values on either side, by equality of the finite and spatial extrema. A finite direct-sum mixture then attains $t$. Existence of a finite-dimensional strategy with value exactly $t$ is $\Sigma^0_1$, by dimension search and quantifier elimination. Every remaining attainable value is an endpoint. Upper endpoints are handled above; lower endpoints are upper endpoints for $-L$. This finite union is $\Sigma^0_2$.

For hardness in all three cases use the winning-probability functional of $\Gclk_e$. Its coefficients are rational, its global bound and finite-dimensional supremum are both one, and attaining the specified value one is precisely perfect spatial play. Theorem~\ref{thm:reduction} and Section~\ref{sec:value} therefore prove Theorem~\ref{thm:bell}. In its promised-bound version every reduction output has valid bound $1$, and the upper predicate restricts to the valid-bound domain, so the same map establishes promise-problem completeness.

\section{The entanglement spectrum and its decision problems}\label{sec:renyi}
A finite entropy bound controls the tail of the Schmidt distribution.
A positive moment gives stronger control, and varying the moment order
reveals a strict ladder of existence languages. We first prove compactness
and the upper predicates, then give the spectral separators. The common
hardness family and the realization of all possible critical indices
follow in Section~\ref{sec:programmable}, using the nested compiler proved
in Appendix~\ref{nested:sec:nested}.

For $0<\alpha<1$ let $\mathcal P_\alpha$ require a perfect pure spatial
strategy with $\Tr h^\alpha<\infty$. Put $\mathcal P_0=\Lq$,
$\mathcal P_1=\LE$, and $\mathcal P_{\rm normal}=\LI$.
For moments of order $\alpha>1$ the finiteness condition imposes no
restriction: $0<\Tr h^\alpha\le1$ for every density, so the R\'enyi
entropy of order $\alpha$ is finite, and the largest eigenvalue is
positive, so the order-infinity entropy is finite too. Thus all these
languages equal $\mathcal P_{\rm normal}$. Here $\mathcal P_{\rm normal}$
is a language label; it does not denote a moment of infinite order. Mixed witnesses can be purified on Bob's side preserving Alice's
marginal. Pure witnesses therefore suffice for these existence languages.

\subsection{Moment budgets are compact and effective}\label{sec:momentupper}
For sorted probabilities with $\sum_i p_i^\alpha\le K$, one has
\begin{equation}\label{eq:momenttail}
 t_D:=\sum_{i>D}p_i\le
 K^{1/\alpha}(D+1)^{-(1-\alpha)/\alpha}.
\end{equation}
Indeed $p_{D+1}^\alpha\le K/(D+1)$ and
$t_D\le p_{D+1}^{1-\alpha}\sum_{i>D}p_i^\alpha$.
This gives uniform tails. Coordinate limits preserve total mass and
converge in $\ell^1$, while Fatou preserves the moment bound. The Schmidt
vectors converge in norm, since
$\|\sqrt p-\sqrt{p'}\|_2^2\le\|p-p'\|_1$. Bounded POVMs have a common
weakly convergent subsequence on fixed copies of $\ell^2$. Their tensor
products converge weakly on finite simple tensors and then on all vectors
by uniform boundedness. Thus all game probabilities converge. The
maximum winning probability $v_G^\alpha(K)$ under the supplied budget
is attained. The same proof uses \eqref{eq:entropytail} at $\alpha=1$.
Trace one alone supplies no uniform tail, as $I_N/N$ demonstrates.

Write $v_{G,D}^\alpha(K)$ for the maximum winning probability with local dimension at most $D$ and moment budget $K$. Normalize the first $D$ probabilities by their retained mass. This
truncation majorizes the full probability vector, so does not increase
the concave moment sum or entropy. Compression of the POVMs therefore
gives
\[
 0\le v_G^\alpha(K)-v_{G,D}^\alpha(K)\le\sqrt{t_D}.
\]
For rational $\alpha$, the finite-dimensional budget constraints are
semialgebraic using nonnegative root variables. For a fixed computable
irrational order, approximate $x^\alpha$ uniformly on $[0,1]$ by rational
polynomials. Compact outer approximations give upper semicomputability.
For lower approximation, move a nonpure boundary spectrum slightly
towards $(1,0,\ldots,0)$: the resource strictly decreases and probabilities
change continuously. Rational spectra and feasible POVMs with strict
slack are dense. At moment budget $K=1$, only a pure marginal is possible
and the value is the finite classical value. The analogous entropy
exception is budget zero. These arguments give computable budget values
for computable orders, and in particular the upper semicomputability
needed here. Equivalently, compact rational outer feasibility tests,
with vanishing rejection slack and uniform constraint error, give
\[
 G\in\mathcal P_\alpha\iff\exists K\ge1\,\forall D\ge1\ R_\alpha(G,K,D)
\]
with $R_\alpha$ decidable. Explicitly, compute rational polynomials $f_D$ with
$\sup_{x\in[0,1]}|f_D(x)-x^\alpha|\le D^{-2}$.
Choose rational bounds $u_D\ge K^{1/\alpha}(D+1)^{-(1-\alpha)/\alpha}$ tending to zero
and rational $\epsilon_D\ge4\sqrt{u_D}$ tending to zero. If $u_D\ge1$, let the stage be true.
Otherwise require ordered $p_i=s_i^2$, $s_i\ge0$, two $D$-dimensional POVM assignments, and
\[
 1-u_D\le\sum_{i=1}^D p_i\le1,\qquad
 \sum_{i=1}^D f_D(p_i)\le K+D^{-1},\qquad
 \operatorname{loss}_G(s,E,F)\le\epsilon_D.
\]
These are rational semialgebraic constraints. A perfect budget-$K$ witness supplies every stage by compression.
Conversely, the true moment of each stage is at most $K+2/D$; the same uniform-tail compactness argument
and Fatou give a normalized perfect limit with moment at most $K$.
The rational-order version is written explicitly in
Section~\ref{sec:programmable}.

\subsection{Tree spectra separate the rungs}
For $b$ orthogonal isometries of common ratio $\nu$ and faithful normal
density $h$, the word-sector argument of Section~\ref{sec:tree} and its $b$-ary version in Appendix~\ref{nested:sec:variants} gives
\[
 h=\tau_{b,\nu}\otimes\rho_{\rm spec},\qquad
 \tau_{b,\nu}=(1-b\nu)\sum_w\nu^{|w|}|w\rangle\langle w|,
 \quad b\nu<1.
\]
One can also see exhaustion by pulling a positive density eigenvector
back through adjoint edges: every pullback multiplies its eigenvalue
by $\nu^{-1}$, so it must terminate before exceeding $\|h\|$.
Consequently
\begin{equation}\label{eq:momenttree}
 \Tr h^\alpha=
 \frac{(1-b\nu)^\alpha}{1-b\nu^\alpha}\Tr\rho_{\rm spec}^\alpha
\end{equation}
when $b\nu^\alpha<1$, and the moment diverges otherwise. The bare-tree
critical index is $\log_2b/n$ at $\nu=2^{-n}$. Its finite compiler is
the $n$-bridge construction with a finite list of $b$ edge labels, as
specified in Appendix~\ref{nested:sec:variants}. For four branches the
normality requirement is $n>2$; for $0<\alpha\le1$,
$n=\lfloor2/\alpha\rfloor+1$ ensures both normality and moment
summability. For larger orders the requirement $n>2$ remains.

For any $0<a<b<1$, choose integers $n,j$ with
$a<\log_2j/n<b$. Such choices exist by taking $n$ large and an integer
between the corresponding exponential bounds. The bare $j$-ary tree
has finite $b$-moment, infinite $a$-moment in every realization, and
finite entropy in its canonical realization. A comb with critical
index one and finite entropy, constructed next, separates every order
below one from entropy. The Catalan game of Section~\ref{sec:sep}
separates entropy from arbitrary normal play. The uniform family of
Section~\ref{sec:programmable} separates finite rank from every
positive rung. Thus
\begin{equation}\label{eq:strictladder}
 \mathcal P_0\subsetneq\mathcal P_a\subsetneq\mathcal P_b
 \subsetneq\mathcal P_1\subsetneq\mathcal P_{\rm normal}
 \quad(0<a<b<1).
\end{equation}
Compression to the positive source corner preserves finite moments;
finite compiler factors preserve them in completeness. These separators
therefore apply to every perfect strategy.

\subsection{Complexity, differences, and critical indices}
The next theorem collects the refinements of Theorem~D. The rungs are
the languages $\mathcal P_0,\mathcal P_\alpha,\mathcal P_1,\mathcal P_{\rm normal}$
defined at the start of this section.

\Needspace{5\baselineskip}
\begin{theorem}[Refinements of Theorem~D]\label{thm:momentrefine}
\leavevmode
\begin{enumerate}[label=(\alph*)]
\item One computable hardness family has a single perfect witness with
all positive moments finite on every YES input, value one on every
input, and no finite-dimensional perfect instance.
\item Every nonempty difference between distinct positive computable
rungs is $D_2(\Sigma^0_2)$-complete; a positive rung minus finite rank
is $\Sigma^0_2$-complete.
\item For a perfect witness with marginal $h$, its critical index is
$\alpha_c(h)=\inf\{\alpha>0:\Tr h^\alpha<\infty\}$. For a spatially perfect game $G$, put
\[
c(G)=\inf\{\alpha_c(h):h\text{ is the marginal of a perfect spatial witness for }G\}.
\]
The possible values of $c(G)$ (an infimum, attained by the realizing
games) are exactly the right-c.e.-in-$0'$ reals in $[0,1]$, already
among games with finite-entropy perfect strategies.
\end{enumerate}
\end{theorem}

\begin{proof}
The construction in Section~\ref{sec:programmable} gives one computable
family $H_e$ such that
\[
 e\in\Adec\iff H_e\in\LI
 \iff H_e\text{ has one perfect state with every positive moment finite}.
\]
Every instance has $\omega_q=1$ and no finite-dimensional perfection.
This proves (a). Together with the upper predicates it proves
$\Sigma^0_2$-completeness for every fixed positive computable rung,
including entropy and normal play, as stated in Theorem~D. The
finite-rank rung is $\Sigma^0_1$-complete.

Exact conjunction and disjunction preserve these resources. For
conjunction tensor the witnesses; moment sums multiply. For disjunction,
the globally matched branch projections reduce the marginal and give
$h=\bigoplus_j p_jh_j$. Hence
$\Tr h^\alpha=\sum_jp_j^\alpha\Tr h_j^\alpha$ and
$S(h)=H(p)+\sum_jp_jS(h_j)$; each positive branch inherits the resource.
Use the full-support branch agreement proved in Section~\ref{sec:sep}.
If $\mathcal P\subsetneq\mathcal Q$ are distinct positive rungs, fix
$G_{\rm sep}\in\mathcal Q\setminus\mathcal P$. Then
\[
 H_e\wedge(G_{\rm sep}\vee H_f)\in\mathcal Q\setminus\mathcal P
 \iff e\in\Adec\text{ and }f\notin\Adec.
\]
All nonempty such differences are therefore $D_2(\Sigma^0_2)$-complete,
and a positive rung minus finite rank is $\Sigma^0_2$-complete, since
no $H_e$ has finite-dimensional perfection. This proves (b). Reverse
differences, and differences between the languages of moment order
greater than one, are empty. The property that NO instances have no
normal perfect strategy holds for $H_e$; it need not hold for every NO
instance of the difference family. Both hardness promises persist under
these operations.

For (c), unused entanglement can increase an individual strategy's
index; the invariant is the infimum $c(G)$. Section~\ref{sec:programmable}
proves that its possible values are exactly the right-c.e.-in-$0'$ reals
in $[0,1]$, and that every such value has a realizing game with a
finite-entropy witness attaining the minimum. The computability
characterization uses \emph{limsup} of computable rational sequences,
following Becher, Bugeaud and Slaman~\cite[Lemma~4]{BBS}.
\end{proof}

\section{A uniform family and programmable critical indices}\label{sec:programmable}
The construction separates the spectrum from the computation that verifies it. Two commuting one-sided shifts first force quarter-plane coordinates and a geometric density. A finite clock and a deterministic tableau then read an integer witness. For programmable critical indices, a comb supplies the desired spectral threshold while an independent quarter-plane register verifies its branching rule. The nested compiler translates the quadratic shift relations into finite game tests.

\subsection{The quarter-plane register and its finite presentation}
Use two modular contractions $S,T$ and impose
\begin{equation}
 S^*S=T^*T=I,\quad ST=TS,\quad S^*T=TS^*.
                                                               \label{eq:nested-A1}
\end{equation}
Name projections $\Pi_x,\Pi_t$ with $SS^*=I-\Pi_x$, $TT^*=I-\Pi_t$.
Name positive contraction counters $C_x,C_t$ and impose their
commutation, $C_x\Pi_x=\Pi_x$, $C_t\Pi_t=\Pi_t$, and
\begin{equation}
 C_xS=\tfrac12SC_x,\ C_xT=TC_x,\quad
 C_tS=SC_t,\ C_tT=\tfrac12TC_t.                               \label{eq:nested-A2}
\end{equation}
All clock and computation labels commute with both counters.
For the clock use centralizer unitaries $Z,J,K,W$ with
\[
 JK=ZKJ,\ [Z,J]=[Z,K]=0,
\]
$Z,J,K$ constant along both shifts, and
\[
 WS=SZW,\ WT=TW,\ W\Pi_x=\Pi_x.
\]
For a projection $E_{\rm ret}$ and a contraction $R_{\rm ret}$ use
\begin{equation}
 (W-I)E_{\rm ret}=0,\qquad
 R_{\rm ret}(W-I)=\tfrac12C_x(I-E_{\rm ret}).                 \label{eq:nested-A3}
\end{equation}
Every row is either centralizer polynomial, degree-one endpoint, or one of the quadratic rows of the translation table in Appendix~\ref{nested:sec:nested}. In particular the mixed row $S^*T=TS^*$ of \eqref{eq:nested-A1} is compiled as $a_S'^*a_T'=ta_Ta_S^*t^*$, in the notation of that appendix; no modular transport is synchronized.

To specify a verifier, supply a finite radius-one deterministic
one-sided cellular automaton, with alphabet $\mathcal A$, interior
rule $F:\mathcal A^3\to\mathcal A$, and left-boundary rule
$G:\mathcal A^2\to\mathcal A$. For every $d$ name four projection
families $P_d,L_d,R_d,N_d$, each summing to identity and commuting
with $C_x,C_t$. Their shift equations are
\begin{equation}
 L_dS=SP_d,\qquad P_dS=SR_d,\qquad P_dT=TN_d.                 \label{eq:nested-A4}
\end{equation}
Thus $L_d(x,t)=P_d(x-1,t)$ for $x>0$, $R_d(x,t)=P_d(x+1,t)$,
and $N_d(x,t)=P_d(x,t+1)$. The unused $L_d(0,t)$ is arbitrary.
Submit the exact finite update polynomials
\begin{equation}
 N_d=(I-\Pi_x)\sum_{F(a,b,c)=d}L_aP_bR_c
              +\Pi_x\sum_{G(b,c)=d}P_bR_c.                   \label{eq:nested-A5}
\end{equation}
Specify every initial track on $\Pi_t$: the return marker from
$E_{\rm ret}$, the head from $\Pi_x$, and the blank work tracks. If
$\iota$ maps these finitely many input bits to an alphabet symbol,
the initial row for $d$ is $\Pi_tP_d=\Pi_t\sum_{\iota(b)=d}\prod_iQ_i^{b_i}$,
where $Q_i^1=Q_i$, $Q_i^0=I-Q_i$. The input projections commute:
they are scalar coordinate markers in the normal representation
proved next. Equivalently one may add their finite commutator rows.
For each rejecting symbol impose $P_d=0$.
Equations \eqref{eq:nested-A4} have modular degree one; \eqref{eq:nested-A5} has modular degree zero,
regardless of its centralizer polynomial degree. This is the precise
admissibility distinction needed for the tableau.

Use the Fibonacci convention $F_1=F_2=1$, $F_{k+2}=F_{k+1}+F_k$. For a total computable predicate $R(e,j,s)$, a finite machine first
scans for the first positive return marker, verifies that its period
is $F_{j+3}$, decodes $j$, and searches for $s$ with $\neg R(e,j,s)$.
Its ordinary finite transition table gives such a cellular automaton:
a symbol records a tape letter and optionally a head with its control
state. A head writes its new letter, and its indicated neighbor receives
the new head at the next step; a boundary rule handles the missing
left neighbor. Invalid head collisions can be assigned a rejecting
symbol. Multiple tapes can be interleaved with finitely many tracks
and finite control, with extra time steps as needed. Only the unique
initialized one-head computation is used by the soundness induction.
This construction is effective in a supplied program for $R$ and in $e$.
Concrete word and row counts depend on that table; the reduction family is parametric. Appendix~\ref{sec:repro} describes the compiled validation instances.

\subsection{Normal classification, the clock, and all positive moments}
Cyclicity gives $\Tr(h_0S^kS^{*k})=2^{-k}$ and the same for $T$.
Hence normality removes both unitary Wold parts. Double commutation
makes the two wandering decompositions compatible and gives
\begin{equation}
 \HH_0=\ell^2(\mathbb N^2)\otimes\KK,\quad
 h_0=\rhoreg\otimes\rhoreg\otimes\rho_{\rm spec},\quad
 \rhoreg=\sum_{x\ge0}2^{-x-1}|x\rangle\langle x|.                \label{eq:nested-A6}
\end{equation}
Indeed the mutually orthogonal spaces $S^xT^t\Pi_x\Pi_t\HH_0$ exhaust the
support; their density blocks are $2^{-x-t}$ times the joint root block.
Normalization makes that block $\rho_{\rm spec}/4$. Equation \eqref{eq:nested-A2} then forces
$C_x=2^{-x}$ and $C_t=2^{-t}$, by induction from each boundary.
Commutation with these counters prevents labels from mixing coordinates
of equal total density weight.

The constant $Z,J,K$ act on $\KK$ and commute with the compact
$\rho_{\rm spec}$. Pick a positive finite-dimensional $\rho_{\rm spec}$ eigenspace and
then a nonzero $Z=zI$ subspace. It is invariant under $J,K$, and
determinants in $JK=zKJ$ give $z^d=1$. Let its finite order be ${d_{\rm clk}}$.
The equations for $W$ force $W(x,t)=Z^x$. Equation \eqref{eq:nested-A3} forces
$E_{\rm ret}(x,t)=I$ exactly when ${d_{\rm clk}}$ divides $x$ on this sector:
its range lies in $\ker(W-I)$; on that kernel the second equation
and $C_x>0$ force it to be identity. On each chosen clock sector the initialized tableau labels
are scalars. Equations \eqref{eq:nested-A4}--\eqref{eq:nested-A5} force each next row to be the unique
classical successor; the induction also proves invariance of the
sector for those projections. Thus rejecting symbols cannot be absent
unless $\forall s\,R(e,j,s)$ for the decoded integer.

Conversely a true witness uses ${d_{\rm clk}}=F_{j+3}$, $a=F_{j+2}$,
$z=\exp(2\pi ia/{d_{\rm clk}})$, $Z=zI_{d_{\rm clk}}$, and Weyl matrices on $\mathbb C^{d_{\rm clk}}$.
With $\rho_{\rm spec}=I_{d_{\rm clk}}/{d_{\rm clk}}$ the honest initialized tableau is diagonal and
all the relations hold. Before the finite compiler factors,
\begin{equation}
 \Tr(h_0^\alpha)=4^{-\alpha}{d_{\rm clk}}^{1-\alpha}(1-2^{-\alpha})^{-2}
 \quad(\alpha>0),\qquad S(h_0)=4+\log_2{d_{\rm clk}}.                    \label{eq:nested-A7}
\end{equation}
This is one common witness for all positive moments.

\paragraph{A bounded reader at every nonreturn.}
Put $\theta=(\sqrt5-1)/2$. If $a$ is a nearest integer to $x\theta$,
then $\delta=x\theta-a$ has nonzero integer norm
$\delta(x\theta'-a)$, where $\theta'=-(1+\sqrt5)/2$.
The second factor has magnitude at most $\sqrt5x+1/2<3x$, giving
$\|x\theta\|>1/(3x)$. Also
$F_k\theta-F_{k-1}=(-1)^{k-1}\theta^k$ and $F_k\theta^k<1$,
so the Fibonacci approximation error is less than $F_k^{-2}$.
These elementary bounds
give, when
$x\le {d_{\rm clk}}/3$,
\[
 \|xa/{d_{\rm clk}}\|>1/(3x)-x/{d_{\rm clk}}^2\ge2/(9x).
\]
For every nonreturn $x\ge2{d_{\rm clk}}/9$, the coprime residue bound gives
$\|xa/{d_{\rm clk}}\|\ge1/{d_{\rm clk}}\ge2/(9x)$. These ranges overlap and cover every
positive nonreturn, including $x>{d_{\rm clk}}$. Concavity of sine on
$[0,\pi/2]$ gives $|z^x-1|\ge4\|xa/{d_{\rm clk}}\|\ge8/(9x)$.
Consequently the canonical reader
\[
 R_{\rm ret}(x)=\begin{cases}2^{-x-1}/(z^x-1),&{d_{\rm clk}}\nmid x,\\0,&{d_{\rm clk}}\mid x\end{cases}
\]
has norm at most $\sup_{x\ge1}9x/2^{x+4}=9/32$. Returns, including
$x=0$, are checked separately and give zero on both sides of \eqref{eq:nested-A3}.

\paragraph{Value one.}
For any $e$, take arbitrarily large Fibonacci periods and the genuine
computation even if it eventually rejects. All equations except the
forbidden-reject row remain exact. The first scan prevents rejection
before time ${d_{\rm clk}}$. Its root probability is at most
$\sum_{t\ge {d_{\rm clk}}}2^{-t-1}=2^{-{d_{\rm clk}}}$, so \eqref{eq:nested-N11} gives
\emph{loss at most $C_{\rm nest}2^{-{d_{\rm clk}}}$}; the constant contains the compilation factors of the row tests. Since $C_{\mathrm{q}}$ and $C_{\mathrm{qs}}$ have the same closure (Section~\ref{sec:prelim}), the finite-dimensional value is one. Finite-dimensional perfection is impossible: $S^*S=I$
would make $S$ unitary, whereas $h_0S=S h_0/2$ and trace cyclicity
would give $1=1/2$.

\subsection{The comb and the independent checker}
The spectral threshold will come from the number of branches attached
at each spine position. A separate geometric register verifies the
branching guard. Keeping its computation time in that separate factor
allows arbitrarily slow total programs without changing the programmed
critical index. First normality establishes coordinates; only then does
the deterministic verifier force the guard on each coordinate fibre.

Let $Q,P,{E_{\rm guard}}$ be centralizer projections, $P\le Q$, and let $\Csp,L$ be
commuting positive contractions, commuting with $Q,P$. Use modular
contractions ${V_{\rm sp}},B_0,B_1$ with
\begin{align*}
 {V_{\rm sp}}^*{V_{\rm sp}}&=Q,& {V_{\rm sp}}{V_{\rm sp}}^*&=Q-P,\\
 B_i^*B_j&=\delta_{ij}{E_{\rm guard}},& B_0B_0^*+B_1B_1^*&=I-Q,\\
 \Csp P&=P/2,& \Csp{V_{\rm sp}}&={V_{\rm sp}}\Csp/2,\\
 \Csp B_i&=B_i\Csp,& LQ&=Q,\qquad LB_i=B_iL/2.
\end{align*}
All modular symbols have ratio $1/2$. These are inner-product, range,
and endpoint rows of the translation table. In particular a range sum
is compiled as a sum $ta_ia_i^*t^*$, together with the transported
range projection. Every range product $XX^*$ used as a root
centralizer label comes with its defining translated row.

Add an independent quarter-plane pair $S,T$ satisfying \eqref{eq:nested-A1}--\eqref{eq:nested-A2}.
For every comb modular symbol $X\in\{{V_{\rm sp}},B_0,B_1\}$ impose both
$XS=SX$, $X^*S=SX^*$ and the corresponding equations with $T$.
For $Q,P,\Csp,L$ impose shift commutation by endpoint rows. These
relations give the tensor decomposition \eqref{eq:nested-A6} with the entire comb
algebra acting on its multiplicity space. The checker has its own
$C_x,C_t,\Pi_x,\Pi_t$ and a separate computation time.

\paragraph{The guard preserves the coordinate fibres.}
Since ${E_{\rm guard}}=B_0^*B_0$, the counter relations and their adjoints give
\begin{equation}
 {E_{\rm guard}}\Csp=B_0^*\Csp B_0=\Csp{E_{\rm guard}},\qquad
 {E_{\rm guard}}L=2B_0^*LB_0=L{E_{\rm guard}}.                                           \label{eq:nested-A8}
\end{equation}
Ordinary and double commutation with checker shifts similarly give
commutation of ${E_{\rm guard}}$ with those shifts and adjoints, so adding
these commutator rows to the presentation is harmless.

The four orthogonal projections $P,{V_{\rm sp}}{V_{\rm sp}}^*,B_0B_0^*,B_1B_1^*$ sum to
identity. Starting from an $h$ eigenvector of eigenvalue $p>0$, each
nonzero pullback through ${V_{\rm sp}}^*,B_i^*$ doubles its eigenvalue. After
finitely many steps it would exceed $\|h\|$ and must vanish.
Thus descendants of root vectors span the support. Since ${V_{\rm sp}}B_i=0$,
surviving words have form $B_w{V_{\rm sp}}^{m-1}P$, with $m\ge1$.
The counter relations force $\Csp=2^{-m}$ and $L=2^{-|w|}$ on their
ranges. Their span is dense, so the joint dyadic eigenspaces
$\HH_{m,\ell}$ exhaust the support.
Equation \eqref{eq:nested-A8} shows ${E_{\rm guard}}$ preserves each such eigenspace.

Name equality projections $E_{\Csp},E_L$ and contraction readers $F_{\Csp},F_L$.
Impose
\begin{equation}
 (\Csp-C_x)E_{\Csp}=0,\quad
 F_{\Csp}(\Csp-C_x)=\tfrac12\Csp C_x(I-E_{\Csp}),                              \label{eq:nested-A9}
\end{equation}
and the analogous rows for $L$. For two unequal dyadic values
$u,v\in(0,1]$, $|uv/[2(u-v)]|\le1$: the larger is at least twice
the smaller. On equal fibers the second equation forces the equality
projection to be identity; on unequal fibers the first forces it to
be zero. The canonical reader is $uv/[2(u-v)]$, with value zero at
equality. Thus the initial checker tracks mark exactly $x=m$ and
$x=\ell$, and receive the bit ${E_{\rm guard}}$. Require every checker label to
commute with $\Csp,L,{E_{\rm guard}},C_x,C_t$, and initialize all tracks and the head.
Its total program computes $k_m$ and rejects exactly when the bit
is different from $\mathbf1_{\{\ell<k_m\}}$. Apply the same finite
\eqref{eq:nested-A4}--\eqref{eq:nested-A5} presentation and forbid rejection.

On a joint $(m,\ell,{E_{\rm guard}})$ fiber the initial symbols are scalars. The
deterministic tableau induction therefore forces the actual finite
computation of this total program. A wrong guard fiber would produce
a nonzero rejecting projection, contradicting faithfulness and the
forbidden row. Hence
\begin{equation}
             {E_{\rm guard}}|_{\HH_{m,\ell}}=\mathbf1_{\{\ell<k_m\}}I.       \label{eq:nested-A10}
\end{equation}
Now, and only now, legal word maps are isometries on the root with
mutually orthogonal ranges. They exhaust the support by the previous
pullback argument, and yield
\begin{equation}
 \begin{gathered}h_{\rm comb}\cong\rhocomb\otimes\rho_{\rm spec},\\
 \rhocomb=\Zcomb^{-1}\sum_{m\ge1,\ |w|\le k_m}
       2^{-(m+|w|)}|m,w\rangle\langle m,w|,\\
 \Zcomb=\sum_{m\ge1}(k_m+1)2^{-m}.\end{gathered}                                \label{eq:nested-A11}
\end{equation}
The complete checker adds $\rhoreg\otimes\rhoreg$, independently of how
long the program takes to verify a particular fiber.

\subsection{Programming and classifying the minimum index}
For a computable rational sequence $a_m$ with $\limsup a_m=r\in[0,1]$,
put
\[
 u_m=\min\!\left(\max(a_m,0),\frac{m}{m+1}\right),\qquad
 k_m=\left\lfloor\frac{mu_m}{1-u_m}\right\rfloor.
\]
Then $0\le k_m\le m^2$,
and rounding changes $k_m/(m+k_m)$ from $u_m$ by less than $1/m$.
Thus its limsup is $r$, and $1\le \Zcomb\le7$.
For $0<\alpha<1$ the moment of \eqref{eq:nested-A11} converges precisely when
\[
              \sum_m2^{(1-\alpha)k_m-\alpha m}<\infty.
\]
The exponent equals $(k_m/(m+k_m)-\alpha)(m+k_m)$.
If $\alpha>r$ the tail decays geometrically; if $\alpha<r$ a
subsequence of terms grows. Therefore $\alpha_c(\rhocomb)=r$,
without a claim about convergence at the endpoint $\alpha=r$.
The entropy is bounded by
\[
 \log_2 7+\sum_{m\ge1}2^{-m}
 \left[(m^2+1)m+\frac{m^2(m^2+1)}2\right]
                         =106+\log_2 7.
\]
Every realization contains the comb factor, so extra multiplicity
cannot lower its critical index. Root multiplicity one gives the
canonical witness attaining minimum index $r$, with finite entropy.

For necessity, restrict the moment order to rational $0<{\beta}<1$.
For sorted Schmidt probabilities with $\sum_i p_i^{\beta}\le K$, use the tail bound \eqref{eq:momenttail} with $\alpha=\beta$ and $D=d_{\rm cut}$.
Choose computable rational upper bounds $\tailbd_{d_{\rm cut}}$ for this tail,
tending to zero, and rational $\epsilon_{d_{\rm cut}}\ge4\sqrt{\tailbd_{d_{\rm cut}}}$,
also tending to zero. For fixed $G,{\beta},K,{d_{\rm cut}}$ let
$\mathcal R(G,{\beta},K,{d_{\rm cut}})$ say there exist two local ${d_{\rm cut}}$-dimensional
POVM assignments and a subnormalized Schmidt vector with ordered
probabilities $p_1,\ldots,p_{d_{\rm cut}}$ such that
\[
 1-\tailbd_{d_{\rm cut}}\le\sum_{i\le {d_{\rm cut}}}p_i\le1,\qquad
 \sum_{i\le {d_{\rm cut}}}p_i^{\beta}\le K,
 \qquad \Pr(\mathrm{reject})\le\epsilon_{d_{\rm cut}}.
\]
When $\tailbd_{d_{\rm cut}}\ge1$ take this stage to be true. At the other stages
this is a decidable semialgebraic feasibility predicate: rational
powers are encoded by nonnegative root variables, and POVM
positivity by principal minors. A perfect witness compresses to
these assignments; its truncated vector differs in norm by at most
$\sqrt{\tailbd_{d_{\rm cut}}}$, giving the displayed slack. Conversely the uniform
tail bound makes these ordered vectors compact in $\ell^2$.
A diagonal subsequence of the bounded POVM matrix entries gives
positive limiting POVMs with the same finite outcome sums. Finite
coordinate tests followed by the uniform vector tail pass their
tensor probabilities to the limit. The limiting vector has norm
one, zero rejection and ${\beta}$-moment at most $K$ by Fatou's lemma.
Consequently the quantifiers are explicitly
\[
 G\in\mathcal P_\beta
 \quad\Longleftrightarrow\quad
 \exists K\in\mathbb N_{\ge1}\ \forall {d_{\rm cut}}\in\mathbb N_{\ge1}\
                         \mathcal R(G,{\beta},K,{d_{\rm cut}}).
\]
For rational $0<a\le1$, $c(G)<a$ is equivalent to this condition for
some rational $0<{\beta}<a$, when $c(G)$ denotes the infimum of
critical indices of perfect witnesses and is finite in $[0,1]$.
Cuts above $1$ are automatic on this domain.
The rational order and integer budget can be paired in one
existential integer, leaving the same $\exists K\,\forall {d_{\rm cut}}$ form.
Consequently the strict upper cut is c.e. in $0'$. Together with the
characterization of Becher, Bugeaud and Slaman \cite[Lemma~4]{BBS} by limsups of computable rationals, this gives exactly the right-c.e.-in-$0'$ indices in
$[0,1]$. The input program computes $k_m$, and contains no oracle.

\section{Semifinite hosts: a critical tree selects a finite depth}\label{sec:selector}
In a diffuse semifinite algebra the Weyl determinant of Section~\ref{sec:clock} is unavailable, so the spatial witness cannot be used. A trace still forces a finite object. A critical binary tree with modular ratio $1/2$ must have a nonzero leaf sector at some finite depth, and a group word chosen by that depth carries the $\Sigma^0_2$ witness. This section builds the resulting reduction and proves the semifinite half of Theorem~B.

\subsection{The critical-tree selector}
Work on a faithful normal support. Fix integers $d\ge2$ and a real $0<\gamma<1$. Let $P,Q$ be centralizer projections, $0\leq C\leq I$ a centralizer contraction, and let $B_1,\ldots,B_d$ satisfy
\begin{align}
 B_i^*B_j&=\delta_{ij}(I-P),&
 \sum_iB_iB_i^*&=I-Q,\label{eq:tree}\\
 \sigma_t^\varphi(B_i)&=d^{-it}B_i,\label{eq:modular}\\
 CQ&=Q,& CB_i&=\gamma B_iC,&[C,P]&=0.\label{eq:depth}
\end{align}
These labels $P,Q,C,B_1,\ldots,B_d$ belong to the selector presentation only.
\begin{lemma}[Critical-tree selector]\label{lem:selector}
For branching $d$, modular ratio $1/d$ and depth ratio $\gamma$, in a normal semifinite realization, $P\ne0$, $\ker C=0$, and $E_n=1_{\{\gamma^n\}}(C)$ satisfy $\sum_{n\geq0}E_n=I$ strongly. Some finite $n$ has $PE_n\ne0$.
\end{lemma}
\begin{proof}
The modular eigenoperators are entire analytic. KMS gives $\varphi(B_iB_i^*)=\varphi(B_i^*B_i)/d$. Summing gives
\begin{equation}\label{eq:defectmass}
 \varphi(P)=\varphi(Q).
\end{equation}
If $P=0$, faithfulness makes $Q=0$, giving a full Cuntz family. Write $\varphi(x)=\tau(hx)$, $h$ injective, $\tau(h)=1$. The covariance is $h^{it}B_i=B_i(h/d)^{it}$, hence $f(h)B_i=B_if(h/d)$ for bounded Borel $f$. For bounded $g\geq0$ with compact support in $(0,\infty)$, $hg(h)$ is positive integrable, and
\[
 \varphi(g(h))=\sum_i\tau(B_i^*hg(h)B_i)
 =\tfrac1d\sum_i\tau(hg(h/d))=\varphi(g(h/d)).
\]
The trace exchange is $\tau(T^{1/2}vv^*T^{1/2})=\tau(v^*Tv)$ for positive integrable $T$. Use indicators of $(d^k,d^{k+1}]$, or monotone approximations to them. All these windows have equal $\varphi$-mass and partition $(0,\infty)$, whose mass is one. Countable additivity is contradicted. Thus $P\ne0$.

Let $z=1_{\{0\}}(C)$. Covariance and its adjoint give $zB_i=B_iz$ by functional calculus at zero. Also $Qz=0$ and $[P,z]=0$. Since $z$ is a centralizer projection, $zMz$ is modularly invariant. The KMS calculation there gives $\varphi(Pz)=0$, hence $Pz=0$. If $z\ne0$, the normalized restriction of $\varphi$ is faithful normal on a semifinite corner with a full critical Cuntz family. The same window argument excludes this. Hence $z=0$.

Put $\mathcal T(X)=\sum_iB_iXB_i^*$. Telescoping gives
\begin{equation}\label{eq:telescope}
 I=\sum_{j=0}^{m-1}\mathcal T^j(Q)+\mathcal T^m(I).
\end{equation}
For a length-$m$ word $v$, $CB_v=\gamma^mB_vC$. Bounded spectral calculus puts its range in $1_{[0,\gamma^m]}(C)\mathcal H$. Thus
\[
 0\leq\mathcal T^m(I)\leq1_{[0,\gamma^m]}(C)\longrightarrow0
\]
strongly by injectivity of $C$. Each $\mathcal T^j(Q)$ is supported in the $\gamma^j$ eigenspace of $C$. These positive terms sum to $I$ and have mutually orthogonal spectral supports. Compressing the sum to each eigenspace gives
\begin{equation}\label{eq:En}
 E_j=\mathcal T^j(Q),\qquad \sum_{j\geq0}E_j=I.
\end{equation}
There is no residual spectral subspace since the sum is $I$. As $[P,C]=0$ and $P\ne0$, some $PE_n$ is nonzero. In particular $Q=E_0$.
\end{proof}

The selector of this section uses $d=2$ and $\gamma=1/2$, with edges reindexed $B_0,B_1$; the coupled construction uses $d=16$ and the same depth ratio.

\subsection{Finite models and a small example}
For words $w$ of length at most $n$, let $B_i$ append $i$ unless $w$ is a leaf, when it vanishes. Let $Q$ select the empty word and $P$ the depth-$n$ words. Put
\begin{equation}\label{eq:treeh}
 C|w\rangle=2^{-|w|}|w\rangle,\qquad
 h_n|w\rangle=\frac{2^{-|w|}}{n+1}|w\rangle.
\end{equation}
Each nonleaf has two distinct children and each nonroot a unique parent. This proves \eqref{eq:tree}, including the cross term. The child-to-parent density ratio is $1/2$, proving \eqref{eq:modular}; the same ratio for $C$ proves \eqref{eq:depth}. Each depth has state mass $1/(n+1)$, whence
\begin{equation}\label{eq:treeentropy}
 \varphi_n(P)=\varphi_n(Q)=\frac1{n+1},\qquad
 S(h_n)=\log_2(n+1)+\frac n2.
\end{equation}
For $n=2$, the seven word densities are $1/3$, two copies of $1/6$, and four copies of $1/12$. Their $C$ labels are $1,1/2,1/4$. The leaf rank is four but its state mass is $1/3$. A finite host's normalized trace gives $\tau(P)\geq1/2$ from \eqref{eq:tree}; this does not bound $\varphi(P)$ away from zero uniformly over states.

\subsection{A finite presentation verifies the depth}\label{sec:group}
Fix a total recursive predicate $\Rdec(e,n,t)$ and $\Adec=\{e:\exists n\,\forall t\,\Rdec(e,n,t)\}$. With $[x,y]=xyx^{-1}y^{-1}$ set
\begin{equation}\label{eq:group}
 \begin{split}
 \Gamma_e=\langle a,b,c\mid {}&
 {r^{\rm grp}}_n^2=1\ (n\geq0),\\[-2pt]
 &{r^{\rm grp}}_n=1\text{ if }\exists t\,\neg\Rdec(e,n,t)\rangle,
 \qquad {r^{\rm grp}}_n=[b,a^nca^{-n}].
 \end{split}
\end{equation}
\begin{lemma}\label{lem:group}
${r^{\rm grp}}_n$ is nontrivial in $\Gamma_e$ exactly when $\forall t\,\Rdec(e,n,t)$; a nontrivial ${r^{\rm grp}}_n$ has order two. Uniformly there is a finite presentation $\mathcal G_e$ and explicit words for embedded $a,b,c$ preserving this property.
\end{lemma}
\begin{proof}
Let $\mathcal D_{\rm good}\subseteq\mathbb N$ be the good integers. Take two copies of $\mathbb V=\bigoplus_{m\in\mathbb Z}\mathbb F_2\varepsilon_m$ and $\mathbb W=\bigoplus_{k\in\mathcal D_{\rm good}}\mathbb F_2z_k$. Define $\mathfrak b(\varepsilon_m,\varepsilon_\ell)=z_{\ell-m}$ if $\ell-m\in\mathcal D_{\rm good}$, zero otherwise. On $\mathbb V\oplus \mathbb V\oplus \mathbb W$ put
\begin{equation}\label{eq:model18}
 (x,y,z)(x',y',z')=(x+x',y+y',z+z'+\mathfrak b(x,y')).
\end{equation}
Bilinearity makes the third coordinate in either association $z+z'+z''+\mathfrak b(x,y')+\mathfrak b(x,y'')+\mathfrak b(x',y'')$. The identity is $(0,0,0)$; the inverse of $(x,y,z)$ is $(x,y,z+\mathfrak b(x,y))$. Simultaneously shifting both $\mathbb V$ indices preserves $\mathfrak b$; adjoin this automorphism as $a$. Put $b=(\varepsilon_0,0,0)$ and $c=(0,\varepsilon_0,0)$. Multiplication gives ${r^{\rm grp}}_n=(0,0,z_n)$ for $n\in\mathcal D_{\rm good}$, identity otherwise. All relators hold, and every good word has a nontrivial order-two image. Bad words are killed by listed relators, proving the assertion.

For precision we give a uniform Craig conversion \cite{Craig} of the c.e.\ relator list to a decidable literal-word list. Chiodo's convention already permits recursive enumeration, so this extra normalization is optional for the cited embedding theorem. Here is a Craig conversion that also preserves the old generator words. Adjoin $d$ with relator $d=1$. For an enumerated relator $r$ at time $t$, replace that occurrence by the literal word $r d^k$, where distinct increasing codes $k$ are chosen effectively with $k\geq t$. Given a word of length $\ell$, membership in the resulting literal relator set is decidable by simulating the first $\ell$ enumeration steps, constructing those coded words of length at most $\ell$, and comparing literally. Use freely reduced old words $r$, which contain no $d$; the final $d^k$ therefore survives free reduction. Include $d$ itself as a relator. Modulo $d=1$, the coded relations are exactly the original $r$. The group is therefore unchanged and the old generator images remain explicit. Higman embedding \cite{Higman}, in the effective uniform formulation of \cite[Theorem 2.2]{Chiodo}, gives a finite presentation with an explicit injective embedding. Injectivity preserves the nontrivial ${r^{\rm grp}}_n$. The model \eqref{eq:model18} need not itself be computable.
\end{proof}

Represent the generators of $\mathcal G_e$ by centralizer unitaries satisfying its finite relators. Add centralizer unitaries $T,L$. Require every group generator and $T$ to commute with $B_0,B_1,C$. Impose
\begin{equation}\label{eq:reader19}
 LQ=cQ,\quad LB_i=B_i aLa^*,\quad [L,C]=[L,P]=0,
\end{equation}
and the centralizer polynomial
\begin{equation}\label{eq:terminal}
 P\bigl(T[b,L]T^*+[b,L]\bigr)=0.
\end{equation}
\begin{lemma}[Level reader]\label{lem:reader}
In every semifinite realization $LE_n=a^nca^{-n}E_n$. The terminal equation then implies that some $n$ is good.
\end{lemma}
\begin{proof}
Group generators commute with $C$, hence with $Q=E_0$. In particular $a^*Q=Qa^*$; this is needed in the induction. For a word $v$ of length $n$, the endpoint equation and commutation of $a$ with the $B_i$ give $LB_vQ=B_v a^nLa^{-n}Q=B_va^nca^{-n}Q$. By \eqref{eq:En} these ranges span $E_n\mathcal H$, proving the level identity. A unitary commuting with $B_i$ also commutes with $B_i^*$: commute its inverse, then take adjoints. It therefore commutes with $P$. Thus $PE_n$ reduces the group generators, $T$ and $L$. Choose a nonzero such sector by Lemma~\ref{lem:selector}. On it \eqref{eq:terminal} becomes $T{r^{\rm grp}}_nT^*=-{r^{\rm grp}}_n$. A bad $n$ has relator ${r^{\rm grp}}_n=I$, giving $I=-I$ on a nonzero space. This contradiction proves goodness.
\end{proof}

\begin{lemma}[Finite realization]\label{lem:hnn}
A good $n$ gives a realization of all relations in a finite von Neumann algebra.
\end{lemma}
\begin{proof}
In $\mathcal G_e\times\langle j\mid j^2=1\rangle$, the subgroup $\langle {r^{\rm grp}}_n,j\rangle$ is $C_2^2$. The map ${r^{\rm grp}}_n\mapsto j{r^{\rm grp}}_n$, $j\mapsto j$ is its automorphism. Form the HNN extension $\Gamma^{\rm HNN}_n$ with stable letter $T$:
\begin{equation}\label{eq:hnn}
 T{r^{\rm grp}}_nT^{-1}=j{r^{\rm grp}}_n,\qquad TjT^{-1}=j.
\end{equation}
The HNN normal-form theorem \cite{HNN} embeds the base group. Consequently $j\ne1$, and $j$ commutes with the base and with $T$, so is central in $\Gamma^{\rm HNN}_n$. This is the group-extension mechanism used in the nonlocal-game setting of \cite{Slofstra}. In $L(\Gamma^{\rm HNN}_n)$, $p_-=(1-\lambda(j))/2$ is central and has trace $1/2$. The corner $M_{\rm HNN}=p_-L(\Gamma^{\rm HNN}_n)$ with trace $2\tau$ is finite. Compressed group unitaries are unitaries of $M_{\rm HNN}$, and \eqref{eq:hnn} becomes $T{r^{\rm grp}}_nT^*=-{r^{\rm grp}}_n$.

Tensor $M_{\rm HNN}$ with the depth-$n$ tree, with density $h_n\otimes1$ relative to matrix trace times the normalized trace of $M_{\rm HNN}$. Group generators and $T$ are constant across vertices. Set $L=a^\ell ca^{-\ell}$ on depth $\ell$. It is a unitary centralizer label commuting with $P,C$. On a nonleaf, $LB_i=B_i aLa^*$ because depth increases by one; both sides vanish on a leaf. At the root $LQ=cQ$. The terminal equation is the sign-flip identity on the leaves and is vacuous elsewhere. The tree relations hold by \eqref{eq:treeh}. Tensor with $R_{\rm hyp}$ and select a factor fibre to obtain a $\II_1$ host. The presentation itself never needs to find the good $n$.
\end{proof}

For a small example, at good depth $n=1$ the tree is $\{\emptyset,0,1\}$ and the blocks of $L$ are $c,aca^{-1},aca^{-1}$. The terminal word is ${r^{\rm grp}}_1=[b,aca^{-1}]$. The model \eqref{eq:model18} with good set $\{1\}$ sends it to $z_1\ne1$, while ${r^{\rm grp}}_0,{r^{\rm grp}}_2$ vanish. The HNN step flips the chosen involution after compression to $j=-1$. One $T$ is not required to flip all good words.

\subsection{One bridge compiles the selector presentation}\label{sec:onebridge}
Use five orthogonal synchronized corners $e_0,m_{00},m_{01},m_{10},m_{11}$ covering $I$, with one FMS copy whose flag is $e_0$ and magic support is their other four corners. Put $e_1=m_{00}$. The actual bridge $v=\btrans_1:e_0\to e_1$ obeys \eqref{eq:bridgecomm}. KMS and the cover give $\varphi(e_0)=1/9$ and $\varphi(m_{ab})=2/9$; here $9=\mathfrak c_1$. These masses are independent of the root model.

\begin{center}
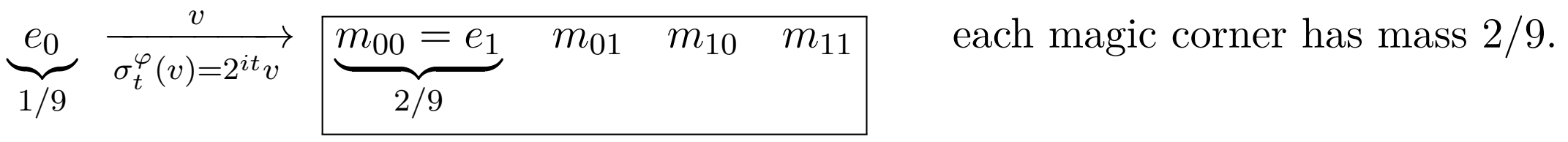
\begin{minipage}{\linewidth}
\centering
\[
 \underbrace{e_0}_{1/9}
 \ \xrightarrow[\sigma_t^\varphi(v)=2^{it}v]{v}\
 \boxed{\underbrace{m_{00}=e_1}_{2/9}\quad
 m_{01}\quad m_{10}\quad m_{11}}
 \qquad\text{each magic corner has mass }2/9.
\]
\captionof{figure}{One calibrated bridge. All copies of transported labels use this same $v$; the box is the four-dimensional magic support.}
\end{minipage}\end{center}

Bind the five corners and the two positive $ZI,IX$ parameters to the qutrit frame. Introduce the two axis unitaries $\mathfrak x,\mathfrak z$ with the explicit formulas of Section~\ref{sec:bridge}. A \emph{typed label} is an operator name together with its required class (projection, positive contraction or unitary). The transported typed label list is exactly
\begin{equation}\label{eq:labels}
 Q,\ C,\ T,\ L,\ \{U_g:g\text{ is a generator of }\mathcal G_e\}.
\end{equation}
Encode a projection by a reflection, a positive contraction by $u_C=2C-I+2i\sqrt{C(I-C)}$, and a unitary directly. A \emph{unitary name} is a formal generator for the chosen unitary encoding, including separate names for separate summands when needed. For each resulting unitary name $\ell$ introduce copies $u_{r,\ell}$, $r=0,1$, with $(u_{r,\ell}-I)(I-e_r)=0$. The \emph{decoded copies} are the actual corner operators recovered from these encodings by the following formulas, respectively: $e_r(I+u_{r,Q})/2$, $e_r(2I+u_{r,C}+u_{r,C}^*)/4$, and $e_ru_{r,\ell}$. Require $u_{r,Q}^2=I$. Introduce the block unitary $V_{1,\ell}$ of \eqref{eq:blockcopies}; since the cover is the whole space it reads
\[
 V_{1,\ell}=e_0u_{0,\ell}+
 \sum_{a,b}\mathfrak x^a\mathfrak z^b e_1u_{1,\ell}\mathfrak z^{-b}\mathfrak x^{-a}.
\]
Refine the retained FMS signs separately for the six frame projections
\[
 F_1,\quad F_2,\quad F_3,\quad G_{12}(1),\quad G_{13}(1),\quad G_{23}(1),
\]
and for every $G_{12}(V_{1,\ell})$. Lemma~\ref{lem:label} first makes $v$ preserve the frame, so $v=I_3\otimes\widetilde v$. The graph-label commutators in the target--source block give
\begin{equation}\label{eq:transport}
 vu_{0,\ell}=u_{1,\ell}v.
\end{equation}
Thus the same actual calibrated transport works for every label. Transport of decoded adjoints and polynomials follows too. $P$ is root-only, encoded by $P_0=e_0(I+u_P)/2$ with $u_P^2=I$ and $(u_P-I)(I-e_0)=0$.

Introduce two centralizer contractions $b_i=e_1b_ie_0$. For compilation encode each as an average of two global unitaries. Extract $B_i=v^*b_i$ on $e_0Me_0$. Their modular eigenvalue is $1/2$ because that of $b_i$ is one. The complete endpoint translation is
\begin{center}\small
\begin{tabular}{@{}ll@{}}
\toprule Root relation & Centralizer polynomial\\\midrule
$B_i^*B_j=\delta_{ij}(I-P)$
& $b_i^*b_j=\delta_{ij}(e_0-P_0)$\\
$\sum_iB_iB_i^*=I-Q$ & $\sum_i b_ib_i^*=e_1-Q_1$\\
$CB_i=\frac12B_iC$ & $C_1b_i=\frac12b_iC_0$\\
$U_gB_i=B_iU_g$ & $(U_g)_1b_i=b_i(U_g)_0$\\
$TB_i=B_iT$ & $T_1b_i=b_iT_0$\\
$LB_i=B_i aLa^*$ & $L_1b_i=b_i a^{[0]}L_0(a^{[0]})^*$\\
\bottomrule
\end{tabular}
\end{center}
Every remaining relation is a root centralizer polynomial: the finite relators of $\mathcal G_e$, $C_0Q_0=Q_0$, $[C_0,P_0]=0$, $[U_g,C_0]=[T_0,C_0]=0$, $L_0Q_0=c^{[0]}Q_0$, $[L_0,C_0]=[L_0,P_0]=0$, and the terminal equation. Expand its root projection as $P=(I+s_P)/2$. Here $a^{[0]},c^{[0]}$ mean the source words for the embedded $a,c$. If $W_{\rm ter}=[b,L]$, its non-self-adjoint polynomial is explicitly
\[
 p=\tfrac12\bigl(TW_{\rm ter}T^*+W_{\rm ter}+s_PTW_{\rm ter}T^*+s_PW_{\rm ter}\bigr).
\]
Use \eqref{eq:G3} for these words and their prefixes. In global scaffold coordinates multiply by $e_0$ and expand $e_0=(I+s_{e_0})/2$ as well. The two off-diagonal blocks impose $p=0$ and $p^*=0$. Expand all other typed polynomials in the same manner, apply graphs and triangles, then Lemma~\ref{lem:rows}. Retain all FMS and refinement questions. Sample all padded question pairs uniformly, reject the listed events, and accept unspecified pairs. This is an effective finite rational Boolean game, denoted $\Gsel_e$.

\paragraph{Soundness and simultaneous completeness.}
Soundness uses the nonzero root corner with normalized state $9\varphi|_{e_0Me_0}$. Equations \eqref{eq:transport} and the table give all root relations as bounded operator identities. For example $C_0v^*=v^*C_1$ gives $C_0B_i=v^*C_1b_i=B_iC_0/2$. Conjugating the range equation by $v^*$ uses $v^*Q_1v=Q_0$. The corner of a semifinite host is semifinite, so Lemmas~\ref{lem:selector} and \ref{lem:reader} prove soundness.

Conversely take any faithful normal root realization $(M,\varphi)$. On $\Mat_5\vN M$ use the state
\begin{equation}\label{eq:scaffoldone}
 \varphi_{\rm sc}=\operatorname{diag}(1,2,2,2,2)/9\ \otimes\varphi.
\end{equation}
Use the identity corner maps $v=|e_1\rangle\langle e_0|\otimes I$, the ideal real FMS effects, and $b_i=|e_1\rangle\langle e_0|\otimes B_i$. The product modular group makes each $b_i$ centralizer because the two modular eigenvalues $2$ and $1/2$ cancel. The whole centralizer is finite, with its faithful trace $\varphi_{\rm sc}$. For any contraction $x$ in a finite von Neumann algebra, the polar partial isometry extends to a unitary $u$: equivalent initial and final projections have equivalent complements in a finite von Neumann algebra. Then
\[
 u_\pm=u(|x|\pm i\sqrt{I-|x|^2}),\qquad x=(u_++u_-)/2.
\]
This supplies the exact encodings of both $b_i$ at once. Use $2Q-I,2P-I$ and the displayed functional-calculus encoding of $C$, copying the same choices to each required corner. Amplify by the scalar qutrit frame. All block labels commute with the ideal FMS signs, so use their actual joint products for refinement. Separate labels use separate questions. Finally apply the standard-form construction in Lemma~\ref{lem:rows}, with Bob's effects defined by its conjugation $\mathcal J_{\rm new}$. Formula \eqref{eq:scaffoldone} and the ideal real FMS effects give exactly the five-dimensional FMS probabilities, including the nonsynchronous signs. All graph and endpoint identities hold simultaneously. A root in a finite von Neumann algebra stays finite; after tensoring with $R_{\rm hyp}$ a $\II_1$ host exists. A $\III_{1/2}$ root stays $\III_{1/2}$ under the finite amplifications.

\subsection{Both uniform promises and the hardness theorem}\label{sec:promises}
\begin{theorem}[Uniform semifinite reduction]\label{thm:selreduction}
Uniformly from a total recursive $\Rdec$ and an input $e$, the compiler gives a finite rational Boolean game $\Gsel_e$ such that
\[
 \Gsel_e\in\perf(\II_1)
 \iff \Gsel_e\in\perf_{\sfhost}
 \iff \exists n\,\forall t\,\Rdec(e,n,t).
\]
For every input $e$, $\Gsel_e$ has a perfect $\III_{1/2}$ strategy and $\omega_q(\Gsel_e)=1$.
\end{theorem}
\begin{proof}
For the equivalence, soundness follows from the compiler and the selector and level-reader lemmas. A good $n$ gives a finite model by Lemma~\ref{lem:hnn}, which the compiler turns into perfect finite-host play and hence a $\II_1$ host.

For the first uniform promise take the uniform KMS representation of $\mathcal O_2$ with generators $s_i$ satisfying $s_i^*s_j=\delta_{ij}I$, $\sum_i s_is_i^*=I$ and $\sigma_t(s_i)=2^{-it}s_i$. Its weak closure is $R_{1/2}$; see the Cuntz/KMS identification in \cite{OP} for the gauge KMS state and \cite[Theorem 4.9 and Example 4.13]{HarrisEmbezzlement} for this factor identification, also used in \cite[Section 3.1]{ReadType}. Set $B_i=s_i$, $P=Q=C=0$, and take all group generators, $T,L$ equal to $I$. Every root equation holds: $LQ=cQ$ and the terminal row are vacuous. In the scaffold \eqref{eq:scaffoldone}, $b_i=|e_1\rangle\langle e_0|\otimes s_i$ is centralizer. Its exact two-unitary completion exists in the finite centralizer, as proved above. The encoding of $C=0$ is $u_C=-I$ on each corner; the zero projections use reflection $-I$ there as well. Their transported block labels commute with the ideal signs. Thus every compiler gadget has a simultaneous perfect realization. The host, including all finite ancillary matrices, is $\III_{1/2}$.

For the second promise take the finite-tree strategy $\mathcal V_N$ with the depth-$N$ finite tree and trivial group labels. Every root equation is exact except the terminal row. Its residual is
\begin{equation}\label{eq:residual}
 {R^{\rm ter}_N}=2P,\qquad
 \varphi_N({R^{\rm ter}_N}^*{R^{\rm ter}_N})=4\varphi_N(P)=\frac4{N+1}.
\end{equation}
Build all graph, prefix, joint and column measurements by their exact formulas even for this nonzero application row. All their structural tests remain exact. Only its final row-zero event can lose, with probability given by \eqref{eq:rowloss}. The lifted root residual is $p=e_0{R^{\rm ter}_N}$, and its graph row has off-diagonal blocks $p,p^*$. The qutrit trace and root mass give
\[
 \varphi_{\rm old}(\Delta_{\rm graph}^2)
 =\tfrac13\varphi_{\rm sc}(p^*p+pp^*)
 =\tfrac{2}{27}\varphi_N({R^{\rm ter}_N}^*{R^{\rm ter}_N}).
\]
The last equality uses traciality on the centralizer. The \emph{padded question count} is the common size of the two question sets after adding always-accepted fresh questions. An \emph{orientation} is a choice of which party receives the row question and which receives the scalar reference question. Let $m_e\geq1$ be the finite number of retained terminal-test copies in the chosen encoding, determined solely by $e$. If $Q_e$ is the padded question count and $\ell_e,d_e$ are the row denominator and slot dimension, then, for each retained terminal-test copy, union bounding its orientations gives a loss of at most $64\ell_e^2/(27d_e^3Q_e^2(N+1))$. Thus the total loss satisfies
\[
 1-\omega(\Gsel_e,\mathcal V_N)
 \le m_e\frac{64\ell_e^2}{27d_e^3Q_e^2}\,\frac1{N+1}=\frac{C^{\rm sel}_e}{N+1}.
\]
Take
\[
 C^{\rm sel}_e=m_e\frac{64\ell_e^2}{27d_e^3Q_e^2}.
\]
Thus $C^{\rm sel}_e$ depends only on the finite compiled specification, denominators, row dimensions and the fixed root mass $1/9$. It is independent of $N$, the density spectrum and the finite tree dimension. Letting $N\to\infty$ proves $\omega_q(\Gsel_e)=1$.
\end{proof}

Choose $\Adec$ to be $\Sigma^0_2$-complete. Theorem~\ref{thm:selreduction} and \eqref{eq:languageequal}, together with Theorem~\ref{thm:upper}, prove $\Sigma^0_2$-completeness of every claimed semifinite existence language in Theorem~B. Furthermore every $\Gsel_e$ is commuting perfect and has quantum value one, and
\[
 \Gsel_e\in\perf_{\co}\setminus\perf_{\sfhost}\iff e\notin\Adec.
\]
This proves $\Pi^0_2$-hardness of $\perf_{\co}\setminus\perf_{\sfhost}$ within that promise, a stronger statement than hardness of intersection with the value-one language. Membership follows from $\perf_{\co}\in\Pi^0_1$ and $\perf_{\sfhost}^c\in\Pi^0_2$: their intersection is $\Pi^0_2$.

Our earlier finite game \cite{Hyperfinite} also has a perfect strategy on a type $\III_{1/2}$ factor and quantum value one, and no perfect strategy whose marginal is normal on an injective von Neumann algebra. The exclusions differ: every $\Gsel_e$ has a perfect strategy on a matrix amplification of the injective factor $R_{1/2}$, and the NO instances exclude every semifinite host instead.

\section{Each type-III subtype has the commuting language}\label{sec:typeIII}
A type-III factor has no nonzero finite projection, so the corner of Section~\ref{sec:cornerupper} does not exist. This section shows that nothing replaces it: every fixed subtype hosts exactly the commuting-perfect games. The state and host may both be chosen.

\begin{backgroundlemma}[Expected-inclusion transfer; \cite{Takesaki,ReadType}]
If $E:N\to M$ is a faithful normal conditional expectation, every
faithful standard-form correlation on $M$ is realized on $N$.
\end{backgroundlemma}
\begin{proof}
For the state $\varphi_N=\varphi_M\circ E$, put
$\mathcal H_{\rm inc}=\overline{M\xi_{\varphi_N}}\subset L^2(N)$.
The map $x\xi_{\varphi_M}\mapsto x\xi_{\varphi_N}$ is an isometry.
Takesaki's expectation theorem \cite{Takesaki} makes $M$ invariant
under the modular group. The Tomita operators agree on the common
core $M\xi$, and their conjugations restrict as $\mathcal J_N|_{\mathcal H_{\rm inc}}=\mathcal J_M$.
Thus for left effects $A$ and right effects $\mathcal J_MB\mathcal J_M$,
\[
 \langle\xi_{\varphi_N},A \mathcal J_NB\mathcal J_N\xi_{\varphi_N}\rangle
 =\langle\xi_{\varphi_M},A \mathcal J_MB\mathcal J_M\xi_{\varphi_M}\rangle.
\]
Positivity, sums and all probabilities are preserved. Support
reduction handles nonfaithful states. This is also
\cite[Lemma 3.1]{ReadType}.
\end{proof}

\begin{lemma}\label{lem:absorb}
For any factor $N_{\rm in}$ with separable predual, $N_{\rm in}\vN R_\infty$ is
$\III_1$, and contains $N_{\rm in}$ with a faithful normal expectation.
\end{lemma}
\begin{proof}
The expectation is $\operatorname{id}\otimes\rho$ for a faithful normal
state $\rho$ on $R_\infty$.
We follow the argument of \cite[Lemma 3.8]{HM}.
For each $0<\lambda<1$, choose $0<\mu<1$ with
$\log\lambda/\log\mu$ irrational.
The injective factors satisfy
$R_\infty\cong R_\lambda\vN R_\mu$ and
$R_\infty\vN R_\lambda\cong R_\infty$.
Consequently $(N_{\rm in}\vN R_\infty)\vN R_\lambda
\cong N_{\rm in}\vN R_\infty$.
The asymptotic-ratio argument of Connes \cite[Section 3.6]{Connes}, used in that proof, gives
$\lambda\in r_\infty(N_{\rm in}\vN R_\infty)$.
Connes's implication \cite[Theorem 3.6.1]{Connes} gives
$r_\infty\setminus\{0\}\subseteq S$.
Every $0<\lambda<1$ therefore belongs to the $S$-invariant, and the
factor is $\III_1$. These steps apply to every factor $N_{\rm in}$ with
separable predual. Hyperfiniteness enters \cite[Lemma 3.8]{HM} only
through its final identification with the injective $\III_1$ factor,
which this lemma does not need.
\end{proof}

\begin{theorem}\label{thm:III}
For each fixed $\lambda\in[0,1]$,
$\perf(\III_\lambda)=\perf_{\co}$.
\end{theorem}
\begin{proof}
The inclusion into commuting perfection is immediate.
For the converse, standardize a commuting-perfect strategy and
choose a factor host $N_{\rm in}$ by central decomposition as in
Section~\ref{sec:prelim}. Put $N_{\rm abs}=N_{\rm in}\vN R_\infty$.
By expected-inclusion transfer and Lemma~\ref{lem:absorb}, this $\III_1$
factor hosts the game. Retain it for $\lambda=1$.

Fix a faithful normal state $\varphi$ on $N_{\rm abs}$.
For a countable subgroup $\Upsilon\subset\mathbb R$, given the discrete
topology, form
$N_\Upsilon=N_{\rm abs}\rtimes_{\sigma^\varphi|_\Upsilon}\Upsilon$.
The Fourier expectation onto $N_{\rm abs}$ is faithful normal, so transfers
the game. It is a factor: $\mathrm T(N_{\rm abs})=\{0\}$ for a $\III_1$ factor, hence
every nonzero modular time is outer.
If a central element has Fourier coefficient $x_g$, commutation
with $N_{\rm abs}$ makes $x_g$ an intertwiner between $\operatorname{id}$ and
$\sigma_g^\varphi$. A nonzero such coefficient for $g\ne0$ would,
by polar decomposition in a factor, implement that automorphism
innerly. Thus only $g=0$ remains, and $x_0$ is scalar.
Faithfulness of Fourier coefficients proves factoriality.

The continuous core is the crossed product by the modular
$\mathbb R$-action; its dual action restricted to the center is the
\emph{flow of weights}. For factors this flow distinguishes the
type-III subtypes \cite{ConnesTakesaki}.
Here is the core computation, with the dual flow specified.
The modular group of the dual state fixes the canonical unitaries
$u_g$ of the discrete crossed product. Writing $\lambda(t)$ for
the continuous-core implementing unitaries, it follows that
\begin{equation}\label{eq:core}
 \begin{split}
 c(N_\Upsilon)
 &\cong c(N_{\rm abs})\rtimes_{\Ad\lambda(g)}\Upsilon\\
 &\cong c(N_{\rm abs})\vN L(\Upsilon).
 \end{split}
\end{equation}
For the second isomorphism the unitaries
$y_g=\lambda(g)^*u_g$ commute with $c(N_{\rm abs})$ and satisfy the group
law of $\Upsilon$; the inverse sends $u_g$ to $\lambda(g)\otimes y_g$.
Since $N_{\rm abs}$ is $\III_1$, $c(N_{\rm abs})$ is a factor, so the center is
$L(\Upsilon)=L^\infty(\widehat\Upsilon)$.
The dual action obeys
$\theta_s(\lambda(t))=e^{-ist}\lambda(t)$ and
$\theta_s(u_g)=u_g$; therefore
$\theta_s(y_g)=e^{isg}y_g$.
Its action on the center is translation by the character
$\chi_s(g)=e^{isg}$.

For $0<\lambda<1$, take
$\Upsilon=t_\lambda\mathbb Z$, $t_\lambda=2\pi/(-\log\lambda)$.
The resulting flow is rotation on the circle with least period
$-\log\lambda$, which is the flow of a $\III_\lambda$ factor \cite{ConnesTakesaki}.
For $\lambda=0$ take $\Upsilon=\mathbb Z+\sqrt2\mathbb Z$.
Its dual is the two-torus, and the flow is
$s\mapsto(e^{is},e^{i\sqrt2s})$.
Fourier characters show ergodicity: an invariant character
indexed by $(m,n)$ requires $m+n\sqrt2=0$, hence $m=n=0$.
Every orbit is a dense one-dimensional subgroup coset of Haar
measure zero (its bounded segments have zero area).
Thus the flow is not transitive. The Connes--Takesaki flow
classification \cite{ConnesTakesaki} gives subtype $\III_0$.
This is the same modular crossed-product mechanism as
\cite[Example 3.9]{HM}, without requiring injectivity of $N_{\rm abs}$.

For clarity non-semifiniteness is a separate conclusion.
The core flow of a semifinite factor is translation on
$L^\infty(\mathbb R)$; it is transitive and admits no invariant
normal probability state. Each of our circle or torus flows
preserves Haar probability, so cannot be that semifinite flow.
For $N_{\rm abs}$ itself non-semifiniteness was already proved by its
$S$-invariant.
\end{proof}
Slofstra's commuting classification \cite{Slofstra}, also proved in \cite[Theorem 6.4]{MNY}, now proves
$\Pi^0_1$-completeness for every fixed subtype, including zero. With Section~\ref{sec:selector} this completes the proof of Theorem~B.

\section{One selector for spatial and semifinite perfection}\label{sec:iionly}
The critical selector can choose the same witness in spatial and semifinite
models. The construction uses two edge families on one scaffold: the
selector is critical, while the four-letter computation tree is subcritical.
The key point is that a nonzero selected terminal sector has positive
intersection with the computation root. This supplies a finite witness
without using a density eigenvector in a diffuse host.

\begin{lemma}[Several edge families on a common chain]\label{lem:multifamily}
Fix $n\ge1$ and finitely many edge families $X_i^{(r)}$. Each family has
its own Gram equations, initial defect and range equation, and its own
within-family endpoint equations with typed centralizer coefficients.
There are no prescribed cross-family Gram equations. If all edges have
modular ratio $2^{-n}$, one finite rational Boolean game compiles these
systems on a common nonzero source corner. It extracts them from every
perfect normal commuting strategy, and every faithful normal realization
extends simultaneously, preserving host type under finite amplification
and finite entropy for spatial realizations.
\end{lemma}
\begin{proof}
Use the corners, activities, bound axes, and common chain
$v=\btrans_n\cdots\btrans_1$ of Lemma~\ref{lem:chain}. Its source mass is
$\mathfrak c_n^{-1}$. For each edge, independently introduce a supported
centralizer contraction $a_i^{(r)}=e_na_i^{(r)}e_0$. A family with initial
defect $D_r$ and range $R_r$ is compiled as
\[
 a_i^{(r)*}a_j^{(r)}=\delta_{ij}(e_0-(D_r)_0),\qquad
 \sum_i a_i^{(r)}a_i^{(r)*}=(R_r)_n.
\]
Only that family's own equations are imposed. Transport the union of
all labels appearing at left endpoints or in range equations, with one
separate graph refinement per encoding and bridge. The common original
signs force $L_nv=vL_0$ for every label in this union. Decode
$X_i^{(r)}=v^*a_i^{(r)}$. Multiplication by $v,v^*$ proves the Gram, range
and endpoint translations in both directions; for example
$L_na_i^{(r)}=a_j^{(r)}(R_{\rm end})_0$ becomes
$L_0X_i^{(r)}=X_j^{(r)}R_{\rm end}$. The centralizer condition and the modular
eigenvalue of $v$ give ratio $2^{-n}$ for every extracted edge.
All equations concern the same faithful corner and state.

Conversely use $\operatorname{diag}(\mu_n)\otimes\varphi_0$ and
$a_i^{(r)}=|e_n\rangle\langle e_0|\otimes X_i^{(r)}$ for every family.
Every lifted contraction lies in the finite centralizer and has its
own two-unitary encoding. Choose the root encodings once and copy them
through the identity corner maps. The resulting block labels commute
with their bridge's canonical signs. Actual joint products give all
separate refinements, followed by the common qutrit and row extensions.
There is one state and one assignment for all families. The construction
adds only finite factors and imposes no mixed orthogonality.
\end{proof}

\subsection{The coupled root presentation}
Use four edges $A_{ab}$ satisfying \eqref{eq:fockrelations} at modular
ratio $1/16$, with root $P_\circ$, and sixteen different edges $B_i$ with
\[
 B_i^*B_j=\delta_{ij}(I-P),\quad \sum_{i=1}^{16}B_iB_i^*=I-Q,
 \quad \sigma_t^\varphi(B_i)=16^{-it}B_i.
\]
For a positive centralizer contraction $C_{\rm sel}$ impose
\[
 C_{\rm sel}Q=Q,\quad C_{\rm sel}B_i=\tfrac12B_iC_{\rm sel},
 \quad[C_{\rm sel},P]=0,\quad[P,A_{ab}]=[C_{\rm sel},A_{ab}]=0.
\]
Lemma~\ref{lem:selector} above uses modular ratio $1/16$ and
depth ratio $1/2$. Retain the coordinate and boundary equations
\eqref{eq:U_x}--\eqref{eq:P_y} and the counter equations
\eqref{eq:C}, with $[C_{\rm ctr},C_{\rm sel}]=0$. Replace the clock by
\begin{equation}\label{eq:equalityreader}
 (2C_{\rm ctr}-C_{\rm sel})P_{\rm ret}=0,\qquad
 T_{\rm inv}(2C_{\rm ctr}-C_{\rm sel})=
 \tfrac18C_{\rm ctr}C_{\rm sel}(I-P_{\rm ret}).
\end{equation}
Here $T_{\rm inv}$ is a general contraction. The tableau has the same
initial and update equations, but its halt prohibition is
$P_{\mathtt H}P=0$. Its machine reads the first positive pulse ${d_{\rm clk}}$, sets
$j={d_{\rm clk}}-1$, and tests $\Rdec(e,j,0),\Rdec(e,j,1),\ldots$, halting at the
first false answer.

Lemma~\ref{lem:multifamily} with $n=4$ compiles the two separate Gram
systems. The transported typed list is exactly
\[
 P_\circ,U_x,U_y,P_x,P_y,C_{\rm ctr},P,C_{\rm sel},Q.
\]
The return projection, its inverse and the tableau projections are
source-only. In particular no relation $A_{ab}^*B_i=0$ is imposed.
Denote this finite game by $F_e$.

\subsection{The terminal sector meets the root}
\begin{lemma}\label{lem:terminalroot}
Let orthogonal isometries $A_1,\ldots,A_r$ have modular ratio
$\nu$ on a faithful normal support. If a centralizer projection $p$
commutes with them and $r\nu<1$, then
\[
 \varphi\left(p\left(I-\sum_i A_iA_i^*\right)\right)
 =(1-r\nu)\varphi(p).
\]
\end{lemma}
\begin{proof}
Commutation with a self-adjoint projection also gives commutation with
the adjoints. KMS gives
$\varphi(A_ipA_i^*)=\nu\varphi(pA_i^*A_i)=\nu\varphi(p)$.
Subtract the sum of these $r$ terms from $\varphi(p)$.
\end{proof}

In a semifinite realization, Lemma~\ref{lem:selector} with $d=16$ gives
some nonzero $p_j=P1_{\{2^{-j}\}}(C_{\rm sel})$. It commutes with all
$A_{ab}$ and their adjoints. Hence
\begin{equation}\label{eq:terminalroot}
 \varphi(p_jP_\circ)=\tfrac34\varphi(p_j)>0.
\end{equation}
Choose $0\ne\xi\in p_jP_\circ\HH$. Its word descendants remain in
$p_j\HH$, are nonzero, and are orthogonal at every fixed word length.
The coordinate inductions identify every finite square on these vectors.

For self-adjoint $X,Y$, with $Y$ injective, the two equations
$(2X-Y)R=0$ and $T(2X-Y)=\kappa XY(I-R)$, $\kappa\ne0$, force
$R=1_{\{0\}}(2X-Y)$. Indeed $R\le K=1_{\{0\}}(2X-Y)$.
For $\zeta\in K\HH$ put $\eta=(I-R)\zeta\in K\HH$. The second equation
gives $XY\eta=0$, and $Y\eta=2X\eta$ gives $2X^2\eta=0$.
Self-adjointness gives $X\eta=0$, then $Y\eta=0$, then $\eta=0$.
Apply this to \eqref{eq:equalityreader}, using injectivity of
$C_{\rm sel}$ from the selector lemma. On the chosen descendants,
$C_{\rm ctr}=2^{-y}$ and $C_{\rm sel}=2^{-j}$. Thus the pulse is
exactly $y=j+1$, including no time-zero pulse.

\begin{lemma}[Definite computation on a terminal sector]\label{lem:terminaltableau}
Suppose $0\ne\xi\in P_\circ\HH$, every $A_w\xi$ belongs to $P\HH$,
and $P_{\rm ret}$ has the specified classical scalar on each $A_w\xi$.
With the coordinate, initial and update equations above, every symbol
projection acts on these descendants as the genuine classical symbol.
If the presentation imposes $P_{\mathtt H}P=0$, that classical
computation cannot halt.
\end{lemma}
\begin{proof}
At each fixed depth, the coordinate recursions give cyclic shifts and
boundary masks on the word descendants. Row zero follows from
\eqref{eq:initial} without its global halt equation. Induct on time at
all spatial coordinates simultaneously. Each factor in the displayed
ordered update product acts as its classical zero or one on the
induction vector: left and right factors refer to the preceding row,
and the pulse is scalar by hypothesis. Exactly the prescribed successor
survives. Multiplication by $U_y$ gives the next-row assertion. The
last-row mask removes wraparound; radius one and blank quiescence make
the right ghost blank throughout the required prefix. This proves the
half-line computation on every sufficiently large square. If a halt
occurred, one nonzero descendant $\eta\in P\HH$ would satisfy
$P_{\mathtt H}\eta=\eta$, contradicting
$P_{\mathtt H}\eta=P_{\mathtt H}P\eta=0$.
\end{proof}

\begin{theorem}[A shared family]\label{thm:sharedfamily}
The family $F_e$ satisfies
\[
 F_e\in\LE\iff F_e\in\LI\iff F_e\in\LU
 \iff\exists j\,\forall t\,\Rdec(e,j,t),\qquad F_e\notin\LJ.
\]
\end{theorem}
\begin{proof}
Semifinite soundness follows from the selected sector, the exact pulse
and Lemma~\ref{lem:terminaltableau}. A false check at that selected $j$
would halt, so $j$ is a witness.

For completeness let $j$ be a witness. On the finite sixteen-letter
tree of words $v$ of length at most $j$, use prefix maps $B_i$, zero
on leaves, with $P$ the leaves, $Q$ the root and
\[
 h_{\rm sel}|v\rangle=(j+1)^{-1}16^{-|v|}|v\rangle,
 \qquad C_{\rm sel}|v\rangle=2^{-|v|}|v\rangle.
\]
Tensor it with the four-letter Fock density $h_F$ of
\eqref{eq:fockdensity}. At selector depth $\ell$, put
$P_{\rm ret}=1_{\{y=\ell+1\}}$ and
\[
 T_{\rm inv}(y,\ell)=
 \begin{cases}[8(2^{\ell+1}-2^y)]^{-1},&y\ne\ell+1,\\
 0,&y=\ell+1.\end{cases}
\]
The nonzero denominator before the factor eight is an integer. The norm
is exactly $1/8$, attained at $\ell=0,y=0$. Negative values occur.
Use the genuine classical tableau for input $\ell$ on each such fibre;
all symbol projections are its diagonal indicator projections. Coordinate
permutations preserve Fock depth, and every other centralizer label is
diagonal. Both edge families have ratio $1/16$. Computations at depths
$\ell<j$ may halt; the terminal equation excludes only depth $j$.
This gives every label and equation simultaneously, also at $j=0$,
where $B_i=0$, $P=Q=C_{\rm sel}=I$.

The selector entropy is $\log_2(j+1)+2j$, so the root product entropy is
\[
 \log_2(j+1)+2j+\log_2(4/3)+4/3<\infty.
\]
The compiler adds finite entropy. The inclusions
$\LE\subseteq\LI\subseteq\LU$ finish the equivalence. In a finite
host the four extracted isometries would have orthogonal range traces
summing to four under a normalized faithful trace, but their sum is at
most the identity. Thus every instance is outside $\LJ$.
\end{proof}

\section{Comparing the type languages}\label{sec:sep}
Equal arithmetical complexity does not mean equal languages. This section shows that the type-I and type-$\II_1$ languages are incomparable and classifies the differences ``perfect in $X$ but not in $Y$'' among six models.

\subsection{A fixed one-bridge separator}\label{sec:shiftseparator}
Apply the complete one-bridge assembly of Section~\ref{sec:onebridge} to the fixed root presentation
\begin{equation}\label{eq:shiftroot}
 A^*A=I,\qquad AA^*=I-P_\circ,\qquad
 \sigma_t^\varphi(A)=2^{-it}A,
\end{equation}
where $P_\circ$ is a centralizer projection. Call the resulting finite rational Boolean game $\Gsh$. Here the transported label list contains just $P_\circ$. Encode its copies by reflections, introduce a centralizer contraction $a=e_1ae_0$, and impose exactly
\[
 a^*a=e_0,\qquad aa^*=e_1-(P_\circ)_1.
\]
Use the six frame refinements and the graph-label refinement for $P_\circ$ from Section~\ref{sec:onebridge}; apply the same graphs, triangles and rational rows. Extraction is $A=v^*a$, and $P=(P_\circ)_0$. Thus $A^*A=e_0$, $AA^*=e_0-P$, and the modular ratio is $1/2$. Conversely every root model of \eqref{eq:shiftroot} extends by that assembly, with $a=|e_1\rangle\langle e_0|\otimes A$.

For its type-I model take $\ell^2(\mathbb N_0)$, $A|m\rangle=|m+1\rangle$, $P_\circ=|0\rangle\langle0|$, and
\[
 h|m\rangle=2^{-m-1}|m\rangle,\qquad
 S(h)=\sum_{m\geq0}(m+1)2^{-m-1}=2\text{ bits}.
\]
The relation $hA=Ah/2$ verifies modular covariance. The scaffold and the finite maximally entangled compiler factors add finite entropy, so $\Gsh$ is perfect spatially with finite entropy.

In any perfect finite-host strategy, extraction takes place in the nonzero finite corner $e_0Me_0$. Its isometry $A$ must be unitary: a faithful finite trace gives $\tau(e_0-AA^*)=0$. Consequently $AA^*=e_0$ forces $P=0$. On the other hand KMS for the original faithful state gives
\[
 \varphi(AA^*)=\tfrac12\varphi(A^*A)=\tfrac12\varphi(e_0),
 \qquad \varphi(P)=\tfrac12\varphi(e_0)>0,
\]
a contradiction. Therefore $\Gsh$ has no finite host and witnesses $\perf(\mathrm I)\not\subseteq\perf(\II_1)$. In particular it has no finite-dimensional perfect strategy. Our earlier binary-scoring game \cite{Gamma} is another finite game that is perfect in infinite dimension with finite entanglement entropy but never perfect in finite dimension.

\subsection{What happens to the Weyl-clock family}\label{sec:weylsemifinite}
The spatial reduction of Sections~\ref{sec:tree}--\ref{sec:reduction} uses the four Fock isometries $A_{ab}$ of \eqref{eq:fockrelations}, with $h_F$ of \eqref{eq:fockdensity}, the coordinate unitaries and boundary projections of \eqref{eq:U_x}--\eqref{eq:P_y}, the Weyl relation \eqref{eq:weyl}, the clock and counter values \eqref{eq:coordinateclock}, and the reader equations \eqref{eq:reader}. Spatial soundness uses finite positive density eigenspaces and a determinant to force a finite-order phase. This determinant is unavailable in a diffuse finite factor.

Here is a normal $\II_\infty$ model of \emph{every} input of that family. Put $\vartheta=(\sqrt5-1)/2$, $\zeta=e^{2\pi i\vartheta}$. In the finite rotation factor $L^\infty(\mathbb T)\rtimes_\vartheta\mathbb Z$, take unitaries $J_{\rm W},K_{\rm W}$ with commutator $\zeta I$ and put $Z_{\rm clk}=\zeta I$. The invariant integral and Fourier coefficient at zero give a faithful finite trace. Irrational rotation is free and ergodic, so a central Fourier coefficient vanishes except at zero and its zero coefficient is constant; the algebra is a factor. It is infinite-dimensional and hence $\II_1$. It is the hyperfinite factor $R_{\rm hyp}$ by amenability of this crossed product and injective-factor classification \cite{Connes76}. The construction itself only needs that it is finite.

Use the tensor density $h_F\otimes1$ on $B(\ell^2(\WW))\vN R_{\rm hyp}$. Set $W=\zeta^y$, $C_{\rm ctr}=2^{-y}$, $P_{\rm ret}=1_{\{y=0\}}$, and
\[
 T_{\rm inv}|_{(x,y)}=
 \begin{cases}0,&y=0,\\
 2^{-y-1}(\zeta^y-1)^{-1},&y>0.
 \end{cases}
\]
The norm bound follows from the quadratic irrationality estimate. For an integer $p$ nearest to $y\vartheta$, the nonzero integer $p^2+py-y^2$ factors as $(p-y\vartheta)(p-y\vartheta')$, $\vartheta'=-(1+\sqrt5)/2$. Since $|p-y\vartheta|\leq1/2$,
\[
 \|y\vartheta\|\geq\frac1{\sqrt5\,y+1/2},\qquad
 |\zeta^y-1|=2\sin(\pi\|y\vartheta\|)
 \geq\frac4{\sqrt5\,y+1/2}>2^{-y-1}.
\]
The sine inequality uses concavity, $\sin(\pi t)\geq2t$ for $0\leq t\leq1/2$. In particular it supplies the explicit all-integer bound $|\sin(\pi y\vartheta)|\geq8/(11y)$, since $\sqrt5<9/4$ and $y\geq1$. For the last inequality it suffices to replace $\sqrt5$ by $9/4$: $2^{y+3}>9y/4+1/2$ for $y=1$, and doubling proves it inductively. Thus $T_{\rm inv}$ is a contraction and \eqref{eq:reader} holds. The binary coordinate recurrences $y\mapsto2y+b$ verify the clock and counter endpoint equations.

In the tableau of Sections~\ref{sec:tree} and \ref{sec:reduction} a waiting front advances one cell per time step, at $x=y$ at time $y$. The first positive-time return pulse $P_{\rm ret}$ freezes it, and the frozen coordinate becomes the unary input of the check of the hardwired predicate; the time-zero pulse is ignored. Here $P_{\rm ret}$ selects $y=0$ only, so no positive-time pulse occurs. The front never freezes, and the decoder never starts a checking computation, irrespective of the hardwired input. Its finite deterministic tableau relations are satisfied by that classical nonhalting tableau in every finite square. The coordinate, boundary and tableau projections are diagonal on word sectors and commute with $h_F\otimes1$. All endpoint labels have the required centralizer type. Lemma~\ref{lem:chain} with $n=4$ applies to exactly the conditional copies and transported labels of Lemma~\ref{lem:four} and to the endpoint rows of Sections~\ref{sec:tree}--\ref{sec:reduction}. Its left-endpoint labels are $P_\circ,U_x,U_y,P_x,P_y,W,C_{\rm ctr},Z_{\rm clk},J_{\rm W},K_{\rm W}$; source-only labels are $P_{\rm ret},T_{\rm inv}$ and the tableau projections. The activity corners and separate refinements coincide with that compiler. Its root mass is $1/121$. The resulting ambient host is $\II_\infty$ for every input. The supplementary exact margin certificate is in Appendix~\ref{app:rotationcertificate}.

Lemma~\ref{lem:chain}, with $n=4$, also extracts the four isometries of every Weyl-clock game in any normal commuting host. Two isometries with orthogonal ranges cannot lie in a nonzero finite corner. Hence every Weyl-clock game has no finite host. This additional fact is independent of the separator $\Gsh$ above.

\subsection{A tracial game with no spatial perfection}
For the other direction use a linear-system game with a perfect commuting strategy and no perfect finite-dimensional strategy, provided by \cite{SlofstraNonclosed}. The solution-group criterion \cite[Theorems 4--5]{CLS} says its distinguished central involution $j$ is nontrivial. The central corner $(1-\lambda(j))L(\Gamma_{\rm sol})/2$ of the solution group $\Gamma_{\rm sol}$ has normalized trace and a perfect left/right regular strategy: equation observables commute in each constraint, square to $I$, and their product is the specified sign because $j=-I$ in the corner. Use their joint spectral projections on the left and the commutant mirrors of variable observables on the right. Consistency and each equation are then exact. A factor fibre is a finite factor host; tensoring with $R_{\rm hyp}$ gives $\II_1$. This is the tracial mechanism in \cite[Proposition 7.4]{ReadType}. Call this game $\Glin$. Gao, Yu and Zhi give an explicit linear-system game with a perfect commuting strategy and finite-dimensional value strictly below one \cite{GYZ}; the same construction places it in $\perf(\II_1)$, and its value gap excludes spatial perfection.

Cleve--Mittal \cite{CM} and \cite[Theorem~5]{CLS} characterize the finite-dimensional case. Spatial perfect play also implies finite-dimensional perfect play for linear-system games; Slofstra \cite{SlofstraNonclosed} uses this implication in the nonclosure construction. Here is the normal spatial argument. Each variable appearing in a constraint has an Alice involution with a self-adjoint Bob mirror on the state vector, by perfect consistency. Equality of those vectors makes Alice's trace-class marginal commute with every such involution. The within-constraint products and parity equations hold on the marginal support, by the zero-loss equations and support separation. A positive density eigenvalue has a nonzero finite-dimensional eigenspace reducing all variable involutions. Restriction to it preserves the products and all constraint equations. Its maximally entangled vector with transposed Bob variables is a finite-dimensional perfect strategy, a contradiction. For a mixed spatial state first purify; the same Alice eigenprojection argument applies. Type-I hosting decomposes into spatial factor fibres, so it is also excluded. Thus $\perf(\II_1)\not\subseteq\perf(\mathrm I)$.

Every type-I factor host tensored with $R_{\rm hyp}$ has type $\II_1$ in finite dimension and type $\II_\infty$ in infinite dimension; a further $B(\ell^2)$ amplification handles the former case. Every $\II_1$ host similarly amplifies to $\II_\infty$. The two witnesses just constructed prove
\[
 \perf(\mathrm I)\ \text{and}\ \perf(\II_1)
 \text{ are incomparable},\qquad
 \perf(\mathrm I),\,\perf(\II_1)
 \subsetneq\perf(\II_\infty).
\]
Equal arithmetical completeness does not imply equal languages.

\subsection{Exact conjunction and disjunction}
We now compare the six languages $\Lq,\LE,\LI,\LJ,\LU,\LC$ of Section~\ref{sec:prelim}. Equation~\eqref{eq:languageequal} identifies $\LU$ with both $\perf(\II)$ and $\perf(\II_\infty)$; Theorem~\ref{thm:III} identifies $\LC$ with each fixed type-III language.
\begin{backgroundlemma}[Exact game operations]
Uniform effective operations $G\wedge G'$ and $G\vee G'$ satisfy, for every $\mathbf X\in\{\Lq,\LE,\LI,\LJ,\LU,\LC\}$,
\[
 G\wedge G'\in\mathbf X\iff G,G'\in\mathbf X,\qquad
 G\vee G'\in\mathbf X\iff G\in\mathbf X\text{ or }G'\in\mathbf X.
\]
The same statements hold for each of the type languages in this paper.
\end{backgroundlemma}
\begin{proof}
First pad question and answer alphabets to common finite alphabets. For an original zero-weight question pair accept everything; preserve its original predicate on positive-weight pairs. Give every padded question pair positive rational weight, accept all pairs involving fresh questions, and reject a padded answer on an original question. This changes no perfect-play language. It permits fixed common question alphabets with full support; no implication below relies on an untested pair.

For conjunction, send both players the same tag and questions for that tagged game, and test its predicate. A perfect strategy restricts to either tag in the same ambient algebra and state. Conversely, tensor two perfect strategies and on a tagged question measure the appropriate factor. Type I is closed under tensor products, as are finite factors, type $\II_\infty$ factors, and semifinite algebras. Amplification gives the stated $\II$ existence languages when needed. For $\LE$ use pure spatial witnesses: marginal entropies add, so their sum is finite. Finite-dimensional witnesses stay finite. Commuting strategies tensor as well. For $\III_\lambda$ use its language equality with $\LC$ from Theorem~\ref{thm:III}; no closure of that subtype under direct tensoring is asserted.

For disjunction, ask the common questions and require each answer to name a branch $s\in\{G,G'\}$ as well as an answer in that branch. Reject mismatched branches and otherwise test the named predicate. A strategy for either game wins by always naming that branch. For soundness let $A_{s,a}^x,B_{s,b}^y$ be the effects and put $P_s^x=\sum_aA_{s,a}^x$, $Q_s^y=\sum_bB_{s,b}^y$. Work first on the faithful standard-form support. The matching argument of Section~\ref{sec:support} makes each branch marginal a projection, with $P_s^x\xi=Q_s^y\xi$ for every $x,y$. Fixing $y$ and using separation makes $P_s^x=P_s$ independent of $x$; likewise $Q_s^y=Q_s$. These are opposite mirrors, $P_s\xi=Q_s\xi=:\xi_s$. At least one $\xi_s$ is nonzero. Since $0\leq A_{s,a}^x\leq P_s$, we have $A_{s,a}^xP_s=P_sA_{s,a}^x=A_{s,a}^x$, and the same holds on Bob's side. Complete each branch POVM to the identity by adding $I-P_s$ to one Alice answer and $I-Q_s$ to one Bob answer. These added effects annihilate $\xi_s$. A losing branch product annihilates $\xi$ and, by the displayed support identities, also $\xi_s$. Normalizing $\xi_s$ therefore wins branch $s$. All effects and the vector remain on the same Hilbert space, in the same ambient $M,M'$; the selected vector need not be faithful. Thus its host type is preserved. This also handles mixed strategies by purification and standardization.

For finite entropy one can work in the original spatial purification. Orthogonality of Bob's branch projections gives $\rho_A=\sum_s p_s\rho_{A,s}$, $p_s=\|\xi_s\|^2$. Entropy concavity, valid also by finite-rank truncation for infinite entropy, gives $\sum_s p_sS(\rho_{A,s})\leq S(\rho_A)<\infty$. Every positive-weight branch has finite entropy. This proves the last model-specific assertion and completes both directions.
\end{proof}

\subsection{Four fixed separators and two uniform families}
Choose both reductions $\Gclk_e$ and $\Gsel_e$ for the same predicate, $e\in\mathrm{FIN}$, and write $\Fin(e)$ for its truth value. Theorems~\ref{thm:reduction} and \ref{thm:selreduction}, including Lemma~\ref{lem:hnn}, give
\begin{align}
 \Gclk_e\in\LE&\iff \Gclk_e\in\LI\iff \Fin(e),
 &&\label{eq:sharedE}\\
 \Gsel_e\in\LJ&\iff \Gsel_e\in\LU\iff \Fin(e),
 &\Gsel_e&\in\LC.\label{eq:sharedJ}
\end{align}
Every $\Gclk_e$ has no finite-dimensional perfect strategy; every $\Gsel_e$ is commuting perfect with $\omega_q=1$. The chained lemma additionally excludes $\Gclk_e$ from $\LJ$ and places it in $\LU$ (Section~\ref{sec:weylsemifinite}). The reductions below use its displayed spatial coordinates and use $\Gsh$ for finite-host exclusion.

Take the fixed one-bridge game $\Gsh$ of Section~\ref{sec:shiftseparator}. Its shift model has root entropy $2$ bits and its finite compiler factors preserve finite entropy. The finite-corner KMS contradiction there proves that it has no finite host. Thus its complete profile uses only the one-bridge assembly.

The tracial solution-group game $\Glin$ is in $\LJ\setminus\LI$; spatial perfection would imply finite perfection. Let $G^{(2)}_{\Cat}$ be the compiled Catalan game of \cite[Theorem C and Sections 3 and 6]{Entropy}. Its perfect spatial strategies have infinite marginal entropy, so it is in $\LI\setminus\LE$. Lemma~\ref{lem:chain} with $n=2$ applies to its nine-corner assembly: activity $e_{g-1}+m_g$, root mass $1/25$, transported encodings $b_+,b_-,p,q,c$ (names of \cite[Section 6]{Entropy}), and the source-only polynomials of \cite[Section 6]{Entropy}. Thus its extracted isometry $A$ has modular ratio $1/4$ also on a commuting normal host. In a finite root corner $A$ would be unitary, and KMS would give $1=1/4$; hence it is outside $\LJ$ as well. Finally choose a fixed NO instance of $\Gsel$ and call it $\Gthree$. Theorem~\ref{thm:selreduction} gives $\Gthree\in\LC\setminus\LU$ and $\omega_q(\Gthree)=1$. Their complete profiles are
\begin{center}\small
\begin{tabular}{@{}lcccccc@{}}
\toprule & $\Lq$ & $\LE$ & $\LI$ & $\LJ$ & $\LU$ & $\LC$\\\midrule
$\Gsh$ &0&1&1&0&1&1\\
$\Glin$ &0&0&0&1&1&1\\
$G^{(2)}_{\Cat}$ &0&0&1&0&1&1\\
$\Gthree$ &0&0&0&0&0&1\\\bottomrule
\end{tabular}
\end{center}
In particular $\Gsh\wedge \Glin$ is in $\LU$ but in neither $\LI$ nor $\LJ$, so even $\LI\cup\LJ\subsetneq\LU$. The remaining inclusions asserted above are strict: $\Gsh$ separates $\Lq$ from $\LE$, $G^{(2)}_{\Cat}$ separates $\LE$ from $\LI$, $\Glin$ separates $\LI$ from $\LU$, and $\Gthree$ separates $\LU$ from $\LC$. Both $\LE$ and $\LI$ are incomparable with $\LJ$.

\Needspace{5\baselineskip}
\subsection{The table, with reductions for every nonempty entry}
\begin{theorem}[Directional differences]\label{thm:differences}
The following table gives row minus column. Each suffix c means computable many-one completeness for the displayed pointclass. Here $D_2$ abbreviates $D_2(\Sigma^0_2)$. Every nonempty cell is complete for its displayed class.
\begin{center}\small
\setlength{\tabcolsep}{5pt}
\begin{tabular}{@{}lcccccc@{}}
\toprule & $\Lq$&$\LE$&$\LI$&$\LJ$&$\LU$&$\LC$\\\midrule
$\Lq$&$\varnothing$&$\varnothing$&$\varnothing$&$\varnothing$&$\varnothing$&$\varnothing$\\
$\LE$&$\Sigma^0_2$-c&$\varnothing$&$\varnothing$&$D_2$-c&$\varnothing$&$\varnothing$\\
$\LI$&$\Sigma^0_2$-c&$D_2$-c&$\varnothing$&$D_2$-c&$\varnothing$&$\varnothing$\\
$\LJ$&$\Sigma^0_2$-c&$D_2$-c&$D_2$-c&$\varnothing$&$\varnothing$&$\varnothing$\\
$\LU$&$\Sigma^0_2$-c&$D_2$-c&$D_2$-c&$D_2$-c&$\varnothing$&$\varnothing$\\
$\LC$&$\Pi^0_1$-c&$\Pi^0_2$-c&$\Pi^0_2$-c&$\Pi^0_2$-c&$\Pi^0_2$-c&$\varnothing$\\\bottomrule
\end{tabular}
\end{center}
\end{theorem}
\begin{proof}
All empty entries follow from the stated inclusions. For upper bounds, $\Lq\in\Sigma^0_1$, $\LE,\LI,\LJ,\LU\in\Sigma^0_2$, and $\LC\in\Pi^0_1$. A difference between two middle languages is therefore in $D_2$. For a middle $\mathbf X$, $\mathbf X\setminus\Lq\in\Sigma^0_2$: write its predicate as $\exists k\forall m$ for $\mathbf X$ and conjoin the $\Pi^0_1$ predicate for nonmembership in $\Lq$. Also $\LC\setminus\Lq\in\Pi^0_1$ and $\LC\setminus\mathbf X\in\Pi^0_2$. It remains to prove the matching lower bounds.

For the four incomparable entries use $\Gclk_e\wedge \Gsh$ as the $\LE$- or $\LI$-side family, and $\Gsel_f\wedge \Glin$ as the $\LJ$-side family. Their disjunction has $\LE$ and $\LI$ coordinates exactly $\Fin(e)$ and $\LJ$ coordinate exactly $\Fin(f)$. Thus its membership in either $\LE\setminus\LJ$ or $\LI\setminus\LJ$ is $\Fin(e)\wedge\neg \Fin(f)$. Swap $e,f$ between the two branches for the reverse differences. The canonical $D_2$ problem reduces in all four cases; in particular $\LI\setminus\LJ$ is $D_2(\Sigma^0_2)$-complete.

For the $\Lq$ column, use $\Gclk_e$ for $\LE$ and $\LI$; use $\Gsel_e\wedge \Glin$ for $\LJ$ and $\LU$. Their larger coordinate is $\Fin(e)$ and their finite-dimensional coordinate is identically zero. Each such entry is therefore $\Sigma^0_2$-hard.

For the four rightward entries of the $\LC$ row, use $\Gclk_e\vee \Gthree$ when $\mathbf X=\LE,\LI$, and $\Gsel_e\vee \Gthree$ when $\mathbf X=\LJ,\LU$. The commuting coordinate is always one and the $\mathbf X$ coordinate is exactly $\Fin(e)$, giving $\Pi^0_2$-hardness. These instances all have $\omega_q=1$, since choosing branch $\Gthree$ approximates perfection in finite dimension. For $\LC\setminus\Lq$, take a $\Sigma^0_1$-complete finite-perfection reduction $G^{\rm MIP}_e$ with $G^{\rm MIP}_e\in\Lq\iff e$ halts, available by perfect completeness in \cite[Theorem 12.6]{MIPRE} and finite-dimensional feasibility. Then $G^{\rm MIP}_e\vee G^{(2)}_{\Cat}$ is always commuting perfect, has $\omega_q=1$ because $G^{(2)}_{\Cat}$ is spatial perfect, and is in $\Lq$ exactly when $e$ halts. Nonhalting reduces to the difference, proving $\Pi^0_1$-hardness even under the promise.

For the two remaining complete nested differences use a shared family. In general suppose $\mathbf Y\subset\mathbf X$ and a computable family $\mathcal F_e$ has both coordinates equal to $\Fin(e)$. Fix $\Gsep\in\mathbf X\setminus\mathbf Y$. For
\begin{equation}\label{eq:sharedcriterion}
 \mathcal D(e,f)=\mathcal F_e\wedge(\Gsep\vee \mathcal F_f)
\end{equation}
the $\mathbf X$ coordinate is $\Fin(e)$ and the $\mathbf Y$ coordinate is $\Fin(e)\wedge \Fin(f)$. The difference coordinate is therefore $\Fin(e)\wedge\neg \Fin(f)$, proving $D_2$-hardness. Apply this first with $(\mathbf X,\mathbf Y)=(\LI,\LE)$, $\mathcal F_e=\Gclk_e$ and $\Gsep=G^{(2)}_{\Cat}$, using \eqref{eq:sharedE}. Apply it again with $(\mathbf X,\mathbf Y)=(\LU,\LJ)$, $\mathcal F_e=\Gsel_e$ and $\Gsep=\Gsh$, using \eqref{eq:sharedJ}. Thus $\LI\setminus\LE$ and $\LU\setminus\LJ$ are $D_2(\Sigma^0_2)$-complete. The required YES witness has finite entropy by Theorem~\ref{thm:reduction}.

For $\LU\setminus\LI$, $\LU\setminus\LE$, and the stronger language
$\LU\setminus(\LI\cup\LJ)$, take the shared family $F_e$ of
Theorem~\ref{thm:sharedfamily} for FIN and fix $H=\Glin\in\LJ\setminus\LI$.
Set $\Lambda(e,f)=F_e\wedge(H\vee F_f)$. Its coordinates are
\[
 \LU(\Lambda)=\Fin(e),\quad
 \LI(\Lambda)=\LE(\Lambda)=\Fin(e)\wedge\Fin(f),\quad
 \LJ(\Lambda)=0.
\]
Each difference therefore has predicate $\Fin(e)\wedge\neg\Fin(f)$.
All three are in $D_2(\Sigma^0_2)$, completing the table and proving
the additional simultaneous exclusion. Replace each $F_e$ by
$F_e\vee\Gthree$ to obtain commuting perfection and $\omega_q=1$
on every input, with the same four lower coordinates. The two
conjuncts then both have value one.
\end{proof}

\begin{corollary}\label{cor:entropyonly}
The language of games with perfect spatial play but no finite-entropy perfect spatial strategy is $D_2(\Sigma^0_2)$-complete. In particular it belongs to neither $\Sigma^0_2$ nor $\Pi^0_2$.
\end{corollary}
\begin{proof}
The identification is $\LI\setminus\LE$ in the theorem. A $D_2$-complete language is hard for both $\Sigma^0_2$ and $\Pi^0_2$ by fixing one of the canonical FIN indices to a YES or NO index. Membership in either pointclass would collapse these two pointclasses, contradicting strictness of the arithmetical hierarchy.
\end{proof}
The corollary concerns exact attainment; it asserts no rate at which entropy must diverge for near-optimal strategies.

For differences between middle models, value-one padding can be made explicit. Replacing $G$ by $G\vee \Gthree$ preserves all lower coordinates and sets the commuting coordinate to one, while forcing $\omega_q=1$. To remove finite-dimensional perfect play as well, use $G\wedge(\Gsh\vee \Glin)$; the second factor has all five infinite-model coordinates equal to one and its $\Lq$ coordinate zero. In particular \eqref{eq:sharedcriterion} for $\LI\setminus\LE$ already has no finite-dimensional perfect instance; adding $\Gthree$ also makes every instance value one. The $\LC$-row hardness families above already have value one, including the separate $\LC\setminus\Lq$ construction.

\begin{corollary}[A natural uniqueness problem]\label{cor:onecomponent}
Deciding whether a graph on $\mathbb N$ with primitive recursive
adjacency has exactly one finite connected component is
$D_2(\Sigma^0_2)$-complete, and hence computably many-one equivalent
to $\LU\setminus\LI$.
\end{corollary}
\begin{proof}
For fixed $j$, at least $j$ finite components is $\Sigma^0_2$:
existentially specify $j$ disjoint nonempty finite connected sets and
universally prohibit an edge leaving any one. Exactly one is at least
one minus at least two.

For hardness, from c.e. sets $W_a,W_b$ construct three stars with centres
$0,1,2$. At stage $s$ add vertices $3s+3,3s+4,3s+5$. Attach the first
to centre zero if $W_a$ gains a new distinct element, otherwise to two.
Attach the second to centre one if either set gains a new element,
otherwise to two. Always attach the third to two. Bounded enumeration
simulation makes adjacency primitive recursive. The three components
are finite precisely under the respective conditions
$W_a$ finite, both $W_a,W_b$ finite, and never. Thus exactly one is
finite precisely when $a\in\mathrm{FIN}$ and $b\notin\mathrm{FIN}$.
The natural statement still has a difference in its logical form.
\end{proof}

\subsection{The refinements of Theorem~B}
The following theorem collects the difference results for the type languages.

\Needspace{5\baselineskip}
\begin{theorem}[Refinements of Theorem~B]\label{thm:Brefine}
Commuting perfection without semifinite perfection is $\Pi^0_2$-complete, with hardness on a family of commuting-perfect games all having $\omega_q=1$. The two languages \emph{semifinite but not spatial} and \emph{semifinite but not finite-entropy spatial} are $D_2(\Sigma^0_2)$-complete, also with no finite-host perfection. Every nonempty directional difference among the six models in Theorem~\ref{thm:differences} is complete for its displayed class.
\end{theorem}
\begin{proof}
The first assertion is proved at the end of Section~\ref{sec:promises}: every $\Gsel_e$ is commuting perfect with $\omega_q(\Gsel_e)=1$, and $\Gsel_e\in\perf_{\co}\setminus\perf_{\sfhost}$ exactly when $e\notin\Adec$. The two languages of the second assertion are $\LU\setminus\LI$ and $\LU\setminus\LE$. The proof of Theorem~\ref{thm:differences} reduces the canonical $D_2(\Sigma^0_2)$-complete problem to both through the family $\Lambda(e,f)$, whose $\LJ$ coordinate is identically zero, so no instance has finite-host perfection. The third assertion is Theorem~\ref{thm:differences}.
\end{proof}

\section{Zero-error codes and perfect play}\label{sec:normal}
A sender uses a finite noisy classical channel once. Beforehand, sender and receiver share an entangled state. The sender encodes one of $m$ messages into the channel input, and the receiver must identify that message with certainty. We ask whether such a code exists. The channel and the message count remain finite; only the shared resource is unrestricted. Zero-error communication depends on exact zeros, rather than on how closely a protocol approaches perfect transmission. This makes attainment decisive, and the question becomes a question about perfect play.

For a finite rational stochastic matrix $N$ and an integer $m\geq1$, write $\ZE_t(N,m)$ for this feasibility problem. The model $q$ uses finite-dimensional tensor products. The model $\qs$ uses a normal density operator on a spatial tensor product, with bounded finite-outcome POVMs. The model $\qc$ uses commuting local operator algebras and a state. The entropy restriction $\finS$ means that both resource marginals have finite von Neumann entropy. We use countably infinite Schmidt decompositions.

\subsection{Channels and trace-class assemblages}
For channel inputs $x,x'$, write $x\sim x'$ when some output $y$ has $N(y\mid x)N(y\mid x')>0$. This is a reflexive relation: it includes $x=x'$. Its restriction to distinct inputs is the confusability graph. Every finite graph $\mathsf X$ has a rational realizing channel $N_{\mathsf X}$: a nonisolated vertex outputs a uniformly random incident edge, and an isolated vertex outputs its own private symbol.

We first extend the finite-dimensional assemblage characterization of \cite{CLMW}. All supports below are closed range supports.

\begin{lemma}[Normal assemblages]\label{lem:assemblage}
An $m$-message normal zero-error code exists if and only if there are positive trace-class operators $R_{i,x}$ and a density $\rho$ such that
\begin{equation}\label{eq:assemblage}
 \sum_xR_{i,x}=\rho\quad(1\leq i\leq m),\qquad
 R_{i,x}R_{j,x'}=0\quad(i\ne j,\ x\sim x').
\end{equation}
The assemblage is realizable with the pure resource $|\sqrt\rho\rangle\!\rangle$. The equivalence includes mixed resources.
\end{lemma}
\begin{proof}
Given a mixed resource $\omega$ and Alice's encoding effect $E_{i,x}$, define
\[
 R_{i,x}=\Tr_A[(E_{i,x}^{1/2}\otimes I)\omega(E_{i,x}^{1/2}\otimes I)].
\]
The sum is Bob's marginal, independently of $i$. For output $y$, Bob must distinguish the positive operators $T_i^y=\sum_xN(y\mid x)R_{i,x}$ perfectly. If $F_i^y$ is a decoding POVM, perfect decoding gives $F_i^y=I$ on $\supp T_i^y$ and $F_j^y=0$ there for $j\ne i$. Thus these supports are orthogonal. Positivity of the sum puts $\supp R_{i,x}$ in $\supp T_i^y$ whenever $N(y\mid x)>0$. This proves the product condition, including equal inputs.

Conversely restrict to $\supp\rho$, so $\rho$ is faithful. Each $0\leq R\leq\rho$ has a unique bounded effect $K$ with
\begin{equation}\label{eq:douglas}
 R=\rho^{1/2}K\rho^{1/2},\qquad 0\leq K\leq I.
\end{equation}
Indeed Douglas' factorization lemma \cite{Douglas} applied to $R\leq\rho$ gives $R^{1/2}=\rho^{1/2}V_0$ for a contraction $V_0$, and $K=V_0V_0^*$. Uniqueness follows from the dense range of $\rho^{1/2}$. The sum identities imply $\sum_xK_{i,x}=I$, by the same dense-range argument.

In a spectral basis for $\rho$, let Alice measure $K_{i,x}^{\mathsf T}$ on the purification $|\sqrt\rho\rangle\!\rangle$. Bob's conditional operator is $R_{i,x}$. Equation~\eqref{eq:assemblage} makes the $T_i^y$ have orthogonal supports, so their support projections, with any unused subspace assigned to one message, form a decoding POVM. This constructs a normal pure code.

Alternatively, any normal mixed perfect code decomposes spectrally into pure perfect codes with the same measurements: every forbidden probability is nonnegative, so it vanishes on every positive-weight component. This observation will also be used for entropy.
\end{proof}

\subsection{Saturation of a clique partition}
Beigi's Lov\'asz-theta bound implies the finite-dimensional clique-cover bound \cite{Beigi}. The following trace-class saturation statement extends \cite[Lemma~5]{Mancinska}.
\begin{lemma}\label{lem:saturation}
If $\mathsf X$ is partitioned into $k$ cliques $\mathcal C_c$, every normal code has at most $k$ messages. If it has exactly $k$, put
\[
 B_{i,c}=\sum_{x\in \mathcal C_c}R_{i,x},\qquad P_{i,c}=\supp B_{i,c}.
\]
On $\supp\rho$ one has
\begin{equation}\label{eq:saturation}
 B_{i,c}=\rho P_{i,c}=P_{i,c}\rho,\qquad
 \sum_iP_{i,c}=I,\qquad \sum_cP_{i,c}=I.
\end{equation}
Each row and column of $(P_{i,c})$ is an orthogonal projection partition.
\end{lemma}
\begin{proof}
For fixed $c$, distinct $i$ give orthogonal $B_{i,c}$, because every pair of inputs in $\mathcal C_c$ is confusable, including identical inputs. Since $B_{i,c}\leq\rho$,
\[
 \sum_i\Tr B_{i,c}\leq\sum_i\Tr(P_{i,c}\rho)\leq1.
\]
Summing over $c$ gives $m\leq k$. If $m=k$, all inequalities are equalities. Faithfulness of $\rho$ gives $\sum_iP_{i,c}=I$. For each $i,c$, positivity and $\Tr[P_{i,c}(\rho-B_{i,c})]=0$ imply $(\rho-B_{i,c})^{1/2}P_{i,c}=0$. Hence $\rho P_{i,c}=B_{i,c}=P_{i,c}\rho$. Summing over $c$ yields $\rho\sum_cP_{i,c}=\rho$; injectivity gives the row sum. A finite family of projections summing to identity is pairwise orthogonal: compress the sum by one member and use positivity of all other compressed members. All traces used here are finite traces of positive trace-class operators.
\end{proof}

A code is \emph{saturating} for this partition if its message count equals the clique bound, $m=k$.

\subsection{The normal correspondence}
Let $G$ have question sets $\mathcal Q_A,\mathcal Q_B$, finite nonempty answer sets, and a Boolean losing predicate on the question pairs of positive sampling probability. Its graph $\Gamma(G)$ has vertices $(\mathrm L,u,a)$ and $(\mathrm R,v,b)$. Distinct answers to the same tagged question form a clique; cross edges are precisely losing tuples. There are no other edges. Set $k=|\mathcal Q_A|+|\mathcal Q_B|$, the number of tagged local questions; Man\v{c}inska's question count $q_{\rm M}$ is our $k$.

For a pure strategy, use Schmidt form $|D\rangle\!\rangle$ with $D\geq0$ and $\Tr D^2=1$. For bounded positive effects, zero probability is equivalent to $EDF^{\mathsf T}=0$, as in Section~\ref{sec:upper}: zero probability is the vanishing of the Hilbert--Schmidt norm of $\sqrt E D\sqrt{F^{\mathsf T}}$; multiplication by the square roots gives one implication, and equality of the closures of the ranges of an effect and its square root gives the other.

\begin{theorem}[Trace-class correspondence]\label{thm:correspondence}
For every finite game $G$,
\[
 \ZE_{\qs}(N_{\Gamma(G)},k)
 \ \Longleftrightarrow\
 G\text{ has a perfect pure spatial strategy with }\Tr D<\infty.
\]
For a specified code marginal $\rho$, the reverse strategy has $D=\rho/\|\rho\|_2$. For a specified game strategy $D$ with finite trace, the forward pure code has marginal $\rho=D/\Tr D$. The game need not be synchronous.
\end{theorem}
\begin{proof}
In the forward construction, store Bob's effects in transposed form, and denote all local effects by $E_{c,a}$ for tagged questions $c$. Put $\rho=D/\Tr D$ and
\[
 R_{c,(c,a)}=\rho^{1/2}E_{c,a}\rho^{1/2},
\]
with zero operators off the designated clique. Each message family sums to $\rho$. For cross losing edges, the perfect-game sandwich $E_{c,a}DE_{c',b}=0$ gives the required zero product. The remaining products between distinct messages vanish by their clique assignments. Lemma~\ref{lem:assemblage} realizes the code. The trace condition is exactly what normalizes its marginal.

For the converse, as in \cite[Theorem~6]{Mancinska}, apply Lemma~\ref{lem:saturation} to an arbitrary saturating code, whether or not it has the displayed form. Aggregate
\[
 T_{c,a}=\sum_iR_{i,(c,a)}.
\]
Each family sums to $\rho$. If $(c,a),(c',b)$ is a cross losing edge, every product $R_{i,(c,a)}R_{j,(c',b)}$ is zero. For $i\ne j$ this is a code condition. For $i=j$ its two factors are supported on the orthogonal row projections $P_{i,c},P_{i,c'}$. Thus $T_{c,a}T_{c',b}=0$.

Use \eqref{eq:douglas} to write $T_{c,a}=\rho^{1/2}E_{c,a}\rho^{1/2}$ with bounded POVMs. The product equality becomes
\[
 \rho^{1/2}E_{c,a}\rho E_{c',b}\rho^{1/2}=0.
\]
The injectivity and dense range of $\rho^{1/2}$ cancel its two outside factors, giving $E_{c,a}\rho E_{c',b}=0$. Use the left-tagged effects on Alice and the transposes of the right-tagged effects on Bob. With $D=\rho/\|\rho\|_2$ these are a perfect spatial strategy. Since $\Tr\rho=1$, $\Tr D=1/\|\rho\|_2<\infty$.

Only the coarse clique projections must commute with $\rho$. The fine outcomes need not do so; this is why the argument retains nonsynchronous games. Restricting every step to finite dimension recovers \cite[Theorem~6]{Mancinska}.
\end{proof}

\begin{remark}
For the game marginal $h=D^2$, the condition $\Tr D<\infty$ is finite R\'enyi entropy of order $1/2$, namely $2\log_2\Tr\sqrt h$. It is stronger than finite von Neumann entropy. Replacing the forward marginal by $D^2$ generally destroys zero-error orthogonality; Appendix~\ref{sec:repro} includes an explicit check on the flagged magic square.
\end{remark}

\subsection{The resource correspondence at every order}\label{sec:codeorders}
Write $h=D^2$ for the game marginal, $Z=\Tr\sqrt h$, and
$\rho_{\rm code}=\sqrt h/Z$ for the code marginal. A normal code first requires
$Z<\infty$. Whenever it exists,
\[
 \Tr\rho_{\rm code}^\beta=Z^{-\beta}\Tr h^{\beta/2},\qquad
 S(\rho_{\rm code})=\log_2Z+(2Z)^{-1}\Tr[\sqrt h\log_2(1/h)].
\]
The resulting correspondence is
\begin{center}\small
\begin{tabular}{@{}ll@{}}\toprule
Code resource & Required game spectrum\\\midrule
Finite rank & finite rank\\
Finite R\'enyi entropy, $0<\beta<1$ & $\Tr h^{\beta/2}<\infty$\\
Finite von Neumann entropy & $\Tr[\sqrt h\log_2(1/h)]<\infty$\\
Finite R\'enyi entropy, $1<\beta\le\infty$ & $\Tr\sqrt h<\infty$\\
Arbitrary normal resource & $\Tr\sqrt h<\infty$\\\bottomrule
\end{tabular}
\end{center}
The logarithmic condition implies summability of the square roots:
only finitely many eigenvalues exceed any fixed positive threshold,
and below such a threshold the logarithm is bounded below.
It cannot be replaced by the half-moment alone. The squared-Catalan
channel of Section~\ref{sec:catalan} has a summable half-moment and
requires infinite code entropy. For $\beta>1$, the four-bridge tree
has finite $\beta/2$-moment but no normal saturating code. At the
spectrum level $p_j\asymp j^{-2}(\log j)^{-4}$ has summable square
roots and a divergent logarithmic half-moment. These examples explain
both endpoints of the table.

\section{Five bridges and the normal row}\label{sec:communication}
We apply the trace-class correspondence to the compiler. The four-bridge games of Section~\ref{sec:reduction} lie exactly at the summability threshold: none of their channels admits a saturating normal code, even on YES inputs. Adding a fifth bridge changes the modular ratio and restores summability. The same five-bridge YES codes have finite marginal entropy. The threshold is visible in the four-letter tree density $h_{F,\nu}=(1-4\nu)\sum_w\nu^{|w|}|w\rangle\langle w|$ of Section~\ref{sec:generaltree}:
\[
 \Tr\sqrt{h_{F,\nu}}
 =\sqrt{1-4\nu}\sum_{\ell=0}^{\infty}(4\sqrt\nu)^\ell,
 \qquad \nu=2^{-n}.
\]
There are four branches per level, so summability requires $n\geq5$.

\subsection{An exact spectrum identity}
The FMS transports need not commute with the row factor $\Mat_{d_{\max}}$, because the sign questions are not synchronized. Removing the row factor in soundness therefore uses the following elementary cancellation instead of a partial trace over an infinite space.

\paragraph{Finite-amplification cancellation.}
If two finite-dimensional tuples of unitaries become unitarily equivalent after $d_{\max}$ identical copies, the original tuples are unitarily equivalent. Indeed they are finite-dimensional representations of the universal $*$-algebra on the tuple. The image in their direct sum is a finite-dimensional $C^*$-algebra. Decompose its two representations into irreducibles; amplification multiplies every multiplicity by $d_{\max}$, so equality of amplified multiplicities cancels $d_{\max}$.

\begin{lemma}[Exact spectrum identity]\label{lem:spectrum}
For every pure perfect strategy of $G^{(n)}$, with full marginal $h$ and Schmidt operator $D=\sqrt h$, there is a normal realization with density $h_0$ such that, including multiplicities,
\begin{equation}\label{eq:specidentity}
 \spec(h)=\spec\left(\operatorname{diag}(\mu_n)\otimes h_0
                    \otimes I_3/3\otimes I_{d_{\max}}/d_{\max}\right).
\end{equation}
Here $h_0$ is the reduced normalized source density after removing the qutrit and row factors; it differs from the source density on the full amplified Hilbert space in Lemma~\ref{lem:four}. Conversely every realization has a perfect strategy with this spectrum. In particular, in the extended nonnegative reals,
\begin{equation}\label{eq:traceidentity}
 \Tr D=c^{\rm amp}_n\Tr\sqrt{h_0},\qquad
 c^{\rm amp}_n=\sqrt{3d_{\max}}\,\frac{1+4\sum_{g=1}^n2^{g/2}}{\sqrt{\mathfrak c_n}}.
\end{equation}
\end{lemma}
\begin{proof}
Restrict to the original marginal support. Synchronization makes the certified scalar matrix algebras commute with the density. The named old projections commute with $\Mat_{d_{\max}}$; the bound parameters and graph units also provide the scalar $\Mat_3$ frame. These two scalar algebras commute, so their exact matrix-unit decompositions give a density $I_3/3\otimes I_{d_{\max}}/d_{\max}\otimes h_{\rm base}$ and factorizations of every named old synchronized projection.

The magic matrix units identify all four density blocks at each bridge. The $\btrans_g$ identify consecutive normalized selected density blocks, with ratio two. The cover therefore implies that the full density has the spectrum in \eqref{eq:specidentity}, where $h_0$ is its normalized source block after removal of both scalar factors. It remains to verify that this $h_0$ is the density of a realization; simply compressing $v^*a_\alpha$ across $\Mat_{d_{\max}}$ would not prove that assertion.

The frame refinements make $v$ act trivially on $\Mat_3$. After removing that frame it is a unitary between the $d_{\max}$-amplifications of the source and far spaces, intertwining their normalized densities and every transported unitary label. In each positive density eigenspace these spaces are finite dimensional. Apply finite-amplification cancellation to their transported unitary tuples. This gives an unamplified unitary for each positive eigenvalue. Their Hilbert direct sum $\widetilde v$ intertwines $h_0$ with the normalized far density, and intertwines all transported labels simultaneously. The densities are faithful, so these eigenspaces exhaust both spaces.

All centralizer contractions $a_\alpha$ are polynomials in old labels and factor across both scalar algebras; write their reduced forms $\widetilde a_\alpha$. Define $\widetilde A_\alpha=\widetilde v^*\widetilde a_\alpha$. The reduced translated equations in Lemma~\ref{lem:four} give every application identity. Intertwining of the normalized densities, together with the source/far weight $2^n$, gives $h_0\widetilde A_\alpha=2^{-n}\widetilde A_\alpha h_0$. This proves the realization assertion. The new creations need not be restrictions of the original creations.

Completeness uses \eqref{eq:scaffold}, followed by the two normalized maximally mixed factors, so has exactly the same spectrum. Taking the square root and the trace of this finite tensor product proves \eqref{eq:traceidentity}; nonnegative spectral sums justify it even when the answer is infinity.
\end{proof}

\subsection{The tree at a general ratio}\label{sec:generaltree}
The tree part of Section~\ref{sec:tree} works for every $0<\nu<1/4$. For clarity we give the density calculation. For a word $w$ in four letters, write $A_w$ for the product of its creations and
\[
 Q_\ell=\sum_{|w|=\ell}A_wA_w^*.
\]
The isometry relations and covariance $h_0A_\alpha=\nu A_\alpha h_0$ imply
\[
 \Tr(h_0Q_\ell)=(4\nu)^\ell,\qquad
 \Tr(h_0P_\circ)=1-4\nu.
\]
The decreasing $Q_\ell$ have a strong limit of zero $h_0$-weight, hence zero by faithfulness. Their successive differences are $\sum_{|w|=\ell}A_wP_\circ A_w^*$. Thus, with $\KK=P_\circ\HH\ne0$,
\begin{equation}\label{eq:treefactor}
 \HH\simeq\ell^2(\WW)\otimes\KK,\quad
 h_0=h_{F,\nu}\otimes\rho_{\rm spec},\quad
 h_{F,\nu}=(1-4\nu)\sum_w\nu^{|w|}|w\rangle\langle w|,
\end{equation}
where $\rho_{\rm spec}$ is a faithful density. The covariance makes every word-sector projection commute with $h_0$, so transporting the root block proves the operator equality. Depth $\ell$ has probability $(1-4\nu)(4\nu)^\ell$; $(4\nu)^\ell$ is the mass at depth at least $\ell$. At $\nu=1/16$ this is \eqref{eq:fockdensity}.

The coordinate, Weyl-clock, bounded period-reader and deterministic tableau equations of Sections~\ref{sec:tree}--\ref{sec:reduction} contain no Fock density ratio. Their proofs use only faithful normality, positive finite-dimensional density eigenspaces, the exact tree decomposition, and a positive weight at every finite depth. All hold in \eqref{eq:treefactor}. Consequently their soundness and canonical completeness hold unchanged for every $\nu=2^{-n}<1/4$. In particular the Weyl determinant still forces a root phase of finite order: on a positive density eigenspace of dimension $d$ on which $Z_{\rm clk}=\zeta$, the relation $J_{\rm W}K_{\rm W}=\zeta K_{\rm W}J_{\rm W}$ gives $\zeta^{d}=1$.

\subsection{Summability and its critical point}
Equation~\eqref{eq:treefactor} gives
\begin{equation}\label{eq:trsqrtF}
 \Tr\sqrt{h_{F,\nu}}
 =\sqrt{1-4\nu}\sum_{\ell=0}^\infty(4\sqrt\nu)^\ell
 =\frac{\sqrt{1-4\nu}}{1-4\sqrt\nu}
\end{equation}
when $\nu<1/16$, and infinity when $1/16\leq\nu<1/4$. For the canonical YES realization $\rho_{\rm spec}=I_{{d_{\rm clk}}}/{d_{\rm clk}}$ with clock order ${d_{\rm clk}}$, Lemma~\ref{lem:spectrum} therefore gives
\begin{equation}\label{eq:YESsum}
 \Tr D=\sqrt{3{d_{\rm clk}}d_{\max}}\,
 \frac{1+4\sum_{g=1}^n2^{g/2}}{\sqrt{\mathfrak c_n}}
 \frac{\sqrt{1-2^{2-n}}}{1-2^{2-n/2}}\quad(n>4).
\end{equation}
For five bridges and the optional eight-bridge construction, the exact coefficients are
\begin{align}
 n=5:&\quad \Tr D=\frac{(81+53\sqrt2)\sqrt7}{2\sqrt{83}}\sqrt{{d_{\rm clk}}d_{\max}},\label{eq:fiveconstant}\\
 n=8:&\quad \Tr D=\frac{(121+60\sqrt2)\sqrt{21}}{2\sqrt{2041}}\sqrt{{d_{\rm clk}}d_{\max}}.\label{eq:eightconstant}
\end{align}
They are approximately $22.65\sqrt{{d_{\rm clk}}d_{\max}}$ and $10.44\sqrt{{d_{\rm clk}}d_{\max}}$.

\begin{remark}[Four bridges are critical]\label{rem:critical}
At $n=4$, every realization has $\Tr\sqrt{h_0}=\infty$, regardless of $\rho_{\rm spec}$, because the per-depth amplitude sum in \eqref{eq:trsqrtF} is constant and nonzero. Lemma~\ref{lem:spectrum} then makes every perfect pure game strategy have infinite amplitude sum. Theorem~\ref{thm:correspondence} implies that the channel of every game $\Gclk_e$ of Theorem~\ref{thm:reduction} has no $k$-message normal code, for every input index. Applying those games directly would therefore give an empty YES set, rather than the desired hardness reduction.
\end{remark}

\subsection{Upper membership}\label{sec:codeupper}
Convert a channel instance $(N,m)$ to a finite game $\mathcal G_{\rm code}(N,m)$. Alice receives $i\in[m]$ and answers a channel input $x$; Bob receives a channel output $y$ and answers $j\in[m]$. All question pairs have positive rational probability. A tuple loses exactly when $N(y\mid x)>0$ and $j\ne i$. With the same resource and measurements,
\begin{equation}\label{eq:codegame}
 \ZE_t(N,m)\ \Longleftrightarrow\
 \mathcal G_{\rm code}(N,m)\text{ has a perfect }t\text{-strategy}.
\end{equation}
Indeed the error probability of the code is the finite sum of these nonnegative expectations weighted by $N(y\mid x)$. It vanishes exactly when every forbidden expectation vanishes. The equivalence preserves marginal entropy as well.

The compact-stratum predicate of Section~\ref{sec:upper} gives $\perf_\qs=\bigcup_\kappa\perf_\qs(\kappa)$ with each stratum $\Pi^0_1$, and mixed strategies reduce to pure perfect components, as in Lemma~\ref{lem:assemblage}. Equation~\eqref{eq:codegame} therefore proves $\ZE_{\qs}\in\Sigma^0_2$. The entropy-budget strata of Section~\ref{sec:finS} prove $\ZE_{\finS}\in\Sigma^0_2$ in the same way. If a mixed perfect resource has finite entropy on one marginal, entropy concavity gives a positive-weight pure perfect component of finite entropy, and for a pure bipartite component the two marginal entropies agree. Hence using one or both finite marginal entropies leads to the same feasibility set.

\subsection{Five bridges for normal hardness}
Let $\Adec=\{e:\exists j\ \forall t\ \Rdec(e,j,t)\}$ for any decidable predicate $\Rdec$. Use the clock presentation of Sections~\ref{sec:tree}--\ref{sec:reduction}, with its computation program hardwiring $e$, and change only its bridge count to five. Denote the resulting game by $G^{(5)}_e$. Its clock and computation mechanism is unchanged. At depth $r$ the four-letter tree carries a $2^r$-by-$2^r$ coordinate square. A root Weyl pair commuting with the root density has a nonzero finite-order phase sector; its determinant on a positive density eigenspace forces this. On a phase of order ${d_{\rm clk}}$ the reader equations \eqref{eq:reader} force $P_{\rm ret}$ to be exactly the kernel projection of $W-I$, so the first positive return is precisely time ${d_{\rm clk}}$. The tableau freezes a waiting front at that time, halts if ${d_{\rm clk}}$ is one or composite, and otherwise checks $\Rdec(e,j,0),\Rdec(e,j,1),\ldots$ for ${d_{\rm clk}}=p_{j+1}$, halting at the first failure. A halt is forbidden by an operator equation, and every depth has positive normal weight.

If the game is perfect, Lemma~\ref{lem:four} extracts this mechanism and therefore a witness $j$. Conversely a witness $j$ uses the canonical tree, a ${d_{\rm clk}}$-dimensional Weyl pair for ${d_{\rm clk}}=p_{j+1}$, the good primitive phase and bounded reader of Section~\ref{sec:clock}, and its nonhalting classical tableau. The reader norm estimate depends only on the phase and the counter, so it holds for every $\nu$. Its density is $h_{F,1/32}\otimes I_{{d_{\rm clk}}}/{d_{\rm clk}}$. This is the simultaneous canonical realization, so Lemma~\ref{lem:four} supplies a perfect game strategy. Formula~\eqref{eq:fiveconstant} makes its Schmidt amplitudes summable.

Now put
\[
 N_e=N_{\Gamma(G_e^{(5)})},\qquad
 k_e=|\mathcal Q_{A,e}|+|\mathcal Q_{B,e}|,
\]
where $\mathcal Q_{A,e},\mathcal Q_{B,e}$ are the question sets of $G^{(5)}_e$. These data are finite and rational and are computed from $e$ without knowing its witness. Theorem~\ref{thm:correspondence} and the preceding two implications give
\[
 \ZE_{\qs}(N_e,k_e)\ \Longleftrightarrow\ e\in \Adec.
\]
This proves $\Sigma^0_2$-hardness and the first part of Theorem~C. No channel in this reduction has a finite-dimensional saturating code: otherwise the finite converse would extract four orthogonal isometries on a nonzero finite-dimensional root space, an impossibility. We make no claim here that a value-one promise is inherited from the source construction.

\subsection{Finite entropy already at five bridges}\label{sec:finiteentropy}
Use exactly the same five-bridge reduction. The forward code marginal is the normalized square root of the canonical game marginal. Its tree factor is
\[
 \widehat h_5=\frac{\sqrt{h_{F,1/32}}}{\Tr\sqrt{h_{F,1/32}}}
 =(1-2^{-1/2})\sum_w2^{-5|w|/2}|w\rangle\langle w|.
\]
At depth $\ell$ there are $4^\ell$ words, so the depth distribution is geometric with ratio $2^{-1/2}$. Its mean is $2^{-1/2}/(1-2^{-1/2})=1+\sqrt2$. Each eigenvalue at depth $\ell$ has negative base-two logarithm $-\log_2(1-2^{-1/2})+(5/2)\ell$. Taking its expectation proves
\begin{equation}\label{eq:fiveentropy}
 S(\widehat h_5)=-\log_2(1-2^{-1/2})+\frac52(1+\sqrt2)
 \approx 7.807\ \text{bits}.
\end{equation}
The full canonical code marginal is
\[
 \operatorname{diag}(\hat\mu_5)\otimes\widehat h_5
 \otimes I_{{d_{\rm clk}}}/{d_{\rm clk}}\otimes I_3/3\otimes I_{d_{\max}}/d_{\max},
 \qquad \hat\mu_{5,i}=\frac{\sqrt{\mu_{5,i}}}{\sum_{i'}\sqrt{\mu_{5,i'}}}.
\]
Thus its entropy is $H(\hat\mu_5)+S(\widehat h_5)+\log_2(3{d_{\rm clk}}d_{\max})$, which is finite. The pure code has the same entropy on both sides. A NO input has no normal code at all, hence no finite-entropy code. The same map therefore proves the finite-entropy half of the first part of Theorem~C.

\begin{remark}[Eight bridges]
With $n=8$, the normalized square root of $h_{F,1/256}$ is exactly $h_{F,1/16}$, with entropy $\log_2(4/3)+4/3$. Its amplitude sum is \eqref{eq:eightconstant}. This is another finite-entropy construction; five bridges already suffice for hardness.
\end{remark}

\subsection{Finite-dimensional zero-error communication}\label{sec:fdrow}
For each fixed dimension, the channel code game in \eqref{eq:codegame} has a rational semialgebraic perfect-strategy predicate: split matrix entries into real and imaginary parts, impose positivity, POVM sums, normalization and zero losing probabilities. Quantifier elimination decides it. Enumerating dimensions gives $\ZE_{\mathrm{q}}\in\Sigma^0_1$.

For hardness, the perfect-completeness reduction of $\mathsf{MIP}^*=\mathsf{RE}$ gives, from a halting instance, a finite game with a perfect finite-dimensional strategy, and from a nonhalting instance a game of entangled value at most $1/2$ \cite[Theorem~12.6]{MIPRE}. Man\v{c}inska's Corollary~7 maps perfect finite-dimensional game feasibility to the saturating one-shot channel threshold \cite{Mancinska}. Thus $\ZE_{\mathrm{q}}$ is $\Sigma^0_1$-complete. This conclusion needs only her finite theorem. Bhattacharyya, Mehta and Zhao leave undecidability of $C^{(1)}_{0,E}$ open in \cite[Section~1.4]{BMZ}. This corollary settles its one-shot threshold problem: it is undecidable and $\Sigma^0_1$-complete already for classical channels. Here $E$ denotes unrestricted finite-dimensional entanglement. The unrestricted asymptotic capacity question remains open.

\section{The commuting row}\label{sec:commuting}
With commuting resources one-shot zero-error feasibility drops to the commuting level. The code game \eqref{eq:codegame} identifies $\ZE_{\qc}$ with perfect commuting play for a particular finite game. Commuting correlations form a compact set, and the NPA hierarchy converges from above to their value \cite{NPA}. Its rational moment relaxations can be decided by real quantifier elimination. A NO instance eventually has an infeasible perfect-value relaxation; conversely a feasible relaxation at every order gives a commuting strategy by compactness of the full moment system. Hence $\ZE_{\qc}\in\Pi^0_1$.

We prove the reduction for the lower bound, including codes with arbitrary commuting resources. Slofstra proved $\Pi^0_1$-hardness for linear system games \cite{Slofstra}; with the NPA bound this is $\Pi^0_1$-completeness, which Mousavi, Nezhadi and Yuen reprove for synchronous games \cite[Theorem~6.4]{MNY}. The source is perfect commuting feasibility for finite synchronous games with positive diagonal question probabilities; see \cite[Theorem~6.4 and Section~1.3]{MNY}. Their synchronous framework and tracial representation are stated in \cite[Theorem~1.6]{MNY}. Such a game has the same question and answer sets on both sides, and distinct answers on an equal-question pair are losing.

\subsection{Normal functionals from a commuting code}
Represent a code on a Hilbert space with vector $\xi$, encoding effects $E_{i,x}$ in an algebra commuting with Bob's von Neumann algebra $\mathcal N$, and decoder effects in $\mathcal N$. Define normal positive functionals
\[
 \phi(T)=\langle\xi,T\xi\rangle,\qquad
 \phi_{i,x}(T)=\langle\xi,E_{i,x}T\xi\rangle\quad(T\in\mathcal N).
\]
The two operators commute, so these functionals are positive and $\sum_x\phi_{i,x}=\phi$. Compress $\mathcal N$ to the support of $\phi$. This makes $\phi$ faithful. It preserves all the functionals, their sums, and decoder tests, since each $\phi_{i,x}\leq\phi$ is supported on the same support. The compressed encoders still commute with the compressed Bob algebra.

When $i\ne j$ and $x\sim x'$, the functionals $\phi_{i,x},\phi_{j,x'}$ have orthogonal supports in $\mathcal N$. To see this, for any common output $y$ their positively weighted functionals are summands of the two distinct-message output functionals. Perfect decoding makes the support of one lie in the eigenvalue-one space of a decoder effect, and that of the other in its kernel. These spectral spaces are orthogonal. Support inclusion for positive summands proves the assertion.

\begin{lemma}[Functional saturation]\label{lem:functionalsaturation}
Suppose the input graph is partitioned into $k$ cliques $\mathcal C_c$ and a commuting code sends $k$ messages. For
\[
 \psi_{i,c}=\sum_{x\in \mathcal C_c}\phi_{i,x},\qquad s_{i,c}=\supp\psi_{i,c},
\]
the $s_{i,c}$ lie in the centralizer
$\mathcal N_\phi=\{s:\phi(sT)=\phi(Ts)\text{ for all }T\in\mathcal N\}$ and satisfy
\[
 \psi_{i,c}(T)=\phi(s_{i,c}T)=\phi(Ts_{i,c}).
\]
Their rows and columns are orthogonal projection partitions.
\end{lemma}
\begin{proof}
For each $c$ the supports are orthogonal in $i$. Domination by $\phi$ gives
\[
 \sum_i\psi_{i,c}(1)\leq\sum_i\phi(s_{i,c})\leq1.
\]
The sum over $c$ is $k$, so every inequality is an equality. Faithfulness gives $\sum_i s_{i,c}=1$. For each $i,c$, $\eta=\phi-\psi_{i,c}$ is positive and $\eta(s_{i,c})=0$. Cauchy--Schwarz for a positive functional gives $\eta(s_{i,c}T)=\eta(Ts_{i,c})=0$. The support property of $\psi_{i,c}$ gives $\psi_{i,c}(s_{i,c}T)=\psi_{i,c}(Ts_{i,c})=\psi_{i,c}(T)$. Combining them proves the formula and centralizer membership.

Summing over $c$ for fixed $i$ gives $\phi((\sum_c s_{i,c}-1)T)=0$ for every $T$. Take $T=(\sum_c s_{i,c}-1)^*$ and use faithfulness to obtain the row sum. A finite sum of projections equal to identity is orthogonal, as in Lemma~\ref{lem:saturation}.
\end{proof}

\subsection{Synchrony makes the fine supports tracial}
Apply the lemma to $\Gamma(G)$, with $k=2|\mathcal Q_A|$, for a synchronous game $G$. Aggregate the fine functionals
\[
 \psi_{c,a}=\sum_i\phi_{i,(c,a)}.
\]
They sum to $\phi$ for each tagged question $c$. Across a cross losing edge their supports are orthogonal: unequal message indices use the code condition, and equal message indices use distinct clique supports in one row of $(s_{i,c})$. Orthogonality passes to finite sums.

Fix an original question $u$. Write $\alpha_a=\psi_{(L,u),a}$, $\beta_a=\psi_{(R,u),a}$, and $p_a=\supp\alpha_a$. For $a\ne b$, synchrony gives $p_a\supp\beta_b=0$. Hence
\[
 \alpha_a(1)\leq\phi(p_a)=\beta_a(p_a)\leq\beta_a(1).
\]
Both outside families sum to one, so these inequalities are equalities for every $a$. The same Cauchy--Schwarz argument used in Lemma~\ref{lem:functionalsaturation}, now for $\phi-\alpha_a$ and $\phi-\beta_a$, gives
\[
 \alpha_a(T)=\beta_a(T)=\phi(p_aT)=\phi(Tp_a).
\]
Faithfulness implies that their supports are the same $p_a$, and that $\sum_a p_a=1$. Thus the fine outcomes are projections in $\mathcal N_\phi$.

The restriction $\tau=\phi|_{\mathcal N_\phi}$ is a faithful normal tracial state. For any original losing tuple $(u,a,v,b)$ the preceding support orthogonality gives $p_{u,a}p_{v,b}=0$. The left/right multiplication construction on $L^2(\mathcal N_\phi,\tau)$ therefore gives a perfect commuting strategy for $G$. This argument starts from an arbitrary saturating code; it assumes neither projective encoders nor a tracial original resource.

Conversely a perfect commuting strategy for synchronous $G$ has its usual tracial projection representation \cite[Theorem~5.5]{PSS}; see also \cite{KPS,MNY}. For each tagged question use its projection partition to encode its designated message in the left algebra of the tracial GNS space; use the right algebra to decode. The conditional functionals are $T\mapsto\tau(p_{c,a}T)$. For distinct designated messages and confusable vertices, their supports are orthogonal by the losing relation. Each output's distinct-message supports therefore give a decoding measurement. This is a commuting $k$-message code for $N_{\Gamma(G)}$.

\begin{theorem}\label{thm:qc}
The one-shot feasibility problem $\ZE_{\qc}$ for finite rational classical channels is $\Pi^0_1$-complete.
\end{theorem}
\begin{proof}
For each finite synchronous game $G$ with positive diagonal question probabilities, the two constructions give the effective equivalence
\[
 G\text{ is commuting-perfect}
 \ \Longleftrightarrow\ \ZE_{\qc}(N_{\Gamma(G)},2|\mathcal Q_A|).
\]
The source problem is $\Pi^0_1$-complete, so this proves hardness. The NPA upper predicate gives membership.
\end{proof}

\section{A channel whose codes require infinite entropy}\label{sec:catalan}
This section proves the second part of Theorem~C. The channel comes from a four-bridge version of the Catalan assembly of \cite{Entropy}. Its game has finite-entropy perfect strategies, but the normalized square-root map of Theorem~\ref{thm:correspondence} turns every one of them into a code with infinite entropy.

\subsection{The squared Catalan density}
Let $\WW_{\rm Cat}$ consist of the empty word and the legal chronological words in an up-step $A$ and a down-step $B$. An up-step is always allowed; a down-step is allowed only at height at least two. Let $u(w)$ count up-steps. The number of words with $j$ up-steps is the Catalan number $\Cat_j$. On $\ell^2(\WW_{\rm Cat})$ let $A,B$ append the corresponding legal letter, with $B$ zero when forbidden. Let $\Ccat$ have eigenvalue $2^{-\mathrm{ht}(w)}$, let $P$ project onto the empty word, and let $Q$ project onto height one.

We require the local relations of \cite[Section~3]{Entropy}:
\begin{align}
 A^*A&=I,& AA^*+BB^*+P&=I,& B^*B&=I-P-Q,\label{eq:Catlocal}\\
 \Ccat P&=P,& \Ccat Q&=Q/2,&0&\leq \Ccat\leq I,\label{eq:Catboundary}\\
 \Ccat A&=A\Ccat/2,& \Ccat B&=2B\Ccat.&&\label{eq:Catcounter}
\end{align}
Here the modular ratio is squared:
\begin{equation}\label{eq:Catsquared}
 h_0 A=Ah_0/16,\qquad[h_0,B]=[h_0,\Ccat]=[h_0,P]=[h_0,Q]=0.
\end{equation}
Here $h_0$ is a faithful normal density. In this section $A,B,P,Q$ denote the Catalan labels just defined. The counter relation $\Ccat A=A\Ccat/2$ is unchanged.

\begin{lemma}[Normal classification at ratio $1/16$]\label{lem:Catclassification}
Every normal realization of \eqref{eq:Catlocal}--\eqref{eq:Catsquared} is, up to a unitary,
\[
 \HH=\ell^2(\WW_{\rm Cat})\otimes\KK_{\rm Cat},\qquad
 h_0=\kcat\otimes\rho_{\rm spec},\qquad
 \kcat=\frac1{8-4\sqrt3}\sum_w16^{-u(w)}|w\rangle\langle w|,
\]
where $\rho_{\rm spec}$ is a faithful normal density. All five local operators act canonically on the first factor and as identity on the second.
\end{lemma}
\begin{proof}
We give the normality argument to check the ratio change in \cite[Lemma~3.1]{Entropy}. The boundary equations put $P,Q$ in the counter eigenspaces one and one half and give $PQ=0$. The range projections $AA^*,BB^*,P$ are orthogonal and sum to identity; $B$ is a partial isometry with initial projection $I-P-Q$. All these projections commute with $\Ccat$.

There is no spectrum of $\Ccat$ in $(1/2,1)$, since there $B$ is isometric and $\Ccat B=2B\Ccat$ would transport it above one. The same reasoning at eigenvalue one gives $1_{\{1\}}(\Ccat)=P$. At eigenvalue one half, $B$ would transport the complement of $Q$ into $P$, orthogonally to its own range, so $1_{\{1/2\}}(\Ccat)=Q$. Iterating $B$ excludes every interval $(2^{-j-1},2^{-j})$. Functional calculus for the bounded intertwining relation justifies these spectral transports.

The zero-counter space reduces the operators and $h_0$. On it $P=Q=0$, so $A,B$ are isometries with orthogonal ranges summing to identity. If its $h_0$-mass is $t$, cyclicity and \eqref{eq:Catsquared} make the two range masses $t/16$ and $t$. Their sum is $t$, forcing $t=0$. Faithfulness removes this space.

Apply legal words to $P\HH$. Their sectors are orthogonal: different last letters have orthogonal ranges; equal last letters cancel with $A^*A=I$ or with $B^*B=I-P-Q$ on a legal predecessor; and a nonempty sector is orthogonal to $P$. Denote the word isometry by $V_w^{\rm Cat}$. Covariance gives
\[
 h_0 V_w^{\rm Cat}=16^{-u(w)}V_w^{\rm Cat}h_0.
\]
The word-sector projections commute with $h_0$, and their closed span reduces all the operators. If its orthogonal complement were nonzero, the compact faithful density there would have a largest positive eigenvalue, with nonzero finite-dimensional eigenspace $L$. The adjoint covariance gives $A^*L=0$. Since $P=0$ on this complement, $BB^*=I$ on $L$. Commutation with $h_0$ makes $L$ reduce $B$, so $B|_L$ is unitary. It also reduces $\Ccat$. Tracing $\Ccat B=2B\Ccat$ on $L$ forces $\Tr_L \Ccat=0$, hence $\Ccat|_L=0$, contrary to the absence of zero-counter mass. Thus the word sectors exhaust the support.

Their density blocks are $16^{-u(w)}V_w^{\rm Cat}(Ph_0P)(V_w^{\rm Cat})^*$. The word count is $\Cat_j$: for $j\geq1$, words with $j$ up-steps and $b$ down-steps, $0\leq b\leq j-1$, number
\[
 \binom{j+b-1}{b}-\binom{j+b-1}{b-1};
\]
reflection of the first bad prefix gives this formula, and summing over $b$ gives $\Cat_j$. The empty word supplies $\Cat_0=1$. The generating function
\[
 F(z)=\sum_{j\geq0}\Cat_jz^j=\frac{1-\sqrt{1-4z}}{2z}
\]
gives $F(1/16)=8-4\sqrt3$. Normalization then gives the asserted tensor product. The same sector identifications fix the local operators.
\end{proof}

Define the critical Catalan density of \cite{Entropy} by
\[
 \hcat=\frac12\sum_w4^{-u(w)}|w\rangle\langle w|.
\]
The generating function yields the exact identities
\begin{equation}\label{eq:Catnumbers}
 \Tr\hcat^2=2-\sqrt3,\qquad
 \kcat=\frac{\hcat^2}{2-\sqrt3},\qquad
 \Tr\sqrt{\kcat}=\frac{\sqrt6+\sqrt2}{2},\qquad
 \frac{\sqrt{\kcat}}{\Tr\sqrt{\kcat}}=\hcat.
\end{equation}
Moreover
\begin{equation}\label{eq:Catentropy}
 S(\hcat)=\sum_{j\geq0}\frac{\Cat_j}{2\,4^j}(1+2j)=\infty,
\end{equation}
since $\Cat_j4^{-j}\sim(\sqrt\pi j^{3/2})^{-1}$. By contrast the squared Catalan factor has
\[
 S(\kcat)=\log_2(8-4\sqrt3)+\frac4{\sqrt3}-2
 \approx0.4094\ \text{bits}.
\]
Indeed, with $F$ as above and $z=1/16$, its mean up-step count is $zF'(z)/F(z)=1/\sqrt3-1/2$; averaging $\log_2 F(z)+4u(w)$ gives the formula. Its multiplicity-weighted probabilities decay geometrically. The canonical perfect strategy of the game $G^{(4)}_{\Cat}$ below therefore has finite entropy: $S(\kcat)$ is its infinite Catalan factor's contribution, and all its other canonical factors are finite dimensional. The channel of $G^{(4)}_{\Cat}$ instead forces the factor $\hcat$ and hence infinite entropy. The communication task itself creates the entropy demand through the square-root map.

\subsection{A four-bridge Catalan assembly}
The Catalan relations are not the four-creation syntax of Definition~\ref{def:admissible}. We therefore specify their own assembly, extending \cite[Section~6]{Entropy} from two bridges to four. Use the seventeen corners, twenty-five frame-bound parameters, cover $\mathfrak c_4=121$, axis equations, support-copy equations and block-unitary transport equations of Lemma~\ref{lem:four}. The transported unitary encodings are those of $B,\Ccat,P,Q$: two unitaries $b_+,b_-$ for $B$, one for the positive contraction $\Ccat$, and the reflections for $P,Q$. Thus $|\Lambda|=5$. For each bridge use the six frame refinements and these five graph-label refinements, on both parties, retaining the original opposite FMS questions.

For $r=0,\ldots,4$, denote the decoded corner copies by $b_r,c_r,p_r,q_r$. Impose the root reflection equations for $p_0,q_0$. Introduce one centralizer contraction $a=(a_++a_-)/2$. The remaining equations are
\begin{align}
 a&=e_4ae_0,&a^*a&=e_0,&aa^*+b_4b_4^*+p_4&=e_4,\label{eq:Catasm1}\\
 b_0^*b_0&=e_0-p_0-q_0,&p_0q_0&=0,&c_0p_0&=p_0,\label{eq:Catasm2}\\
 c_0q_0&=q_0/2,&c_4a&=ac_0/2,&c_0b_0&=2b_0c_0.\label{eq:Catasm3}
\end{align}
The range equation is imposed at the far corner because $aa^*$ lies under $e_4$; conjugation by $v$ below moves it to the source.

Let $v={\btrans}_4{\btrans}_3{\btrans}_2{\btrans}_1$ be the bridge chain. Extraction on the source gives
\[
 A=v^*a,\qquad B=b_0,\qquad \Ccat=c_0,\qquad P=p_0,\qquad Q=q_0.
\]
Conjugate the far range equation by $v$, using the actual transported copies. This gives $AA^*+BB^*+P=e_0$. The other local equations follow directly; for example
\[
 \Ccat A=c_0v^*a=v^*c_4a=v^*ac_0/2=A\Ccat/2.
\]
Since $a$ is a centralizer contraction and $hv=16vh$, its source density has $h_0A=Ah_0/16$. The other decoded labels commute with it. After removal of the finite scalar factors, use exactly the finite-amplification cancellation proof of Lemma~\ref{lem:spectrum}, now for these five transported unitary labels. It gives a realization of \eqref{eq:Catlocal}--\eqref{eq:Catsquared} and the same spectrum identity with $\mu_4$.

Conversely use the canonical Catalan operators and density $\kcat$ in $\operatorname{diag}(\mu_4)\otimes\kcat$, put $a=|e_4\rangle\langle e_0|\otimes A$, and use identity transports. The ratio makes $a$ commute with the scaffold density. The blockwise polar encodings, actual graphs, label refinements and finite Boolean-row extension used in Lemma~\ref{lem:four} realize all the displayed equations simultaneously. Thus they compute a finite rational game $G^{(4)}_{\Cat}$ with a perfect normal strategy.

\subsection{Every normal code has infinite entropy}
Set $N=N_{\Gamma(G^{(4)}_{\Cat})}$ and let $m$ be the number of tagged local questions of $G^{(4)}_{\Cat}$. The canonical strategy just constructed has finite amplitude sum, by \eqref{eq:Catnumbers} and its finite scaffold and scalar factors. Theorem~\ref{thm:correspondence} therefore gives a normal $m$-message code.

Consider any normal code, including a mixed resource, and let $\rho_B$ be its Bob marginal. The assemblage converse and Theorem~\ref{thm:correspondence} construct a pure perfect game strategy with
\[
 h_{\rm game}=\frac{\rho_B^2}{\Tr\rho_B^2},\qquad
 \rho_B=\frac{\sqrt{h_{\rm game}}}{\Tr\sqrt{h_{\rm game}}}.
\]
The assembly spectrum identity and Lemma~\ref{lem:Catclassification} give
\[
 h_{\rm game}\simeq
 \operatorname{diag}(\mu_4)\otimes\kcat\otimes\rho_{\rm spec}
 \otimes I_3/3\otimes I_{d_{\max}}/d_{\max}.
\]
Here $\simeq$ means unitary equivalence of positive operators. Trace summability forces $\Tr\sqrt{\rho_{\rm spec}}<\infty$. Taking the normalized square root gives
\begin{equation}\label{eq:codemarginal}
\begin{gathered}
 \rho_B\simeq\operatorname{diag}(\hat\mu_4)\otimes\hcat\otimes\rho_{\rm spec,code}
                 \otimes I_3/3\otimes I_{d_{\max}}/d_{\max},\\
 \hat\mu_{4,i}=\frac{\sqrt{\mu_{4,i}}}{\sum_j\sqrt{\mu_{4,j}}},\quad
 \rho_{\rm spec,code}=\frac{\sqrt{\rho_{\rm spec}}}{\Tr\sqrt{\rho_{\rm spec}}}.
\end{gathered}
\end{equation}
The Catalan factor is exactly $\hcat$; the full marginal also includes the finite factors and the possible spectator $\rho_{\rm spec,code}$. Equation~\eqref{eq:Catentropy} gives $S(\rho_B)=\infty$ for every code.

For a pure resource its two marginal spectra coincide, so Alice's entropy is infinite as well. For a mixed resource $\omega=\sum_jt_j|\psi_j\rangle\langle\psi_j|$, each positive-weight component is a perfect code under the same encoding and decoding measurements. This follows by applying positivity to each forbidden error expectation. Each such pure component has infinite entropy on both marginals. Extended entropy concavity gives
\[
 S(\omega_A)\geq t_j S((\psi_j)_A)=\infty,
\]
and similarly for Bob. This proves the second part of Theorem~C for all normal codes.

\section{Discussion and open problems}\label{sec:discussion}
The hierarchy separates several ways to ask for certainty. Finite-dimensional perfection has a finite existential witness. Commuting perfection is controlled by compactness. Value one asks for arbitrarily accurate finite witnesses. Spatial and semifinite perfection combine an unbounded finite choice with an infinite verification. In the spatial reduction the chosen object is the order of a clock phase; in the semifinite reduction it is the depth of a critical tree. Without a trace, as in every type-III factor, the choice disappears and the commuting level returns.

\subsection{Entropy budgets}
The entropy restriction does not lower the complexity. Each YES instance
$\Gclk_e$ of the spatial reduction has a finite-entropy witness.
Corollary~\ref{cor:entropybeaver} shows that the least sufficient integer entropy budget, maximized over games of bit length at most $n$, is a function of Turing degree $0''$ that eventually dominates every total $0'$-computable function. Corollary~\ref{cor:schmidtbeaver} proves the same profile for the least sufficient Schmidt stratum. Finite-entropy
compactness also controls entire prescribed tables by
Corollary~\ref{cor:entropybudget}. The finite-entropy and arbitrary-spatial
decision sets have the same completeness degree, although the sets
themselves differ: the entropy-game construction in \cite{Entropy} has
spatial perfection and no finite-entropy perfect strategy, and
Corollary~\ref{cor:entropyonly} classifies their difference.

\subsection{Prescribed tables and escaping mass}
The largest-Schmidt compactness argument of Theorem~\ref{thm:strata} does not directly control a prescribed correlation table without a uniform tail bound. For a concrete obstruction, put
\[
 D_{{\rm esc},n}=\operatorname{diag}(1,\underbrace{n^{-1/2},\ldots,n^{-1/2}}_{n\text{ entries}},0,\ldots).
\]
Then $\Tr D_{{\rm esc},n}^2=2$ and $D_{{\rm esc},n}\to|0\rangle\langle0|$ in operator norm. Alice measures $|0\rangle\langle0|$ and its complement; Bob has one trivial outcome. Every normalized vectorization of $D_{{\rm esc},n}$ gives Alice probabilities $(1/2,1/2)$. The normalized limit gives $(1,0)$. The largest-amplitude normalization preserves zero equations while changing nonzero normalized probabilities.

For the finite-entropy realization class, Corollary~\ref{cor:entropybudget}
proves the $\Sigma^0_2$ upper bound for exact rational or real-algebraic
tables.

\subsection{Uniform membership and algebraic correlations}
Two questions remain about prescribed spatial correlations. Is membership of real-algebraic tables in $C_{qs}$ recursively enumerable uniformly over finite scenarios? Must every spatially perfect game have an algebraic perfect table? The two answers cannot both be yes: enumerating algebraic tables, checking their losing zeros, and dovetailing their membership searches would enumerate the YES instances of our $\Sigma^0_2$-complete family. The fixed-scenario remark of Fu, Miller and Slofstra \cite{FMS} does not by itself settle the required uniformity.

\subsection{Zero-error capacities}\label{sec:capacity}
Theorem~C concerns one use of a channel. A tensor power of the channel of a game has the strong-product confusability graph, which is not the confusability graph of the AND-repeated game, so one-shot hardness does not transfer to the entanglement-assisted zero-error capacity. Whether this capacity is computable, and whether normal entanglement can give a strictly larger capacity than finite-dimensional entanglement, remain open. For the squared-Catalan channel of Section~\ref{sec:catalan}, normal entanglement already reaches the clique-partition bound of Lemma~\ref{lem:saturation} in one use; whether finite-dimensional entanglement reaches the same capacity asymptotically is unknown.

\subsection{Other questions}
The type languages are existence languages: the host and its normal state may be chosen together. They do not decide whether a prescribed factor and state realize a correlation, or determine the least resources of a strategy. Natural questions ask which syntactic classes force finite-entropy spatial perfection whenever spatial perfection is possible, and how quantitative entropy requirements interact with the difference classification.

Can the game compiler be made substantially smaller? Can the period reader be realized by a simpler predicate? What are the optimal effective bounds for prescribed tables when one supplies an entropy budget or a computable bound on the Schmidt tail? A classification of unrestricted prescribed-table membership would require additional control of the normalization that escapes in the example above.

\section*{How this work was done}\label{sec:methods}
\addcontentsline{toc}{section}{How this work was done}
Opus 5.5, GPT-6.1 Sol and GPT-6 assisted with derivations, checking, programming and writing. The authors selected the questions, directed the work and take responsibility for the paper. The semifinite-but-not-spatial completion, the moment ladder and the nested compiler were derived with models from one developer and checked by a model from the other; these checks were automated.

Exact and numerical checks of the finite identities, counts and examples used in the proofs are in the repository,
\begin{center}\codearchive,\end{center}
and Appendix~\ref{sec:repro} describes their scope. Appendix~\ref{app:leanA} describes the formal verification of Theorem~A, together with the selected formalized lemmas for Theorems~B--D.

\appendix
\section{The nested assembly}\label{nested:sec:nested}
\subsection{Input, output, and the scope of the calibration}
An input is a finite real rational $*$-polynomial presentation in typed
centralizer labels and contractions $X_1,\ldots,X_s$. A label has type
projection, unitary, positive contraction, or contraction. The modular
symbols have the prescribed common ratio
\begin{equation}
 h_0X_i=\tfrac12X_ih_0.                                      \label{eq:nested-N0}
\end{equation}
Each monomial contains at most two modular symbols, counting adjoints;
centralizer coefficients can occur before, between, and after them.
Centralizer polynomial equations and degree-one endpoint equations are
included. A normal realization has a nonzero Hilbert space and a faithful
trace-class density $h_0$ of trace one, with all identities on its support.
The ratio in \eqref{eq:nested-N0} is part of the input convention; it is not an operator to be
synchronized. Section~\ref{nested:sec:variants} treats the common ratio $2^{-n}$.
Arbitrary unrelated real ratios are outside this stated lemma: the
calibration used here supplies powers of two.

The proof uses two calibrated levels so that an entire inner activity,
including off-diagonal lifted contractions, is transported by one outer
map. A centralizer corner unitary then identifies equal-density corners.
Products of the resulting modular operators can be translated into actual
centralizer products. The whole-activity refinement is what prevents
independent gauges from changing those products.

\begin{theorem}[Nested compiler]\label{nested:thm:nested}
Every such presentation effectively determines a finite rational Boolean
game. A perfect normal spatial strategy extracts a normal realization
on a source corner of mass $1/81$. Every normal realization has one
simultaneous perfect extension using a $25$-corner scaffold, a qutrit
frame, and a finite row ancilla. The extension preserves finiteness of
every positive moment and of entropy. With types, centralizer conditions,
and modular covariance exact, constructed approximate realizations obey
\begin{equation}
 1-\omega\le C_{\rm nest}\sum_j\|r_j\|_{2,h_0}^2,\qquad
 \|r\|_{2,h_0}^2=\Tr(h_0r^*r),                              \label{eq:nested-N1}
\end{equation}
where $C_{\rm nest}$ depends only on the finite compiled presentation.
\end{theorem}

\subsection{Corners, parameters, and activities}
Begin with nine synchronized corner projections
\[
 f,\ m_{00},m_{01},m_{10},m_{11},\ E_{00},E_{01},E_{10},E_{11}.
\]
Name a synchronized activity ${\Pact_{\rm in}}$ and impose
\begin{equation}
 {\Pact_{\rm in}}=f+\sum_{ab}m_{ab},\qquad {\Pact_{\rm in}}+\sum_{ab}E_{ab}=I.
                                                               \label{eq:nested-N2}
\end{equation}
The inner conditional FMS has flag $f$, activity ${\Pact_{\rm in}}$, and selected corner
$e=m_{00}$. The outer conditional FMS has flag \emph{the whole ${\Pact_{\rm in}}$},
activity $I$, and selected corner $E=E_{00}$. Retain every original
question, both orientations of each event binding, and every inactive
answer. The inner inactive event is $I-{\Pact_{\rm in}}$. The outer inactive event is
zero. In particular its column partition has flag ${\Pact_{\rm in}}$, four $E_{ab}$
answers, and a zero inactive answer. No new inner flag is substituted
for ${\Pact_{\rm in}}$ in that partition.

The frame-bound parameter list is exactly the nine corners, ${\Pact_{\rm in}}$, and
the four positive $ZI,IX$ events, two for each bridge: fourteen
parameters. Use the parameter-binding rows of Section~\ref{sec:projection}.
Set $M_{\rm in}={\Pact_{\rm in}}-f$ and $M_{\rm out}=I-{\Pact_{\rm in}}$. The four unitary axis labels
are equated to $I-2M+2E_{\mathrm{axis},+}$ for their respective bridge.
Let $r_{ab}=x_{\rm in}^az_{\rm in}^be$ and
$R_{ab}=x_{\rm out}^az_{\rm out}^bE$ denote the actual magic corner maps.
These are centralizer expressions in the bound parameters and axes.
The two calibrated transports satisfy
\begin{equation}
 u^*u=f,\ uu^*=e,\ hu=2uh;\qquad
 {\btrans}^*{\btrans}={\Pact_{\rm in}},\ {\btrans}{\btrans}^*=E,\ h{\btrans}=2{\btrans}h.                                 \label{eq:nested-N3}
\end{equation}
The soundness argument below derives the source mass $\Tr(hf)=1/81$ from these relations and the cover.

\subsection{The complete transport list and the whole-activity block}
Encode a projection $P$ by $2P-I$, a positive contraction $C$ by a
unitary with $C=(2I+U+U^*)/4$, a general contraction by the average of
two unitaries, and a unitary directly. Identities in a supported corner
use that corner's identity. The outer transported list consists of:
\begin{enumerate}
\item the two inner projections $f,e$, regarded as projections in ${\Pact_{\rm in}}$;
\item all source label encodings that occur at any endpoint (transporting
all source labels is an allowed deterministic choice);
\item both unitary encodings of every lifted contraction $a_i=e a_i f$;
\item the finitely many derived coefficients specified below when a
coefficient occurs inside a quadratic word.
\end{enumerate}
For every encoding name $k$, introduce global unitaries $U_k,V_k$ with
\[
 (U_k-I)(I-{\Pact_{\rm in}})=0,\qquad (V_k-I)(I-E)=0.
\]
A root-supported encoding additionally fixes $I-f$. Its transported
copy will then fix $E-f'$. Introduce a global unitary ${V^{\rm blk}_k}$ and impose
\begin{equation}
 {V^{\rm blk}_k}={\Pact_{\rm in}} U_k+\sum_{ab}R_{ab}(E V_k)R_{ab}^*.                    \label{eq:nested-N4}
\end{equation}
There is no $I-{\Pact_{\rm in}}-M_{\rm out}$ term since the outer activity is $I$.
In \eqref{eq:nested-N4}, $\Pact_{\rm in}U_k$ acts on the entire five-corner space ${\Pact_{\rm in}}\HH$, including
off-diagonal blocks between $f$ and $e$. Replacing it by a sum of five
unrelated diagonal encodings would lose the required common transport.
The identity $(U_k-I)(I-{\Pact_{\rm in}})=0$ gives $U_k(I-{\Pact_{\rm in}})=I-{\Pact_{\rm in}}$.
Left multiplication by $U_k^*$ gives $U_k^*(I-{\Pact_{\rm in}})=I-{\Pact_{\rm in}}$;
taking adjoints gives $(I-{\Pact_{\rm in}})U_k=I-{\Pact_{\rm in}}$. Thus $[U_k,{\Pact_{\rm in}}]=0$.
The same proof gives $[V_k,E]=0$. Consequently $\Pact_{\rm in}U_k$ is a
unitary on ${\Pact_{\rm in}}\HH$, and each $R_{ab}(EV_k)R_{ab}^*$ is unitary
on its own magic corner. Their supports are orthogonal and cover $I$,
so \eqref{eq:nested-N4} is block diagonal for
${\Pact_{\rm in}}\HH\oplus\bigoplus_{ab}E_{ab}\HH$ and is unitary.

For $f$ use $U_f=I-2{\Pact_{\rm in}}+2f$, and for $e$ use
$U_e=I-2{\Pact_{\rm in}}+2e$; their outer unitary names determine
\[
 f'=\tfrac12 E(I+V_f),\qquad e'=\tfrac12 E(I+V_e).
\]
For each bridge separately refine its retained signs for the six projections
\[
 F_1,\quad F_2,\quad F_3,\quad G_{12}(1),\quad G_{13}(1),\quad G_{23}(1).
\]
For the outer bridge also refine them separately for every $G_{12}({V^{\rm blk}_k})$.
The unchanged opposite questions are retained. There are six refinements at the inner bridge and $6+|\Lambda|$ at
the outer bridge, per retained question and per party.

Label refinement (Lemma~\ref{lem:label}) makes every
retained sign commute with the six frame projections. In the frame
coordinates these generate $\Mat_3\otimes I$: the $F_i$ give the diagonal
matrix units and $F_iG_{ij}(1)F_j=e_{ij}\otimes I/2$ gives the
others. Their commutant is $I_3\otimes B(\HH_{\rm base})$.
Because the corners are frame-bound, the sign and its $(c_0,f)$ block
are frame scalar. Thus ${\btrans}=I_3\otimes\widetilde {\btrans}$ (and likewise $u$).
Writing the outer sign as $I_3\otimes\widetilde{\mathcal X}_0$,
the $(1,2)$ frame block of its commutator with
\[
 G_{12}({V^{\rm blk}_k})=\tfrac12(F_1+F_2+e_{12}\otimes {V^{\rm blk}_k}+
                                      e_{21}\otimes {V^{\rm blk}_k}^*)
\]
is $[\widetilde{\mathcal X}_0,{V^{\rm blk}_k}]/2$. On the outer sign support,
with flag space ${\Pact_{\rm in}}\HH$ and selected magic space $E\HH$, the bridge
lemma gives
\[
 \widetilde{\mathcal X}_0=
 \begin{pmatrix}0&\widetilde {\btrans}^*\\\widetilde {\btrans}&0\end{pmatrix},
 \qquad
 {V^{\rm blk}_k}=\begin{pmatrix}\Pact_{\rm in}U_k|_{{\Pact_{\rm in}}\HH}&0\\0&EV_k|_{E\HH}\end{pmatrix}.
\]
Here $c_0$ means the selected outer magic direction $E$ and $f$
means its flag, which is the \emph{whole} ${\Pact_{\rm in}}$ rather than the inner $f$.
The $(c_0,f)$ block of the commutator is exactly
$\widetilde {\btrans}(\Pact_{\rm in}U_k)-(EV_k)\widetilde {\btrans}$, so
\begin{equation}
                 {\btrans}(\Pact_{\rm in}U_k)=(EV_k){\btrans},                            \label{eq:nested-N5}
\end{equation}
with the same ${\btrans}$ for every $k$, including the two
encodings of a contraction. Here $\Pact_{\rm in}U_k$ may mix all
five inner corners: if $\Pact_{\rm in}U_k=(U_{rs})$ in inner-corner
coordinates, \eqref{eq:nested-N5} says $EV_k={\btrans}(U_{rs}){\btrans}^*$.
Multiplying \eqref{eq:nested-N5} by the unitary adjoints gives transport also for
$U_k^*,V_k^*$. Decoding the same polynomials in these unitaries
therefore transports projections, positive contractions and averages
of two unitaries. Hence $f'={\btrans}f{\btrans}^*$, $e'={\btrans}e{\btrans}^*$,
and, with ${\btrans}_0={\btrans}f$, ${\btrans}_1={\btrans}e$,
\begin{equation}
                  a_i'={\btrans}_1a_i{\btrans}_0^*,                          \label{eq:nested-N6}
\end{equation}
including both encodings of each $a_i$ rather than just its diagonal
compressions. Both sides are actual operators; neither is an
independently gauged copy.

Introduce a centralizer contraction $t$ by two global unitaries and impose
\begin{equation}
 t=f'te,\qquad t^*t=e,\qquad tt^*=f'.                         \label{eq:nested-N7}
\end{equation}
The corner maps $t$ and $a_i$ are synchronized centralizer expressions;
$u,{\btrans},{\btrans}_0,{\btrans}_1$ are extracted transports and are never graph words.
Set $\HH_0=f\HH$ and $h_0=81fhf|_{\HH_0}$. The extracted contractions are
\begin{equation}
                         X_i={\btrans}_0^*ta_i.                      \label{eq:nested-N8}
\end{equation}
The soundness argument below shows that $h_0$ is a faithful trace-one density and that \eqref{eq:nested-N8} satisfies \eqref{eq:nested-N0}.

\subsection{All quadratic translations and arbitrary coefficients}
Put $\ell={\btrans}_0^*t{\btrans}_1^*$. It is a corner unitary from $e'\HH$ to $f\HH$.
Direct multiplication, using only \eqref{eq:nested-N6}--\eqref{eq:nested-N7}, gives
\begin{align}
 X_i^*X_j&=a_i^*a_j,\label{nested:eq:innernew}\\
 {\btrans}_0X_iX_j^*{\btrans}_0^*&=ta_ia_j^*t^*,\label{nested:eq:rangenew}\\
 X_iX_j&=\ell a_i'ta_j.\label{nested:eq:productnew}
\end{align}
For the last identity use $a_i{\btrans}_0^*={\btrans}_1^*a_i'$; the intermediate
projections are $e,f',e'$ and cancel by their stated source/range rules.
The following table is consequently an equivalence in both directions.
\begin{center}\small
\begin{tabular}{@{}ll@{}}
\toprule Root identity & Centralizer identity\\\midrule
$X_i^*X_j={R_{\rm tar}}$ & $a_i^*a_j={R_{\rm tar}}$ at $f$\\
$\sum_iX_iX_i^*={R_{\rm tar}}$ & $\sum_i ta_ia_i^*t^*={R_{\rm tar}}'$ at $f'$\\
$X_iX_j=X_kX_l$ & $a_i'ta_j=a_k'ta_l$\\
$X_i^*X_j=X_kX_l^*$ & $a_i'^*a_j'=ta_ka_l^*t^*$\\
$LX_i=X_jR$ & $L'ta_i=ta_jR$\\\bottomrule
\end{tabular}
\end{center}
Here ${R_{\rm tar}}'={\btrans}_0{R_{\rm tar}}{\btrans}_0^*$ and $L'={\btrans}_0L{\btrans}_0^*$ are obtained from the
transported encodings. The mixed quadratic row uses
${\btrans}_0a_i^*a_j{\btrans}_0^*=a_i'^*a_j'$. Conjugation by ${\btrans}_0$, or multiplication
by $\ell$, is injective on the specified corner. Thus these are
operator equivalences rather than implications inferred from traces.
In particular they give the displayed quadratic identities.

For completeness of the admissible class, here is a finite procedure for
interior coefficients. Give $X$ degree $1$ and $X^*$ degree $-1$.
In a density eigenbasis, a degree-$d$ word has matrix entries only where
the output eigenvalue is $2^{-d}$ times the input eigenvalue. Distinct
degrees are disjoint and orthogonal in $\|\cdot\|_{2,h_0}$; split a
polynomial into degrees $-2,-1,0,1,2$. Take adjoints of negative degrees.
For centralizer coefficients ${K_L},{K_M},{K_R}$:
\begin{enumerate}
\item Degree one: ${K_L}X_i{K_R}$ lifts to ${K_L}'ta_i{K_R}$.
\item Degree zero: ${K_L}X_i{K_M}X_j^*{K_R}$ lifts to
${K_L}'ta_i{K_M}a_j^*t^*{K_R}'$ at $f'$. A term ${K_L}X_i^*{K_M}X_j{K_R}$ first becomes
${K_L}a_i^*t^*{K_M}'ta_j{K_R}$ at $f$; transport that whole supported centralizer
expression to $f'$ through its own derived encoding.
\item Degree two: put $H_{K_L}=t^*{K_L}'t$ at $e$ and transport it to
$H_{K_L}'={\btrans}_1H_{K_L}{\btrans}_1^*$ at $e'$. Then
\[
             {K_L}X_i{K_M}X_j{K_R}=\ell H_{K_L}'a_i'{K_M}'ta_j{K_R}.
\]
Indeed $\ell H_{K_L}'={K_L}\ell$, by multiplication and $tt^*=f'$.
\end{enumerate}
A derived coefficient with a known rational norm bound $K$ is encoded
after division by $\max(1,K)$, and its defining equation is added as a
centralizer polynomial. Such a bound follows by multiplying and summing
the declared contraction bounds. Only the two displayed kinds of derived
coefficients are needed for degree two. Define the base labels and $t$
first, the original transports next, and then the derived coefficients
and their transports. Definitions involve no newly transported derived
coefficient. Adding their
separate refinements constrains the original
${\btrans}$, and never extracts a fresh gauge. This is a finite, non-circular
assembly for the whole stated class.

\subsection{Soundness and the Boolean game}
Start with a normal strategy, purify without querying the environment,
and reduce to the original marginal supports by facts (N1)--(N8) of Section~\ref{sec:compiler}. Perfect matching precedes every use of projectivity. Boolean
row tests give the rational identities on these supports. Graphs and
triangles then give actual common unitary words. Parameter bindings
identify the actual FMS events; they do not replace them by free labels.
Apply the two conditional bridge theorems to those bound events. The
frame refinements apply first; the block refinements then give \eqref{eq:nested-N5}.
Decode labels, impose \eqref{eq:nested-N7}, and apply \eqref{eq:nested-N8} and the translation procedure.
This order uses neither the intended application nor its spectral
classification in extracting the scaffold.

All cancellations of a coefficient operator use its dense range on the
original local support, as proved in (N1)--(N8). On a mixed-state cut
$A\mid BE$ the coefficient map need not be injective. All graph calculations are
bounded block identities. The source density is faithful since $f$ commutes with $h$ and
$h$ is faithful on the reduced support.

Write $h=h_A$ on $P_A\HH_A$. The synchronized
corners are projections commuting with $h$; their sum identities
make them orthogonal. Since $hu=2uh$ and $h{\btrans}=2{\btrans}h$, cyclicity of the
trace of a trace-class operator times bounded operators gives
\[
 \begin{aligned}\Tr(he)&=\Tr(huu^*)=\Tr(u^*hu)=2\Tr(hf),\\
 \Tr(hE)&=\Tr(h{\btrans}{\btrans}^*)=\Tr({\btrans}^*h{\btrans})=2\Tr(h{\Pact_{\rm in}}).\end{aligned}
\]
The centralizer magic-corner maps give equal masses to the four
inner selected copies and, separately, the four outer selected
copies. Thus $\Tr(h{\Pact_{\rm in}})=9\Tr(hf)$ and
$1=\Tr(h{\Pact_{\rm in}})+4\Tr(hE)=9\Tr(h{\Pact_{\rm in}})$, proving $\Tr(hf)=1/81$.
The bridge lemma holds also for a zero activity, so no nonvanishing sector
is assumed here. It follows that $\HH_0=fP_A\HH_A\ne0$ and
$h_0=81fhf|_{\HH_0}$ is faithful, trace class and trace one.

A supported unitary commutes with its support by the calculation
above. On $\HH_0$ the four decoding types are therefore respectively
$U$, $(I+U)/2$ with $U^2=I$, $(2I+U+U^*)/4$, and
$(U_++U_-)/2$. They are respectively unitary, projection, positive
contraction, and contraction, and each commutes with $h_0$.
The supported $a_i=e a_i f$ is a contraction; \eqref{eq:nested-N7} makes $t$ a
partial isometry from $e$ onto $f'$, and ${\btrans}_0$ is one from $f$ onto
$f'$. Hence $X_i={\btrans}_0^*ta_i$ maps $\HH_0$ into itself with norm at
most one. Since $h{\btrans}_0^*={\btrans}_0^*h/2$ and all other factors centralize $h$,
\[
 h_0X_i=81fh{\btrans}_0^*ta_i
 =\tfrac{81}{2}{\btrans}_0^*hta_i
 =\tfrac{81}{2}{\btrans}_0^*ta_ifhf=\tfrac12X_ih_0.
\]
Transport of decoded labels and their adjoints, followed by the
bounded corner multiplications in Section~\ref{nested:sec:nested}, now gives every
imposed application identity on this actual $\HH_0$ in both
directions. The degree splitting uses the density eigenbasis on
this support, never an inverse of $h$.

Bob's original marginal support is treated in the same way: matched
coarse effects are its projective mirrors, the bridge and refinement
lemmas apply on that support, and the same cyclicity gives source
mass $1/81$. Mirroring products reverses their order, as in (N4a);
no extra mirror equation is imposed. Alice's extracted realization
already proves soundness; the symmetric support argument justifies
all Bob-side bindings and refinements, including on a purified
mixed-state cut with its original Bob marginal.

Expand the final finite centralizer polynomials in their real rational
unitary words, cancel adjacent inverse tokens, and close under proper
prefixes.
Add all graphs, placement triangles, product triangles,
inverse triangles, parameter bindings, and the off-diagonal polynomial
rows. Apply the Boolean-row construction already proved above. Retain
all FMS and refinement questions, pad the two question sets, and sample
every pair uniformly with positive rational probability $Q^{-2}$.
Accept every unspecified answer pair. Perfection gives zero probability
for each individual rejecting rule, proving soundness.

\subsection{One simultaneous completeness model}
Let $\mu=\operatorname{diag}(1,2,2,2,2)/9$. On
$\mathbb C^5_{\rm out}\otimes\mathbb C^5_{\rm in}\otimes\HH_0$ take
\begin{equation}
                     h_{\rm sc}=\mu\otimes\mu\otimes h_0.     \label{eq:nested-N9}
\end{equation}
Outer index $0$ is the inner activity; inner index $0$ is $f$ and
inner index $1$ is $e$. The selected outer index is $1$. Thus
\begin{align*}
 f&=|0,0\rangle\langle0,0|\otimes I,&
 e&=|0,1\rangle\langle0,1|\otimes I,\\
 f'&=|1,0\rangle\langle1,0|\otimes I,&
 e'&=|1,1\rangle\langle1,1|\otimes I,\\
 {\btrans}&=|1\rangle\langle0|\otimes I_5\otimes I,&
 t&=|1,0\rangle\langle0,1|\otimes I,\\
 a_i&=|0,1\rangle\langle0,0|\otimes X_i,&
 a_i'&=|1,1\rangle\langle1,0|\otimes X_i.
\end{align*}
The density weights at $e$ and $f'$ are both $2h_0/81$, so $t$
centralizes $h_{\rm sc}$. Commutation of $a_i$ with $h_{\rm sc}$ is
exactly $2h_0X_i=X_ih_0$. All the copies use the identity outer
identifications. The inner and outer FMS effects are the real canonical
ones on their respective five-dimensional factors, with the inner
copy inactive off outer index $0$. Their compressed sign vectors are
$|f,f\rangle+(-1)^{xy}|c_x,r_y\rangle$, up to the common positive
amplitude and spectator factors. The FMS sign completeness used here
is the explicit calculation at the end of the proof of
Lemma~\ref{lem:four}.

Every positive density eigenspace is finite-dimensional. For a
centralizer contraction $T$, its polar partial isometry extends on
each such eigenspace to a unitary $V$. Then
\begin{equation}
 U_\pm=V\bigl(|T|\pm i\sqrt{I-|T|^2}\bigr),\qquad
                         T=(U_++U_-)/2.                     \label{eq:nested-N10}
\end{equation}
Direct sums give bounded centralizer unitaries. On a prescribed
support such as ${\Pact_{\rm in}}$, do this inside that support and extend by identity
outside. For a root positive contraction use
$U=2C-I+2i\sqrt{C(I-C)}$, and for a projection use $2P-I$.
Encode source choices once and transport precisely those choices into
$E$; \eqref{eq:nested-N4} now holds simultaneously for every label, even a lifted
$a_i$ with off-diagonal inner blocks. Encode the global $t$ by \eqref{eq:nested-N10}.

Append $I_3/3$ and all graph, triangle, and parameter-binding
projections. For every refinement use actual joint products: its
block unitary commutes with its own FMS signs. Distinct labels require
no mutual commutation. Finally append $I_{d_{\max}}/d_{\max}$ and the
Boolean-row effects. Alice uses these operators and Bob their
transposes in a product density eigenbasis. Facts (N1)--(N8) prove every
matching, calibration, orthogonality, and row-zero test with this one
state and assignment.

Both nested bridges have magic density twice the
flag density. For the inner one the flag spectator density is $h_0$
(up to its scalar outer weight); for the outer one it is
$\mu\otimes h_0$. Denote either normalized spectator by $\rho$,
and the total flag mass by $\lambda$. The canonical magic directions
satisfy $\langle c_x,r_y\rangle=\epsilon_{xy}/\sqrt2$, where
$\epsilon_{xy}=(-1)^{xy}$. Compressing $\Omega_h$ to the two sign
supports gives, omitting only the unchanged frame and row factors,
\[
 \Omega_{xy}=\sqrt\lambda\bigl(|f,f\rangle+
            \epsilon_{xy}|c_x,r_y\rangle\bigr)\otimes\Omega_\rho.
\]
Alice's sign observable exchanges $f,c_x$ and Bob's exchanges $f,r_y$.
Applying the swaps gives
$(\mathcal X_x\otimes I)\Omega_{xy}
=\epsilon_{xy}(I\otimes\mathcal Y_y)\Omega_{xy}$.
Their spectral sign answers therefore never violate $ab=\epsilon_{xy}$.
The other column/row projections give the retained answers and the
calibration rules exactly. The inner measurements have inactive
effect $I-{\Pact_{\rm in}}$ off outer index $0$; that sector is matched on the two
sides. The outer bridge is active everywhere.

Under $E\HH\cong\mathbb C^5_{\rm in}\otimes\HH_0$, let
$\widehat U_k$ be the entire chosen unitary on ${\Pact_{\rm in}}\HH$.
Identity outer identifications make
\[
                    {V^{\rm blk}_k}=I_{\rm out}\otimes\widehat U_k.
\]
Every outer FMS effect, including its noncentralizer sign effects,
acts on the outer factor alone and hence commutes with ${V^{\rm blk}_k}$,
even when $\widehat U_k$ has off-diagonal inner blocks. The inner
bridge has only the six frame refinements, which commute because
the frame is a separate scalar factor. Thus every labelling
refinement is the actual positive joint product of commuting
projections, sums to the original coarse measurement, and has the
required graph marginal. Transposition in the product density eigenbasis preserves
these commutators and products on Bob's side. All centralizer
matching and orthogonality tests follow from \eqref{eq:nested-N8}, and the appended
row factor gives all Boolean row tests simultaneously with the same
state. This proves the claimed nested completeness assignment.

The complete marginal is a tensor product of $h_0$ with a fixed
finite-dimensional density $\eta$. Thus
$\Tr[(\eta\otimes h_0)^\alpha]=\Tr(\eta^\alpha)\Tr(h_0^\alpha)$
for every $\alpha>0$, and its entropy is $S(\eta)+S(h_0)$.
Conversely a finite full moment gives a finite moment of the extracted
realization: the source corner $f$ commutes with the marginal density $h$, so
$\Tr[(hf)^\alpha]=\Tr(h^\alpha f)\le\Tr h^\alpha$, and normalization by the
mass $1/81$ gives $\Tr h_0^\alpha\le81^\alpha\Tr h^\alpha$. Likewise the block
decomposition $h=hf\oplus h(I-f)$ gives $S(h_0)\le81\,S(h)$.
These observations give the stated resource preservation.

\subsection{The loss constant, including the normalization factors}
Keep exact structural hypotheses and choose all derived coefficient
encodings from their actual expressions, so their defining rows stay
exact. A homogeneous source residual $r$ has a lifted centralizer
residual $p$. For a degree-one or degree-two lowering residual the
lift has input corner $f$, so
\[
 \|p\|_{2,h_{\rm sc}}^2=\frac1{81}\|r\|_{2,h_0}^2.
\]
A degree-zero residual transported to $f'$ instead has factor $2/81$.
For negative degrees take the adjoint; if $h_0r=2^{-d}rh_0$,
cyclicity gives $\|r^*\|_{2,h_0}^2=2^{-d}\|r\|_{2,h_0}^2$.
Therefore each chosen translation has factor $\beta_j/81$, with
$\beta_j\le4$. Splitting degrees preserves the sum of squared norms
because their density matrix entries have disjoint ratios.

For the qutrit row $\Delta_j$ the two off-diagonal blocks are $p_j,p_j^*$.
Since $p_j$ is in the centralizer, trace cyclicity gives
\[
 \Tr(h_{\rm old}\Delta_j^2)=\frac{2\beta_j}{243}\|r_j\|_{2,h_0}^2.
\]
Let $L_j$ clear the rational row denominators and let $d_j$ be its
padded power-of-four column dimension. Equation~\eqref{eq:rowloss}
then gives, for one oriented row-zero test,
\begin{equation}
 Q^{-2}p(\Pi^{\rm row}_0,\Pi^{\rm ref}_{d_j})=
 \frac{8L_j^2\beta_j}{243d_j^3Q^2}\|r_j\|_{2,h_0}^2.
                                                               \label{eq:nested-N11}
\end{equation}
Sum these probabilities; overlapping rules can only reduce total loss.
If both orientations are retained, include both in the sum (equivalently
multiply the maximum constant by two). One may therefore take the
maximum displayed coefficient, with the explicit orientation
multiplicity, as $C_{\rm nest}$. It is independent of the Hilbert space,
density spectrum, witness period, and computation time. If just one
forbidden projection has root probability $p$, its residual squared norm
is $p$, and \eqref{eq:nested-N1} reads loss at most $C_{\rm nest}p$. This proves the
completeness estimate.

\section{The bridge-count and branching variants}\label{nested:sec:variants}
\begin{corollary}[Common dyadic ratio and arbitrary finite branching]
For every fixed $n\ge1$, replace the inner bridge by an $n$-bridge chain
and the outer bridge by an $n$-bridge chain with first flag the entire
inner activity. The same degree-two compiler works with common ratio
$2^{-n}$. If $d_n=1+4n$ and
$Z_n=1+4\sum_{g=1}^n2^g$, the scaffold has $d_n^2$ scalar corners and
source mass $Z_n^{-2}$. Any finite number $b$ of creation labels may
be used. For a full $b$-ary isometry presentation a normal realization
requires $b2^{-n}<1$ and contains the weighted Fock factor with weights
$(1-b2^{-n})2^{-n|w|}$.
\end{corollary}
\begin{proof}
The explicit chain corner list is $e_0,m_{g,ab}$, $1\le g\le n$,
with $e_g=m_{g,00}$, activity $e_{g-1}+\sum_{ab}m_{g,ab}$, and cover
$e_0+\sum_{g,ab}m_{g,ab}=I$. The bridge transports ${\btrans}_g$ have ratio
$2$; their product has ratio $2^n$. Use the block equation \eqref{eq:nested-N4} at
each link, including its inactive identity term when the link activity
is not $I$. Transport the same encoding list through every link.
This gives the actual common product transport, by induction on $g$.
The inner chain density is
\[
 Z_n^{-1}\operatorname{diag}(1;2^{\oplus4};4^{\oplus4};\ldots;
 (2^n)^{\oplus4})\otimes h_0.
\]
Take two such factors. The inner last selected corner and the outer
last selected copy of the source both have density $2^nh_0/Z_n^2$,
so the same $t$ construction applies. Replace ${\btrans}_0,{\btrans}_1$ by restrictions
of the common outer chain product. Every multiplication in the proof
above remains valid, and the common quadratic factor still exists.
The finite mass factors in \eqref{eq:nested-N11} change accordingly and remain
presentation-dependent constants.

The number $b$ affects only the finite list of lifted $a_i$ and its
isometry, orthogonality, and range rows. No bridge has to possess $b$
magic corners: its magic square still has four. If $A_i^*A_j=\delta_{ij}I$
and $hA_i=2^{-n}A_ih$, cyclicity gives total range mass $b2^{-n}$.
Pull a positive density eigenvector backwards through nonzero $A_i^*$;
its eigenvalue increases by $2^n$ and so the process must terminate.
The defect and all its word descendants exhaust the support. The word
sectors are orthogonal, and their densities are $2^{-n|w|}$ times
the root density. Normalization forces $b2^{-n}<1$ and gives the
stated factor. Its $\alpha$-moment converges precisely when
$b2^{-n\alpha}<1$. This proves both the $n\ne4$ and $b\ne4$ variants.
\end{proof}

\paragraph{Exactly $n$ bridges for the bare-tree presentations of Section~\ref{sec:renyi}.}
The bare-tree and degree-one endpoint presentations need only one
chain, hence exactly $n$ bridges, rather than the two chains needed
for the quadratic extension. Here is its complete specialization.
Use the preceding $1+4n$ corner list and cover, bind those corners and
the $2n$ positive axis events to the frame, and set
$v={\btrans}_n\cdots {\btrans}_1$. For each transported encoding use the
copies $u_{g,\ell}$ fixing $I-e_g$ and the block equation
\eqref{eq:blockcopies}. Refine that link for the six frame projections and every
$G_{12}(V_{g,\ell})$. This gives the same calibrated transport of every
label through $v$, with no independent link gauges. Introduce
$b_i=e_nb_ie_0$ and extract $A_i=v^*b_i$. The input rows are exactly
\[
 b_i^*b_j=\delta_{ij}e_0,\qquad
 \sum_i b_ib_i^*=e_n-P_n,\qquad L_nb_i=b_jR_0,
\]
plus the root centralizer polynomial rows. These are respectively
the isometry/orthogonality, range, and endpoint translations; multiply
by $v,v^*$ to obtain both implications. The cover fixes source mass
$Z_n^{-1}$. The one-chain density displayed in the proof, the same
finite-block encodings, and actual joint refinements give simultaneous
completeness. The one-oriented loss constant is
\[
 \max_j\frac{8L_j^2\beta_j}{3Z_nd_j^3Q^2},\qquad
 \beta_j=1\ \hbox{for source and endpoint rows},\quad
 \beta_j=2^n\ \hbox{for range rows}.
\]
For the two-chain quadratic version replace $Z_n$ by $Z_n^2$ and use
$\beta_j\le2^{2n}$. Any finite $b$ is handled just by the displayed
finite $i,j$ lists. Thus the one-chain $n$-bridge compiler used in Section~\ref{sec:renyi} and the nested degree-two extension are both specified.

\section{Software checks}\label{sec:repro}
The finite checks are in the directory \texttt{verification/} of the repository below, and the Lean development of Appendix~\ref{app:leanA} is in \texttt{lean/}. The proofs in this paper do not depend on the finite checks. They test finite instances of the identities, constants and constructions; \texttt{verification/DETAILS.md} records every check and count.

\centerline{\codearchive}

\paragraph{Checked exactly.} Exact rational or algebraic arithmetic checks
\begin{itemize}
\item the tree, clock and reader identities on ten finite models (depths $0$--$4$, clock orders $2$ and $3$), where exactly the four top-depth isometry rows fail, as \eqref{eq:fourdefects} predicts;
\item a compiled four-bridge instance with one modular isometry and no transported labels ($748$ rational rows, $27\,991$ questions per party), and the worked tableau of Table~\ref{tab:toy};
\item the scaffold constants $4n+1$, $6n+1$ and $\mathfrak c_n$, the five- and eight-bridge constants \eqref{eq:fiveconstant} and \eqref{eq:eightconstant}, the code entropy \eqref{eq:fiveentropy} and the Catalan identities \eqref{eq:Catnumbers};
\item finite binary trees, reader stand-ins and chain scaffold masses for the type rows, and the Boolean profiles of every reduction in Theorem~\ref{thm:differences};
\item the finite selector models of Section~\ref{sec:iionly}, and the tree spectra and comb arithmetic of Sections~\ref{sec:renyi}--\ref{sec:programmable};
\item five literal instances of the nested compiler of Appendix~\ref{nested:sec:nested} (three predicates covering both decision branches and two comb programs, with $84\,402$ to $103\,474$ rational rows each), every row evaluated in the rational free group ring, and a synthetic presentation with $518\,601$ rows that exercises every translation route;
\item the rotation margin of Appendix~\ref{app:rotationcertificate}.
\end{itemize}

\paragraph{Checked in floating point.} With explicit tolerances: the truncated Fock model with its Weyl pair and tableau; good primitive phases for every prime below $4000$ (Lemma~\ref{lem:goodphase} is the proof for all primes); the trace-class correspondence on the magic square and the flagged magic square, including the negative control for the marginal $D^2$ mentioned after Theorem~\ref{thm:correspondence}; the reader bounds and reductions of Section~\ref{sec:iionly}; the Fibonacci clock models of orders $3$, $5$ and $8$; randomized group-law and nested transport identities.

\paragraph{Not checked.} None of the finite checks proves infinite-dimensional existence or extraction, compactness, the general compiler, the Higman and HNN embedding theorems, the classification of type-III factors, or any hardness reduction. No full reduction game is written out; the questions of the full instance in Section~\ref{sec:reduction} are only counted. The finite group stand-ins do not represent the groups $\mathcal G_e$ of Section~\ref{sec:group}.

\paragraph{How to run.} In \texttt{verification/}, run \texttt{python -B run\_all.py -{}-output-dir DIR} with \texttt{DIR} new or empty; add \texttt{-{}-full-counts} to repeat both question counts. The runner regenerates the nested instances, compares counts and hashes with the shipped records, and fails if any check fails. Its \texttt{README.md} lists dependencies and maps the program labels to the equations of this paper. The paper itself builds with \texttt{python -B build.py} in \texttt{paper/}.

\paragraph{Lean.} Besides the proof of Theorem~A, a set of local Lean results checks selected steps of the written proofs. Table~\ref{tab:leancore} shows representative entries. WHOLE means that the stated local assertion is checked in full, and STEP that one step of the cited result is checked. Theorems B--D are not checked beyond these local lemmas.

\begin{center}\footnotesize
\captionof{table}{Scope of selected Lean results; WHOLE refers only to the stated local assertion.}\label{tab:leancore}
\begin{tabular}{@{}>{\raggedright\arraybackslash}p{.36\linewidth}>{\raggedright\arraybackslash}p{.49\linewidth}>{\raggedright\arraybackslash}p{.09\linewidth}@{}}
\toprule Local result & Paper reference (stable label) & Extent\\\midrule
Good primitive phase, every prime & Lemma~\ref{lem:goodphase} (\texttt{lem:goodphase}) & WHOLE\\
Printed admissibility of the Theorem~A application tables & Definition~\ref{def:admissible} (\texttt{def:admissible}); table~\eqref{eq:labelset} & WHOLE\\
Bridge component identity & Equation~\eqref{eq:bridgevector} (\texttt{eq:bridgevector}) & WHOLE\\
Magic-square rigidity and matrix units & Lemma~\ref{lem:bridge} (\texttt{lem:bridge}) & STEP\\
Scalar Fock entropy series & Equation~\eqref{eq:fockentropy} (\texttt{eq:fockentropy}) & WHOLE\\
Conditional clock and reader implications & Equations~\eqref{eq:det}, \eqref{eq:coordinateclock}--\eqref{eq:D} (\texttt{eq:det}, \texttt{eq:coordinateclock}, \texttt{eq:reader}, \texttt{eq:return}, \texttt{eq:D}) & STEP\\
\bottomrule
\end{tabular}
\end{center}

\section{An exact rotation margin}\label{app:rotationcertificate}
For $\vartheta=(\sqrt5-1)/2$ and $\zeta=e^{2\pi i\vartheta}$, use the lower bound of Section~\ref{sec:weylsemifinite}.
Put $\gamma_y=2^{y+1}|\zeta^y-1|$. The global minimum of $\log_2 \gamma_y$ is attained at $y=1$, and
\begin{equation}\label{eq:margin}
 2.89<\log_2 \gamma_1<2.9,\qquad
 \log_2 \gamma_1=2.898452049757\ldots.
\end{equation}
Here is a certificate, including the infinite tail. For $y\geq3$, the preceding bound gives $\gamma_y\geq2^{y+3}/(\sqrt5\,y+1/2)>8$: at $y=3$ use $\sqrt5<9/4$, and the ratio increases because $2(\sqrt5\,y+1/2)>\sqrt5\,(y+1)+1/2$. For $y=2$ the exact certificate described below gives $\gamma_2>8$. For $y=1$ it gives $1<\gamma_1<8$ and $\gamma_{1,-}^{100}>2^{289}$, $\gamma_{1,+}^{10}<2^{29}$ for rational bounds $\gamma_{1,-}<\gamma_1<\gamma_{1,+}$. To construct them, enclose $\sqrt5$ by integer square roots at scale $10^{80}$, and $\pi$ by $16\arctan(1/5)-4\arctan(1/239)$ with 60 alternating terms and the next-term remainder. Evaluate $\sin(\pi\vartheta)$ at the rational endpoints using 40 alternating terms and the next term; sine is decreasing on this interval in $(\pi/2,\pi)$. Round outward to denominator $10^{40}$ and compare the displayed integer powers. For $y=2$, use $|\sin(2\pi\vartheta)|=\sin(\pi(2\vartheta-1))$; its argument is in $(0,\pi/2)$, where sine increases, and the same series gives $16\sin(\pi(2\vartheta-1))>8$. The exact-fraction routine in \texttt{verification/types/} carries out these steps. It certifies the $y=1,2$ enclosures and the rational base and coefficient inequality of the tail estimate. Together with the induction above these prove the minimum over all positive integers. The weaker positive margin of Section~\ref{sec:weylsemifinite} is all that contractivity requires.

\section{Results used from other papers}\label{app:companions}
Tables~\ref{tab:imported} and \ref{tab:imported-continued} list every load-bearing result imported from other papers, with the place where it is used. Results restated with proof here are marked as such.

\begin{center}\footnotesize

\begin{minipage}{\linewidth}
\captionof{table}{Imported results.}\label{tab:imported}
\begin{tabular}{@{}>{\raggedright\arraybackslash}p{.40\linewidth}>{\raggedright\arraybackslash}p{.27\linewidth}>{\raggedright\arraybackslash}p{.25\linewidth}@{}}
\toprule Result & Source & Used in\\\midrule
Perfect completeness of $\mathsf{MIP}^*=\mathsf{RE}$ & \cite[Theorem~12.6]{MIPRE} & Sections~\ref{sec:upper}, \ref{sec:sep}, \ref{sec:fdrow}\\
Commuting perfection is $\Pi^0_1$-complete & \cite{Slofstra}; \cite[Theorem~6.4]{MNY} & Sections~\ref{sec:typeIII}, \ref{sec:commuting}\\
Finite-dimensional value one is $\Pi^0_2$-complete & \cite[Theorem~6.9]{MNY} & Table~\ref{tab:landscape}\\
Convergent moment hierarchy & \cite{NPA} & Sections~\ref{sec:upper}, \ref{sec:commuting}\\
Flagged magic-square geometry & \cite{Mancinska}; square \cite{Mermin,Peres} & Section~\ref{sec:bridge}\\
Normal bridge, label refinement, rational rows, row loss & \cite[Lemmas~4.1, 4.2, Theorem~B, Eq.~(32)]{Entropy}; restated with proof & Section~\ref{sec:compiler}\\
Catalan game, local relations and two-bridge assembly & \cite[Theorem~C, Sections~3 and 6, Lemma~3.1]{Entropy} & Sections~\ref{sec:sep}, \ref{sec:catalan}\\
Standard-form reduction of commuting strategies & \cite[Lemma~1.6]{ReadType}; restated & Section~\ref{sec:support}\\
Commuting calibrated bridge & \cite[Lemma~2.2]{ReadType}; proof recalled & Section~\ref{sec:bridgecomm}\\
Expected-inclusion transfer & \cite{Takesaki}; \cite[Lemma~3.1]{ReadType}; proof given & Section~\ref{sec:typeIII}\\
Tracial mechanism for linear-system games & \cite[Proposition~7.4]{ReadType} & Section~\ref{sec:sep}\\
Cuntz gauge KMS state and its factor & \cite{OP}; \cite[Theorem~4.9, Example~4.13]{HarrisEmbezzlement}; \cite[Section~3.1]{ReadType} & Section~\ref{sec:promises}\\
Bounded centralizer perturbation & \cite[Lemma~4.5]{PT} & Section~\ref{sec:bridgecomm}\\
Effective Higman embedding & \cite{Higman}; \cite[Theorem~2.2]{Chiodo}; \cite{Craig} & Section~\ref{sec:group}\\
HNN normal form & \cite{HNN} & Section~\ref{sec:group}\\
\bottomrule
\end{tabular}
\end{minipage}
\end{center}

\begin{center}\footnotesize

\begin{minipage}{\linewidth}
\captionof{table}{Imported results. (continued)}\label{tab:imported-continued}
\begin{tabular}{@{}>{\raggedright\arraybackslash}p{.40\linewidth}>{\raggedright\arraybackslash}p{.27\linewidth}>{\raggedright\arraybackslash}p{.25\linewidth}@{}}
\toprule Result & Source & Used in\\\midrule
Asymptotic ratio set and $S$-invariant & \cite[Section~3.6, Theorem~3.6.1]{Connes} & Section~\ref{sec:typeIII}\\
Absorption argument & \cite[Lemma~3.8]{HM} & Section~\ref{sec:typeIII}\\
Flow of weights classification & \cite{ConnesTakesaki} & Section~\ref{sec:typeIII}\\
Injective factor classification & \cite{Connes76} & Section~\ref{sec:weylsemifinite}\\
Linear-system games: nonclosure, solution groups, spatial implies finite & \cite{SlofstraNonclosed}; \cite[Theorems~4--5]{CLS}; \cite{CM} & Section~\ref{sec:sep}\\
Assemblage characterization of zero-error codes & \cite{CLMW} & Section~\ref{sec:normal}\\
Clique saturation, square-root map, Corollaries~7 and~8 & \cite[Lemma~5, Theorem~6, Corollaries~7--8]{Mancinska} & Sections~\ref{sec:normal}, \ref{sec:fdrow}\\
Douglas factorization & \cite{Douglas} & Section~\ref{sec:normal}\\
Synchronous tracial representation & \cite[Theorem~5.5]{PSS}; \cite{KPS,MNY} & Section~\ref{sec:commuting}\\
Limit lemma & Shoenfield (1959) \cite{Shoenfield} & Section~\ref{sec:entropybudgets}\\
Right-c.e.-in-$0'$ reals as computable limsups & \cite[Lemma~4]{BBS} & Section~\ref{sec:programmable}\\
\bottomrule
\end{tabular}
\end{minipage}
\end{center}

The dependence is one way. This paper uses the entropy-game paper \cite{Entropy} and the type-game paper \cite{ReadType}; neither uses any result proved here.

\section{Lean verification of Theorem A}\label{app:leanA}
Theorem A is proved in Lean~4 with no hypotheses; the proof uses only
Lean's three standard axioms. The comparator challenge \nolinkurl{lean/Challenge.lean}
states Theorem A as follows, with \nolinkurl{TheoremAStatement} as defined
in \nolinkurl{lean/Sigma2/Statements.lean}:
\begin{flushleft}\leftskip1.5em\footnotesize\ttfamily\frenchspacing
\detokenize{theorem theoremA_paper : TheoremAStatement}\\[4pt]
\detokenize{def TheoremAStatement : Prop :=}\\
\hspace*{1em}\detokenize{SigmaTwoComplete SpatialPerfect} $\wedge$ \detokenize{SigmaTwoComplete EntropyPerfect} $\wedge$\\
\hspace*{1em}\detokenize{PromiseHard SpatialPerfect PromisedGame} $\wedge$ \detokenize{PromiseHard EntropyPerfect PromisedGame}
\end{flushleft}
Here the languages consist of natural-number codes of finite rational
Boolean two-player games, and invalid codes belong to none of them.
\texttt{SpatialPerfect} is the set of games with a perfect normal spatial
strategy (perfection is attained, not merely value one), and
\texttt{EntropyPerfect} the set of games with such a strategy whose two
reduced density operators have finite entropy; Section~\ref{sec:prelim}
explains why the one-marginal entropy convention gives the same language.
\texttt{SigmaTwoComplete} means $\Sigma^0_2$ membership and hardness under
total computable many-one reductions. \texttt{PromiseHard} says that each
hardness reduction can be chosen so that every output lies in
\texttt{PromisedGame}: it has finite-dimensional value one and no perfect
finite-dimensional strategy. The same statement is proved as
\nolinkurl{theoremA} in \nolinkurl{lean/Sigma2/Main.lean} and as
\nolinkurl{theoremA_paper}, through the printed admissibility
predicate, in \nolinkurl{lean/Sigma2/MainPaper.lean}. The theorem
\nolinkurl{theoremA_shared_family} in \nolinkurl{lean/Sigma2/SharedFamily.lean}
proves that a single total computable map is a promised reduction for both
languages, which is the family of Theorem A.

\subsection{What is checked}
The first column uses this paper's numbering. The Lean names
\nolinkurl{lemma_4_4_strong_of_obligations} and
\nolinkurl{lemma_3_1_compact_sandwiches} refer to Lemma~\ref{lem:four} at
$n=4$ and to Lemma~\ref{lem:sandwich}. Several theorems in the table take
as an argument a compiler specification, of type
\nolinkurl{CompilerSpecification.Specification}. The final theorems supply
it from the proved obligations \nolinkurl{CompilerObligations.obligations_all},
so it is not a hypothesis of any final theorem. The final theorems have no hypotheses.
The formal $\Sigma^0_2$ upper bounds of Theorem~A use no decision procedure for real arithmetic.
Each stage is a conjunction of polynomial equations and non-strict inequalities. Every satisfiable finite stage or finite entropy relaxation has a
solution with all coordinates in $[-1,1]$: amplitudes and probabilities lie
in $[0,1]$, and the formal stages encode each effect as $E=Y^*Y$, where a Gram factor $Y$ of an effect in a POVM has entries of modulus
at most one. On this box each constraint is Lipschitz in the maximum norm (each coordinate moved by at most the spacing), with a constant
computed from its coefficients. A stage therefore fails exactly when, for
some $k$, at every point of the grid of spacing $1/(k+1)$ some constraint is
violated by more than its Lipschitz constant times $1/(k+1)$. Soundness is
the Lipschitz estimate and completeness is compactness of the box. Each
test is a finite exact rational computation, so the stages are uniformly
$\Pi^0_1$ (\nolinkurl{UpperGridCertificate.boxTruth_iff_cert},
\nolinkurl{Upper.finite_stage_piOne}, \nolinkurl{Upper.entropy_relaxation_piOne}).

\begin{center}\footnotesize
\begin{minipage}{\linewidth}
\setlength{\tabcolsep}{4pt}
\begin{tabular}{@{}>{\raggedright\arraybackslash}p{.17\linewidth}>{\raggedright\arraybackslash}p{.44\linewidth}>{\raggedright\arraybackslash}p{.33\linewidth}@{}}
\toprule Printed result & Lean name & Checked form\\\midrule
Theorem A &
\nolinkurl{Sigma2.theoremA};
\nolinkurl{Sigma2.theoremA_paper};
\nolinkurl{Sigma2.theoremA_shared_family} &
Full: both completeness clauses, both promised-hardness clauses, and one
computable hardness family for both languages.\\[4pt]
Lemma 4.5, $n=4$ &
\nolinkurl{Sigma2.CompilerContract.lemma_4_4_strong_of_obligations};
\nolinkurl{Sigma2.CompilerSpecification.strong_compiler};
\nolinkurl{Sigma2.CompilerSpecification.four_bridge_compiler};
\nolinkurl{Sigma2.CompilerObligations.obligations_all} &
Partial: four-bridge exact compilation, with all seven compiler
obligations proved; not general $n$ or commuting hosts.\\[4pt]
Theorem 3.1, membership clauses; Lemma 3.2 &
\nolinkurl{Sigma2.Upper.uniform_strata_piOne};
\nolinkurl{Sigma2.Upper.spatial_union};
\nolinkurl{Sigma2.Upper.spatial_sigmaTwo};
\nolinkurl{Sigma2.Upper.lemma_3_1_compact_sandwiches} &
Partial: the first two clauses of Theorem 3.1; Lemma 3.2.\\[4pt]
Section 3.2 and the relaxations of Proposition 3.3 &
\nolinkurl{Sigma2.UpperGridCertificate.boxTruth_iff_cert};
\nolinkurl{Sigma2.UpperGridCertificate.boxTruth_piOne};
\nolinkurl{Sigma2.UpperBoxStages.stage_boxTruth};
\nolinkurl{Sigma2.UpperBoxStages.entropy_boxTruth};
\nolinkurl{Sigma2.Upper.finite_stage_piOne} &
The finite predicates are uniformly $\Pi^0_1$ by exact rational grid
certificates; quantifier elimination is not used.\\[4pt]
Proposition 3.3; Corollary 3.4; equation (5) &
\nolinkurl{Sigma2.UpperEntropyAssembly.perfect_of_sequence};
\nolinkurl{Sigma2.Upper.entropy_tail};
\nolinkurl{Sigma2.Upper.entropy_stage_slack};
\nolinkurl{Sigma2.Upper.entropy_scheduled_iff};
\nolinkurl{Sigma2.Upper.entropy_relaxation_piOne};
\nolinkurl{Sigma2.Upper.entropy_sigmaTwo} &
Partial: perfect-game compactness and effective finite-entropy
membership; not prescribed-table compactness, entropy infima or optimal
entropy.\\[4pt]
Sections 2 and 4, normal toolkit &
\nolinkurl{Sigma2.Preliminaries.normal_pure_perfect_equivalence} &
Partial: the normal/pure perfection equivalence, not all of
(N1)--(N8).\\[4pt]
Lemma 4.1 &
\nolinkurl{Sigma2.Compiler.lemma_4_1_normal_bridge};
\nolinkurl{Sigma2.Compiler.calibrated_bridge_vector} &
Partial: the spatial support interface and component identity; not
commuting hosts.\\
\bottomrule
\end{tabular}
\end{minipage}
\end{center}

\begin{center}\footnotesize
\begin{minipage}{\linewidth}
\setlength{\tabcolsep}{4pt}
\begin{tabular}{@{}>{\raggedright\arraybackslash}p{.17\linewidth}>{\raggedright\arraybackslash}p{.44\linewidth}>{\raggedright\arraybackslash}p{.33\linewidth}@{}}
\toprule Printed result & Lean name & Checked form\\\midrule
Lemma 4.2 &
\nolinkurl{Sigma2.Compiler.lemma_4_2_label_refinement};
\nolinkurl{Sigma2.Compiler.joint_refinement_iff} &
Coarse-sign identification and the converse commuting refinement.\\[4pt]
Lemma 4.3; equation (19) &
\nolinkurl{Sigma2.Compiler.lemma_4_3_boolean_rows};
\nolinkurl{Sigma2.Compiler.row_loss_formula} &
Conditional: perfect games, with the \texttt{RowsHold} and
\texttt{NamedBinaryProjections} premises; (19) only as a trace
identity under a slot-sum hypothesis, not an event probability.\\[4pt]
Definition 4.4 &
\nolinkurl{Sigma2.Presentation};
\nolinkurl{Sigma2.Presentation.admissible};
\nolinkurl{Sigma2.NormalRealization};
\nolinkurl{Sigma2.Compiler.Compiled};
\nolinkurl{Sigma2.Compiler.ApproxCompiled};
\nolinkurl{Sigma2.PaperAdmissible};
\nolinkurl{Sigma2.PaperAdmissibility.application_paper_admissible};
\nolinkurl{Sigma2.PaperAdmissibility.saturation_bridge} &
\texttt{Presentation.admissible}, the domain of the compiler theorems,
omits two transport conditions of the printed definition.
\texttt{PaperAdmissible} is the printed definition; it holds for the
Theorem A input tables, and every admissible presentation saturates to a
\texttt{PaperAdmissible} one with equivalent normal realizations and the
same compiled game.\\[4pt]
Section 4.7, equations (28)--(29) &
\nolinkurl{Sigma2.CompilerSpecification.approximate_compilation};
\nolinkurl{Sigma2.CompilerApproxBase.approxCompiled_of_base15} &
Partial: an existential dimension-independent loss bound under exact
structural hypotheses; not the explicit coefficient (29), robust
extraction, or preservation of an input witness or its finite
dimension.\\[4pt]
Section 5.1, equations (30)--(33) &
\nolinkurl{Sigma2.Tree.fock_decomposition};
\nolinkurl{Sigma2.Tree.fock_density};
\nolinkurl{Sigma2.Tree.fock_entropy};
\nolinkurl{Sigma2.Tree.no_finite_realization} &
Rooted Fock structure, weights, entropy and finite-dimensional
obstruction.\\[4pt]
Lemma 5.1 &
\nolinkurl{Sigma2.Tree.lemma_5_1_definite_computation};
\nolinkurl{Sigma2.ReductionSoundBridge.nonreject_of_data} &
Classical tableau on a root-clock sector, with its concrete data from
the row-extraction modules.\\
\bottomrule
\end{tabular}
\end{minipage}
\end{center}

\begin{center}\footnotesize
\begin{minipage}{\linewidth}
\setlength{\tabcolsep}{4pt}
\begin{tabular}{@{}>{\raggedright\arraybackslash}p{.17\linewidth}>{\raggedright\arraybackslash}p{.44\linewidth}>{\raggedright\arraybackslash}p{.33\linewidth}@{}}
\toprule Printed result & Lean name & Checked form\\\midrule
Section 6, equations (40)--(46); Lemma 6.1 &
\nolinkurl{Sigma2.Clock.finite_phase_sector};
\nolinkurl{Sigma2.Clock.coordinate_clock};
\nolinkurl{Sigma2.Clock.exact_return};
\nolinkurl{Sigma2.Clock.lemma_6_1_good_primitive_phase};
\nolinkurl{Sigma2.Clock.bounded_reader} &
Conditional: \nolinkurl{finite_phase_sector}, \nolinkurl{coordinate_clock} and \nolinkurl{exact_return} take \nolinkurl{hP : P.admissible} and \nolinkurl{D : Clock.Data R}; the last two also take a root vector, and exact return a primitive-phase eigenvector. \nolinkurl{bounded_reader} assumes a primitive phase and its separation bound; Lemma~6.1 supplies a suitable phase and bound for every prime. \nolinkurl{Reduction.realization_sound} supplies the clock data for Theorem~A.\\[4pt]
Section 7; Theorem 7.1 &
\nolinkurl{Sigma2.Reduction.effective_program};
\nolinkurl{Sigma2.Reduction.application_admissible};
\nolinkurl{Sigma2.Reduction.presentation_codes_computable};
\nolinkurl{Sigma2.Reduction.realization_sound};
\nolinkurl{Sigma2.Reduction.realization_complete};
\nolinkurl{Sigma2.CompiledReduction.theorem_7_1_reduction} &
Effective input construction, both perfection equivalences and absence
of any finite-dimensional perfect output.\\[4pt]
Section 7.1, equations (50)--(54) &
\nolinkurl{Sigma2.CompiledFockApproximation.normal_approximation};
\nolinkurl{Sigma2.CompiledFockApproximation.finite_approximation};
\nolinkurl{Sigma2.CompiledValue.four_defect_approximation};
\nolinkurl{Sigma2.CompiledValue.value_one};
\nolinkurl{Sigma2.CompiledValue.reduction_promised} &
Partial: geometric approximation and the promise on every input; the
$Q,L,d$ of \nolinkurl{four_defect_approximation} are existential, not the
actual row counts or denominators.\\[4pt]
Other local results &
Thin core, \nolinkurl{lean/THIN-CORE.md} &
Local statements only; see Table~\ref{tab:leancore}.\\
\bottomrule
\end{tabular}
\end{minipage}
\end{center}

\noindent\textbf{Not formalized:} Theorems B--D beyond their thin-core
lemmas, including the general semifinite and type classification and the
communication classification; Theorem~\ref{thm:bell} beyond its
winning-functional identities, and the last two clauses of
Theorem~\ref{thm:strata}; the general $n$-bridge and commuting-host
compiler; the economical ten-label schedule, its size counts and
equation~\eqref{eq:valueconstant} with actual row denominators (the
formal compiler transports all typed labels); the reduction of strategies
on arbitrary Hilbert spaces to $\ell^2(\mathbb N)$ by compression;
real quantifier elimination, which the formal proof does not need.

\noindent\textbf{Conventions:} both local spaces are $\ell^2(\mathbb N)$,
and normal states are countable ensembles of unit vectors in
$\ell^2(\mathbb N\times\mathbb N)$, their completed tensor product. A
finite-dimensional strategy is a unit vector supported on finitely many
basis vectors, with POVMs on $\ell^2(\mathbb N)$. Every game has at least
one question and one answer for each party.

\subsection{Trust base}
The proofs use Lean 4.35.0 and the Mathlib commit
\begin{quote}\small
\nolinkurl{56a005855ea17e0b0958f73ec801ae140b0a1679};
\end{quote}
every pin is in \nolinkurl{lean/}. Every declaration of the
\texttt{Sigma2} library, including every theorem in the table, uses no
axioms beyond Lean's three standard axioms:
\begingroup\scriptsize
\begin{verbatim}
'Sigma2.theoremA' depends on axioms: [propext, Classical.choice, Quot.sound]
'Sigma2.theoremA_paper' depends on axioms: [propext, Classical.choice, Quot.sound]
'Sigma2.theoremA_shared_family' depends on axioms: [propext, Classical.choice, Quot.sound]
\end{verbatim}\endgroup
The proof of Theorem A passes two independent kernel checkers,
\texttt{leanchecker} and \texttt{nanoda}. The comparator challenge
\nolinkurl{lean/Challenge.lean} imports only Mathlib and defines the
statement vocabulary transparently. It states Theorem A as displayed
above, the shared hardness family \nolinkurl{theoremA_shared_family}, and
the printed admissibility definition \nolinkurl{paper_admissible_iff}.
\nolinkurl{lean/Solution.lean} imports the proofs, and
\nolinkurl{lean/comparator.json} permits only the three standard axioms.
From \texttt{lean/}, on Linux with bubblewrap and with the dependencies
placed as below, one command checks all three:
\texttt{lake comparator -{}-config comparator.json}.
Records of the comparator run and the kernel checks are in
\nolinkurl{lean/evidence/}.

\subsection{Reproduce}
With the Lean 4.35.0 toolchain first on the path and the pinned
dependency sources placed under \nolinkurl{lean/vendor/} as described in
\nolinkurl{lean/README.md}, run from \texttt{lean/} (the first line is
PowerShell syntax):
\begin{flushleft}\leftskip1.5em\ttfamily\frenchspacing
\$env:LEAN\_NUM\_THREADS = \textquotesingle 1\textquotesingle\\
lake exe cache get\\
lake build\\
lake env lean -{}j1 -{}M6144 Audit.lean
\end{flushleft}
The first line selects one Lean thread, and \texttt{lake exe cache get}
fetches the Mathlib build cache. \texttt{lake build} rebuilds every project
module from source. It audits the axioms of every declaration of the
library, including private declarations, and rejects admitted proofs and
additional axioms; it audits the proofs of the two final Theorem A theorems and
checks that every Lean name in the table exists. The last command prints
the axioms of every theorem in the table; \nolinkurl{lean/AXIOMS.txt} is
the expected output.

\bibliographystyle{plain}
\bibliography{references}

\begin{thebibliography}{10}

\bibitem{AaronsonBB}
Scott Aaronson.
\newblock The {Busy Beaver Frontier}.
\newblock Technical Report TR20-115, Electronic Colloquium on Computational
  Complexity, 2020.

\bibitem{BPR}
Saugata Basu, Richard Pollack, and Marie-Fran{\c c}oise Roy.
\newblock {\em Algorithms in Real Algebraic Geometry}.
\newblock Springer, second edition, 2006.

\bibitem{BBS}
Ver{\'o}nica Becher, Yann Bugeaud, and Theodore~A. Slaman.
\newblock The irrationality exponents of computable numbers.
\newblock {\em Proceedings of the American Mathematical Society},
  144(4):1509--1521, 2016.

\bibitem{Beigi}
Salman Beigi.
\newblock Entanglement-assisted zero-error capacity is upper bounded by the
  {Lov\'asz} theta function.
\newblock {\em Physical Review A}, 82:010303(R), 2010.
\newblock arXiv:1002.2488v2.

\bibitem{BMZ}
Archishna Bhattacharyya, Arthur Mehta, and Yuming Zhao.
\newblock On the undecidability of quantum channel capacities.
\newblock \url{https://arxiv.org/abs/2601.22471v3}, 2026.
\newblock In Proceedings of the 67th Annual IEEE Symposium on Foundations of
  Computer Science (FOCS 2026), to appear.

\bibitem{CZ}
Ranyiliu Chen and Yuming Zhao.
\newblock A non-robust quantum correlation self-test, 2026.
\newblock arXiv:2609.25117v2.

\bibitem{Chiodo}
Maurice Chiodo.
\newblock On torsion in finitely presented groups.
\newblock {\em Groups, Complexity, Cryptology}, 6(1):1--8, 2014.
\newblock \url{https://doi.org/10.1515/gcc-2014-0001}; arXiv:1107.1489v4.

\bibitem{CLS}
Richard Cleve, Li~Liu, and William Slofstra.
\newblock Perfect commuting-operator strategies for linear system games.
\newblock {\em Journal of Mathematical Physics}, 58:012202, 2017.
\newblock arXiv:1606.02278.

\bibitem{CM}
Richard Cleve and Rajat Mittal.
\newblock Characterization of binary constraint system games.
\newblock In {\em Automata, Languages, and Programming (ICALP 2014)}, pages
  320--331, 2014.
\newblock arXiv:1209.2729.

\bibitem{Coladangelo3322}
Andrea Coladangelo.
\newblock The {$I_{3322}$} {Bell} inequality requires infinite dimensions.
\newblock \url{https://arxiv.org/abs/2609.06038}, 2026.

\bibitem{CS}
Andrea Coladangelo and Jalex Stark.
\newblock An inherently infinite-dimensional quantum correlation.
\newblock {\em Nature Communications}, 11:3335, 2020.
\newblock Preprint arXiv:1804.05116.

\bibitem{Connes}
Alain Connes.
\newblock Une classification des facteurs de type {III}.
\newblock {\em Annales scientifiques de l'\'Ecole Normale Sup\'erieure},
  6(2):133--252, 1973.

\bibitem{Connes76}
Alain Connes.
\newblock Classification of injective factors: Cases {II$_1$}, {II$_\infty$},
  {III$_\lambda$}, $\lambda\ne1$.
\newblock {\em Annals of Mathematics}, 104(1):73--115, 1976.

\bibitem{ConnesTakesaki}
Alain Connes and Masamichi Takesaki.
\newblock The flow of weights on factors of type {III}.
\newblock {\em Tohoku Mathematical Journal}, 29(4):473--575, 1977.
\newblock Parts 1 and 2.

\bibitem{Craig}
William Craig.
\newblock On axiomatizability within a system.
\newblock {\em The Journal of Symbolic Logic}, 18(1):30--32, 1953.

\bibitem{CLMW}
Toby~S. Cubitt, Debbie Leung, William Matthews, and Andreas Winter.
\newblock Improving zero-error classical communication with entanglement.
\newblock {\em Physical Review Letters}, 104:230503, 2010.
\newblock arXiv:0911.5300.

\bibitem{Douglas}
Ronald~G. Douglas.
\newblock On majorization, factorization, and range inclusion of operators on
  {Hilbert} space.
\newblock {\em Proceedings of the American Mathematical Society},
  17(2):413--415, 1966.

\bibitem{Douglas3322}
Seth Douglas.
\newblock The {I3322} quantum value is attained spatially but not in finite
  dimension.
\newblock Zenodo, \url{https://doi.org/10.5281/zenodo.21782008}, 2026.
\newblock All versions, first released August 2026; arXiv:2609.05555.

\bibitem{Hyperfinite}
Seth Douglas and Nidhal Mghirbi.
\newblock A finite game that normal hyperfinite strategies approach but never
  win.
\newblock \url{https://doi.org/10.5281/zenodo.23816141}, 2026.
\newblock Version 1.0.0, preprint.

\bibitem{ReadType}
Seth Douglas and Nidhal Mghirbi.
\newblock Finite games that read the type of a von {N}eumann algebra.
\newblock \url{https://doi.org/10.5281/zenodo.23931154}, 2026.
\newblock Version 1.1.0, preprint.

\bibitem{CommunicationPrecursor}
Seth Douglas and Nidhal Mghirbi.
\newblock One-shot zero-error communication with normal entanglement is
  {$\Sigma^0_2$}-complete.
\newblock Zenodo, \url{https://doi.org/10.5281/zenodo.23924868}, 2026.

\bibitem{TypePrecursor}
Seth Douglas and Nidhal Mghirbi.
\newblock Perfect play has different arithmetical complexity in semifinite and
  type {III} algebras.
\newblock Zenodo, \url{https://doi.org/10.5281/zenodo.23925091}, 2026.

\bibitem{Gamma}
Seth Douglas and Nidhal Mghirbi.
\newblock Perfect play in a binary-scoring nonlocal game requires infinite
  dimension but only finite energy.
\newblock \url{https://doi.org/10.5281/zenodo.23772890}, 2026.
\newblock Version 2.0.0, preprint.

\bibitem{Entropy}
Seth Douglas and Nidhal Mghirbi.
\newblock Perfect play in a finite nonlocal game requires infinite entanglement
  entropy.
\newblock \url{https://doi.org/10.5281/zenodo.23932017}, 2026.
\newblock Version 1.1.0, preprint.

\bibitem{SpatialPrecursor}
Seth Douglas and Nidhal Mghirbi.
\newblock Perfect play with spatial quantum strategies is
  {$\Sigma^0_2$}-complete.
\newblock Zenodo, \url{https://doi.org/10.5281/zenodo.23910986}, 2026.

\bibitem{Fritz}
Tobias Fritz.
\newblock Curious properties of free hypergraph {$C^*$}-algebras.
\newblock {\em Journal of Operator Theory}, 83(2):423--445, 2020.
\newblock arXiv:1808.09220.

\bibitem{FMS}
Honghao Fu, Carl~A. Miller, and William Slofstra.
\newblock The membership problem for constant-sized quantum correlations is
  undecidable.
\newblock {\em Communications in Mathematical Physics}, 406:96, 2025.
\newblock arXiv:2101.11087v5.

\bibitem{GYZ}
Minbo Gao, Tianshi Yu, and Lihong Zhi.
\newblock An explicit counterexample to {Tsirelson's} problem via a linear
  system game.
\newblock \url{https://arxiv.org/abs/2610.10248}, 2026.

\bibitem{HM}
Uffe Haagerup and Magdalena Musat.
\newblock Classification of hyperfinite factors up to completely bounded
  isomorphism of their preduals.
\newblock {\em Journal f{\"u}r die reine und angewandte Mathematik},
  630:141--176, 2009.
\newblock arXiv:0706.3463.

\bibitem{HarrisGraph}
Samuel~J. Harris.
\newblock Universality of graph homomorphism games and the quantum coloring
  problem.
\newblock {\em Annales Henri Poincar\'e}, 25(10):4321--4356, 2024.
\newblock arXiv:2305.18116.

\bibitem{HarrisEmbezzlement}
Samuel~J. Harris.
\newblock Self-testing of exact entanglement embezzlement.
\newblock \url{https://arxiv.org/abs/2605.22713v2}, 2026.

\bibitem{Higman}
Graham Higman.
\newblock Subgroups of finitely presented groups.
\newblock {\em Proceedings of the Royal Society of London. Series A},
  262(1311):455--475, 1961.

\bibitem{HNN}
Graham Higman, Bernhard~H. Neumann, and Hanna Neumann.
\newblock Embedding theorems for groups.
\newblock {\em Journal of the London Mathematical Society}, s1-24(4):247--254,
  1949.

\bibitem{MIPRE}
Zhengfeng Ji, Anand Natarajan, Thomas Vidick, John Wright, and Henry Yuen.
\newblock {MIP}$^*$ = {RE}.
\newblock {\em Annals of Mathematics}, 2026.
\newblock Accepted; arXiv:2001.04383v4.

\bibitem{KPS}
Se-Jin Kim, Vern~I. Paulsen, and Christopher Schafhauser.
\newblock A synchronous game for binary constraint systems.
\newblock {\em Journal of Mathematical Physics}, 59:032201, 2018.
\newblock arXiv:1707.01016.

\bibitem{KrohnGrimberghe}
Artus Krohn-Grimberghe.
\newblock Certified exact identification of the {I3322} quantum value.
\newblock \url{https://arxiv.org/abs/2609.29693}, 2026.

\bibitem{Lin}
Junqiao Lin.
\newblock $\mathsf{MIP}^{co}=\mathsf{coRE}$.
\newblock \url{https://arxiv.org/abs/2510.07162}, 2025.

\bibitem{Imitation}
Martino Lupini, Laura Man\v{c}inska, Vern~I. Paulsen, David~E. Roberson,
  Giannicola Scarpa, Simone Severini, Ivan~G. Todorov, and Andreas Winter.
\newblock Perfect strategies for non-signalling games.
\newblock {\em Mathematical Physics, Analysis and Geometry}, 23:7, 2020.
\newblock arXiv:1804.06151.

\bibitem{Mancinska}
Laura Man\v{c}inska.
\newblock The right entanglement for perfect nonlocality and zero-error
  communication.
\newblock \url{https://arxiv.org/abs/2610.06512v1}, 2026.

\bibitem{MR}
Laura Man\v{c}inska and David~E. Roberson.
\newblock Quantum homomorphisms.
\newblock {\em Journal of Combinatorial Theory, Series B}, 118:228--267, 2016.
\newblock Preprint ``Graph Homomorphisms for Quantum Players'',
  arXiv:1212.1724.

\bibitem{MRV}
Laura Man\v{c}inska, David~E. Roberson, and Antonios Varvitsiotis.
\newblock On deciding the existence of perfect entangled strategies for
  nonlocal games.
\newblock {\em Chicago Journal of Theoretical Computer Science}, 2016(5), 2016.
\newblock arXiv:1506.07429.

\bibitem{Mermin}
N.~David Mermin.
\newblock Simple unified form for the major no-hidden-variables theorems.
\newblock {\em Physical Review Letters}, 65:3373--3376, 1990.

\bibitem{MNY20}
Hamoon Mousavi, Seyed~Sajjad Nezhadi, and Henry Yuen.
\newblock On the complexity of zero gap {MIP}$^*$.
\newblock In {\em 47th International Colloquium on Automata, Languages, and
  Programming (ICALP 2020)}, volume 168 of {\em LIPIcs}, pages 87:1--87:12,
  2020.
\newblock arXiv:2002.10490.

\bibitem{MNY}
Hamoon Mousavi, Seyed~Sajjad Nezhadi, and Henry Yuen.
\newblock Nonlocal games, compression theorems, and the arithmetical hierarchy.
\newblock In {\em Proceedings of the 54th Annual ACM SIGACT Symposium on Theory
  of Computing}, pages 1--11, 2022.
\newblock Full version arXiv:2110.04651v2.

\bibitem{NPA}
Miguel Navascu\'es, Stefano Pironio, and Antonio Ac\'in.
\newblock A convergent hierarchy of semidefinite programs characterizing the
  set of quantum correlations.
\newblock {\em New Journal of Physics}, 10:073013, 2008.
\newblock arXiv:0803.4290.

\bibitem{Nezhadi}
Seyed~Sajjad Nezhadi.
\newblock {\em Quantum Games, Graphs, and G\"odel}.
\newblock PhD thesis, University of Maryland, 2025.

\bibitem{OP}
Dorte Olesen and Gert~K. Pedersen.
\newblock Some {$C^*$}-dynamical systems with a single {KMS} state.
\newblock {\em Mathematica Scandinavica}, 42:111--118, 1978.

\bibitem{PS}
Connor Paddock and Simon Schmidt.
\newblock A commuting operator self-test for exact entanglement embezzlement.
\newblock \url{https://arxiv.org/abs/2609.38083}, 2026.

\bibitem{PalVertesi}
K{\'a}roly~F. P{\'a}l and Tam{\'a}s V{\'e}rtesi.
\newblock Maximal violation of a bipartite three-setting, two-outcome {Bell}
  inequality using infinite-dimensional quantum systems.
\newblock {\em Physical Review A}, 82:022116, 2010.

\bibitem{PSS}
Vern~I. Paulsen, Simone Severini, Daniel Stahlke, Ivan~G. Todorov, and Andreas
  Winter.
\newblock Estimating quantum chromatic numbers.
\newblock {\em Journal of Functional Analysis}, 270(6):2188--2222, 2016.

\bibitem{Pauwels}
Jef Pauwels.
\newblock The quantum supremum of the {$I_{3322}$} {Bell} inequality is not
  attained in finite dimension.
\newblock \url{https://arxiv.org/abs/2608.29734}, 2026.

\bibitem{PT}
Gert~K. Pedersen and Masamichi Takesaki.
\newblock The {Radon--Nikodym} theorem for von {N}eumann algebras.
\newblock {\em Acta Mathematica}, 130:53--87, 1973.

\bibitem{Peres}
Asher Peres.
\newblock Incompatible results of quantum measurements.
\newblock {\em Physics Letters A}, 151:107--108, 1990.

\bibitem{Rado}
Tibor Rad\'o.
\newblock On non-computable functions.
\newblock {\em Bell System Technical Journal}, 41(3):877--884, 1962.

\bibitem{Shoenfield}
Joseph~R. Shoenfield.
\newblock On degrees of unsolvability.
\newblock {\em Annals of Mathematics}, 69(3):644--653, 1959.

\bibitem{SlofstraNonclosed}
William Slofstra.
\newblock The set of quantum correlations is not closed.
\newblock {\em Forum of Mathematics, Pi}, 7:e1, 2019.
\newblock arXiv:1703.08618.

\bibitem{Slofstra}
William Slofstra.
\newblock Tsirelson's problem and an embedding theorem for groups arising from
  non-local games.
\newblock {\em Journal of the American Mathematical Society}, 33:1--56, 2020.
\newblock arXiv:1606.03140.

\bibitem{Takesaki}
Masamichi Takesaki.
\newblock Conditional expectations in von {N}eumann algebras.
\newblock {\em Journal of Functional Analysis}, 9(3):306--321, 1972.

\bibitem{VLSWW}
Lauritz van Luijk, Alexander Stottmeister, Reinhard~F. Werner, and Henrik
  Wilming.
\newblock Embezzlement of entanglement, quantum fields, and the classification
  of von {N}eumann algebras.
\newblock \url{https://arxiv.org/abs/2401.07299v4}, 2024.

\end{thebibliography}
\end{document}